\documentclass[11pt,twoside]{unswthesis}
\usepackage{amsfonts}
\usepackage{amssymb}
\usepackage{amsmath}
\usepackage{epsf}
\usepackage{graphicx}
\usepackage{epsf}
\begin{document}

\pagestyle{empty}

\unitlength=1cm
\thispagestyle{empty}
\begin{flushright}
\end{flushright}
\begin{center}
{\bf {\Large Universit\'e de Strasbourg }}\\
\vspace{0.3cm}
{\bf {\large LE COLL\`EGE DOCTORAL}}\\
\vspace{1.cm}
{\bf {\huge Habilitation thesis }}\\
{\bf {\Large (Th\`ese d'habilitation \`a diriger des recherches) }}\\
\vspace{0.9cm}
\vspace{0.8cm}
{\rm   Sp\'ecialit\'e : Physique th\'eorique}\\
\vspace{0.4cm}
by\\
\vspace{0.4cm}
{\bf {\Large Rimantas LAZAUSKAS}}\\
\vspace{0.4cm}

\begin{picture}(16,0.5)
\put(0,0.1){\line(1,0){16}}
\put(0,0){\line(1,0){16}}
\end{picture}\\
\vspace{0.5cm}
{\Huge {\begin{tabular}[t]{c} Application of the complex scaling method \\
in quantum scattering theory \end{tabular} }} \\
\vspace{0.4cm}
\begin{picture}(16,0.5)
\put(0,0.1){\line(1,0){16}}
\put(0,0){\line(1,0){16}}
\end{picture}\\
\vspace{0.3cm}
\vspace{1.0cm}
\begin{center}
\begin{tabular}[t]{cl}
Composition du Jury \hspace{0.5cm}&
\begin{tabular}[c]{|rll}
\vspace{-0.0cm} {\bf M.}&{\bf DUFOUR}, & {\large Garant de Th\`ese}\\
\vspace{-0.0cm} {\bf D.}&{\bf BAYE}, & {\large Rapporteur} \\
\vspace{-0.0cm} {\bf N.}&{\bf BARNEA}, & {\large Rapporteur}    \\
\vspace{-0.0cm} {\bf M.}&{\bf GATTOBIGIO}, & {\large Rapporteur}\\
\vspace{-0.0cm} {\bf C.}&{\bf BECK}, & {} \\
\vspace{-0.0cm} {\bf J.}&{\bf CARBONELL}, & {}    \\
\vspace{-0.0cm} {\bf P.A.}&{\bf HERVIEUX} & {}
\end{tabular}
\end{tabular}
\end{center}
\vspace{2.0cm}
\begin{flushright}
Travail pr\'epar\'e au sein de l'Institut Pluridisciplinaire Hubert Curien\\
\vspace{-0.1cm} 23, rue du Loess\\
\vspace{-0.1cm} 67037 Strasbourg cedex 2 \end{flushright}
\end{center}


\begin{flushright}
\textit{To the memory of \\ \textbf{Claude GIGNOUX}\\}
\end{flushright}
\vspace{2 cm}

I owe a lot to Claude Gignoux, who was my cosupervisor during the
PhD thesis in Grenoble, almost 20 years ago. First of all Claude
was an exemplary person - modest and shy, but at the same time
very open minded, always available to help or motivate a young
student.  And certainly he was an extraordinary physicist, due to
his shyness quite little renown abroad. Now very few persons know
that the first numerical solution of 3-body Faddeev equations has
been realized by Claude, during his PhD.  Solution of 4-body
Faddeev-Yakubovsky equations has also been pioneered by Claude and
Jaume Carbonell (my PhD supervisor) long time ago in Grenoble.

\vspace{0.5 cm}

 Finally, my adventure with complex scaling method has been
strongly influenced by Claude. In the end of 2003 I was finalizing
my PhD, whereas Claude was taking retirement. Claude's approach
was quite straightforward -- without any ceremonies he took all
his office notes, notebooks, archives and was ready to throw them
in to rubbish bin. Luckily I was passing by his office and could
save some of them. Sometime latter listing these old notes of
Claude I found his very valuable remarks on the possible
implementation of the complex scaling method for solving
scattering problems. It took me a while to test these ideas, which
eventually turned into gold!

\newpage

\tableofcontents
\bibliographystyle{ieeetr}
\newpage 

\renewcommand{\thepage}{\roman{page}} \setcounter{page}{1} 
\cleardoublepage
\renewcommand{\thepage}{\arabic{page}} \setcounter{page}{1} %
\setcounter{chapter}{0} \pagestyle{fancyplain} \mainmatter

\chapter{Introduction}

\pagestyle{fancyplain}

\bigskip

\bigskip There is a  countless number of problems in quantum mechanics, which
require very accurate numerical solutions. Few-body systems are
the perfect example, as these systems develop individual
characters depending on the number of the constituent particles.
The existence of striking differences in neighboring few-body
systems is a well established phenomenon, which is mostly related
to the correlated motion, the feat that few-body systems are
usually far from saturation and the presence of Pauli principle.
This individual behavior requires a very specific and accurate
treatment, whereas the approximate solutions based on restricted
model space (mean field, Born-Oppenheimer approximation, etc.)
often fail to describe the few-body systems.

The last two decades have witnessed decisive progress in physics
by ab initio calculations. Nevertheless would they be variational,
coupled-cluster methods, No-core shell model, Monte-Carlo or
lattice techniques, they are mostly limited to the bound state
problems. On the other hand, rigorous solution of the particle
collisions, incorporating elastic, rearrangement and breakup
channels, for a long time remained limited to the three-body
case~\cite{Glockle_bible,Arnas_bench}. The main difficulty is
related to the fact that, unlike the bound state wave functions,
scattering wave functions are not localized. Therefore, the
solution of the scattering problem in configuration space implies
to solve the problem of multidimensional integro-differential
equations subject to extremely complex boundary conditions. This
problem constitutes an important challenge both in advancing the
formal as well as the numerical aspects of the few-body
collisions.

There is a rising interest in applying bound-state-like methods to
handle non-relativistic scattering problems. Indeed, the very
first idea of using bound state solutions to solve many-body
scattering problems dates back to E.~P.~Wigners R-matrix
theory~\cite{Wigner:47}. In this approach, the scattering
observables were obtained from the configuration space solutions
in the interaction region, which were expanded in squared
integrable basis functions and thus without imposing the
appropriate boundary conditions. While this technique is still
very popular, it requires  nevertheless an important numerical
effort related to the inversion of the full Hamiltonian matrix. It
also fails to address the possibility to break the system in more
than two clusters. In a recent review~\cite{PPNP}, written with my
colleagues, we have weighted and analyzed such methods. From a
long-term perspective, the complex scaling technique seems the
most promising one.

The complex scaling~(CS) technique has also a fancy history. A
very similar approach to CS has been introduced already during the
World War II by D.~R.~Hartree \textit{et al.}~\cite{HMN46,Co83} in
the study of the radio wave propagation in the atmosphere.
D.~R.~Hartree with his team were searching to determine the
complex eigenvalues of the second order differential equations. In
practice, this problem is equivalent to the one encountered when
aiming to determine the positions of the resonant states in
quantum two-particle collisions. Nevertheless these works have not
been continued after the war, whereas the original work by
D.~R.~Hartree has not been widely publicized, being presented only
as a scientific report in a review with a limited outreach. In the
late sixties J.~Nuttall and H.~L.~Cohen~\cite{NC69} proposed a
very similar technique to treat the generic scattering problem for
short range potentials. Few years later J.~Nuttall even employed
this method to solve a three-nucleon scattering problem above the
breakup threshold~\cite{MDN72}. Nevertheless due to an unlucky
mismatch, these pioneering works of J.~Nuttall \textit{et al.}
have also been interrupted. Actually, the action of the CS
operator on short-ranged potentials transforms them into
complicate oscillating structures, which are not easy to handle or
interpret. On the contrary, the CS operation is trivial for the
Coulomb potential, although in this case Nuttal's method is not
applicable directly. Based on J.~Nuttall's \textit{et al.} works
and the later mathematical foundation of E.~Baslev and
J.~.M.~Combes~\cite{BC71} the original method of Hartree has been
recovered in order to calculate resonance eigenvalues in atomic
physics~\cite{Ho83,Mo98}. The efficiency of the CS technique to
calculate positions of the atomic resonances nourished some
efforts to apply this method also in calculating resonances
dominated by short ranged interactions.
Surprisingly the pioneering works of J.~Nuttall's \textit{et al.}~\cite%
{MDN72} on the scattering remained without pursue for a long time.
To avoid the complications related to the CS transformation of
short-ranged potentials one may construct transformations acting
only beyond the physical domain of interaction, thus leading to
the exterior complex scaling method~\cite{ecsm_SIMON1979211}.
Exterior complex scaling method has proved to be efficient and
competitive in determining resonance positions, however
applications of this method to the  quantum collisions problems
remains very limited. This is due to the fact that the exterior
complex scaling, unlike the original complex scaling method,
contains several serious deficiencies both from the formal as well
as practical point of view.

Only recently, a variant of the complex scaling method based on
the spectral
function formalism has been presented by K.~Kat\={o}, B.~Giraud et al.~\cite%
{MKK01,GK03,GKO04} and applied in the works of K.~Kat\={o} \textit{et al.}~%
\cite{MKK01,SMK05,AMKI06,KKMTI10,KKWK11,KKMI13}. This variant has
been mostly applied in describing radiative decay reactions, due
to the presence of an external perturbation. CS method is well
adapted to solve this kind of problems. Indeed, as such processes
are due to the disintegration of compact objects (bound states),
they are described by the wave functions containing only outgoing
waves in the asymptotes (decay products propagate from the mutual
center of mass -- position of the original compact state). The CS
operation transforms outgoing waves into exponentially bound
functions, which rends problem tractable using square integrable
basis.

From this point of view, few-body collisions remains the most
complicated case. Within a time-independent formalism, collision
describing wave function involves both an incoming wave and
outgoing waves. An incoming wave originates from the plane wave,
describing original setup, where the projectile approaches a
target and continues without scattering. On the contrary, outgoing
waves represents all the possible scattering events, modifying the
original state. As aforementioned, CS efficiently transforms
outgoing waves into exponentially bound functions, however the
incoming waves are transformed into exponentially diverging
functions. These diverging functions should be treated with a
special care.

The revival of the Nuttal's work on collisions by CS method
started with a work of A.~T.~Kruppa\textit{\ et al.}~\cite{KSK07},
where it has been demonstrated how the collisions containing
residual Coulomb interaction can be addressed for two-particle
case. During the last few years I have
realized series of studies developing CS method in few-body collisions~\cite%
{LC11,La12}. These efforts will be highlighted in this
'Habilitation \`{a} diriger des recherches'. In what follows I
will summarize its tentative contents.

After a short introduction of the CS method, the limits of its
applicability will be addressed. The natural limitation arises
from the possibility to apply CS operator on the interaction. CS
is an analytical transformation in coordinate space, thus the
potential should be analytic function of coordinates. Practically,
this limitation can be overcame if analytic basis functions are
used, whereas matrix elements of the CS potential are evaluated
using contour rotation technique. Other straightforward limitation
is due to the need in keeping the product of the incoming wave
times the potential compact. As a consequence, this requirement
translates into a condition for the potential to be exponentially
bound and the upper bound of the complex scaling parameter to be
used in the calculations. Some particular functional forms of the
potential may acquire singularity poles in the complex plane,
which may render the numerical calculations unstable. Finally, as
I have demonstrated in one of my first studies on CS~\cite{LC11},
for the collisions involving more than two particles some more
stringent constrains are present. These constrains arise from the
fact that the incoming wave and the residual target-projectile
interaction contain respectively diverging and converging regions
which does not perfectly overlap in the multidimensional
N-particle space.

Next the outreach of the CS method will be reviewed. I have tested
dozens of different potentials as well as several numerical
techniques for which CS method turns to be efficient, accurate but
also an easy to implement tool. In particular, I have demonstrated
that CS also works for optical potentials, which simulates effects
of the absorbtion by the target. It has also been demonstrated
that for some short-range potentials, which are not exponentially
bound, CS method may still be successful.

As the first important test of CS method I have performed
calculations in a three-nucleon sector. Namely, neutron and proton
scattering on deuteron, based on simplistic nucleon-nucleon
interaction model, was considered. For this case, accurate
calculations exist realized using conventional approach (i.e.
imposing physical boundary conditions). The obtained results both
for the breakup as well as for the elastic scattering amplitudes
were surprisingly accurate and has been achieved using very
limited numerical resources~\cite{LC11}, even compared to much
more technically complex conventional calculations. At the same
time I have demonstrated that repulsive Coulomb interaction could
be treated in 3-body collisions within CS method. Its treatment
requires some minor approximations, which consist in neglecting
long-ranged Coulomb polarization terms.

Next challenge was to explore the aptitude of CS method in a more
general 3-body systems. To this aim, I have considered the problem
of deuteron
scattering on $^{12}$C nucleus in its ground state. In these calculations, $%
^{12}$C nucleus was considered as a single object by describing
interaction between the projectile nucleons and $^{12}$C nucleus
with a phenomenological optical potential. A realistic
neutron-proton potential has been used to describe interaction
between the nucleons composing projectile (deuteron).
Dynamics of the reaction included elastic d+$^{12}$C, neutron transfer to p+$%
^{13}$C as well as deuteron's breakup n+p+$^{12}$C channels. These
calculations have been compared with an alternative conventional
approach based on a description of the reaction dynamics in
momentum-space~\cite{Deltuva_3B}. Very accurate results have been
obtained for the elastic, the transfer and the breakup reaction
cross sections~\cite{DFL12}. Thus once again proving efficiency of
CS method this time for a 3-different particle system, which
comprise optical potential and relatively strong Coulomb
repulsion.

More recently, CS approach has been generalized to treat
four-nucleon reactions in the cases where both three-cluster and
four-nucleon breakup channels are present. Once again, very
reliable results have been obtained in describing p+$^3$He and
n+$^3$H collisions~\cite{La12,PhysRevC.72.034003}.

My last adventure with CS method led to develop approach
appropriate to describe collisions involving three charged
particles. It is worth noticing that for a long time it has been
believed that CS technique is not appropriate for the scattering
process dominated by  long-range interactions. A novel method has
been developed, which combines complex scaling, distorted wave and
Faddeev-Merkuriev equation formalisms~\cite{Me80}. For a moment,
this formalism has been tested in studying three realistic
Coulombic problems: electron scattering on ground states of
Hydrogen and Positronium atoms as well as a $e^+$+H(n=1)
$\leftrightarrow$ p+Ps(n=1) reaction. Accurate results were
obtained in a wide energy region, extending beyond the atom
ionization threshold.

This research project summarizes my recent activity in developing
a very promising method to describe few-particle scattering
problem. I intend to demonstrate the efficiency of the CS method
in describing complicated scattering process involving N$>$2
particle systems, where the conventional scattering theory methods
requiring explicit treatment of the boundary conditions fail or
become technically overcomplicated. These developments opens the
way for describing complex many-particle reactions, involving
multiple transfer, rearrangement and breakup channels.

\chapter{Theory}

\section{Coordinates}

Our ability to solve any physical problem strongly relies on a
proper choice of the relevant degrees of freedom. In this context,
the few-body physics makes no exception. A proper selection of a
coordinate set may essentially reduce complexity of the problem or
in contrary rend it unsolvable. One should first think hard when
trying to make an optimal choice for the coordinates, by adapting
it to each particular problem as well as to the available
numerical/analytical tools. The selected coordinates should
describe efficiently the system, must be easy to handle when
evaluating matrix elements (economically evaluate integrals in
multi-dimensional space), to express different Hamiltonian terms,
like kinetic or potential energies, etc..

\begin{itemize}
\item \textit{Single particle coordinates}%
\begin{equation*}
\vec{r}_{1},\vec{r}_{2},\vec{r}_{3},..,\vec{r}_{N}
\end{equation*}
constitute the simplest and the most used coordinate set. One of
the main assets of this set is the presence of the simple
expression for a kinetic-energy term:
\begin{equation}
H_{0}=-\sum_{i=1}^{N}{\frac{\hbar ^{2}}{2m_{i}}}\Delta _{\vec{r}_{i}}.
\end{equation}
In the last expression, $m_{i}$ denotes the mass of the particle
$i$. Other very important aspect of this coordinate set is related
with the simplicity in performing systems wave function's
(anti)symmetrization procedure. Nevertheless this set has also a
serious drawback, since it does not allow to separate explicitly
the center of mass degrees of freedom for multiparticle $N>2$
systems.

In this work I will outline only two other types of coordinate sets, which
will be applied in the following applications

\item \textit{Perimetric coordinates} for a three-body system are defined  as%
\begin{eqnarray}
u &=&r_{12}+r_{31}-r_{23},  \notag \\
v &=&r_{12}+r_{23}-r_{31},  \label{eq_perim_cor} \\
z &=&r_{23}+r_{31}-r_{12}, \\
\vec{R}
&=&{\frac{m_{1}\vec{r}_{1}+m_{2}\vec{r}_{2}+m_{3}\vec{r}_{3}}{M}}
,\notag
\end{eqnarray}%
where $M=m_1+m_2+m_3$ is the total mass of the system, with $r_{ij}=\left\vert \overrightarrow{r}_{i}-\overrightarrow{r}%
_{j}\right\vert$. One needs to supplement these radial coordinates
with three angles $\left( \alpha ,\beta ,\gamma \right) $
describing the orientation of the triangle, made by three
particles placed at its vertices, in space. These coordinates vary
in the interval $\left[ 0,\infty \right]$. They satisfy
automatically the triangular conditions and results into simple
Jacobian. The great asset of this set is that it locates the cusps
of a three-particle wave function at the origin of the
coordinates. At the same time if (as example) particle $1$ recedes
from pair $\left( 23\right) $ coordinate $u$ starts growing with
the separation distance, thus allowing a proper approximation of
the systems wave functions behavior in the asymptote region. For a
total angular momentum (L=0), the wave function of the system
becomes independent of the Euler's angles whereas the matrix
elements of the kinetic energy operator between the states $\psi
_{i}$ and $\psi _{j}$ may be expressed as
\begin{eqnarray}
\left\langle \psi _{i}\left\vert H_0\right\vert \psi _{j}\right\rangle
&=&2\int_{0}^{\infty }du\int_{0}^{\infty }dv\int_{0}^{\infty }dz  \notag \\
&&\times\left\{ \left[ \frac{u(v+z)(u+v+z)}{m_{1}}+\frac{uz(z+u)}{m_{2}}+%
\frac{uv(u+v)}{m_{3}}\right] \frac{d\psi _{i}}{du}\frac{d\psi _{j}}{du}%
\right.  \notag \\
&&+\left[ \frac{vz(v+z)}{m_{1}}+\frac{v(u+z)(u+v+z)}{m_{2}}+\frac{vu(u+v)}{%
m_{3}}\right] \frac{d\psi _{i}}{dv}\frac{d\psi _{j}}{dv}  \notag \\
&&+\left[ \frac{vz(v+z)}{m_{1}}+\frac{uz(z+u)}{m_{2}}+\frac{z(u+v)(u+v+z)}{%
m_{3}}\right] \frac{d\psi _{i}}{dz}\frac{d\psi _{j}}{dz} \\
&&-\frac{vz(v+z)}{m_{1}}\left[ \frac{d\psi _{i}}{dv}\frac{d\psi _{j}}{dz}+%
\frac{d\psi _{i}}{dz}\frac{d\psi _{j}}{dv}\right] -\frac{uz(u+z)}{m_{2}}%
\left[ \frac{d\psi _{i}}{du}\frac{d\psi _{j}}{dz}+\frac{d\psi _{i}}{dz}\frac{%
d\psi _{j}}{du}\right]  \notag \\
&&\left. -\frac{uv(u+v)}{m_{3}}\left[ \frac{d\psi _{i}}{du}\frac{d\psi _{j}}{%
dv}+\frac{d\psi _{i}}{dv}\frac{d\psi _{j}}{du}\right] \right\}.  \notag
\label{eq_h0_perim}
\end{eqnarray}
It is possible to extend this expression to $L>0$
case~\cite{Baye_perimc}, however not considered in this work.

Perimetric coordinates are very efficient in handling 3-body bound
state problems, related with central interactions, which diverge
at the origin (like Coulomb). Unfortunately,  angular momentum
algebra operations become quite involved for this coordinate set.
Other important drawback of this, otherwise very handy set, is
absence of a simple generalization to $N>3$ systems.
\begin{figure}[th]
\centering
\includegraphics[width=90mm]{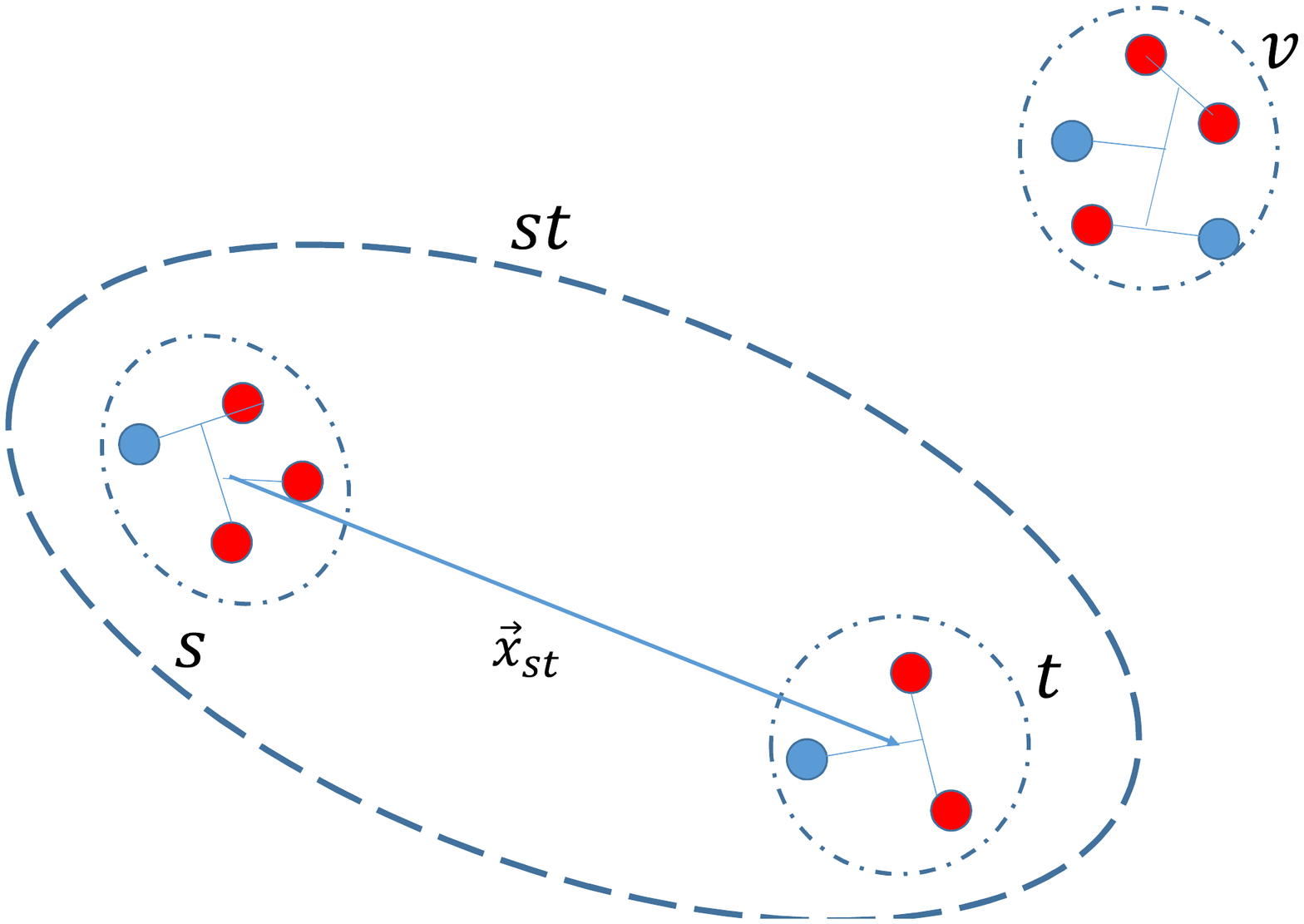}
\caption{ Jacobi coordinate $\vec{x}_{st}$ joining two multiparticle
clusters $s$ and $t$ to form a cluster $st$. }
\label{fig:clusters_st}
\end{figure}

\item \textit{Jacobi coordinates} are the most practical choice to
formulate the multiparticle scattering problem. This set
automatically separates center-of-mass degrees of freedom but also
it allows to separate asymptotes of diverse collisions channels,
related with creation of different multiparticle clusters. Jacobi
coordinates are generalized to the systems containing arbitrary
number of particles, they  present simple and flexible scheme  to
break multiparticle system into separate clusters. One constructs
Jacobi coordinates by systematically dividing the system in
clusters and their subclusters; a coordinate connecting two
clusters $\left( s\right) $ and $\left( t\right) $ is expressed
using a general formulae:
\begin{equation}
\vec{x}_{st}=\sqrt{\frac{2m_{s}m_{t}}{m(m_{s}+m_{t})}}\left( \vec{r}_{t}-%
\vec{r}_{s}\right),
\end{equation}%
where $m_{s}$ and $m_{t}$ are the masses of the clusters, while $%
\overrightarrow{r}_{s}$ and $\overrightarrow{r}_{t}$ are
respective positions of their center-of-masses. A mass factor $m$
of free choice is introduced into the former expression in order
to retain the proper units of the distances. When studying systems
of identical particles it is convenient to identify this mass with
the mass of a single particle. In terms of Jacobi coordinates the
free Hamiltonian is expressed as:
\begin{equation}
H_{0}=-\sum_{\left(st\right) \subset P}{\frac{\hbar ^{2}}{m}}\Delta _{\vec{x}%
_{st}}-{\frac{\hbar ^{2}}{2M}}\Delta _{\vec{R}},
\end{equation}
with $\vec{R}$ denoting the center-of-mass position and $M$ the
total mass of the system. The sum runs over all the possible
branches of the tree $\left( st\right) \subset P$ (as example, see
Figure~\ref{fig:clusters_st} ), breaking multiparticle system into
separate clusters until all the clusters are broken into single
particles. Throughout this work Jacobi coordinates will be mostly
employed and therefore I pay more attention to this type of
coordinates in the following subsections.
\end{itemize}

\subsection{3-body Jacobi coordinates} 

\begin{figure}[th]
\centering
\includegraphics[width=90mm]{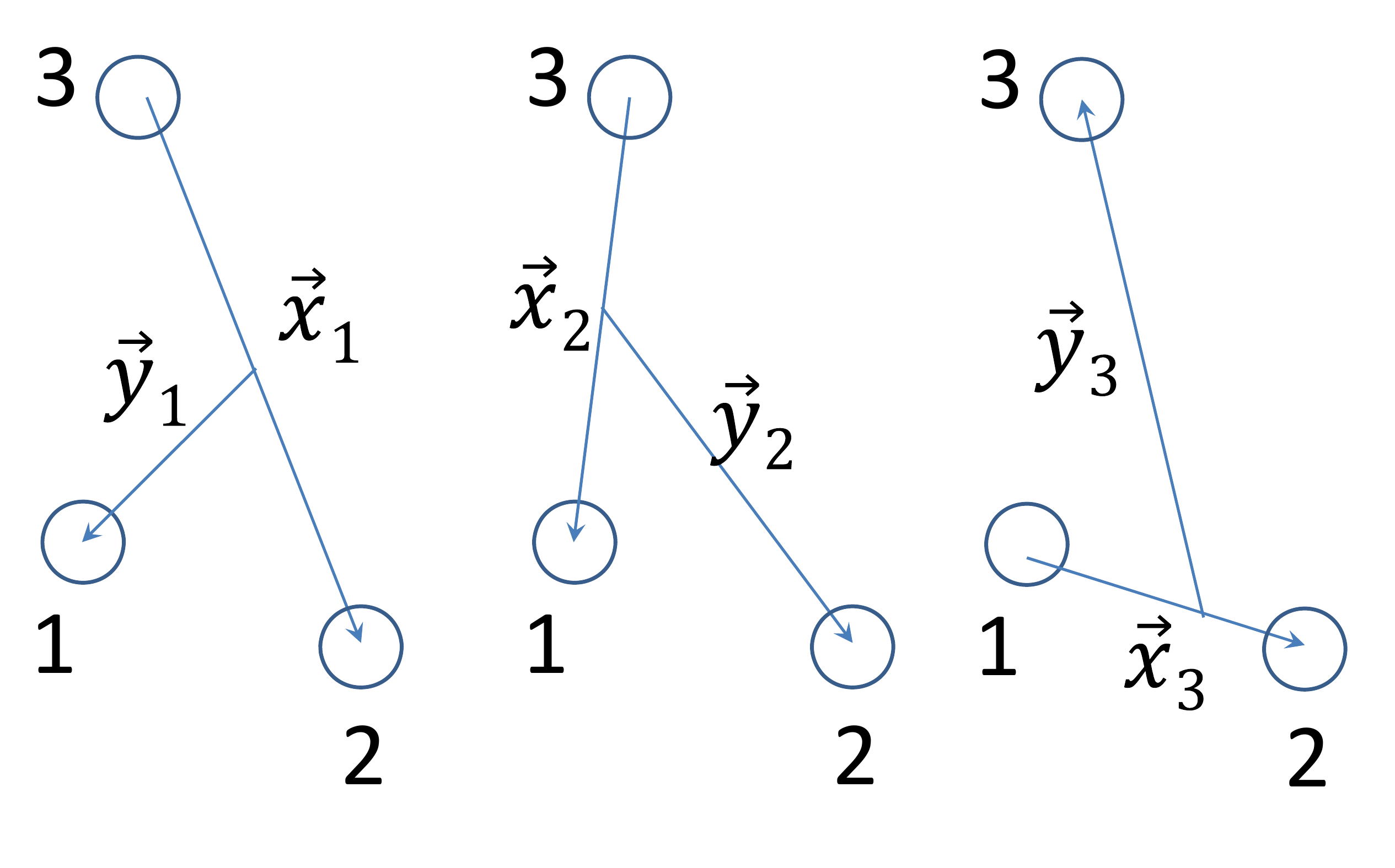}
\caption{Three possible sets of 3-body Jacobi coordinates }
\label{fig:3b_coord}
\end{figure}
To each sequence $\alpha\equiv(\alpha \beta \gamma )\supset (
\beta \gamma )$ one may associate two Jacobi coordinates, see
Fig.\ref{fig:3b_coord}:
\begin{equation}
\vec{x}_{\alpha }=\sqrt{\frac{2m_{\beta }m_{\gamma }}{(m_{\beta }+m_{\gamma
})m}}(\vec{r}_{\gamma }-\vec{r}_{\beta }),\qquad \vec{y}_{\alpha }=\sqrt{%
\frac{2m_{\alpha }(m_{\beta }+m_{\gamma })}{Mm}}\left[ \vec{r}_{\alpha }-{%
\frac{m_{\beta }\vec{r}_{\beta }+m_{\gamma }\vec{r}_{\gamma }}{m_{\beta
}+m_{\gamma }}}\right] ,
\end{equation}
where, as before, $m$ is some constant having dimension of a mass
conveniently chosen to retain the standard distance units for the
relative coordinates. By index $\alpha $ one considers a chain of
partition $\alpha\equiv(\alpha \beta \gamma )\supset ( \beta
\gamma )$. This set is supplemented by the center of-mass
coordinate
\begin{equation}
\vec{R}={\frac{m_{1}\vec{r}_{1}+m_{2}\vec{r}_{2}+m_{3}\vec{r}_{3}}{M}}.
\end{equation}

By performing cyclic permutation three independent sets of Jacobi
coordinates (or partition chains) are obtained, namely:
$1\equiv(123 )\supset ( 23 )$; $2\equiv(123)\supset (31)$ and
$3\equiv(123)\supset (12)$. Any of these three sets constitutes a
complete coordinate base in configuration space. Equivalent
adjacent coordinate pairs
may be established in the momentum space, defined by:%
\begin{equation}
\vec{p}_{\alpha }=-i\hbar \frac{\partial }{\partial \vec{x}_{\alpha }};\;%
\vec{q}_{\alpha }=-i\hbar \frac{\partial }{\partial \vec{y}_{\alpha }},
\end{equation}
and given by:
\begin{equation}
\vec{p}_{\alpha }=\sqrt{\frac{m_{\beta }m_{\gamma }}{2(m_{\beta }+m_{\gamma
})m}}(\vec{k}_{\gamma }-\vec{k}_{\beta }),\qquad \vec{q}_{\alpha }=\sqrt{%
\frac{m_{\alpha }(m_{\beta }+m_{\gamma })}{2Mm}}\left[ \vec{k}_{\alpha }-{%
\frac{m_{\beta }\vec{k}_{\beta }+m_{\gamma }\vec{k}_{\gamma }}{m_{\beta
}+m_{\gamma }}}\right] ,
\end{equation}
where $\vec{k}_{\alpha }$ represents momentum of the particle $\alpha$.

\subsection{Relations between different coordinate sets}

The three Jacobi coordinate sets are equivalent, they describe the
same configuration of three particles in configuration (momentum)
space. Therefore these coordinates are related and one may easily
establish relation between these coordinate sets. Indeed, there
exist an orthogonal transformation:
\begin{eqnarray}
\vec{x}_{\alpha } &=&c_{\alpha \beta }\vec{x}_{\beta }+s_{\alpha \beta }\vec{%
y}_{\beta }, \\
\vec{y}_{\alpha } &=&-s_{\alpha \beta }\vec{x}_{\beta }+c_{\alpha \beta }%
\vec{y}_{\beta },
\end{eqnarray}
satisfying orthonormality condition:
\begin{equation}
c_{\alpha \beta }^{2}+s_{\alpha \beta }^{2}=1,
\end{equation}
and
\begin{equation}
c_{\alpha \beta}=-\sqrt{\frac{m _\alpha m_\beta }{(M-m
_\beta)(M-m _\alpha)}}; \; s_{\alpha \beta }=\epsilon _{\alpha \beta }\sqrt{%
1-c_{\alpha \beta }^{2}}=\epsilon _{\alpha \beta }\sqrt{\frac{M
m_{\gamma}}{(M-m _{\beta })(M-m _{\alpha })}},
\end{equation}
where $\epsilon _{\alpha \beta }=(-1)^{\beta -\alpha }sign{(\beta -\alpha )}$
with $sign(\beta -\alpha )$ representing the sign of the subtraction $(\beta
-\alpha )$. I.e. $\epsilon _{21}=\epsilon _{32}=\epsilon _{13}=+1=-\epsilon
_{12}=-\epsilon _{23}=-\epsilon _{31}$ and:
\begin{equation}
c_{\alpha \beta }=c_{\beta \alpha };\; s_{\alpha \beta }=-s_{\beta \alpha }.
\end{equation}
The modules of the Jacobi coordinates are expressed:
\begin{eqnarray}
x_{\beta }(x_{\alpha },y_{\alpha },u_{\alpha }) &=&[c_{\beta \alpha
}^{2}x_{\alpha }^{2}+s_{\beta \alpha }^{2}y_{\alpha }^{2}+2s_{\beta \alpha
}c_{\beta \alpha }x_{\alpha }y_{\alpha }u_{\alpha }]^{1/2},  \notag \\
y_{\beta }(x_{\alpha },y_{\alpha },u_{\alpha }) &=&[s_{\beta \alpha
}^{2}x_{\alpha }^{2}+c_{\beta \alpha }^{2}y_{\alpha }^{2}-2s_{\beta \alpha
}c_{\beta \alpha }x_{\alpha }y_{\alpha }u_{\alpha }]^{1/2}, \\
u_{\beta }(x_{\alpha },y_{\alpha },u_{\alpha })
&=&{\frac{1}{x_{\beta }y_{\beta }}}\left[ (c_{\beta \alpha
}^{2}-s_{\beta \alpha }^{2})x_{\alpha }y_{\alpha }u_{\alpha
}-s_{\beta \alpha }c_{\beta \alpha }(x_{\alpha }^{2}-y_{\alpha
}^{2})\right],  \notag
\end{eqnarray}%
with $u_{i}=\cos \alpha _{i}=\hat{x}_{i}.\hat{y}_{i}$.

\subsection{4-body Jacobi coordinates\label{sec_4b_jac_coor}}

\bigskip
\begin{figure}[th]
\centering
\includegraphics[width=150mm] {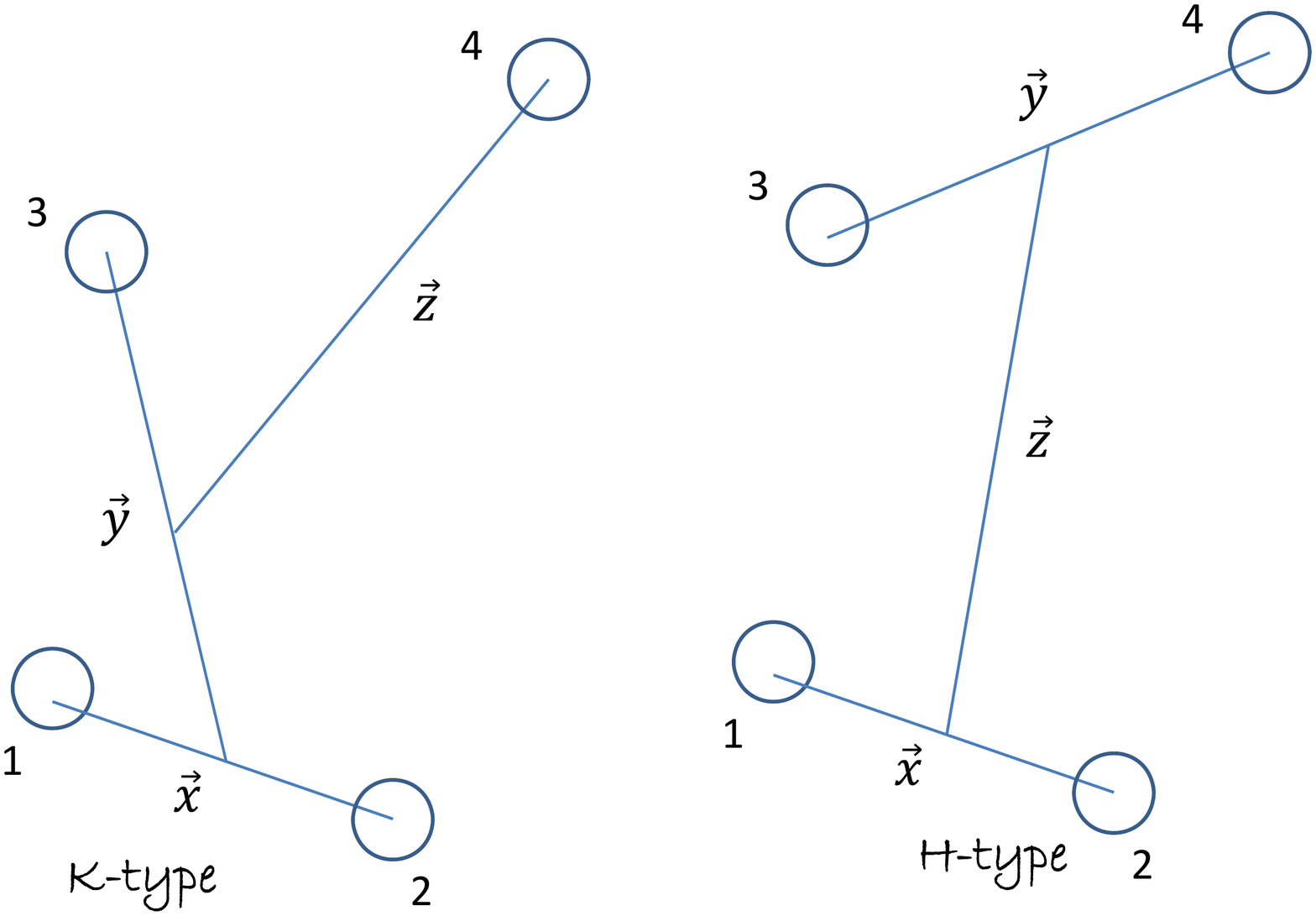}
\caption{4-particle Jacobi coordinate sets proper to describe FY
components, denoted in this work as $K_{12,3}^4$ and
$H_{12}^{34}$, represented by the partition chains $(1234)\supset
(123) \supset(12)$ and $(1234)\supset (12)(34) \supset(12)$
respectively.} \label{fig:4b_coord}
\end{figure}
For a four body system one can construct 48 sets of Jacobi
coordinates, since there are 2 types of partitions, see Fig.
~\ref{fig:4b_coord} and furthermore there are 4! possible
rearrangements of the 4 particles. Definitions of these
coordinates are as follows:
\begin{equation}
\begin{tabular}{ll}
K-type partition \textit{(ij,k)l} & $\left\{
\begin{tabular}{ll}
$\vec{x_{ij}}$ & $=\sqrt{2\mu _{ij}}(\vec{r}_{j}-\vec{r}_{i})\smallskip $ \\
$\vec{y_{ij,k}}$ & $=\sqrt{2\mu _{ij,k}}(\vec{r}_{k}-\frac{m_{i}\vec{r}%
_{i}+m_{j}\vec{r}_{j}}{m_{i}+m_{j}})\smallskip $ \\
$\vec{z_{ijk,l}}$ & $=\sqrt{2\mu _{ijk,l}}(\vec{r}_{l}-\frac{m_{i}\vec{r}%
_{i}+m_{j}\vec{r}_{j}+m_{k}\vec{r}_{k}}{m_{i}+m_{j}+m_{k}})\smallskip $%
\end{tabular}
\right.  ,\smallskip  $ \\
H-type partition \textit{(ij)(kl)} & $\left\{
\begin{tabular}{ll}
$\vec{x_{ij}}$ & $=\sqrt{2\mu _{ij}}(\vec{r}_{j}-\vec{r}_{i})\smallskip $ \\
$\vec{y_{kl}}$ & $=\sqrt{2\mu _{kl}}(\vec{r}_{l}-\vec{r}_{ki})\smallskip $
\\
$\vec{z_{ij,kl}}$ & $=\sqrt{2\mu _{ij,kl}}(\frac{m_{k}\vec{r}_{k}+m_{l}\vec{r%
}_{l}}{m_{k}+m_{l}}-\frac{m_{i}\vec{r}_{i}+m_{j}\vec{r}_{j}}{m_{i}+m_{j}})$%
\end{tabular}
\right. .$%
\end{tabular}%
\end{equation}
In the last formulaes the undimensional terms $\mu
_{ij,kl}=\frac{(m_i+m_j)(m_k+m_l)}{m(m_i+m_j+m_k+m_l)}$,
representing reduced mass of the clusters $(ij)$ and $(kl)$ were
employed.

 Relation between the different sets of the Jacobi coordinates
is less trivial than in a three-body case. It is convenient to
express it in a matrix form:
\begin{equation}
\left(
\begin{array}{c}
\vec{x^{\prime }} \\
\vec{y^{\prime }} \\
\vec{z^{\prime }}%
\end{array}
\right) =\left[ M_{3\times 3}\right] \left(
\begin{array}{c}
\vec{x} \\
\vec{y} \\
\vec{z}%
\end{array}
\right) .
\end{equation}
Due to the orthogonality of the Jacobi coordinates and the fact that the norm $%
\rho ^{2}=x^{2}+y^{2}+z^{2}$ is conserved the coordinate
transformation matrices M are unitary. In practice it is
convenient however to split the task in two steps, as:
\begin{equation}
(\vec{x}\vec{y}\vec{z})\longrightarrow(\vec{x}\vec{y}^{(i)
}\vec{z}^{\prime })\longrightarrow(\vec{x}^{\prime
}\vec{y}^{\prime }\vec{z}^{\prime }).
\end{equation}
During each of these steps only two vectors are manipulated, thus
requiring only transformation operation similar to 3-body case. In
the first step an intermediate vector $\vec{y}^{(i) }$ is
introduced for the convenience. The practical realization of
passage between different sets of coordinates is explained in more
details in Appendix B of the~\cite{These_Rimas_03}.

\subsection{General transformation of the Jacobi coordinates}

Transformation between any two Jacobi coordinates sets, describing
$N$-particle system, is far from trivial and consist of
multiplication with a matrix of the size  $( N-1)\times(N-1):$
\begin{equation}
\left(
\begin{array}{c}
\vec{x}^{\prime } \\
\vec{y}^{\prime } \\
\vec{z}^{\prime } \\
.. \\
\vec{w}^{\prime }%
\end{array}%
\right) =\left[ M_{N-1\times N-1}\right] \left(
\begin{array}{c}
\vec{x} \\
\vec{y} \\
\vec{z} \\
.. \\
\vec{w}%
\end{array}%
\right) .
\end{equation}
Nevertheless in analogy with a 4-body case, this operation might
be split into multiple three-body type coordinate transformation
steps, which involves only coupling of two different vectors at
the time. I.e.:
\begin{eqnarray}
\left(
\begin{array}{c}
\vec{x}^{\prime } \\
\vec{y}^{\prime } \\
\vec{z}^{\prime } \\
.. \\
\vec{w}^{\prime }%
\end{array}%
\right) &=&\left(
\begin{array}{cccc}
\left[ M_{2\times 2}\right] _{xy} &  &  &  \\
& 1 &  &  \\
&  & ... &  \\
&  &  & 1%
\end{array}%
\right) \left(
\begin{array}{c}
\vec{x} \\
\vec{y}^{\prime \prime } \\
\vec{z}^{\prime } \\
.. \\
\vec{w}^{\prime }%
\end{array}%
\right) \\
&=&\left(
\begin{array}{cccc}
1 &  &  &  \\
& \left[ M_{2\times 2}\right] _{yz} &  &  \\
&  & ... &  \\
&  &  & 1%
\end{array}%
\right) \left(
\begin{array}{c}
\vec{x} \\
\vec{y} \\
\vec{z}^{\prime \prime } \\
.. \\
\vec{w}^{\prime }%
\end{array}%
\right) \\
&=&..=\left(
\begin{array}{cccc}
1 &  &  &  \\
& 1 &  &  \\
&  & ... &  \\
&  &  & \left[ M_{2\times 2}\right] _{vw}%
\end{array}%
\right) \left(
\begin{array}{c}
\vec{x} \\
\vec{y} \\
\vec{z} \\
.. \\
\vec{w}%
\end{array}%
\right).
\end{eqnarray}
Expressions of the $2x2$ matrix $\left[ M_{2\times 2}\right]
_{xy}$ coefficients are obtained from the relations given for
3-body Jacobi coordinate transformations, by considering total
masses of the clusters involved in transforming coordinates.

\section{Faddeev-Yakubovsky equations\label{sec:FY_eq}}

The Schr\"{o}dinger equation is the fundamental equation of
physics describing quantum mechanical behavior. The properties as
well as the evolution of an isolated system may be established
from the set of the energy conserving physical solutions of the
time-independent Schr\"{o}dinger equation. Nevertheless one should
be cautious that this equation suffers from severe formal as well
as practical anomalies in describing many-body scattering
problems, starting from the 3-body case. The main difficulty is
related with a lack of tools to account for the rich variety of
the  N-body asymptotic states and our inability to impose the
proper boundary conditions, constraining the solutions of the
Schr\"{o}dinger equation to the physical ones. As will be
demonstrated in the next section, the complex scaling (CS) method
provides an efficient remedy and may be employed to solve
scattering problems starting from the Schr\"{o}dinger equation.
Nevertheless in order to get a better insight into a few-particle
scattering problem it is of great benefit to develop a
mathematically proper formalism. This feat has been achieved by
L.D.~Faddeev in the late sixties, related to the three-particle
problems~\cite{Faddeev:1960su} dominated by the short-ranged
interactions. Just a few years later Faddeev's revolutionary work
has been generalized to any number of particles by
O.A.~Yakubovsky~\cite{yakubovsky:67}. Finally, there exist also
modification of the three-body Faddeev equations, allowing to
treat long-ranged pairwise interactions, proposed by
S.P.~Merkuriev~\cite{Me80}.

In what follows I will briefly highlight the derivation of the
Faddeev-Yakubovsky equations in configuration space.

\subsection{The 3-body scattering and channels}

There are four possible types of the reaction channels in a three-particle
system. One can specify three different types of the binary channels
\begin{eqnarray}
&&1+(23),  \notag \\
&&2+(31), \\
&&3+(12) , \notag
\end{eqnarray}%
which should be supplemented with a so-called three-body breakup
channel:
\begin{equation}
1+2+3.
\end{equation}%
In principle, by taking any of these four configurations as an
initial state after the particles interact (collide) the system
may end in any of the four available configurations with a certain
probability. By virtue of Quantum Mechanics all these processes
happen simultaneously and must be encoded in the systems wave
function! Moreover a system of any two particles may possess
several bound states and thus there may exist many asymptotic
states within each of the 3-existing binary particle
configurations.

We start from the standard Schr\"{o}dinger equation considering a
three particle system interacting by the short-ranged binary
potentials, for simplicity of the notation we denote $V_{1}\equiv
\mathcal{V}_{23};\quad
V_{2}\equiv \mathcal{V}_{31};\quad V_{3}\equiv \mathcal{V}_{12}$%
\begin{equation}
(E-H_{0}-V_{1}-V_{2}-V_{3})\Psi =0,
\end{equation}%
where as usual $E$ denotes systems total energy, $H_{0}$ is the
kinetic energy operator and $\Psi$ - the total systems wave
function. From the total wave function $\Psi $ three different
wave function components are constructed:
\begin{equation}
\begin{array}{cc}
F_{i}=(E-H_{0})^{-1}V_{i}\Psi; & i=(1,2,3).%
\end{array}%
\end{equation}%
By substitution the last relation in to Schr\"{o}dinger equation it is easy
to check, that%
\begin{equation}
\Psi =F_{1}+F_{2}+F_{3}.
\end{equation}

The functions $F_{1},F_{2},F_{3}$ are called Faddeev components.
In the configuration space region where particle $1$ goes away the
interaction terms   vanish $V_{2}\equiv 0$ \& $V_{3}\equiv 0$,
thus forcing: $F_{2}\rightarrow 0$ \& $F_{3}\rightarrow 0$. In
this region the component $F_{1}$ fully absorbs the behavior of
the systems wave function. Therefore Faddeev component $F_{1}$
contains the complete asymptote of the systems wave function, when
particle 1 goes away, in such a way separating the asymptote
related to the binary 1+(23) particle channels from the ones
belonging to 2+(31) and 3+(12) configurations.

Instead of working with a single wave function $\Psi $ and a
single Schr\"{o}dinger equation, one may formulate a set of
coupled equations for the wave function
components $F_{i}$. This feat is realized in a set of three Faddeev equations:%
\begin{eqnarray}
(E-H_{0})F_{1} &=&V_{1}(F_{1}+F_{2}+F_{3}),  \notag \\
(E-H_{0})F_{2} &=&V_{2}(F_{1}+F_{2}+F_{3}), \\
(E-H_{0})F_{3} &=&V_{3}(F_{1}+F_{2}+F_{3}).  \notag
\end{eqnarray}

One may easily remark that adding three Faddeev equations one recovers Schr%
\"{o}dinger equation for the total systems wave function $\Psi$.

By employing Jacobi coordinates, one may easily separate and drop
the dependence on the center of mass degrees of freedom. Then,
like a total systems wave function $\Psi$, its Faddeev components
$F_{i}$ are  functions in six-dimensional space $R^6$, defined by
the Jacobi coordinates $\vec{x}$ and $\vec{y}$. It is natural to
associate $F_{i}$ to its proper Jacobi coordinate set. For
example $F_{1}$ may be expressed as a either function of $\left( \vec{x}_{1},%
\vec{y}_{1}\right)$, or $\left( \vec{x}_{2},\vec{y}_{2}\right)$, or finally $%
\left( \vec{x}_{3},\vec{y}_{3}\right)$. However it is much more
convenient
and makes more sense to express $F_{1}$ as a function of $\left( \vec{%
x}_{1},\vec{y}_{1}\right) $, since once expressed in its proper
coordinate set, Faddeev components maintain the simplest
structural behavior.

\subsection{Boundary conditions}

Differential equations should be supplemented with appropriate
boundary conditions in order to limit their possible solutions to
the physical ones. In this sense, and in particular when related
to the scattering problem, the benefits of the Faddeev components
becomes obvious.

The physical wave functions should be integrable and free of the
contact singularities, therefore they are expressed using regular
functions. This feat might be conveniently imposed by:
\begin{eqnarray}
\left. F_{i}(\vec{x}_{i},\vec{y}_{i})\right\vert _{x_{i}\rightarrow 0}
&\rightarrow &f(\widehat{x}_{i},\vec{y}_{i}),  \notag \\
\left. F_{i}(\vec{x}_{i},\vec{y}_{i})\right\vert _{y_{i}\rightarrow 0}
&\rightarrow &f(\vec{x}_{i},\widehat{y}_{i}),
\end{eqnarray}%
or in a more practical form:
\begin{eqnarray}
x_{i}\left. F_{i}(\vec{x}_{i},\vec{y}_{i})\right\vert _{x_{i}=0} &=&0,
\notag \\
y_{i}\left. F_{i}(\vec{x}_{i},\vec{y}_{i})\right\vert _{y_{i}=0}
&=&0. \label{eq:3b_bc_orig}
\end{eqnarray}

It is easy to formulate the 'external' boundary conditions for a
bound state problem. Bound state wave functions are compact
(square integrable), thus corresponding Faddeev components must
vanish in the far asymptotes:
\begin{eqnarray}
\left. F_{i}(\vec{x}_{i},\vec{y}_{i})\right\vert _{x_{i}\rightarrow \infty }
&=&0,  \notag \\
\left. F_{i}(\vec{x}_{i},\vec{y}_{i})\right\vert
_{y_{i}\rightarrow \infty } &=&0.  \label{eq:3b_bsxy_inf}
\end{eqnarray}%
In practice, one may prefer to limit the solution of the
differential equations to some finite region in space. In this
case one may require numerical solutions to vanish at the borders
of some large enough box, reducing the former conditions to:
\begin{eqnarray}
\left. F_{i}(\vec{x}_{i},\vec{y}_{i})\right\vert _{x_{i}=x_{\max }} &=&0,
\notag \\
\left. F_{i}(\vec{x}_{i},\vec{y}_{i})\right\vert _{y_{i}=y_{\max }} &=&0.
\end{eqnarray}

For the scattering problems, the regularity condition at the origin eq.(\ref%
{eq:3b_bc_orig}) remains valid. However the 'external' boundary
conditions turn to be much more complicated than for the bound
state problems. Nevertheless, like in a 2-body case, they should
represent a combination of the outgoing spherical wave and the
incoming plane wave. Moreover, as pointed out above, the Faddeev
components are built to separate different binary channels. By
limiting ourselves to the scattering problems arising from a
binary initial channel (initial state describes scattering of two
clusters), one may notice that the far asymptotes of the Faddeev
components should include~\cite{Merkuriev_boundcon}:

\begin{itemize}
\item An incoming plane wave part due to initial channel
$b_{j}^{(in)}$, if this wave is proper to the considered Faddeev
component. Since, by virtue of Faddeev equations, asymptotes of
the binary channels are separated into the appropriate Faddeev
components.

\item The outgoing spherical waves of the binary channels proper to the
considered Faddeev component.

\item If a 3-particle breakup is energetically accessible, i.e.
systems total energy in the center of mass frame is positive,
Faddeev components will also
include the outgoing 3-particle waves.\footnote{%
It is possible to formulate the boundary conditions including the
breakup for the case when particles are not charged and with some
approximations for the case when two particles are charged. Still
one should mention that Faddeev equations by themselves does not
provide specific framework to handle breakup asymptotes.}

By considering a system of non-charged particles, interacting by
short-range interactions, the aforementioned conditions can be summarized~\cite{Merkuriev_boundcon}:%
\begin{eqnarray}
\left. F_{i}(\overrightarrow{x}_{i},\overrightarrow{y}_{i})\right\vert
_{x_{i}\rightarrow \infty } &=&\mathcal{A}_{b_{j}^{(in)}}^{(i)}(\widehat{x}%
_{i},\widehat{y}_{i},\frac{x_{i}}{y_{i}})\frac{\exp (i\sqrt{\frac{m}{\hbar
^{2}}E}R)}{R^{\frac{5}{2}}}, \\
\left. F_{i}(\overrightarrow{x}_{i},\overrightarrow{y}_{i})\right\vert
_{y_{i}\rightarrow \infty } &=&\sum\limits_{b_{i}}\varphi _{b_{i}}(%
\overrightarrow{x}_{i})\left( \delta _{b_{i},b_{j}^{(in)}}\exp (i%
\overrightarrow{q}_{_{b_{i}}}\cdot \overrightarrow{y}_{i})+\mathcal{A}%
_{b_{i},b_{j}^{(in)}}(\widehat{y}_{i})\frac{\exp (i\left\vert
\overrightarrow{q}_{_{b_{i}}}\right\vert y_{i})}{y_{i}}\right)  \notag \\
&+&\mathcal{A}_{b_{j}^{(in)}}^{(i)}(\widehat{x}_{i},\widehat{y}_{i},\frac{%
x_{i}}{y_{i}})\frac{\exp (i\sqrt{\frac{m}{\hbar ^{2}}E}R)}{R^{\frac{5}{2}}}.
\end{eqnarray}
Here the first equation is a simple consequence of the fact that all
two-body wave functions vanish in their far asymptotes, the remaining term
contains an asymptote of the three-particle breakup. Terms $\mathcal{A}%
_{b_{i},b_{j}^{(in)}}(\widehat{y}_{i})$ and $\mathcal{A}%
_{b_{j}^{(in)}}^{(i)}(\widehat{x}_{i},\widehat{y}_{i},\frac{x_{i}}{y_{i}})$
describe binary and breakup amplitudes respectively. Binary amplitude $%
\mathcal{A}_{b_{i},b_{j}^{(in)}}(\widehat{y}_{i})$ describes transition from
the initial binary channel $b_{j}^{(in)}$ to one of the open binary channels
$b_{i}$, which is proper to Faddeev component $F_{i}$. Concerning the
breakup amplitude, one should note that $\mathcal{A}_{b_{j}^{(in)}}^{(i)}(%
\widehat{x}_{i},\widehat{y}_{i},\frac{x_{i}}{y_{i}})$ represents
only a part of the full amplitude, incorporated in a particular
Faddeev component $i$. The three breakup amplitude components
related to the same initial binary channel $b_{j}^{(in)}$ should
be added in order to retrieve a full breakup amplitude. In the
last equation summation is run over all available bound states
$b_{i}$ in the binary-particle cluster associated with the
component $i$. The momenta $\overrightarrow{q}_{_{b_{i}}}$ satisfy
energy
conservation condition:%
\begin{equation}
q_{_{b_{i}}}=\sqrt{\frac{m}{\hbar ^{2}}%
(E+E_{b_{j}^{(in)}}^{(2b)}-E_{b_{i}}^{(2b)})},
\end{equation}
where $E_{b_{i}}^{(2b)}$ denotes the 2-particle binding energy associated
with a channel $b_{i}$.

It is possible to generalize the last expressions for the systems
containing two charged particles. In this case, the free waves
should be replaced by their generalized expressions, built by
taking into account Coulomb interaction. Analytic expressions of
the breakup waves are not known for a case of charged particles.
One may still formulate approximate ones, based on semiclassical
approximations, if two of three particles are
charged~\cite{Merkuriev_3bcoulbc}.

\end{itemize}

\subsection{Faddeev-Merkuriev equations\label{sec_FM_equations}}

In the eighties, the original Faddeev equations, destined to solve
three-body problems governed by short-range interactions, have
been developed by S.P.~Merkuriev~\cite{Me80} to treat Coulombic
systems. Merkuriev proposed to split Coulomb potential $V_{\alpha
}$ into two parts (short and long range), $V_{\alpha }=V_{\alpha
}^{s}+V_{\alpha }^{l}$, by means of some cut-off function $\chi
_{\alpha }$.
\begin{equation}
V_{\alpha }^{s}(x_{\alpha },y_{\alpha })=V_{\alpha }(x_{\alpha
})\chi _{\alpha }(x_{\alpha },y_{\alpha });\qquad V_{\alpha
}^{l}(x_{\alpha },y_{\alpha })=V_{\alpha }(x_{\alpha })[1-\chi
_{\alpha }(x_{\alpha },y_{\alpha })].
\end{equation}%
Using the last identity the set of three Faddeev equations is
rewritten:
\begin{equation}
(E-H_{0}-V_{\alpha }-W_{\alpha })\Psi _{\alpha }=V_{\alpha
}^{s}\sum_{\alpha \neq \beta =1}^{3}\Psi _{\beta };\qquad
W_{\alpha }=V_{\beta }^{l}+V_{\gamma }^{l} . \label{MFE}
\end{equation}%
Here $E$ is a center of mass energy and $H_{0}$ is the free
Hamiltonian of a three-particle system.  In these equations the term $%
W_{\alpha }$ represents a non-trivial long-range three-body
potential. This term includes the residual interaction between a
projectile particle $\alpha $ and a target composed of particles
$\left( \beta \gamma \right)$. In order to obtain a set of
equations with compact kernels and which efficiently separate the
wave function asymptotes of different binary particle channels, the function $%
\chi_{\alpha }$ should satisfy certain conditions~\cite{Me80}. To
satisfy these conditions Merkuriev proposed a cut-off function in
a form:
\begin{equation}
\chi _{\alpha }(x_{\alpha },y_{\alpha })=\frac{2}{1+\exp \left[ \frac{%
(x_{\alpha }/x_{0})^{\mu }}{1+y_{\alpha }/y_{0}}\right] },
\end{equation}%
with parameters $x_{0},y_{0}$ and $\mu $, which can be
parametrized differently in each channel $\alpha$. A constrain
$\mu>2$ should be however respected, while the choice of $x_{0}$
and $y_{0}$ remains arbitrary. From the physics perspective a
parameter $x_{0}$ is associated with the effective size of the
2-body interaction; it makes therefore sense to associate this
parameter with a size of two-body bound state. On the other hand
the parameter $y_{0}$ is associated with a size of three-body
region, where the three-particle overlap is important.

Faddeev-Merkuriev (FM) equations, as formulated in eq.(\ref{MFE}),
project the wave function's asymptotes of the $\alpha $-$\left(
\beta \gamma \right)$ particle  channels to the component $\Psi
_{\alpha }$. The total systems wave function is
recovered by adding the three FM components $\Psi (\vec{x},\overrightarrow{y%
})=\Psi _{1}(\vec{x},\overrightarrow{y})+\Psi _{2}(\vec{x},\overrightarrow{y}%
)+\Psi _{3}(\vec{x},\overrightarrow{y})$. Similarly, by adding up
three equations eq.(\ref{MFE}), formulated for each component
$\Psi_{\alpha }$, the Schr\"{o}dinger equation is recovered.


In order to solve FM equations numerically, it is convenient to
express
each FM component $\Psi _{a}$ in its proper set of Jacobi coordinates $(\vec{%
x}_{\alpha},\vec{y}_{\alpha})$. Further it is practical to employ
partial waves to express the angular dependence of these
components:
\begin{equation}
\Psi _{\alpha }(\vec{x}_{\alpha },\vec{y}_{\alpha
})=\sum\limits_{l_{x},l_{y}}\frac{f_{\alpha
,l_{x},l_{y}}^{(LM)}(x_{\alpha },y_{\alpha })}{x_{\alpha
}y_{\alpha }}\left\{ Y_{l_{x}}(\widehat{x}_{\alpha })\otimes
Y_{l_{y}}(\widehat{y}_{\alpha })\right\} _{LM}, \label{eq_PW_exp}
\end{equation}%
here $\vec{l}_{x}$ and $\vec{l}_{y}$ are partial
angular momenta associated with the Jacobi coordinates $\vec{x}%
_{\alpha }$ and $\vec{y}_{\alpha }$ respectively. Naturally, the
total angular momentum $\vec{L}=\vec{l}_{x}+\vec{l}_{y}$ of the
system should be
 conserved.

Let select  an initial scattering state $\widetilde{\Psi
}_{a}^{(in)}$,  associated with a Jacobi coordinate set $\alpha $
(this feat will be expressed by the Kroneker $\delta _{\alpha ,a}$
function). The scattering
state $(a)$ is defined by a particle $\alpha$, which with momentum $%
q_{\alpha }=\frac{m_{e}}{\hbar ^{2}}\sqrt{E-E_{a}}$ impinges on a
bound particle pair $\left( \beta \gamma \right) $. This bound
state is defined by
a proper angular momentum quantum number $l_{x}^{(a)}$ and binding energy $%
E_{a}$. The relative angular momentum quantum number $l_{y}^{(a)}$
should satisfy triangular conditions,  related with the angular
momenta conservation
condition  $\vec{l}_{x}^{(a)}+\vec{l}_{y}^{(a)}=%
\vec{L}$. Then
\begin{equation}
\Psi _{\alpha }^{(a)}(\vec{x}_{\alpha },\vec{y}_{\alpha })=%
\widetilde{\Psi }_{a}^{(in)}(\vec{x}_{\alpha },\vec{y}_{\alpha
})\delta _{\alpha ,a}+\widetilde{\Psi }_{\alpha }^{(a)}(\vec{x}_{\alpha },%
\vec{y}_{\alpha }). \label{eq_inwave_sep}
\end{equation}%
The standard procedure is with a term $\widetilde{\Psi }%
_{a}^{(in)}(\vec{x}_{\alpha },\vec{y}_{\alpha })$ to separate a
free incoming wave of particle $\alpha$ with respect to a bound
pair of particles $(\alpha,\beta)$. Nevertheless Coulomb field of
particle $\alpha$ easily polarizes and excites the target,
resulting into long-range coupling between different target
configurations~\cite{Gailitis,Hu_PRL}. As a result, the scattering
wave function in its asymptote may approach a free-wave solution
very slowly and reach it only in far asymptote, beyond the region
covered by the numerical calculation. It might be useful to
represent incoming wave function by distorted waves, which
describe more accurately asymptotic solution. It is, the incoming
wave may be generalized to satisfy a 3-body Schr\"{o}dinger
equation:
\begin{equation}
(E-H_{0}-V_{\alpha }-\widetilde{W}_{\alpha })\widetilde{\Psi }%
_{a}^{(in)}\equiv 0 , \label{Sc_eq_aux_pot}
\end{equation}%
with some auxiliary long-range potential $\widetilde{W}_{\alpha }(\vec{x}%
_{\alpha },\vec{y}_{\alpha })$. This potential is exponentially
bound in $x_{\alpha }$ direction and therefore does not contribute
to particle recombination process. Nevertheless it may couple
different target states. Such an auxiliary potential can be
conveniently expressed by employing a separable expansion:
\begin{equation}
\widetilde{W}_{\alpha }(\vec{x}_{\alpha },\vec{y}_{\alpha
})=\sum_{a,b}\left\vert \varphi _{a,l_{x}}(\vec{x}_{\alpha
})\right\rangle \lambda _{ab}(y_{\alpha })\left\langle \varphi
_{b,l_{x}}(\vec{x}_{\alpha })\right\vert .\label{eq_aux_pot}
\end{equation}%
Radial amplitudes representing a distorted incoming wave
$\widetilde{\Psi }_{a}^{(in)}(\vec{x}_{\alpha },\vec{y}_{\alpha
})$ satisfy  standard boundary condition:
\begin{eqnarray}
\frac{1}{x_\alpha}\widetilde{f}_{\alpha
,l_{x},l_{y}}^{(in,a)}(x_{\alpha },y_{\alpha
}\rightarrow \infty )&=&\varphi _{a,l_{x}}(x_{\alpha })\widehat{j}%
_{l_{y}}(q_{a}y_{\alpha })\delta
_{l_{y},l_{y}^{(a)}}  \\
&+&\sum_{b}\delta _{\alpha
,b}\widetilde{A}_{b,a}(E)\sqrt{\frac{q_{b}}{q_{a}}}\varphi
_{b,l_{x}}(\vec{x}_{\alpha })\exp (iq_{b}y_{\alpha }-il_{y}\pi
/2)\delta _{l_{y},l_{y}^{(b)}} , \nonumber
\end{eqnarray}
where $\widetilde{A}_{b,a}(E)$ is the scattering amplitude due to
the
auxiliary long-range potential $\widetilde{W}_{\alpha }(\vec{x}_{\alpha },%
\vec{y}_{\alpha })$.  Equation~(\ref{Sc_eq_aux_pot}) is easy to
solve numerically using close coupling expansion~\cite{Massey289}.
Close coupling procedure allows to eliminate dependence on
$\vec{x}_{\alpha }$, thus leading to a standard 2-body coupled
channel problem.
By solving  eq.(\ref{Sc_eq_aux_pot}), the incoming wave $\widetilde{\Psi }%
_{a}^{(in)}(\vec{x}_{\alpha },\vec{y}_{\alpha })$ is obtained
numerically and may be further employed to solve the three-body FM
equations. By inserting
expressions~(\ref{eq_inwave_sep}-\ref{Sc_eq_aux_pot}) into
original FM equation~(\ref{MFE}), one obtains:
\begin{equation}
(E-H_{0}-V_{\alpha }-W_{\alpha })\widetilde{\Psi }_{\alpha
}^{(a)}=V_{\alpha
}^{s}\sum_{\alpha \neq \beta =1}^{3}\left( \widetilde{\Psi }_{\beta }^{(a)}+%
\widetilde{\Psi }_{\beta }^{(in)}\delta _{\beta ,a}\right) +(W_{\alpha }-%
\widetilde{W}_{\alpha })\widetilde{\Psi }_{a}^{(in)}\delta
_{\alpha ,a} . \label{eq_FM_inh}
\end{equation}

The FM amplitude $\widetilde{f}_{\alpha
,l_{x},l_{y}}^{(a)}(x_{\alpha },y_{\alpha }),$ associated with the
component $\widetilde{\Psi }_{\alpha }^{(a)}(\vec{x}_{\alpha
},\vec{y}_{\alpha })$, in the asymptote contains only outgoing
waves. It may contain two-types of them: ones representing binary
process where a particle $\alpha $ is liberated but a pair of
particles $\left( \beta \gamma \right) $ remains bound and
outgoing waves representing the breakup of the system into three
free particles:
\begin{eqnarray}
\frac{1}{x_\alpha}\widetilde{f}_{\alpha
,l_{x},l_{y}}^{(a)}(x_{\alpha },y_{\alpha}&\rightarrow& \infty
)=\sum_{b}\delta _{\alpha
,b}\overline{A}_{b,a}^{{}}(E)\sqrt{\frac{q_{b}}{q_{a}}}\varphi_{b,l_{x}^{(b)}}(\vec{x}_{\alpha}) \exp(iq_{b}y_{\alpha}-il_{y}^{(b)}\pi /2) \nonumber\\
&+&A_{a,l_{x},l_{y}}(E,\frac{x_{\alpha }}{y_{\alpha
}},\sqrt{x_{\alpha }^{2}+y_{\alpha }^{2}})\exp
(i\sqrt{\frac{m_{e}}{\hbar ^{2}}E(x_{\alpha }^{2}+y_{\alpha
}^{2})}).
\end{eqnarray}
The amplitude $\overline{A}_{b,a}(E)$ represents transition
between
the distorted binary channels, whereas the amplitude $A_{a,l_{x},l_{y}}(E,\frac{x_{\alpha }%
}{y_{\alpha }},\sqrt{x_{\alpha }^{2}+y_{\alpha }^{2}})$ is set to
describe three-particle breakup process. These amplitudes can be
extracted from the solution $\widetilde{\Psi }_{\alpha }^{(a)}$ of
the FM equations by applying Green's theorem. In this study, we
will concentrate only on the scattering
amplitudes related to the rearrangement reactions. The amplitude $\overline{A}%
_{b,a}(E)$ is given by:
\begin{eqnarray}
\overline{A}_{b,a}(E) &=&\sqrt{q_{a}q_{b}}\frac{m}{\hbar
^{2}}\left\{
\left\langle \Psi ^{(a)}\left\vert E-H_{0}\right\vert \widetilde{\Psi }%
_{b}^{(in)}\right\rangle -\left\langle \widetilde{\Psi }_{b}^{(in)}\left%
\vert E-H_{0}\right\vert \Psi ^{(a)}\right\rangle \right\}   \\
&=&\sqrt{q_{a}q_{b}}\frac{m}{\hbar ^{2}}\left\langle \Psi
^{(a)}\left\vert \sum_{\alpha }\left\{ \left( V_{\alpha }+\widetilde{W}%
_{\alpha }\right) \delta _{\alpha ,b}-V_{\alpha }\right\}
\right\vert \widetilde{\Psi }_{b}^{(in)}\right\rangle .
\label{eq_greens_th}
\end{eqnarray}

The total scattering amplitude is given by:
\begin{equation}
A_{b,a}(E)=\overline{A}_{b,a}(E)+\widetilde{A}_{b,a}(E).
\end{equation}
In terms of this full amplitude, partial scattering cross section
for a process $b\rightarrow a$ and a partial wave L is defined by:
\begin{equation}
\sigma _{ab}^{L}(E)=\frac{2\pi a_{0}^{2}}{\frac{m_{\alpha
}(m_{\beta
}+m_{\gamma })}{(m_{\alpha }+m_{\beta }+m_{\gamma })m}q_{a}^{2}}%
(2L+1)\left\vert A_{a,b}(E)\right\vert ^{2} .
\end{equation}
One may also define total inelastic cross section for a collision
$(a)$:
\begin{equation}
\sigma _{a,inel}^{L}(E)=\frac{\pi a_{0}^{2}}{2\frac{m_{\alpha
}(m_{\beta
}+m_{\gamma })}{(m_{\alpha }+m_{\beta }+m_{\gamma })m}q_{a}^{2}}%
(2L+1)\left( 1-\left\vert 1+2iA_{a,a}(E)\right\vert ^{2}\right) .
\end{equation}

\subsection{The four-body FY equations}


The derivation of the four-body Faddeev-Yakubovsky equations
starts by defining three-body like FY components:
\begin{equation}\label{eq_3b_FY_com}
\begin{array}{cc}
  \psi_{ij}=G_0 V_{ij} \Psi & (i<j).
\end{array}
\end{equation}
Here $G_0$ denotes a free four-body Green's function, while
$V_{ij}$ denotes binary potential between the particles $i$ and
$j$. Naturally, there exist six different three-body like FY
components for a four body system. By combining the three-body
like FY components, one may define two types of FYCs, denoted as
the components of the type-K and the type-H and given by:
\begin{equation}\label{eq_4b_FY_com}
\begin{array}{cc}
  K_{ij,k}^l=G_{ij} V_{ij}(\psi_{jk}+\psi_{ik}) & (i<j); \\
  H_{ij}^{kl}=G_{ij} V_{ij}\psi_{kl} & (i<j; \quad k<l).
\end{array}
\end{equation}
 By permuting particle indexes one
may construct 12 independent components of the type-K as well as 6
independent components of the type-H. The asymptotes of the
components $K_{ij,k}^l$ and $H_{ij}^{kl}$ incorporate the  3+1 and
the 2+2 particle channels respectively, see
Fig.~\ref{fig:4b_coord}.

In this work only systems of four identical nucleons will be
considered. Within the isospin formalism neutrons and protons are
treated as isospin-degenerate states of the same particle,
nucleon. FY components which differ by the order of the particle
indexing are related due to the symmetry of particle permutation.
There remain only two independent FYCs, which are further denoted
$K \equiv K_{12,3}^4$ and $H \equiv H_{12}^{34} $ by omitting
their indexing. FY equations for a case of the four identical
particles read~\cite{These_Rimas_03,Lazauskas_4B}:
\begin{eqnarray}
\left( E-H_{0}-V_{12}\right) K &=&V_{12}(P^{+}+P^{-})\left[
(1+Q)K+H\right],
\notag \\
\left( E-H_{0}-V_{12}\right) H &=&V_{12}\tilde{P}\left[
(1+Q)K+H\right] , \label{FY1}
\end{eqnarray}
where $H_0$ is a kinetic energy operator, whereas $V_{ij}$
describes the interaction between $i$-th and $j$-th nucleons. FYCs
may be converted from one coordinate set to another by using the
particle permutation operators, which are summarized as follows:
$P^{+}=(P^{-})^{-1}\equiv P_{23}P_{12}$, $Q\equiv - P_{34}$ and
$\tilde{P} \equiv P_{13}P_{24}=P_{24}P_{13}$, where $P_{ij}$
indicates operator permuting particles $i$ and $j$.

In terms of the FYCs, the total wave function of an $A=4$ system
is given by:
\begin{equation}  \label{FY_wave_func}
\Psi =\left[ 1+(1+P^{+}+P^{-})Q\right] (1+P^{+}+P^{-})K
+(1+P^{+}+P^{-})(1+\tilde{P})H.
\end{equation}

Each FY component $F=(K,H)$ is considered as a function, described
in its proper set of Jacobi coordinates, defined in the
section~\ref{sec_4b_jac_coor}.

Angular, spin and isospin dependence of these components is
described using the tripolar harmonics $\mathcal{Y}_{\alpha
}(\hat{x},\hat{y},\hat{z})$, i.e:
\begin{equation}
\langle \vec{x}\vec{y}\vec{z}|F\rangle =\sum_{\alpha
}\;{\frac{F_{\alpha }(xyz)}{xyz}}\;\mathcal{Y}_{\alpha
}(\hat{x},\hat{y},\hat{z}).  \label{KPW}
\end{equation}%
The quantities $F_{\alpha }(xyz)$ are called the regularized FY
amplitudes, where the label $\alpha $ holds for a set of 10
intermediate quantum numbers describing a given four-nucleon
quantum state $\left( J^{\pi },T,T_{z}\right) $. By using the
LS-coupling scheme the tripolar harmonics are defined for
components $K$ and $H$ respectively by
\begin{eqnarray}
\mathcal{Y}_{\alpha_K }&\equiv  &\left\{ \left[ \left( l_{x}l_{y}
\right)_{l_{xy}}l_{z}\right]_L \left[ \left(\left(
s_{1}s_{2}\right)_{s_{x}}s_3\right)_{S3}s_{4}\right]_S\right\}
_{J^{\pi }\cal{M}} \otimes \left[ \left(\left(
t_{1}t_{2}\right)_{t_{x}}t_3\right)_{T3}t_{4}\right]_{T\cal{T}_Z},
\label{PWD} \\
\mathcal{Y}_{\alpha_H }&\equiv &\left\{ \left[ \left(
l_{x}l_{y}\right)_{l_{xy}}l_{z}\right]_L \left[\left(
s_{1}s_{2}\right)_{s_{x}}\left(
s_{3}s_{4}\right)_{s_{y}}\right]_S\right\} _{J^{\pi
}\cal{M}}\otimes \left[ \left( t_{1}t_{2}\right)_{t_{x}}\left(
t_{3}t_{4}\right)_{t_{y}}\right]_{T\cal{T}_Z} .
\end{eqnarray}

The next step is to separate the incoming plane wave of the two
colliding clusters from $K$ (or $H$) partial components:
\begin{eqnarray}\label{In_wave_sep}
K(\vec{x},\vec{y},\vec{z})=K^{out}(\vec{x},\vec{y},\vec{z})+K^{in}(\vec{x},\vec{y},\vec{z}), \\
H(\vec{x},\vec{y},\vec{z})=H^{out}(\vec{x},\vec{y},\vec{z})+H^{in}(\vec{x},\vec{y},\vec{z}).
\end{eqnarray}
The expansion of the incoming plane wave in the tripolar harmonics
provides:
\begin{eqnarray}
F_{\alpha_K }^{in}(x,y,z)=\delta_{3+1}\kappa^{(3)}_{\alpha_K}(x,y) \cdot \hat{j}_{ l_z}(q_3z)/q_3, \\
F_{\alpha_H
}^{in}(x,y,z)=\delta_{2+2}\kappa^{(22)}_{\alpha_H}(x,y) \cdot
\hat{j}_{ l_z}(q_{22}z)/q_{22},
\end{eqnarray}
here $\delta_{3+1}$=1 and $\delta_{2+2}=0$ if one considers the
incoming state of one particle projected on the bound cluster of 3
particles (like $n+^3H$). Alternatively, $\delta_{3+1}$=0 and
$\delta_{2+2}=1$ if one considers the incoming state of 2+2
particle clusters (like $^2H+^2H$). The functions
$\kappa^{(3)}_{\alpha_K}(x,y)$ and $\kappa^{(22)}_{\alpha_H}(x,y)$
represent regularized Faddeev amplitudes of the corresponding
bound state wave functions containing 3 and 2+2 particle clusters
respectively. The $q^2_3=\frac{m}{\hbar^2}(E-\epsilon_3)$ and
$q^2_{22}=\frac{m}{\hbar^2}(E-\epsilon_{2}-\epsilon_{2}$) are the
momenta of the relative motion of the free clusters.  Here we
suppose that the system possesses only one three-particle and only
one two-particle bound states with the binding energies equal
$\epsilon_3$ and $\epsilon_{2}$ respectively. By inserting
eq.~(\ref{In_wave_sep}) into eq.~(\ref{FY1}) one may rewrite FY
equations in their driven form:
\begin{eqnarray}
\left( E-H_{0}-V_{12}\right) K^{out}-V_{12}(P^{+}+P^{-})\left[
(1+Q)K^{out}+H^{out}\right] &=&V_{12}(P^{+}+P^{-})\left[
(1+Q)H^{in}+QK^{in}\right],
\notag \\
\left( E-H_{0}-V_{12}\right) H^{out}-V_{12}\tilde{P}\left[
(1+Q)K^{out}+H^{out}\right] &=&V_{12}\tilde{P}\left[
(1+Q)K^{in}\right]. \label{FY_drive}
\end{eqnarray}
One may note that the $K^{out}$ and $H^{out}$ components in the
asymptote contain only various combinations of the outgoing waves.
If the breakup into three or four clusters is energetically
allowed, the FY components of both types retain parts of the
outgoing waves describing breakup. In addition, the $K^{out}$
components fully absorb outgoing waves representing the 3+1
particle channels, whereas the $H^{out}$ components  fully absorb
the outgoing waves corresponding to 2+2 particle channels. In the
asymptote, where at least one particle recedes from the others,
they take the following forms:

{\begin{footnotesize}
\begin{eqnarray}
K^{out}(\vec{x},\vec{y},\vec{z})=A_{31}(\hat{z})\psi^{(3)}(\vec{x},\vec{y})
\frac{\exp(iq_{3}z)}{|z|}+
 A^K_{211}(\hat{y},\hat{z})\psi^{(2)}(\vec{x})\frac{\exp(iq_{2}X)}{|X|^{5/2}}+A^K_{1111}(\hat{x},\hat{y},\hat{z})\frac{\exp(iq_{1}R)}{|R|^{4}},
\notag \\
H^{out}(\vec{x},\vec{y},\vec{z})=A_{22}(\hat{z})\psi^{(22)}(\vec{x},\vec{y})
\frac{\exp(iq_{3}z)}{|z|}+
 A^H_{211}(\hat{y},\hat{z})\psi^{(2)}(\vec{x})\frac{\exp(iq_{2}X)}{|X|^{5/2}}+A^H_{121}(\hat{x},\hat{z})\psi^{(2)}(\vec{y})\frac{\exp(iq_{2}Y)}{|Y|^{5/2}},\notag\\
 +A^H_{1111}(\hat{x},\hat{y},\hat{z})\frac{\exp(iq_{1}R)}{|R|^{4}},
\label{Out_wave_assymp}
\end{eqnarray}
\end{footnotesize}}
where terms $A$ represent various types of amplitudes of
scattering in two, three and four clusters. Wave functions
$\psi^{(3)}$, $\psi^{(22)}$ and $\psi^{(2)}$  represent various
cluster bound states and thus are exponentially bound.

\section{The complex scaling method}

\begin{figure}[th]
\centering
\includegraphics[width=150mm] {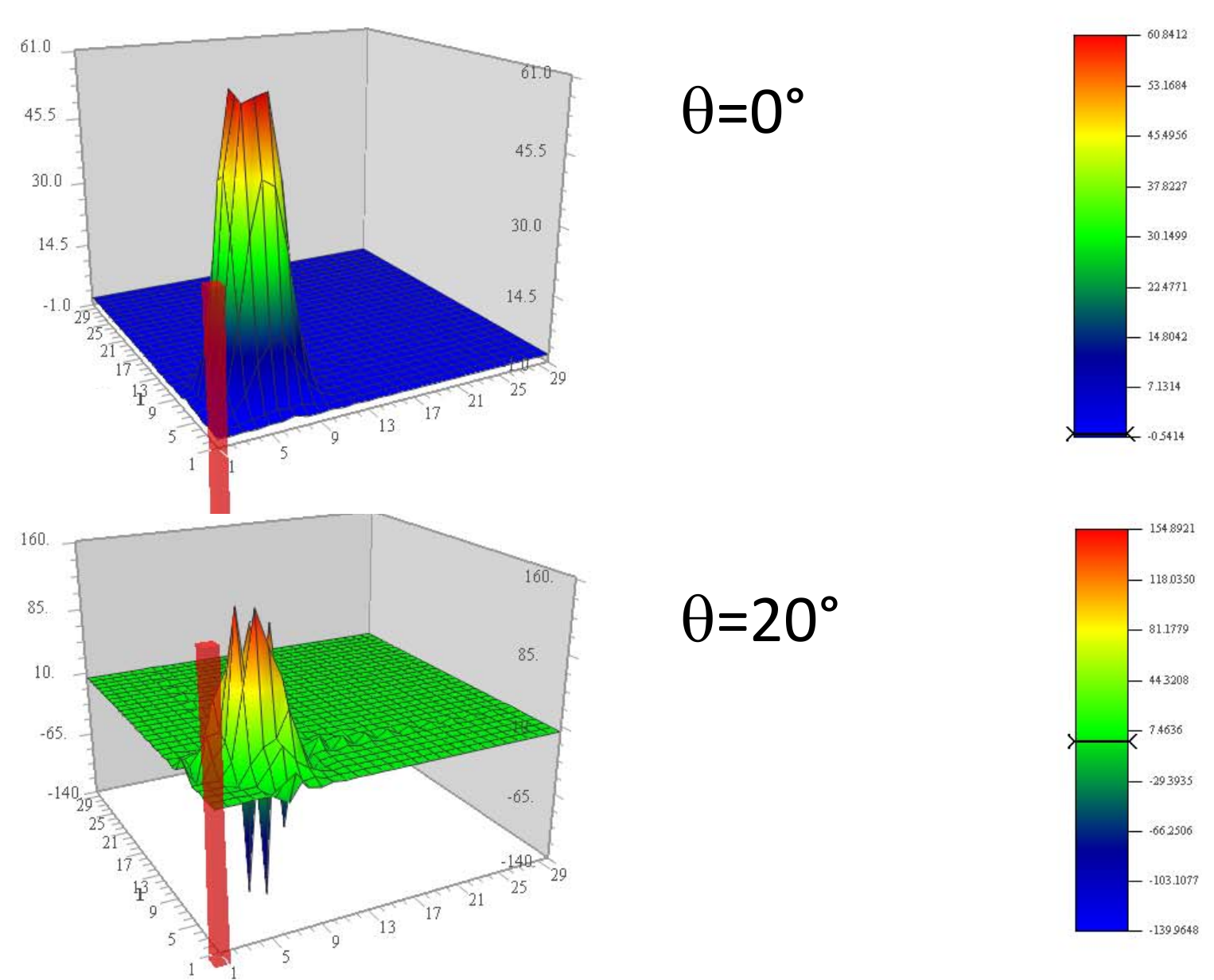}
\caption{Real parts of the potential energy matrix elements for
INOY np interaction in $3S_1$ wave. The projection is made on
Lagrange-Laguerre basis of 30 functions. The z-axis reflects the
size of the matrix elements in MeV. In the upper figure matrix
elements for the original potential, in the bottom figure for the
CS transformed potential with $\protect\theta=20^\circ$.}
\label{fig:CS_pot_nl}
\end{figure}

A method very similar to the complex scaling~(CS) has been
introduced already during the World War II by D.R.~Hartree et
al.~\cite{HMN46,Co83} in relation to the study of the radio wave
propagation in the atmosphere. D.R.~Hartree et al. were solving
second order differential equations for the complex eigenvalues.
In practice, this problem is equivalent to the one of finding
S-matrix pole positions, in relation with the resonant states of
quantum two-particle collision process. In the late sixties
J.~Nuttal and H. L.~Cohen~\cite{NC69} proposed a very similar
method to treat
 scattering problems, dominated by the short range
potentials. Few years later J.~Nuttal even employed this method to
solve a three-nucleon scattering problem above the breakup
threshold~\cite{MDN72}. Nevertheless these pioneering works of
J.~Nuttal have been abandoned, while based on J.~Nutall's work and
the mathematical foundation of E.~Baslev and J.
M.~Combes~\cite{BC71} the original method of D.R.~Hartree has been
recovered
in order to calculate resonance eigenvalues in atomic physics~\cite%
{Ho83,Mo98}. Such an omission is mostly due to the fact that short
range potentials may earn highly nontrivial structures after the
complex scaling transformation is applied, see
figures~\ref{fig:CS_pot_nl} and~\ref{fig_local_sc_pot} (or refer
to~\cite{LC05_3n,REB81} for more details). On the other hand this
transformation does not affect the radial form of the Coulomb
potential.

Only recently a variant of the complex scaling method based on the spectral
function formalism has been presented by K.~Kat\={o}, B.~Giraud et al.~\cite%
{MKK01,GK03,GKO04} and applied in the works of K. Kat\={o} et al.~\cite%
{MKK01,SMK05,AMKI06,KKMTI10,KKWK11,KKMI13}. This variant will be described
in detail in the next subsection. On the contrary in the later works of
A.T.~Kruppa et al.~\cite{KSK07} as well as in the works of J.~Carbonell and
R.L.~\cite{LC11,La12} the original idea of J.~Nuttal and H.L.~Cohen is
further elaborated.

\bigskip

\subsection{The complex scaling operator}

Numerous problems in quantum mechanics are related to isolated
systems, which are subject to energy conservation. Usually this
kind of problems can be reformulated  in a time-independent frame
by factoring out the time-dependent part of their wave-function.
When considering non-relativistic dynamics a time-independent
formalism leads to solve a generalized N-body Schr\"{o}dinger
equation with an eventually present inhomogeneous term:
\begin{equation}
(E-\widehat{H}_{0}-\sum_{i}\widehat{V}_{i})\Psi =I.
\label{eq:schr_drv_base}
\end{equation}%
In the last equation $E$ is systems total energy, $\Psi $ its wave
function, or at least its non-trivial part\footnote{%
As it will be demonstrated, it is possible to rewrite the problem
in such a way that the wave function $\Psi $ contains only
outgoing waves.}. $H_{0}$ denotes kinetic energy operator, whereas
$\widehat{V}_{i}$ denotes operators representing potential energy
terms. For sake of simplicity one may express the total Hamiltonian as $\widehat{H}=%
\widehat{H}_{0}+\sum_{i}V_{i}.$ Eventually on the right hand of
the last equation an inhomogeneous term $I$ is present. An
inhomogeneous term appears in diverse scattering problems and may
be straightforwardly related to an initial state, which for the
problems related with a realistic experiment is supposed to be
known a-priori (predefined by the experimental setup) and
therefore represents a trivial part of the problem. In this case
$\Psi $ represents wave function's behavior in relation with a
final state, describing  distribution of the reaction products.
Since the reaction products should evolve from the
collision-center (closely localized area, where particles are
supposed to hit each other), their distribution should be
described by the outgoing spherical waves -- waves evolving in all
the directions from the collision area. The wave function $\Psi $
thus carries key information about the considered system, in
particular in its far-asymptote information about the particle
distribution after reaction takes place is encoded and thus is
straightforwardly related with the experimental observables. The
main asset of the complex scaling method is due to simple and
efficient treatment of the outgoing waves.

Problems of finding bound or resonant states, particle collisions
or reactions due to an impact of an external probe might be
presented in the general form of eq.~(\ref{eq:schr_drv_base}).
Nevertheless direct solution of the last equation presents a
formidable task already for a three-particle systems. Additional
complications arise due to the fact that most of the computational
methods in quantum mechanics have been developed for the Hermitian
operators. However the physical Hamiltonians are Hermitian only
when they operate on bounded (square integrable) functions. Wave
functions describing resonant states or particle collisions does
not meet the last criteria. Nevertheless as will be demonstrated
here, an extension of the variational principle and of the other
well-known theorems in quantum mechanics to the non-Hermitian
operators can be made by carrying out similarity transformations
$\widehat{S}$, which converts outgoing scattered waves, $\phi
^{out}$, into square integrable functions. That is,
\begin{equation}
E\left( \widehat{S}\Psi \right) -\left( \widehat{S}\widehat{H}\widehat{S}%
^{-1}\right) \left( \widehat{S}\Psi \right) =\widehat{S}I,
\end{equation}%
such that
\begin{equation}
\widehat{S}\phi ^{out}(r\rightarrow \infty )\rightarrow 0,
\end{equation}%
and $\widehat{S}\phi ^{out}(r)$ is in the Hilbert space although
$\phi ^{out}(r)$ is not. The complex-scaling operator, to be
defined below, is only one example of a vast set of similarity
transformations for which the last equation is satisfied. However
the simplicity of the complex-scaling operator and its conformity
with the existing numerical methods makes it unexcelled in the
practical applications.

The complex-scaling (CS) operator is defined as
\begin{equation}
\widehat{S}=\exp (i\theta r\frac{\partial }{\partial r}),
\end{equation}%
such that
\begin{equation}
\widehat{S}f(r)=f(re^{i\theta }).
\end{equation}

As already mentioned, of particular interest is the action of this
operator on the outgoing scattered waves
\begin{equation}
\widehat{S}\phi ^{out}(r\rightarrow \infty )\propto \exp
(ikre^{i\theta }),
\end{equation}
in this equation $k$ denotes the scattering momentum.

CS transformation of the Hamiltonian is also rather trivial. For a
sake of clarity, and without loss of generality, let us consider
an one-dimensional radial Hamiltonian. When the potential is
dilation analytic, the complex-scaled Hamiltonian is simply:
\begin{eqnarray}
H_{l}^{\theta } &=&\widehat{S}\widehat{H}\widehat{S}^{-1} \\
&=&-\frac{\hbar ^{2}}{2\mu }\frac{d^{2}}{e^{2i\theta
}dr^{2}}+\frac{\hbar ^{2}}{2\mu }\frac{l(l+1)}{e^{2i\theta
}r^{2}}+V(re^{i\theta }).
\end{eqnarray}

From the last expression it follows that the kinetic energy
operator is simply scaled by the factor $e^{-2i\theta }$ after the
CS transformation is applied:
\begin{equation}
T_{ij}^{\theta }=\frac{1}{e^{2i\theta }}T_{ij}.
\end{equation}

Calculation of the potential energy matrix is more complicated,
but still rather standard. For the local potential one has:
\begin{equation}
V_{ij}^{\theta }=\left\langle f_{i}\left\vert \widehat{S}\widehat{V}\widehat{%
S}^{-1}\right\vert f_{j}\right\rangle =\int_{0}^{\infty
}f_{i}(r)V(re^{i\theta })f_{j}(r)dr, \label{eq:me_cspot}
\end{equation}

If the potential is non-local $V(r,r\prime )$:
\begin{equation}
V_{ij}^{\theta }=\int_{0}^{\infty }e^{i\theta
}f_{i}(r)V(re^{i\theta },r^{\prime }e^{i\theta })f_{j}(r^{\prime
})drdr^{\prime }.
\end{equation}
One may refer to the section~\ref{sec:cs_pot_trnsf} for a more
detailed discussion on the CS transformation of the potential
energy.

\subsection{Bound states \label{sec:cs_bound_state}}

In quantum mechanics, bound states are defined as localized
solutions of the Schr\"{o}dinger equation, without a source term
$I\equiv 0$. These states
appear as the poles of the S-matrix on a positive imaginary momentum axis $%
k_{bs}=i\left\vert k_{bs}\right\vert =\sqrt{\frac{mE_{bs}}{\hbar
^{2}}}$ (see figure~\ref{fig_cauchy}). Bound state wave functions
in their asymptotes involve only outgoing waves and thus:
\begin{equation}
\phi _{bs}(r\rightarrow \infty )\propto \exp (ik_{bs}r)=\exp (-\left\vert
k_{bs}\right\vert r).
\end{equation}%
By virtue of the last equation, bound state wave functions are
exponentially bound and belong to the Hilbert space. The action of
the CS operator on a bound state wave function gives:
\begin{equation}
\widehat{S}\phi _{bs}(r\rightarrow \infty )\propto \exp (-\left\vert
k_{bs}\right\vert re^{i\theta })=\exp (-\left\vert k_{bs}\right\vert r\cos
\left( \theta \right) )\exp (-i\left\vert k_{bs}\right\vert r\sin \left(
\theta \right) ).
\end{equation}%
This function remains in the Hilbert space as long as a CS angle
satisfy $mod(\theta -\pi /2,2\pi )<\pi /2$ \footnote{In practice
it is convenient to limit the complex scaling angles to $0\leq
\theta <\pi /2$.} Naturally one may solve CS Schr\"{o}dinger
equation for $\overline{\phi }_{bs}^{\theta }=\widehat{S}\phi
_{bs}:$
\begin{equation}
E_{bs}\overline{\phi }_{bs}^{\theta }-\left( \widehat{S}\widehat{H}\widehat{S%
}^{-1}\right) \overline{\phi }_{bs}^{\theta }=0,  \label{eq:2b_cs_se}
\end{equation}%
to determine bound state energies $E_{bs}$ and their wave function
representations $\overline{\phi }_{bs}^{\theta }$ due to CS
transformation. As long as CS angle satisfies the condition
$mod(\theta -\pi /2,2\pi )<\pi /2,$ the last equation might be
solved using techniques based on Hilbert space methods by
expanding $\overline{\phi }_{bs}^{\theta }$ with a square
integrable basis function set.

Obviously CS transformation does not bring any added value in solving bound
state problem by itself, since the CS Schr\"{o}dingers equation (\ref%
{eq:2b_cs_se}) is more complicated than a non-transformed one.
After CS transformation the structure of a bound state wave function $\overline{%
\phi }_{bs}^{\theta }(r)$ becomes more complicated than its original image $%
\phi _{bs}(r)$, gaining additional oscillating factor $\exp
(-i\left\vert k_{bs}\right\vert r\sin \left( \theta \right) )$ in
the far asymptote. Nevertheless, as it will be demonstrated in the
following, if one wish to apply CS method to solve scattering
problems, CS images of the bound state wave functions are needed
as an input in constructing initial state wave function. To this
aim it turns to be numerically advantageous to solve
eq.~(\ref{eq:2b_cs_se}) and determine $\overline{\phi
}_{bs}^{\theta }(r)$,
than try to construct $\overline{\phi }_{bs}^{\theta }$ using $\widehat{S}%
\phi _{bs}(r)$ relation.

\subsection{Resonant states\label{sec:res_states}}

In this study I will restrict to the resonant states related with
the S-matrix poles appearing in the 4$^{th}$ energy quadrant.
Two-particle resonant state wave functions are defined by the
outgoing wave solutions of the two-body Schr\"{o}dinger equation.
It is
\begin{equation}
\phi_n(r\rightarrow\infty )\propto\exp (i k_n^{res}r),
\end{equation}
with  $k_n^{res}=\sqrt{\frac{m}{\hbar^2}E_n^{res}}$ representing
momentum of a resonant state $n$. It is of particular interest  to
express an action of the complex scaling operator on a wave
function of a resonant state:
\begin{eqnarray}
\widehat{S}\phi _{n}^{res}(r &\rightarrow &\infty )\propto \exp
(ik_{n}^{res}re^{i\theta })=\exp (i\left\vert k_{n}^{res}\right\vert
re^{i\left( \theta -\vartheta ^{res}\right) }) \\
&=&\exp (i\left\vert k_{n}^{res}\right\vert r\cos \left( \theta
-\vartheta ^{res}\right) )\exp (-\left\vert k_{n}^{res}\right\vert
r\sin \left( \theta -\vartheta ^{res}\right) )
\label{eq:res_state_wf}
\end{eqnarray}
Thus one may easily see that if the condition
\begin{equation}
\text{mod}(\theta -\vartheta ^{res},2\pi )<\pi  \label{eq:2b_res_cn}
\end{equation}%
is satisfied, the complex-scaled resonance wave functions become
exponentially convergent.

It is of interest to see how a CS transforation affects the
spectra of the Hamiltonian. According to the Aguilar, Balslev and
Combes theorem~\cite{BC71}, see fig.~\ref{Fig_eigen_csm}):
\begin{enumerate}
\item The bound state poles remain unchanged under the transformation.

\item The cuts are now rotated downward making an angle of
$2\theta $ with a real axis.

\item The resonant poles are \textquotedblleft exposed\textquotedblright\ by
the cuts once the \textquotedblleft rotational angle\textquotedblright\ $%
\theta $ is greater than $-\frac{1}{2}Arg(E_{res})$, where
$E_{res}$ is the complex resonance energy.
\end{enumerate}

\bigskip

It is easy to prove this theorem for the short-range potentials.
In such a case, the asymptotic behavior of the scattering states
is given by:
\begin{equation}
\phi ^{scatt}(r\rightarrow \infty )=A(k)e^{-ikr}+B(k)e^{ikr}
\end{equation}%
where as usual center-of-mass kinetic energy (E) in terms of
momentum (k) is expressed
\begin{equation}
E=\frac{\hbar ^{2}}{2\mu }k^{2},
\end{equation}%
The energy takes any real positive value (provided that the
threshold energy is taken as zero). The complex-scaled scattering
states are given by
\begin{equation}
\widehat{S}\phi ^{scatt}(r\rightarrow \infty
)=A(k)e^{-ikre^{i\theta }}+B(k)e^{ikre^{i\theta }}.
\end{equation}%
One can see that these wave functions diverge if $\theta <\pi $,
since the real part of the exponential factor $e^{-ikre^{i\theta
}}$is positive. The only bounded non-divergent (not square
integrable) functions are obtained when $k$ gets complex values,
\begin{equation}
k=\left\vert k\right\vert e^{-i\theta }
\end{equation}%
and therefore (when the threshold is taken as the zero reference energy)
\begin{equation}
E=\left\vert E\right\vert e^{-i2\theta }
\end{equation}

According to the Aguilar, Balslev and Combes theorem~\cite{BC71}
(ABC theorem), in order to find the resonant states one should
simply solve an eigenvalue problem for a complex Hamiltonian:
\begin{equation}
H_{l}^{\theta }\widetilde{\phi }_{n}^{\theta }(r)=E_{n}^{\theta }\widetilde{%
\phi }_{n}^{\theta }(r),
\end{equation}%
keeping in mind that the resonant eigenvalues are
\textquotedblleft exposed\textquotedblright\ by the cuts of the
rotated continuum states, that is $\theta
>-\frac{1}{2}Arg(E_{R})$.

The complex analog to the variational principle provides the
formal justification to the use of the computational techniques
that originally were developed for the bound state problems. The
Rayleigh quotient
\begin{equation}
E_{n}^{\theta }=\frac{(\phi \left\vert H_{l}^{\theta }\right\vert \phi )}{%
(\phi \left\vert \phi \right) }
\end{equation}%
provides a stationary approximation to the true complex eigenvalue $%
E_{n}^{\theta }$ when $\phi $ is a c-normalizable eigenfunction of $%
H_{l}^{\theta }$, which is close to exact solution $\widetilde{\phi }%
_{n}^{\theta }(r).$ This means that the calculated eigenvalues,
corresponding to some resonant state, will stabilize around the
exact solution without providing any bound (upper, lower) for the
eigenenergy.

In practice, the convergence of the calculated resonance
eigenvalues might be improved by either increasing the  size of
the eigenfunction basis (or density of the wave-function
discretization points), or by increasing the complex
scaling angle beyond its critical value $\theta =-\frac{1}{2}Arg(E_{R})$%
\footnote{It is important to note, as will be demonstrated in the
following section that there may exist potential depending maximal
value of the CS angle $\theta p,$ beyond which one is not able to
realize CS transformation of the potential.}. The physical
resonance eigenvalues frequently appear close to the thresholds
and therefore the values of $\left\vert k_{n}^{res}\right\vert $
in eq.(\ref{eq:res_state_wf}) are usually small, resulting slow
decaying exponent for the CS resonant wave functions. At the same
time, due to the presence of the exponent $\exp (i\left\vert
k_{n}^{res}\right\vert r\cos \left( \theta -\vartheta
^{res}\right) )$ these asymptotes might be strongly oscillating.
This demonstrates that much of the care should be taken in
describing  the far-extending parts  of the resonant wave
functions.

These general developments might be easily extended to the
problems related with a few-particle resonant states. One must
simply keep in mind that a few-particle resonance wave function
might involve more than one outgoing wave, related with a presence
of more than one scattering threshold. Therefore very similar
condition, as one formulated for a 2-body case in
eq.(\ref{eq:2b_res_cn}), should be validated relative to each open
threshold.  Furthermore one should be aware of the possible
appearance of the discretized continuum pseudostates, associated
with a presence of the resonant states in the multiparticle
subsystems (see fig.~\ref{fig_cauchy}). In the momentum manifold
these pseudostates align along the lines starting from a resonant
subsystem's momentum and are bent by angle $\theta$ relative to
the real axis.
\subsection{Extended completeness relation}

\begin{figure}[th]
\centering
\includegraphics[width=90mm]{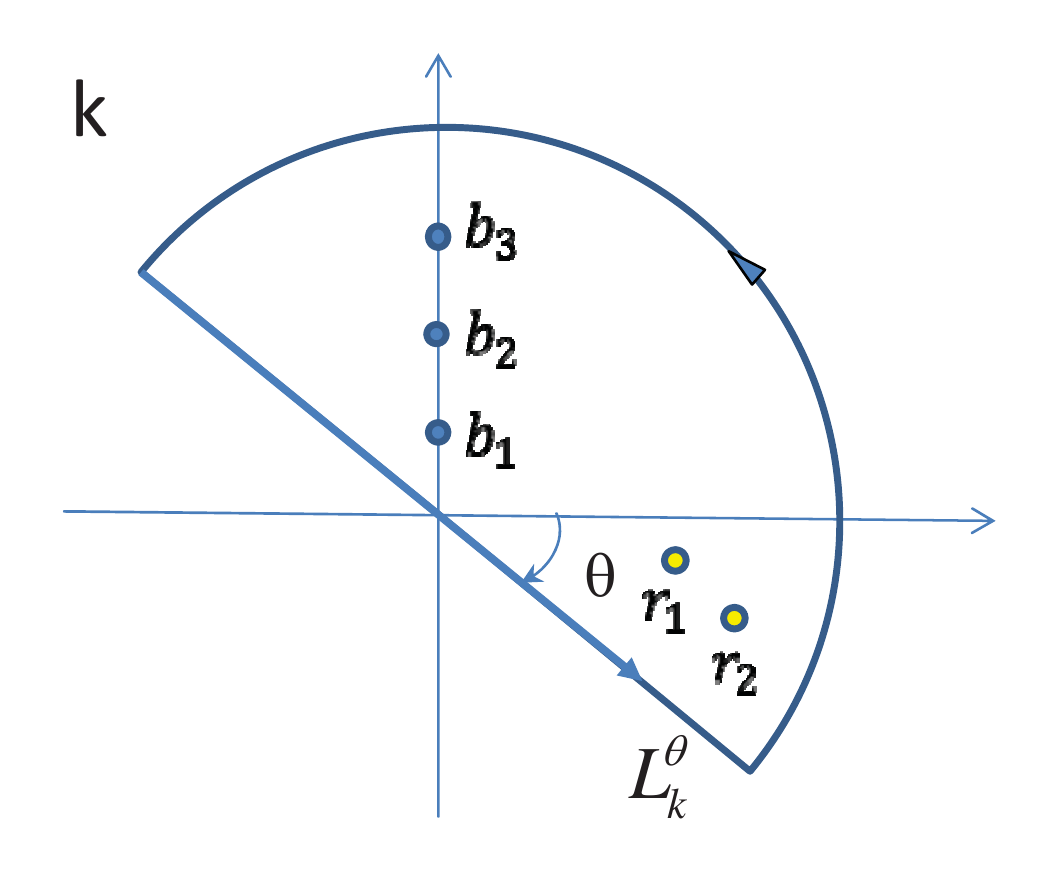}
\caption{The Cauchy integral contour in the momentum plane for the
completeness relation of the complex scaled Hamiltonian. The $b_1,b_2,..$
and $r_1,r_2,..$ represent the bound and resonant poles respectively. }
\label{fig_cauchy}
\end{figure}

The complex eigenvalues obtained for a complex-scaled Hamiltonian
have a very
physical interpretation. In the work of K.~Kat\={o}, B.~Giraud et al.~\cite%
{MKK01,GK03,GKO04} the completeness relation of
T.~Berggren~\cite{Be73} has been proved for the complex scaled
Hamiltonian solutions representing bound, resonant as well as
single- and coupled-channel scattering states. This completeness
relation can be formulated for the Cauchy integral contour in the
momentum plane as demonstrated in fig.~\ref{fig_cauchy}, as:
\begin{equation}
\mathbf{1}=\sum_{b}\left\vert \chi _{b}^{\theta }\right) \left( \chi
_{b}^{\theta }\right\vert +\sum_{r}^{n_{r}^{\theta }}\left\vert \chi
_{r}^{\theta }\right) \left( \chi _{r}^{\theta }\right\vert
+\int_{L_{k}^{\theta }}dk_\theta\left\vert \chi _{k_{\theta }}\right) \left(
\chi _{k_{\theta }}\right\vert,  \label{eq:GF_CSM}
\end{equation}
here $\chi _{b}^{\theta }$ and $\chi _{r}^{\theta }$ are the
complex scaled bound and resonant state wave-functions
respectively. Only the resonant states encircled by a semicircle
rotated by an angle $\theta $ must be considered. Remaining
continuum states $\chi _{k}^{\theta }$ are located on the rotated
momentum axis $L_k^\theta$ (see figure~\ref{fig_cauchy}). One
should mention that the definition of the complex scaled bra- and
ket-states for a non-Hermitian $H^\theta$ is different from one
defined for Hermitian Hamiltonians. For the complex scaled
Hamiltonian $H^\theta$ one express a bra-state as bi-conjugate
solution of the equivalent ket-state. In practice, for the
discrete (resonant and bound) states we can use the same wave
functions for the bra- and ket-states; for the continuum states,
the wave function of a bra-state is given by that of the
equivalent ket-state divided by the S-matrix.

Using the former completeness relation, one may construct the
complex scaled Green's function as
\begin{equation}
\mathcal{G^{\theta }}(E,\mathbf{r},\mathbf{r^{\prime }})=\sum_{b}\frac{%
\left\vert \chi _{b}^{\theta }(\mathbf{r})\right) \left( \chi _{b}^{\theta }(%
\mathbf{r^{\prime }})\right\vert }{E-E_{b}}+\sum_{r}^{n_{r}^{\theta }}\frac{%
\left\vert \chi _{r}^{\theta }(\mathbf{r})\right) \left( \chi _{r}^{\theta }(%
\mathbf{r^{\prime }})\right\vert }{E-E_{r}}+\int_{L_{k}^{\theta }}dk_{\theta
}\frac{\left\vert \chi _{k_{\theta }}(\mathbf{r})\right) \left( \chi
_{k_{\theta }}(\mathbf{r^{\prime }})\right\vert }{E-E_{\theta }},
\label{eq:Greens_func}
\end{equation}%
where $E_{b}$ and $E_{r}=(E_{R}-\frac{i}{2}\Gamma )$ are the energy
eigenvalues of the bound and relevant resonant states respectively.
Variables $\mathbf{r}$ reflect all the internal coordinates of the
multiparticle system under consideration.

\bigskip
\begin{figure}[th]
\centering
\includegraphics[width=70mm] {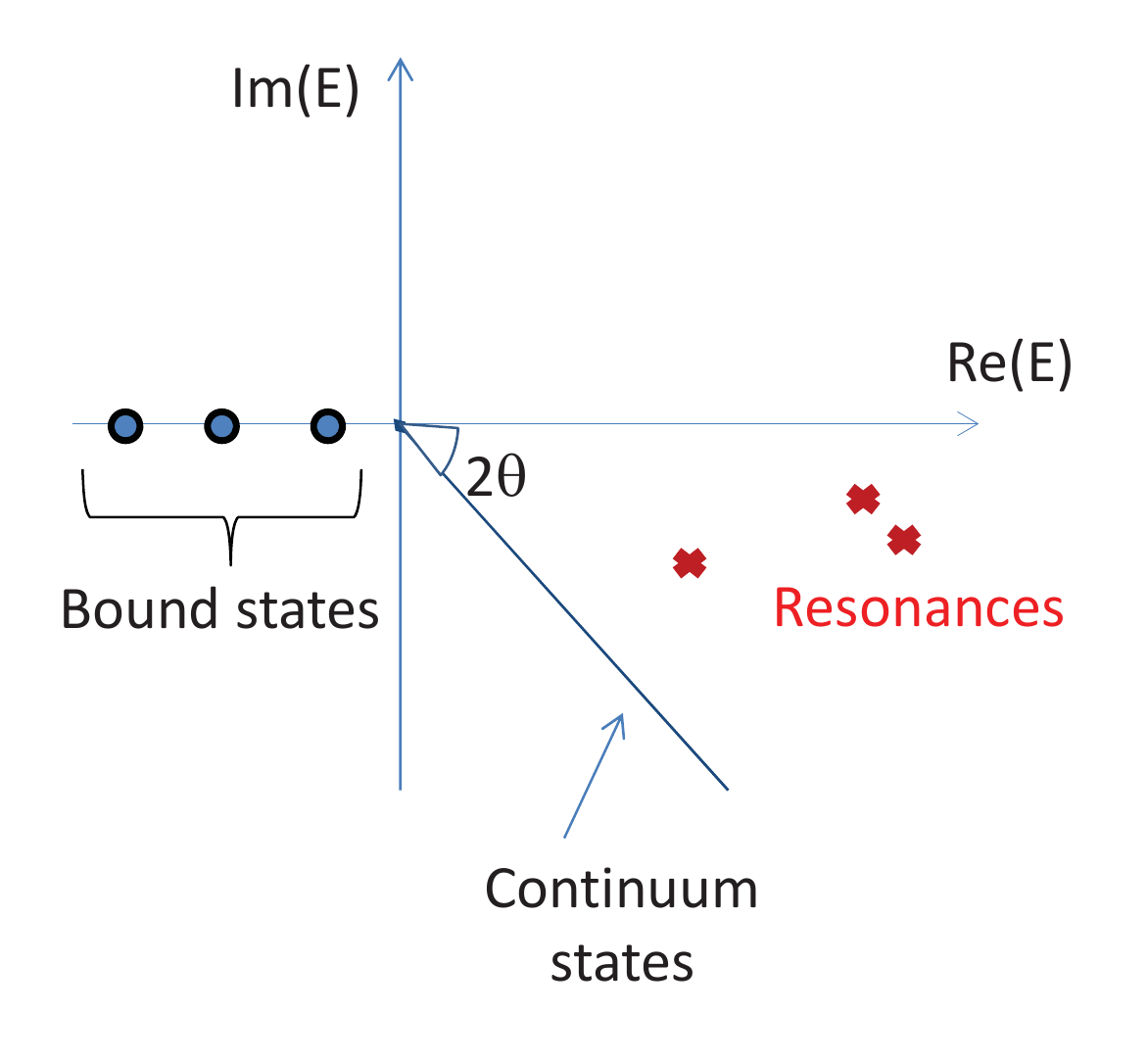} 
\includegraphics[width=90mm] {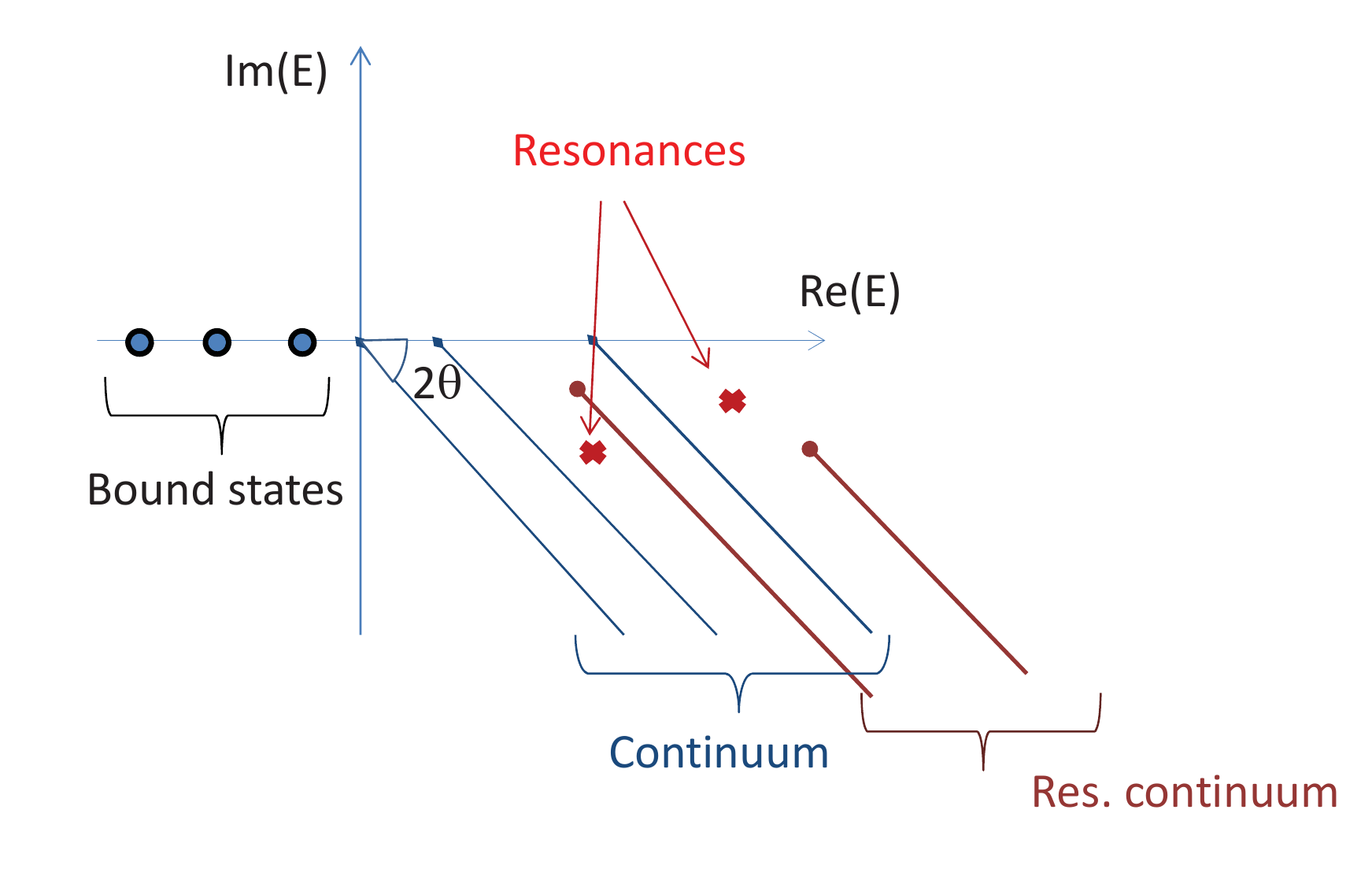}
\caption{Schematic representation of the eigenvalues of the
complex scaled Hamiltonian, $H^{\protect\theta }$ , according to
the theorem of Aguilar, Balslev and Combes~\protect\cite{BC71}.
For a two-body system (left panel) bound states are obtained as
negative real energy eigenvalues, continuum-pseudostates are
rotated by angle $2\protect\theta $, resonant states inside
$2\protect\theta $ branch may also be obtained. For a many-body
system (right panel) several rotated continuum branches exist
associated with bound and resonant thresholds in its subsystems.}
\label{Fig_eigen_csm}
\end{figure}

For the sake of simplicity, the contour depicted in the
figure~\ref{fig_cauchy} represents the simplest 2-body case. Still
all of the presented relations remain valid for the many-body
system; one only should keep in mind that the obtained spectra may
have a much more complicated structure. Following the ABC
theorem~\cite{BC71} the eigenvalues of the complex-scaled two-body
Hamiltonian, which are associated with the bounded wave function,
splits into three categories: bound state eigenvalues situated on
the negative horizontal energy axis, the pseudo-continuum states
scattered along the positive energy axis rotated by angle $2\theta
$ and eigenvalues representing the resonant states whose
eigenergies satisfies the relation -$arg(E)<2\theta $ (see left
panel of figure~\ref{Fig_eigen_csm}). For the many-body system,
bound states will be situated on the horizontal part of the energy
axis, situated below the lowest systems separation into
multiparticle clusters threshold (see figure~\ref{Fig_eigen_csm}).
Pseudo-continuum states will scatter along the $2\theta $-lines
projected
from each possible separation threshold. In addition, one will have $2\theta $%
-lines projected from the "resonant thresholds", where one or more
sub-clusters are resonant. Finally, many-body resonance
eigenvalues will manifest as discrete points inside the semicircle
making angle $2\theta $ with real energy axis and derived from the
lowest threshold.

\subsection{Reactions due to external probes\label{sec:react_exprb}}

There is a vast group of problems in physics where a system is
initially in a bound state and is excited to the continuum by a
perturbation. In particular, it concerns reactions led by
electro-magnetic and weak probes. For these reactions one is led
to evaluate the strength (or  the response) function, which in the
lowest order perturbation theory is provided by
\begin{equation}
S(E)=\sum_{\nu }\left\vert \left\langle \Psi _{\nu }\left\vert \hat{O}%
\right\vert \Psi _{0}\right\rangle \right\vert ^{2}\delta (E_{\nu }-E_{0}-E),
\end{equation}%
where $\hat{O}$ is the perturbation operator which induces a
transition from a bound-state $\Psi _{0},$ with a ground-state
energy $E_{0}$, to a state $\Psi _{\nu }$ with an energy $E_{\nu
}$. Both wave functions are solutions of the same Hamiltonian :
$H$ . The energy is measured from some standard value, e.g., a
particle-decay threshold energy. When the excited state is in the
continuum, the label $\nu $ is continuous and the sum must be
replaced by an integration. The final state wave function $\Psi
_{\nu }$ may have complicate asymptotic behavior in configuration
space if it represents a continuum state. On the other hand the
expression may  be rewritten by avoiding summation over the final
states
\begin{eqnarray}
S(E) &=&\left\langle \Psi _{0}\left\vert \hat{O}^{\dag }\delta (H-E_{\nu })%
\hat{O}\right\vert \Psi _{0}\right\rangle \\
&=&-\frac{1}{\pi }Im\left\langle \Psi _{0}\left\vert \hat{O}^{\dag }G(E_{\nu
}+i\varepsilon )\hat{O}\right\vert \Psi _{0}\right\rangle =-\frac{1}{\pi }%
Im\left\langle \Psi _{0}\left\vert \hat{O}^{\dag }\right\vert \Phi
_{\nu }\right\rangle,  \label{Strenght_func}
\end{eqnarray}%
with
\begin{equation}
(H-E_{\nu })\Phi _{\nu }=\hat{O}\Psi _{0}.
\end{equation}%
The right hand side of the former equation is compact, damped by
the bound-state  wave function $\Psi _{0}$. The wave function
$\Phi _{\nu }$ in its asymptote will contain only outgoing waves.
Therefore the last inhomogeneous equation might be readily solved
using complex scaling techniques
\begin{equation}
(H^{\theta }-E_{\nu })\overline{\Phi }_{\nu }^{\theta
}=\hat{O}^{\theta }\Psi _{0}^{\theta }. \label{eq:extern_probe}
\end{equation}%
To do so, one should construct the CS inhomogeneous term, present
in  the right hand side of the last equation. A practical way to
obtain complex-scaled bound state wave functions $\Psi
_{0}^{\theta }$ is to  solve bound state problem for the
complex-scaled Hamiltonian
\begin{equation}
(H^{\theta }-E_{0})\Psi _{0}^{\theta }=0,
\end{equation}
as explained in the section \ref{sec:cs_bound_state}.

In order to solve eq.~(\ref{eq:extern_probe}), one projects it on
a chosen square-integrable basis ($f_i$) employed to expand wave
function $\Psi _{0}^{\theta }$:
\begin{equation}
\Psi _{0}^{\theta }(r)=\sum_i c_i^\theta f_i(r)
\label{eq:linal_ex},
\end{equation}
where naturally the expansion coefficients $c_i^\theta$ are
complex numbers. This procedure leads to a standard linear algebra
problem:
\begin{equation}
\left( \left[ E\right] -\left[ \hat{H^{\theta }}\right] \right)
c^\theta=w^{in,\theta }. \label{eq:ep_lin_alg_probe}
\end{equation}
here $\left[ E\right] ,\left[ \hat{H^{\theta }}\right] $ are the
same matrices as for CS resonances problem, representing
projection of norm matrix and Hamiltonian. Vector $w^{in,\theta }$
represents projection of the inhomogenious term $\hat{O}^{\theta
}\Psi _{0}^{\theta }$ on a chosen basis $f_{i}(r)$.

 There are two distinct ways to solve the linear algebra problem
eq.~(\ref{eq:ep_lin_alg_probe}) and evaluate the associated
strength function eq.~(\ref{Strenght_func}). The first one, and
probably the most practical one, relies on the direct solution of
the linear algebra problem. Once the coefficients $c_i^\theta$ are
obtained, it makes no difficulty to calculate the strength
function of eq.~(\ref{Strenght_func}):
\begin{equation}
S(E)=-\frac{1}{\pi} Im \sum_i c_i^\theta w_i^{in,\theta }.
\end{equation}
One may keep in mind that complicated few-particle problems may
lead to  linear algebra problems of very considerable size, where
Hamiltonian matrix largely exceeds storage capacities of the
available hardware. To confront this problem, iterative linear
algebra methods exist~\cite{Saad:2003}, which allows to find the
solution by avoiding storage of the  matrix.

\subsubsection{Complex scaled Green's function method \label{CS_GF}}

Alternative solution of a linear algebra problem
eq.~(\ref{eq:ep_lin_alg_probe}) relies on the spectral expansion,
widely employed in the
works~\cite{MKK01,SMK05,AMKI06,KKMTI10,KKWK11,KKMI13}. In this
case, the solution of the linear algebra problem is expanded in
the
eigensolutions $\widetilde{\phi }_{i}^{\theta }(r)$ of the Hamiltonian matrix $\left[ \hat{H^{\theta }}\right]$:%
\begin{eqnarray}
\overline{\Phi }_{\nu }^{\theta }(\mathbf{r}) &=&\sum_{i=1}^{N}a_{i}%
\widetilde{\phi }_{i}^{\theta }(r) \\
a_{i} &=&\frac{\left( \widetilde{\phi }_{i}^{\theta }(r)\right. \left\vert
w^{in,\theta }\right) }{E-E_{i}^{\theta }}
\end{eqnarray}

Finally, strength function is obtained via:
\begin{equation}
S(E)=-\frac{1}{\pi }Im\left\langle \Psi _{0}^{\theta }\left\vert (\hat{O}%
^{\dag })^{\theta }\right\vert \overline{\Phi }_{\nu }^{\theta
}\right\rangle .
\end{equation}

By inserting the last relation into the eq.~(\ref{Strenght_func}),
one finally gets:
\begin{eqnarray}  \label{Strenght_func_gf}
S(E) &=&S_b(E)+S^\theta_r(E)+S^\theta_k(E), \\
S_b(E)&=&-\frac{1}{\pi }Im\sum_{b}\frac{\left( \Psi_{0}^{\theta }\right\vert(%
\hat{O}^\dag)^\theta\left\vert \chi_{b}^{\theta }\right) \left( \chi
_{b}^{\theta }\right\vert \hat{O}^\theta \left\vert \Psi_{0}^{\theta }\right)%
}{E-E_b}, \\
S^\theta_r(E)&=&-\frac{1}{\pi }Im\sum_{r}^{n_{r}^\theta}\frac{\left(
\Psi_{0}^{\theta }\right\vert(\hat{O}^\dag)^\theta\left\vert
\chi_{r}^{\theta }\right) \left( \chi _{r}^{\theta }\right\vert \hat{O}%
^\theta \left\vert \Psi_{0}^{\theta }\right)}{E-E_r}, \\
S^\theta_k(E)&=&-\frac{1}{\pi }Im\int_{L^\theta_k}\frac{\left(
\Psi_{0}^{\theta }\right\vert(\hat{O}^\dag)^\theta\left\vert \chi_{k_\theta
}\right) \left( \chi _{k_\theta}\right\vert \hat{O}^\theta \left\vert
\Psi_{0}^{\theta }\right)}{E-E_\theta}.
\end{eqnarray}
In practice (numerical solution), one works with a finite basis;
then, the last term containing integration is replaced by a sum
running over all the complex eigenvalues, representing continuum
pseudo states. All the eigenvalues are obtained as solutions of
the complex scaled Hamiltonian with a pure outgoing wave boundary
condition -- exponentially converging ones due to the complex
scaling.

The obtained total strength function $S(E)$ should be independent
of the angle $\theta$, employed in the calculation. Furthermore
the strength function component $S_b(E)$ as well as its partial
components due to contribution of the separate bound
states are also independent of $\theta$. The partial components of the $%
S^\theta_r(E)$, corresponding to narrow resonant states, also turn
to be independent of $\theta$, as long as the angle $\theta$ is
large enough to encircle these resonances. However if a resonance
is large enough and is not encircled by the contour $L^\theta_k$
its contribution to the strength function is reabsorbed by the
pseudo-continuum states in the $S^\theta_k(E)$ term. This feature
has been clearly demonstrated in the ref.~\cite{AMKI06} for a
chosen 2-body example.

\begin{figure}[th]
\centering
\includegraphics[width=80mm] {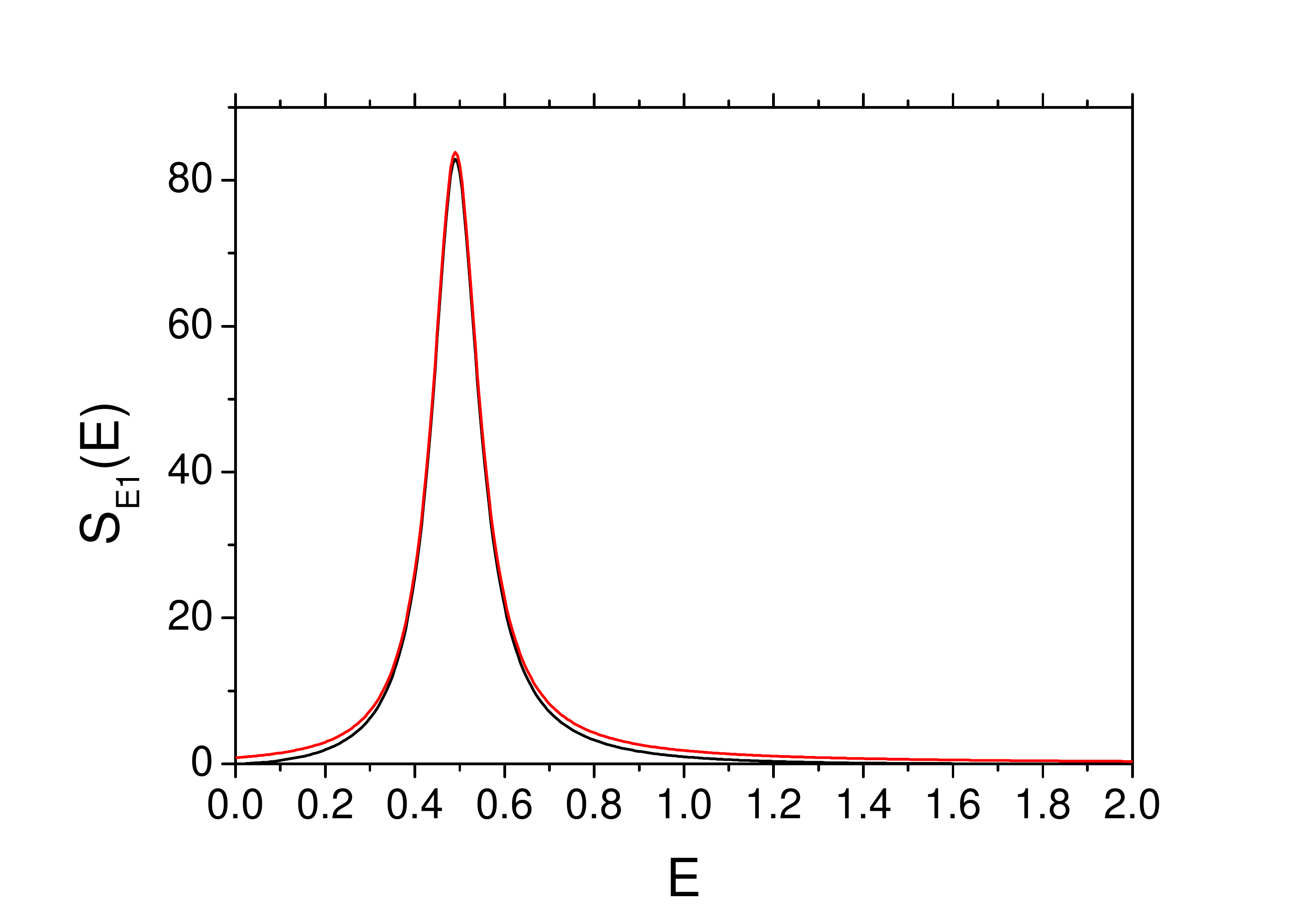} 
\includegraphics[width=80mm] {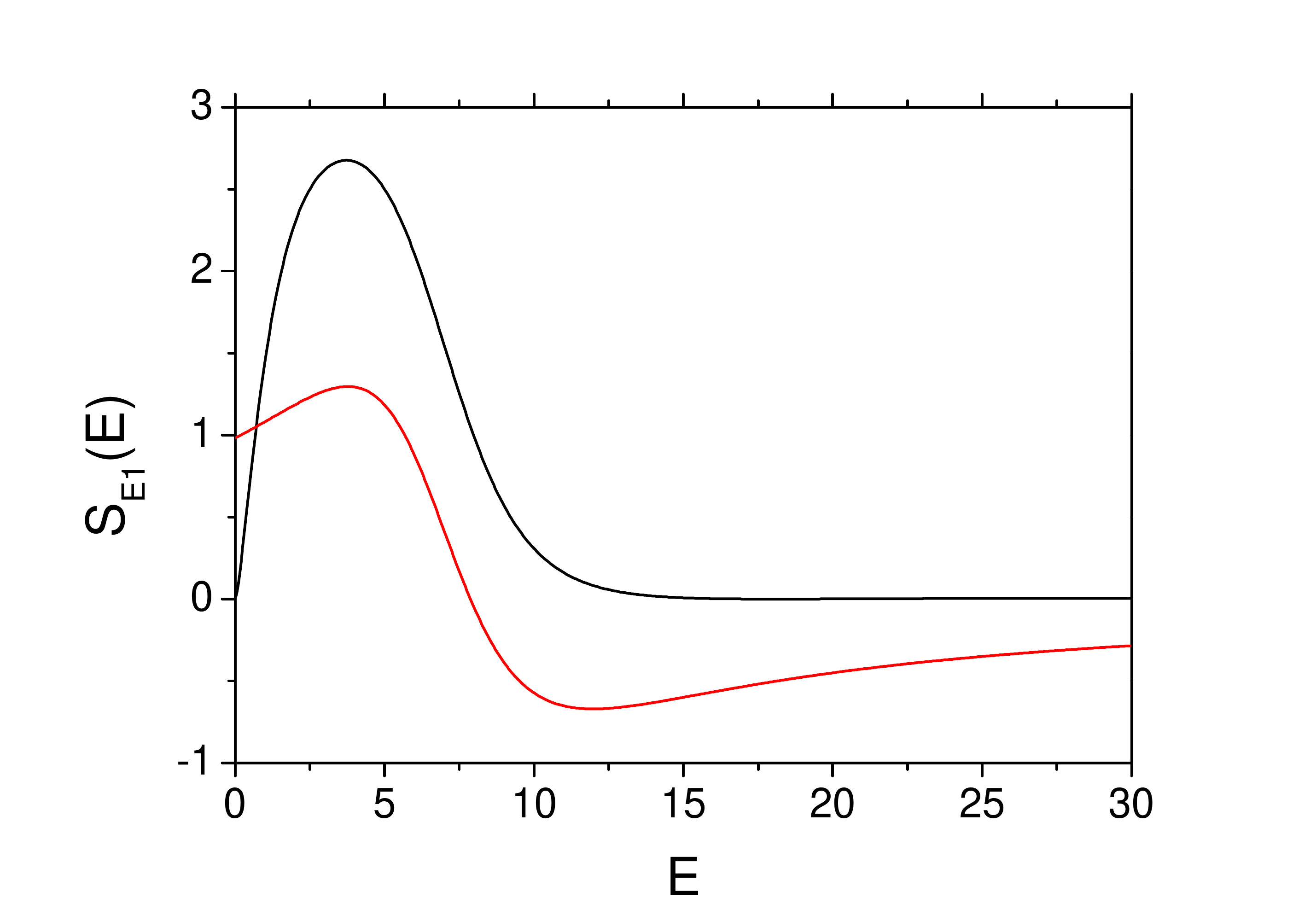}
\caption{Dipole-response functions (black line) with separated
contribution of the resonance obtained via complex scaling (red
line). In the left panel
results are presented for a potential containing a narrow resonance at $%
E^{res}=0.48-0.064i$, whereas right panel is for a potential
supporting resonance at $E^{res}=6.56-3.9i$.}
\label{fig_impact_res}
\end{figure}

Another instructive example is provided in
figure~\ref{fig_impact_res}, comparing contributions to E1
strength function by a narrow and broad resonances. One may see
that narrow resonance carries most of the strength. Contribution
of the broad resonances is comparable to the one of the continuum.
Furthermore, while the full strength function is a positive
quantity, the partial contributions of the resonant or continuum
states may contain regions in energy with negative contribution to
the total strength function. However once all the partial
contributions are summed positive value of the total strength
function should be recovered.

Relation~(\ref{Strenght_func_gf}) offers an unique feature to
separate the contributions of the resonant and bound states in the
strength function, providing clear physical interpretation of the
various components in the strength function.


\section{Complex-scaling method for the collisions}

In the previous sections I have demonstrated how efficient CS
method could be in handling problems dominated by the outgoing
wave functions. Particle collisions turns to be slightly more
complicated case, since their wave functions contain the incoming
waves $\Psi _{a}^{in}$, associated with the initial
projectile-target states, and highly untrivial outgoing waves
$\Psi _{a^{\prime }}^{out}$ representing various possible reaction
channels:
\begin{equation}
\Psi ^{scatt}(k_{a})_{\rho \rightarrow \infty }=\Psi
_{a}^{in}(k_{a})+\sum\limits_{a^{\prime }}f_{a^{\prime }a}(k_{a})\Psi
_{a^{\prime }}^{out}(k_{a})
\end{equation}
Nevertheless, once again, the problem might be reformulated in a
way suitable for CS method, as have been demonstrated for the
first time by J.~Nuttal and H.I.~Cohen~\cite{NC69}.

\subsection{Scattering, two-body problem}

\subsubsection{Short range, exponentially bound, interactions}

The idea of J.~Nuttal and H.I.~Cohen~\cite{NC69} can  be briefly
formulated as follows. The Schr\"{o}dinger equation is recast into
its inhomogeneous (driven) form by splitting the wave function
into the sum $\Psi (r)=$ $\Psi
^{out}(r)+\Psi ^{in}(r)$, where an incident (free) $\Psi ^{in}(\mathbf{r}%
)=\exp (i\mathbf{k}\cdot \mathbf{r})$ wave is separated. A
remaining untrivial part of the systems wave function $\Psi
^{out}(\mathbf{r})$ describes scattered waves and may be found by
solving a second-order differential equation with an inhomogeneous
term:
\begin{equation}
\lbrack E-\hat{H_{0}}-V(\mathbf{r})]\Psi ^{out}(\mathbf{r})=V(\mathbf{r}%
)\Psi ^{in}(\mathbf{r}).  \label{Schro_2B}
\end{equation}%
The scattered wave is represented in the asymptote  by an outgoing
wave $\Psi ^{out}\sim \exp (ikr)/r,$ where $k=\sqrt{2\mu E}/\hbar
$ is the wave number for the relative motion. If one scales all
the particle coordinates by a
constant complex factor, i.e. $r_{i}^{\theta }=e^{i\theta }r_{i}$ with $%
Im(e^{i\theta })>0$, the corresponding scattered wave $\overline{\Psi }%
^{out,\theta }(\mathbf{r})$ will vanish exponentially $\sim \exp (-kr\sin
\theta )$ as particle separation $r$ increases. Moreover if the interaction
is of short range -- exponentially bound with the longest range $\eta ^{-1}$
-- then after complex scaling the right hand side of eq.~(\ref{Schro_2B})
also tends to zero at large $r$, if :
\begin{equation}
\tan \theta <\eta /k.  \label{Cond_shrp}
\end{equation}%
From here we introduce the notation $f^{\theta }(r)=f(re^{i\theta })$ for
the complex-scaled functions. The complex scaled driven Schr\"{o}dinger
equation becomes:
\begin{equation}
\lbrack E-e^{-i2\theta }\hat{H_{0}}-V^{\theta }(\mathbf{r})]\overline{\Psi }%
^{out,\theta }(\mathbf{r})=V^{\theta }(\mathbf{r})\overline{\Psi }%
^{in,\theta }(\mathbf{r}).  \label{Schro_2B_cs}
\end{equation}%
If the condition in eq.~(\ref{Cond_shrp}) is satisfied, the former
inhomogeneous equation may be solved by using a compact basis to expand $%
\overline{\Psi }^{out,\theta }(\mathbf{r})$, thus by employing
standard bound-state techniques:
\begin{equation}
\overline{\Psi }^{out,\theta
}(\mathbf{r})=\sum_{i=1}^{N}c_{i}^\theta f_{i}(r),
\end{equation}
with $c_{i}^\theta$ denoting complex expansion coefficients, while
$f_{i}(r)$ is a function from the conveniently chosen compact
basis. After projecting equation on the basis states $f_{i}(r)$,
as previously,  one gets linear algebra problem to be solved:
\begin{equation}
\left( \left[ E\right] -\left[ \hat{H^{\theta }}\right] \right)
c^\theta=v^{in,\theta } \label{eq:linal_sc},
\end{equation}%
here $\left[ E\right] ,\left[ \hat{H^{\theta }}\right] $ are the
same square matrices
representing projection of norm matrix and Hamiltonian, whereas vector $%
v^{in,\theta }$ denotes projection of the inhomogenious term $V^{\theta }(%
\mathbf{r})\overline{\Psi }^{in,\theta }(\mathbf{r})$ on a chosen basis $%
f_{i}(r)$.

As discussed in a previous section, there are two mathematically
equivalent ways to solve the last set of linear equations in order
to obtain vector $c^\theta$, which contains
coefficients $c^\theta_{i}$ representing projection of the function $\overline{\Psi }%
^{out,\theta }(\mathbf{r})$:
\begin{itemize}
\item Solve linear-algebra problem, formulated in
eq.~(\ref{eq:linal_sc})
 \item Use spectral expansion of
the last equation into eigensolutions of matrix $\left[
\hat{H^{\theta }}\right]$.
In this case:%
\begin{eqnarray}
\overline{\Psi }^{out,\theta }(\mathbf{r}) &=&\sum_{i=1}^{N}a_{i}\widetilde{%
\phi }_{i}^{\theta }(r) \\
a_{i} &=&\frac{\left( \widetilde{\phi }_{i}^{\theta }(r)\right.
\left\vert v^{in,\theta }\right) }{E-E_{i}^{\theta }}.
\end{eqnarray}
There are no need to repeat the arguments of the previous section
reflecting the advantages of two different methods. It worths only
mentioning that spectral expansion formalism allows to use the
same dataset of the eigensolutions to obtain results on the bound,
resonant states as well as particle collisions or reactions due to
the external probes.

From the obtained CS representation of the scattered wave function
$\overline{\Psi }^{out,\theta }(\mathbf{r})$ there are three ways
to extract scattering observables.

\item The most straightforward way is based on the analysis of the
asymptotic behavior of the outgoing waves. In this case the
scattering amplitude $f_{k}(\hat{r})$ is extracted in a similar
way as the asymptotic normalization coefficient from the
bound-state wave function, that is, by matching asymptotic
behavior of the solution:
\begin{equation}
\overline{\Psi }^{out,\theta }(\mathbf{r\rightarrow \infty })=f_{k}(\widehat{%
k})e^{-i\theta }\exp (ikre^{i\theta })/r.
\end{equation}

\item Another well known alternative is to use the integral
relations, which one gets after applying the Green's
theorem~\cite{KSK07,LC11,HK12}. For a simple case of two-particle
scattering this gives: \begin{small}
\begin{eqnarray}
f_{k}(\widehat{k}) &=&-\frac{1 }{E_{cm}}e^{i3\theta }\int (\Psi
^{in\ast })^{\theta }(\mathbf{r})V^{\theta }(\mathbf{r})\left[ \overline{%
\Psi }^{out,\theta }(\mathbf{r})+\overline{\Psi }^{in,\theta }(\mathbf{r})%
\right] d^{3}r \\
&=&-\frac{1 }{E_{cm}}e^{i3\theta }\int (\Psi ^{in\ast })^{\theta }(%
\mathbf{r}){V^{\theta }}(\mathbf{r})\overline{\Psi }^{out,\theta }(\mathbf{r}%
)d^{3}r-\frac{1 }{E_{cm}}\int (\Psi ^{in}(\mathbf{r}))^{\ast }V(%
\mathbf{r}){\Psi }^{in}(\mathbf{r})d^{3}r. \label{integral_2B}
\end{eqnarray}
\end{small}%
Where $E_{cm}=\frac{\hbar^2 k^2}{2\mu}$ is center-of-mass energy
of the colliding particles. In the second relation one has
separated the Born term, which may be evaluated without performing
complex scaling. The $(\Psi ^{in\ast })^{\theta }(\mathbf{r})$
term is obtained by applying complex-scaling operation on the
bi-conjugate function $(\Psi ^{in}(\mathbf{r}))^{\ast }$. The
radial part of the former function  coincides with  one of the
$(\Psi ^{in})^{\theta }(\mathbf{r})$, whereas complex-conjugation
is applied only on angular functions (spherical harmonics).
Therefore manipulations involving bi-conjugate functions is
straightforward.

If the spectral expansion is used, the scattering amplitude
$f_{k}(\widehat{k})$ is obtained as a sum of the separate
contributions: a Born term, contributions from bound, resonant and
discretized continuum states obtained as eigensolutions of $\left[
\hat{H^{\theta }}\right]$.

\item Finally, the scattering phaseshifts may be extracted using
continuum level density (CLD) formalism. One starts with the CLD
definition:
\begin{equation}
\Delta (E)=-\frac{1}{\pi }\text{Im}\left( \text{Tr}[G(E)-G_{0}(E]\right) ,
\label{eq:cld_0}
\end{equation}%
where $G(E)=(E-H)^{-1}$ and $G_{0}(E)=(E-H_{0})^{-1}$ denote full
and free Green's functions, respectively. In principle, the former
expression may be generalized to the scattering of two composit
clusters. Then $H_{0}$, besides the kinetic energy, should include
interactions inside separate clusters, whereas $H$ includes all
the interaction terms in two-cluster system. Thus CLD express the
effect from the interactions connecting two clusters. When the
eigenvalues of $H$ and $H_{0}$ are obtained approximately
($\epsilon _{i} $ and $\epsilon _{i}^{0}$ respectively) within the
framework including finite number of the basis functions (N), the
discrete CLD is defined:
\begin{equation}
\Delta (E)_{N}=\sum_{i}\delta (E-\epsilon _{i})-\sum_{j}\delta (E-\epsilon
_{j}^{0}).  \label{eq:cld_1}
\end{equation}%
The CLD is related to the scattering phaseshift as:
\begin{equation}
\Delta (E)=\frac{1}{\pi }\frac{d\delta (E)}{dE}  \label{eq:cld_2}
\end{equation}%
and thus one can inversely calculate the phaseshift~($\delta $) by
integrating the last equation obtained as a function of energy. These
equations are difficult to apply for real Hamiltonians, as one will
necessarily confront the singularities present in eqs.~(\ref{eq:cld_0}-\ref%
{eq:cld_1}). However by using CS expressions for the Green's
functions, these singularities are avoided and replaced by the
smooth Lorentzian functions. By plugging in CS Green's function
expression~(\ref{eq:Greens_func}) into eq.~(\ref{eq:cld_1}) and
after some simple algebra one gets:
\begin{equation}
\Delta (E)_{N}=\overline{\rho }_{N}^{\theta }(E)-\overline{\rho }%
_{0,N}^{\theta }(E)  \label{cld_3}
\end{equation}%
and
\begin{eqnarray}
\overline{\rho }_{N}^{\theta }(E) &=&\sum_{b}^{n_{b}}\delta (E-E_{b})+\frac{1%
}{\pi }\sum_{r}^{n_{r}^{\theta }}\frac{Im(E_{r})}{\left[ E-Re%
(E_{r})\right] ^{2}+\left[ Im(E_{r})\right] ^{2}}\notag\\
&+&\frac{1}{\pi }%
\sum_{k}^{N-n_{r}^{\theta }-n_{b}}\frac{Im(E_{k}^{\theta })}{\left[ E-%
Re(E_{k}^{\theta })\right] ^{2}+\left[ Im(E_{k}^{\theta })%
\right] ^{2}},  \label{eq:cld_rhop} \\
\overline{\rho }_{0,N}^{\theta }(E) &=&\frac{1}{\pi
}\sum_{k}^{N}\frac{
Im(E_{0,k}^{\theta })}{\left[ E-Re(E_{0,k}^{\theta })\right] ^{2}+%
\left[ Im(E_{0,k}^{\theta })\right] ^{2}}.
\end{eqnarray}%
In the last expression $E_{b},E_{r}$ and $E_{k}^{\theta }$ are the
eigenvalues of the full CS Hamiltonian $H^{\theta }$, representing
bound,
resonant and continuum states respectively. The term $\overline{\rho }%
_{0,N}^{\theta }$ is equivalent to $\overline{\rho }_{N}^{\theta
}(E)$ only obtained for a free CS Hamiltonian $H_{0}^{\theta }$;
this term contains
only pseudo-continuum states $(E_{0,k}^{\theta })$ aligned along $2\theta $%
-lines pointing out from the scattering thresholds (see figure~\ref%
{Fig_eigen_csm}).

By plugging the last two relations into eq.~(\ref{eq:cld_2}) and
integrating
it over the energy it is easy to get an expression for the phaseshifts:%
\begin{equation}
\delta (E)=n_{b}\pi +\delta _{r}(E)+\delta _{k}(E)
\label{eq:cld_b0},
\end{equation}

with
\begin{eqnarray}
\delta _{r}(E) &=&\sum_{r}^{n_{r}^{\theta }}\arctan \left[ \frac{Re%
(E_{r})-E}{Im(E_{r})}\right] ,\label{eq:cld_b1} \\
\delta _{k}(E) &=&\sum_{k}^{N-n_{r}^{\theta }-n_{b}}\arctan \left[ \frac{%
Re(E_{k}^{\theta })-E}{Im(E_{k}^{\theta })}\right]
-\sum_{k}^{N}\arctan \left[ \frac{Re(E_{0,k}^{\theta })-E}{Im%
(E_{0,k}^{\theta })}\right] \label{eq:cld_b2}
\end{eqnarray}

One may see that in these expression total phaseshift is obtained
as a sum from the separated contributions of  bound $n_{b}\pi $,
resonant $\delta _{r}(E)$ and  continuum $\delta _{k}(E)$ states.
According to the Levinson theorem bound states simply contribute
in providing shift of the phase at the origin by $n_{b}\pi .$
Contribution of each resonant state to the phaseshift might be
uniquely separated and they should not depend on the CS parameter
$\theta$ as long as calculations are numerically converged. If the
CS angle is able to "expose" all the resonant states one gets also
angle independent definition for the overall contribution of the
continuum states to the total phaseshift. If some resonances are
not exposed, their contribution to the phaseshift are compensated
by the appropriate change in the continuum contribution $\delta
_{k}(E)$~\cite{MYO20141}.
\end{itemize}

\subsubsection{Presence of a long-range interaction}

Let us consider a case where particle interaction apart
short-range part includes an
additional long-range term $V(\mathbf{r})=V_{s}(\mathbf{r})+V_{l}(\mathbf{r})$, where $V_{s}(%
\mathbf{r})$ is exponentially bound, whereas $V_{l}(\mathbf{r})$
is long-ranged. CS method can be generalized to treat this problem
if for the long-range term $V_{l}(\mathbf{r})$ the incoming wave
solution $\Psi _{l}^{in}(\mathbf{r})$ is analytic and can be
extended in to the complex r-plane~\cite{KSK07,Elander_CSM,LC11}.
Then one is left to solve the equivalent driven Schr%
\"{o}dinger equation:
\begin{equation}
\lbrack E-e^{-i2\theta }\widehat{H_{0}}-V^\theta(\mathbf{r})]\overline{\Psi }%
_{s}^{sc}(\mathbf{r})=V^\theta_{s}(\mathbf{r})(\Psi_{l}^{in})^\theta(\mathbf{%
r}).  \label{Schro_2B_lr}
\end{equation}%
The inhomogeneous term on the right hand side of the former
equation is moderated by the short-range interaction term,
therefore it is exponentially bound if the condition
eq.(\ref{Cond_shrp}) is fulfilled by the short range potential
$V_{s}(\mathbf{r})$. Perfect example is related with a presence
of the Coulomb interaction $V_l(%
\mathbf{r})=\frac{\hbar ^{2}\eta }{\mu r}$ . For this case the
incoming wave solution is well known and is usually expressed by the regular Coulomb functions $(\Psi_{l}^{in})^\theta(\mathbf{%
r})\equiv F_{l}(\eta,kr e^{i\theta})$.

One may establish a relation equivalent to the
eq.(\ref{integral_2B}) in order to determine the
long-range-modified short-range interaction amplitude
$f_{k,s}(\widehat{k})$ :
\begin{equation}
 f_{k,s}(\widehat{k})
=-\frac{1}{E_{cm}}e^{i3\theta }\int (\Psi
_{l}^{in\ast })^{\theta }(\mathbf{r})V_{s}^{\theta }(\mathbf{r})\overline{%
\Psi }_{s}^{sc}(\mathbf{r})d^{3}r -\frac{1}{E_{cm}}\int (\Psi _{l}^{in}(\mathbf{r}))^{\ast }V_{s}(%
\mathbf{r}){\Psi }_{l}^{in}(\mathbf{r})d^{3}r. \label{integraal}
\end{equation}

The total scattering amplitude $f_{k}(\widehat{k})$ is a sum of a
short-range one and the scattering amplitude due to the long-range
term alone $f_{k,l}(\widehat{k})$, known analytically:
\begin{equation}
f_{k}(\widehat{k})=f_{k,s}(\widehat{k})+f_{k,l}(\widehat{k}).
\end{equation}%
%
%
%
%
%
%


\subsubsection{\protect\bigskip Short-range, exponentially non-bound,
interactions}

It is natural to pose a question about application of the CS
method to describe scattering governed by short range
interactions, decaying faster than 1/r$^{3},$ but which are not
exponentially bound. From the formal point of view CS method, as
described in two previous subsections, is not applicable for this
case. On the other hand one may imagine solving a problem for a
modified potential
\begin{equation}
\widetilde{V}(r)=f(r)V(r),
\end{equation}%
where $f(r)$ is some analytic function, which is very close to $1$
in the space region where the potential energy is important
compared to the kinetic energy term, while this function makes
$\widetilde{V}(r)$ vanish exponentially in the far asymptote.
Depending on the choice of the function
$f(r)$ one may rend scattering observables provided by the potential $%
\widetilde{V}(r)$ very close to ones obtained by the original potential $%
V(r)$. On the other hand one has no formal obstacles to apply CS
method in solving scattering problem related to the potential
$\widetilde{V}(r)$. Such a phenomenon has been already considered
by J.~Nutall~\cite{PhysRevA.12.486}. One may see that if basis of
exponentially bound functions is used to solve
eq.~(\ref{Schro_2B}) or eq.~(\ref{Schro_2B_lr}), in this case
basis by itself partly fulfills function of the regulator $f(r)$.
Furthermore they have demonstrated that calculated scattering
phases spiral around the exact value once one increases the basis
size; it may approach very close to the exact value but when the
basis is further increased the calculated phases start to recede
from the exact ones continuing the spiral movement.
In~\cite{PhysRevA.29.2933} it has been suggested to use Pad\'{e}
summation technique to gain accuracy from the approximately
calculated phases which spiral around the exact value. For set of
2-body potentials they have demonstrated convergence of the
Pad\'{e} series and thus possibility to get very accurate
evaluation of the scattering phaseshifts.

\section{Example of the solution on a finite grid\label{sec:2b_sc_splines}}

To test the applicability of our approach we consider a system of
two nucleons with a mass $\frac{\hbar ^{2}}{m}=41.47$
MeV.fm$^{2}$, where the  strong part of the nucleon-nucleon (NN)
interaction is described by the spin-dependent S-wave MT I-III
potential, formulated in~\cite{MT13} and parameterized
in~\cite{MT13p}:
\begin{equation}
V_{S}(r)=-A_{S} {\exp (-1.55r)\over r} +1438.72{\exp (-3.11r)\over
r},
\end{equation}%
where $V_{S}(r)$ is in MeV and $r$ is in fm units. The attractive
Yukawa strength is given by $A_{s=0}=-513.968$ MeV.fm and
$A_{s=1}=-626.885$ MeV.fm for the two-nucleon interaction in spin
singlet and triplet states respectively.

MT I-III potential has been chosen for two reasons. On one hand it
is a widely employed potential for which accurate benchmark
calculations exist. On the other hand this potential, being a
combination of the attractive and repulsive Yukawa terms, reflects
well the structure of the realistic nucleon-nucleon interaction:
it is strongly repulsive at the origin but posses a narrow
attractive well situated at $r\approx 1$ fm. Note that many
numerical techniques fail to treat potentials like MT I-III, which
include a repulsive core.

\begin{figure}[tbp]
\centering
\hspace{-0.5cm}\includegraphics[width=85mm]{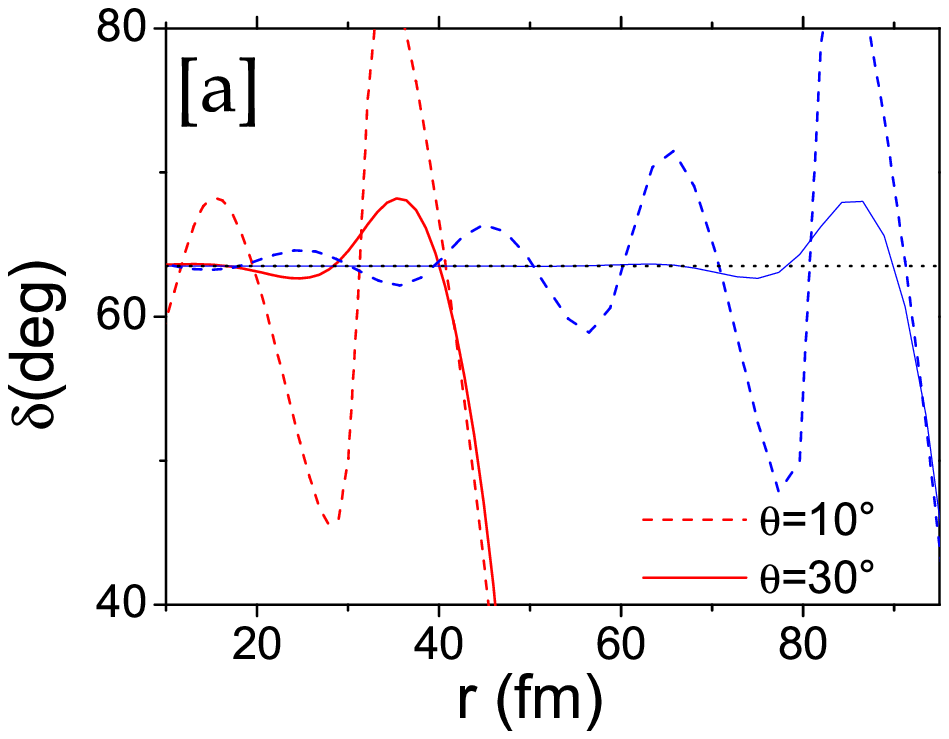}\hspace{-0.5cm}
\includegraphics[width=85mm]{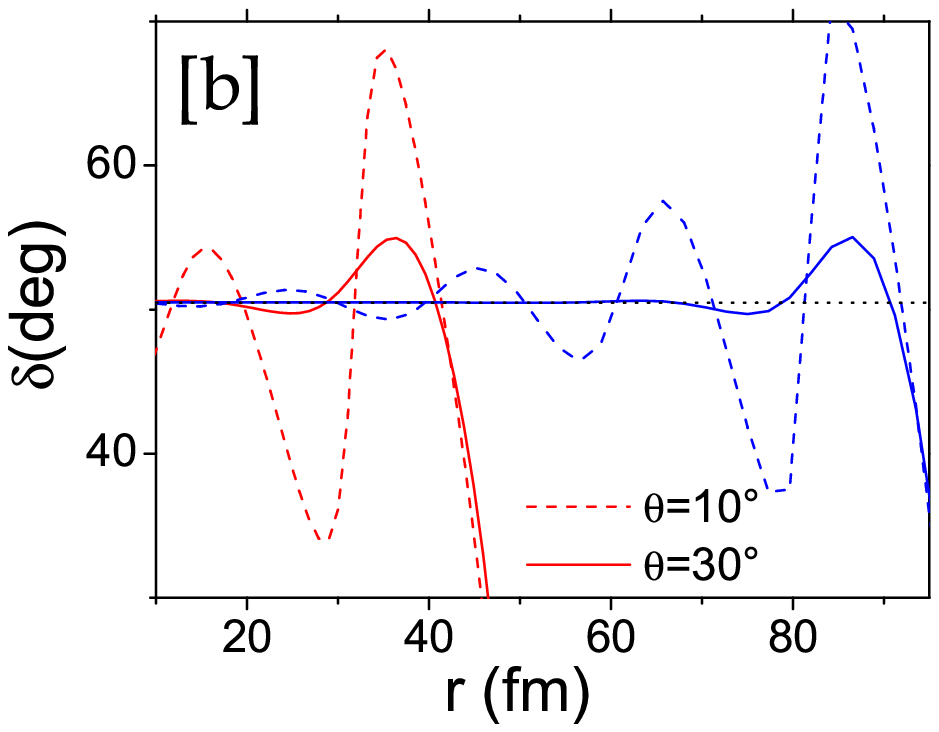}
\caption{$^1$S$_0$ NN phaseshifts at $E_{cm}$=1 MeV  and extracted
locally by calculating logarithmic derivatives of the wave
function. Calculations were performed with cut-off imposed at
$r_{max}$=50 (in red, curves diverging close to 50 fm) and 100 fm
(in blue, curves diverging close to 100 fm) using a complex
rotation angle $\protect\theta$ =10$^\circ$ (dashed lines) and $%
\protect\theta =$30$^\circ$ (solid line). The pure strong
interaction result is presented in the left figure (a) and
calculations including repulsive Coulomb interaction for pp-pair
are presented in the right figure (b). They are compared to the
exact results indicated by a dotted horizontal line.}
\label{fig:rr_dep}
\end{figure}

We have first considered a two-body case. In
figure~\ref{fig:rr_dep} we present our results for the NN
$^{1}$S$_{0} $  phaseshifts at $E_{cm}=$1 MeV. Two calculation
sequences have been performed by forcing  $\overline{\psi
}_{l}^{sc}$ to vanish at the border of the numerical grid set at
$r_{max}=50$ fm (in red)  and $r_{max}=100$ fm (in blue)
respectively, whereas the complex scaling angle $\theta$ has been
chosen to be 10$^\circ$ (dashed lines) and 30$^\circ$ (solid
lines). The phaseshifts are extracted by calculating logarithmic
derivative of the wave function at a given distance and adjusting
it to proper asymptotic behavior, including complex scaled Bessel
or Coulomb functions. As one can see, the extracted phaseshifts
oscillate with $r$. This oscillatory behavior is due to the
premature enforcement of $\overline{\psi }_{l}^{sc}(r)$ to vanish
at the border of the grid $r_{\max }$. The phaseshifts extracted
close to $r_{\max }$ are strongly affected by the cut-off and are
thus not reliable. The amplitude of the close-border oscillations
is sizeably reduced by either increasing $r_{\max }$ or $\theta $,
i.e. by reducing the sharpness of the numerical cut-off. The
extracted phaseshifts corresponding  to the calculation with
$r_{\max }=100$ fm and $\theta =$30$^\circ$ are stable in a rather
large window, which starts at $r\sim 5$ fm (right outside the
interaction region) and extends up to $r\sim 70$ fm. Beyond this
value the effect due to cut-off sets in. In the stability region
the extracted phaseshifts agree well with the ''exact'' results
(dotted line), obtained by solving scattering problem using the
standard (i.e. not complex rotated) boundary condition technique.

\begin{figure}[h!]
\centering
\hspace{-0.5cm}\includegraphics[width=85mm]{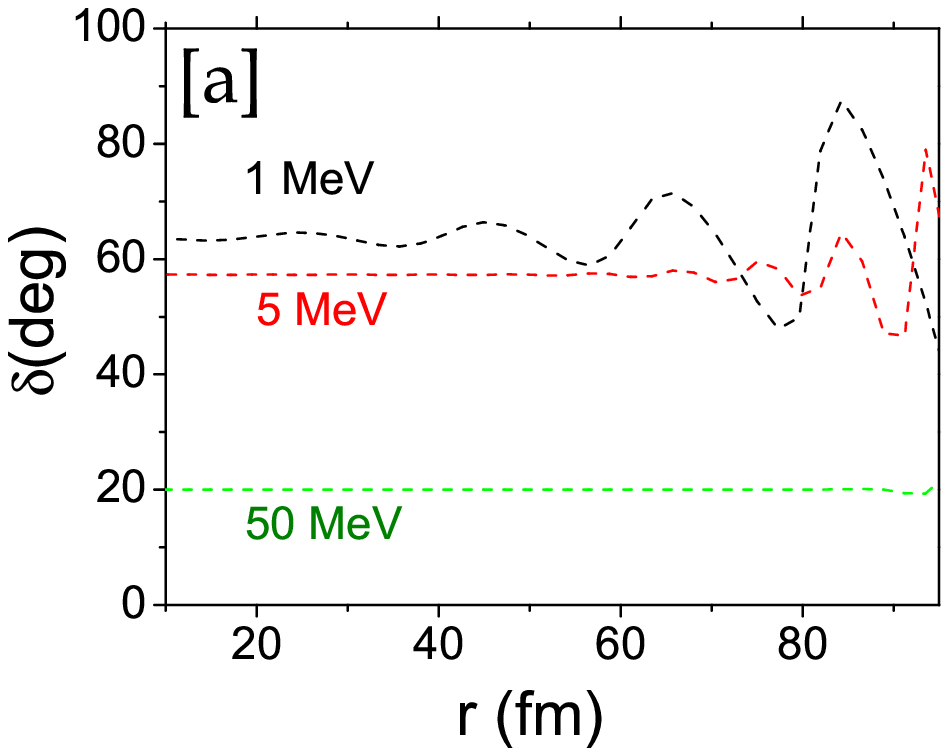}\hspace{-0.5cm}
\includegraphics[width=85mm]{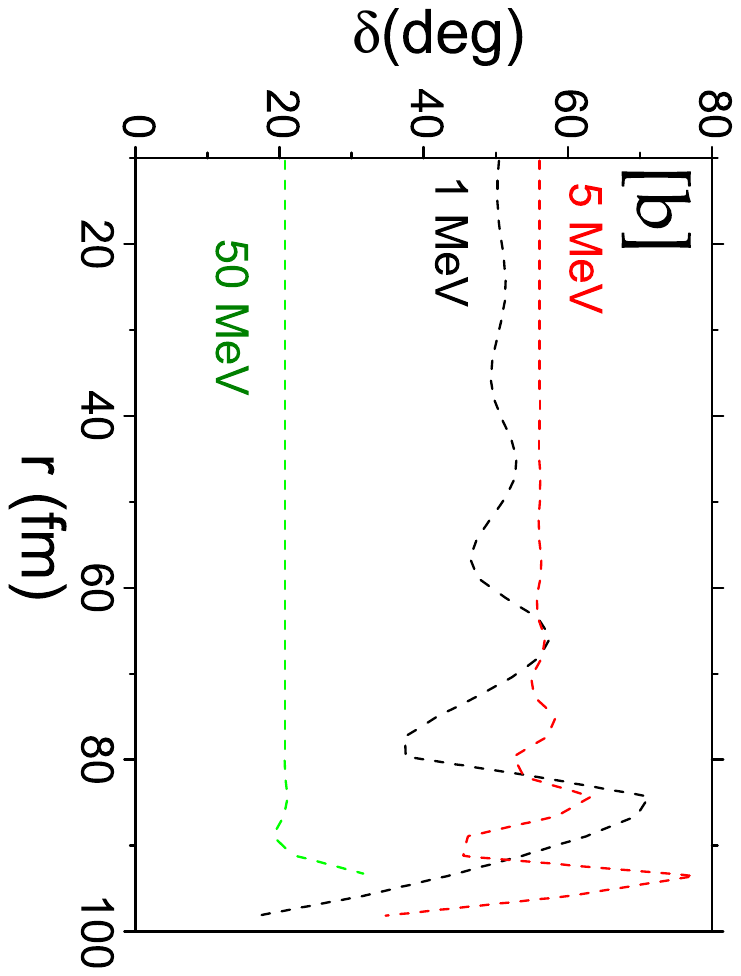}
\caption{ $^1$S$_0$ NN phaseshifts calculation at $E_{cm}$=1, 5
and 50 MeV. Calculations were performed with a cut-off imposed at
$r_{max}$= 100 fm using the complex rotation angle
$\protect\theta$ =10$^\circ$. The pure strong interaction result
are presented in the left figure (a), and those including
repulsive Coulomb interaction for pp-pair are presented in the
right figure (b).} \label{fig:r_dep_en}
\end{figure}

In figure~\ref{fig:r_dep_en} we have compared the  NN
$^{1}$S$_{0}$ phaseshifts at different energies --   $E_{cm}$=$1$,
$5$ and $50$ MeV -- by fixing $r_{\max }=100$ fm and $\theta$
=10$^\circ$.
 One can see that when increasing the energy, the  effect of the cut-off reduces, sizeably improving the stability of the extracted
phaseshifts. The inclusion of the repulsive Coulomb term does not
have any effect on the quality of the method.

\begin{figure}[h!]
\centering
\includegraphics[width=14cm]{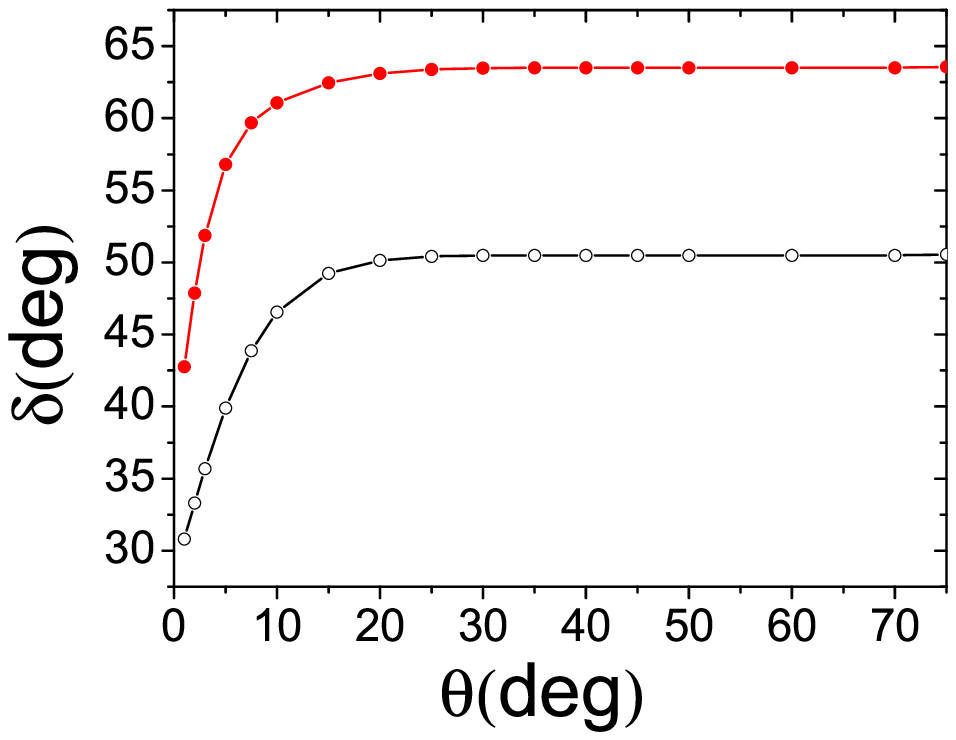}\hspace{1.cm}
\caption{Dependence of the calculated NN  $^1$S$_0$ phaseshift
using integral expression as a function of the complex rotation
angle. The grid was limited to r$_{max}$=100 fm. The upper curve
corresponds to Coulomb-free case and the bottom one includes
Coulomb.}\label{fig:integral_2B}
\end{figure}

One may improve considerably the accuracy of the phaseshifts by
using the integral relation given in eq.~(\ref{integral_2B}). The
results are displayed  in tables~\ref{tab_2B_1M}, \ref{tab_2B_50M}
and in figure~\ref{fig:integral_2B}. The phaseshifts converge to a
constant value by either  increasing the cut-off radius r$_{\max
}$ or the complex rotation angle. A spectacular accuracy of five
digits is easily reached. One should notice however that the use
of very large values of $\theta$ should be avoided, due to the
fact that the function $\overline{\psi }_{l}^{sc}(r)$ as well as
the complex scaled potential $V(re^{i\theta})$ might become very
steep and rapidly oscillating, see the discussion in the next
section. At higher energy, the function $\overline{\psi
}_{l}^{sc}(r)$ vanishes faster and thus one may easily achieve
convergence by employing smaller values of r$_{\max }$ and/or
$\theta$.

\begin{table}[tbp]
\begin{center}
\caption{Calculation of the scattering phaseshifts using integral
expressions at $E_{cm}=$1 MeV }
\label{tab_2B_1M}%
\vspace{0.5cm}
\begin{tabular}{|c|cccc|cccc|}\hline
& \multicolumn{4}{c|}{MT I-III} & \multicolumn{4}{c|}{MT
I-III+Coulomb}\\$r_{max}$ (fm)& 5$^\circ$ & 10$^\circ$ &
30$^\circ$ & 50$^\circ$ & 5$^\circ$ & 10$^\circ$ & 30$^\circ$ &
50$^\circ$ \\\hline\hline
 10 & 44.420 & 49.486 & 55.790 & 56.676
& 33.999 & 36.390 & 41.528 & 43.805
\\
25 & 34.704 & 44.211 & 62.654 & 63.743 & 24.772 & 34.910 & 50.693
& 50.698
\\
50 & 56.812 & 61.083 & 63.482 & 63.512 & 39.895 & 46.546 & 50.487
& 50.491
\\
100 & 66.502 & 63.822 & 63.512 & 63.512 & 55.463 & 50.811 & 50.491
& 50.491
\\
150 & 62.497 & 63.485 & 63.512 & 63.512 & 49.317 & 50.474 & 50.491
& 50.491
\\
\hline exact & \multicolumn{4}{c|}{63.512} &
\multicolumn{4}{c|}{50.491}\\\hline
\end{tabular}
\end{center}
\end{table}

\begin{table}[tbp]
\begin{center}
\caption{Calculation of the scattering phaseshifts using integral
expressions at $E_{cm}=$50 MeV }
\label{tab_2B_50M}%
\vspace{0.5cm}
\begin{tabular}{|c|cccc|cccc|}\hline
$r_{max}$ (fm)& \multicolumn{4}{c|}{MT I-III} &
\multicolumn{4}{c|}{MT I-III+Coulomb}\\

& 3$^\circ$ & 5$^\circ$ & 10$^\circ$ & 30$^\circ$ & 3$^\circ$ &
5$^\circ$ & 10$^\circ$ & 30$^\circ$ \\\hline\hline 10 & 19.400 &
19.719 & 19.923 & 19.605 & 19.795 & 20.245 & 20.610 & 20.313
\\
25 & 20.788 & 20.135 & 20.027 & 20.032 & 21.530 & 20.864 & 20.755
& 20.760
\\
50 & 20.014 & 20.026 & 20.027 & 20.027 & 20.734 & 20.754 & 20.755
& 20.755
\\
100 & 20.027 & 20.027 & 20.027 & 20.027 & 20.755 & 20.755 & 20.755
& 20.755
\\\hline
exact & \multicolumn{4}{c|}{20.027} &
\multicolumn{4}{c|}{20.755}\\ \hline
\end{tabular}
\end{center}
\end{table}

\subsection{General remarks about the complex scaling method\label{SA_GF}}

\subsubsection{Spectral decomposition vs solution of the linear equation}

As demonstrated in ref.~\cite{KSK07}, and briefly discussed in the
section \ref{CS_GF}, there are two approaches to solve linear
algebra problems, arising from the solution of a system of
differential equations with an inhomogenious term, as generalized
in eq.~(\ref{eq:schr_drv_base}). They are: direct solution of the
linear algebra problem or the method based on the spectral
expansion of the linear algebra matrix. These two methods are
fully equivalent, if accurately solved they provide results which
coincide up to numerical round-off error.

It should be noted that a full spectral decomposition of the $%
H^{\theta }$ is required to express CS Green's function in
eq.~(\ref{eq:Greens_func}) and to evaluate the scattering
amplitudes. The scattering amplitude, except in the case of
resonant scattering, is not determined by one or a few dominant
eigenvalues\footnote{One should notice however, if one tries to
approximate the phaseshifts using only few eigenvalues, which are
closest to the scattering energy, then the
CLD formalism may provide better convergence than the relations~(\ref{eq:cld_2}-%
\ref{eq:cld_rhop}).}. This may turn out to be a crucial obstacle
in applying CS Green's function method in studying many-body
systems, since the resulting algebraic eigenvalue problem becomes
too large to be fully diagonalized. In this case the original
prescription of J. Nuttall \textit{et al.},  based on direct
solution of the linear algebra problem, turns to be strongly
advantageous. The last prescription requires solution of the
linear algebra problem eq.(\ref{eq:linal_sc}) at chosen energy
points, allowing one to solve a resulting large-scale problem by
iterative methods (requiring no explicit storage of the matrix
elements).

On the other hand CS Green's function formalism provides clear
physical interpretation of the scattering observables in terms of
 bound, resonant and continuum states. Furthermore, the same
input of eigenvalues and eigenvectors may be used to approximate
CS Green's function expression and then describe different
processes in a chosen N-body system: bound states, resonant
states, particle collisions or reactions induced by an external
perturbations. In such a way a solid framework may be constructed
to study correlations between the different physical observables.

\subsubsection{CLD versus integral relation}

\begin{figure}[th]
\centering
\includegraphics[width=81mm] {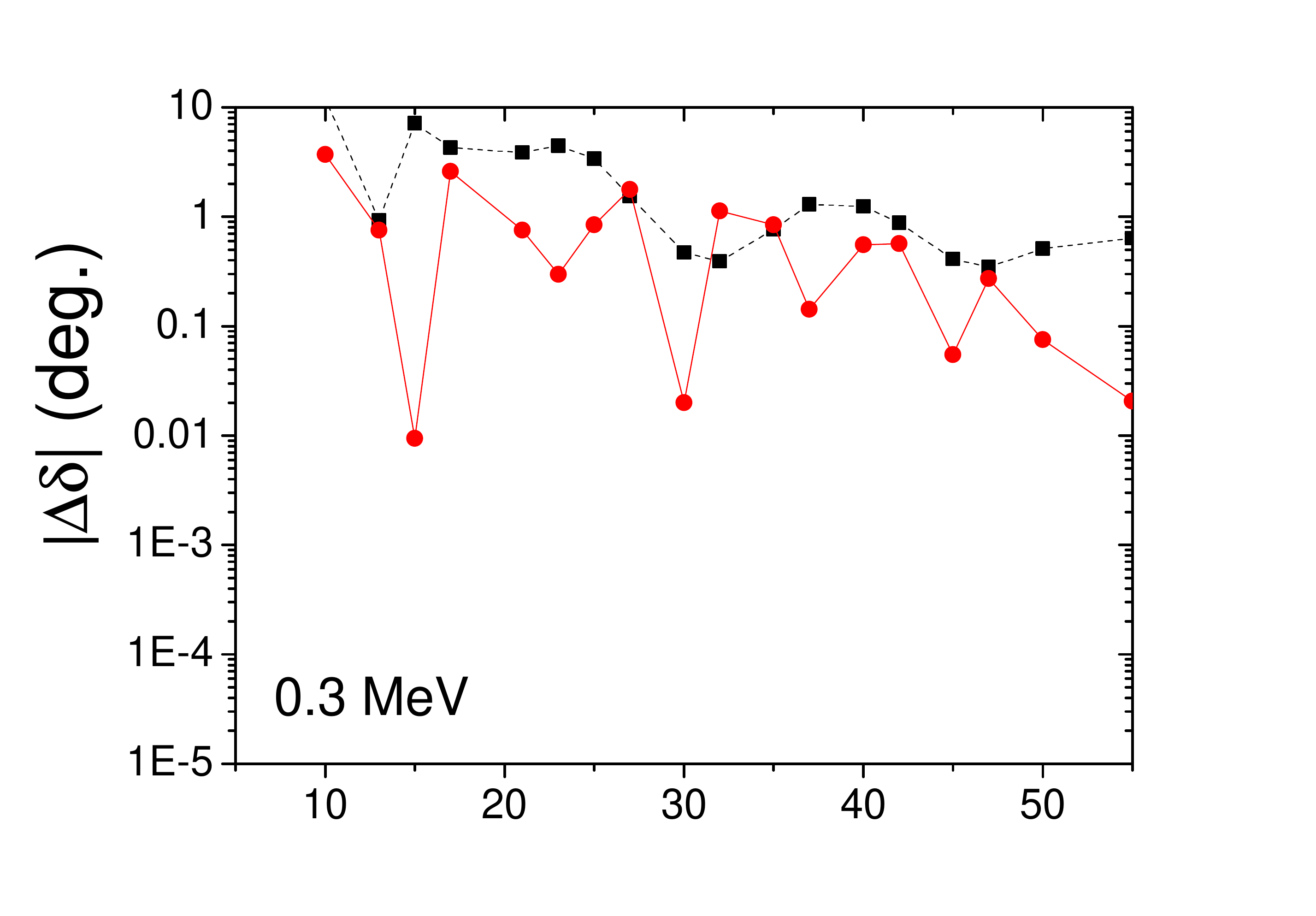} 
\includegraphics[width=81mm] {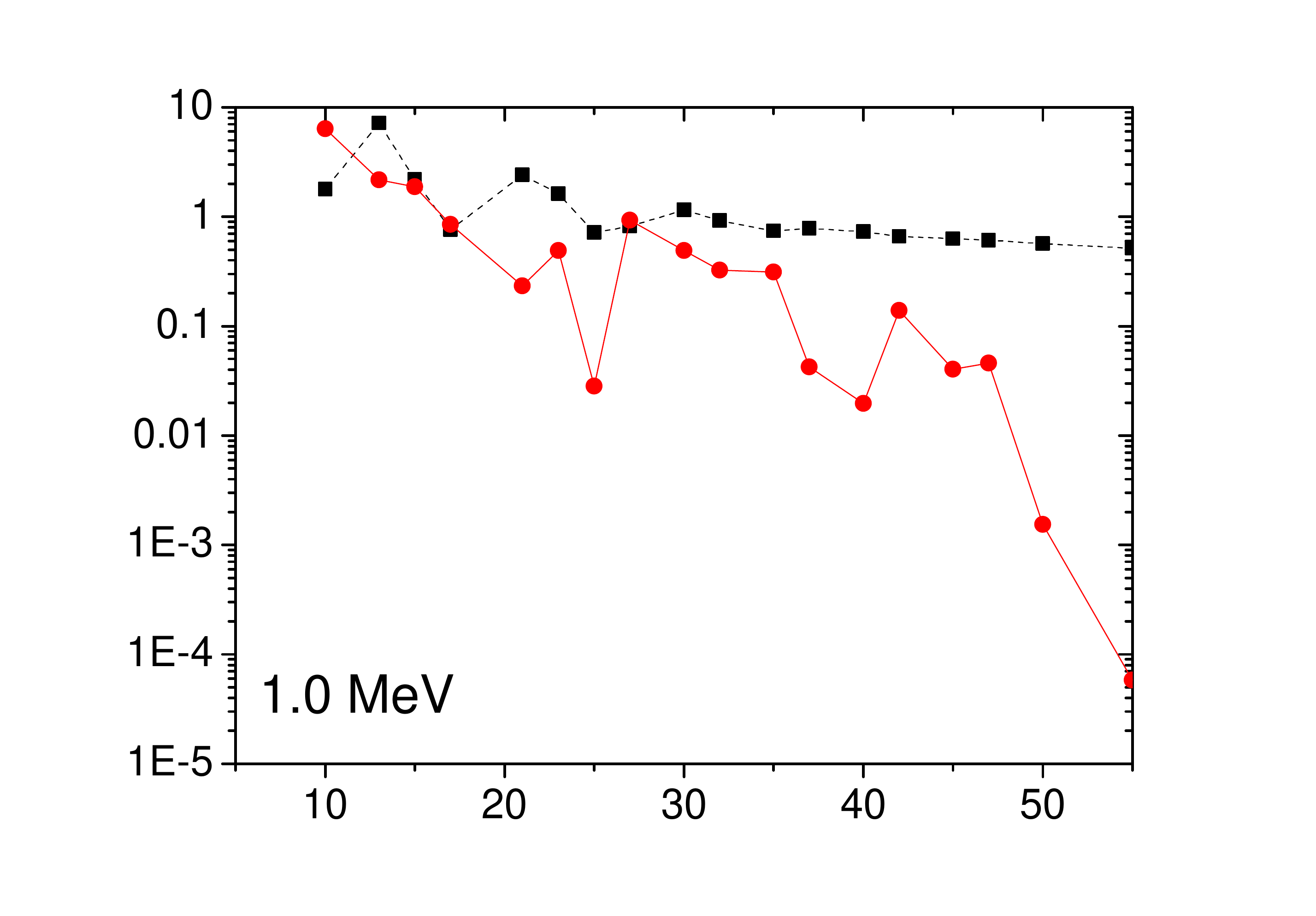} \includegraphics[width=81mm]
{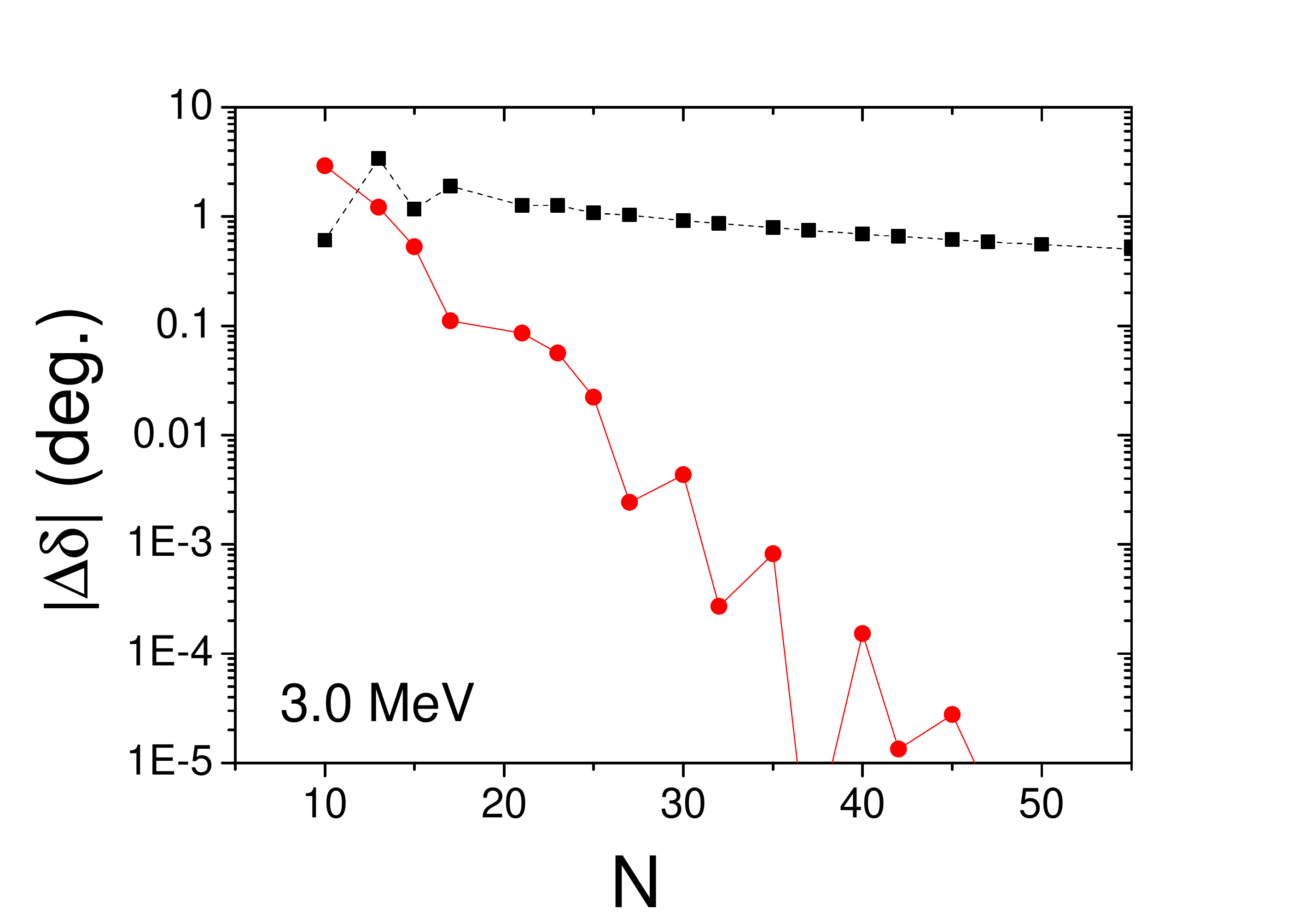} \includegraphics[width=81mm] {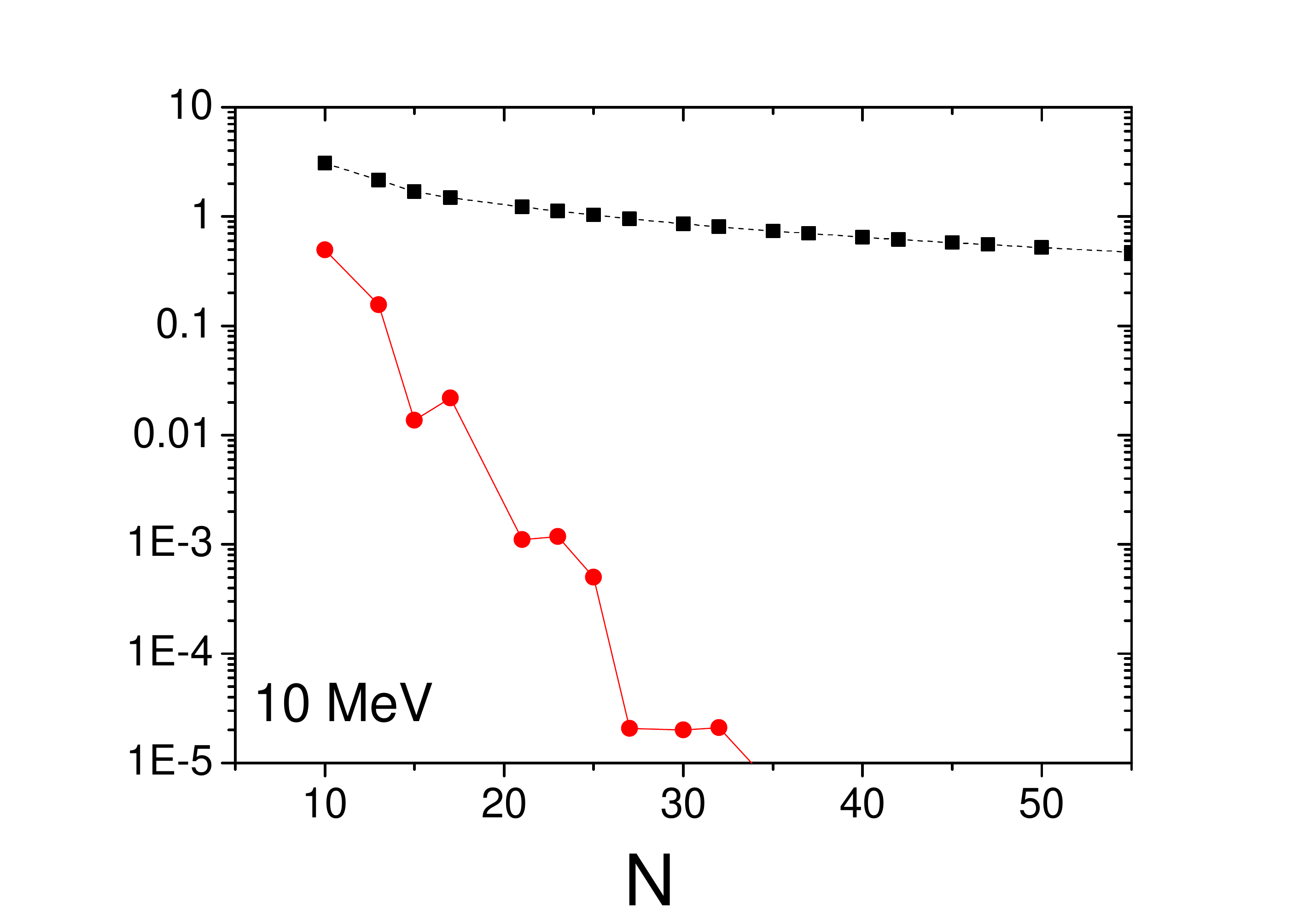}
\caption{Comparison of the absolute errors in extracting S-wave phaseshifts
by using CLD and the integral relation formalisms. Results for four
different energies are compared as a function of the Lagrange-Laguerre basis
functions used to perform the calculations. Complex scaling angle has been
set to $\protect\theta=15^\circ$.}
\label{fig_cld_vs_int}
\end{figure}

One disposes also three principally different methods to extract
the scattering phaseshifts: by analyzing the shape  of the wave
function's asymptote, employing integral relation or  relation
based on CLD expression. I would not discuss in detail  the first
method, success of it strongly depends on the  choice of the wave
function's region to extract phaseshifts and thus it strongly
relies on the skillfulness of a person performing calculations.
One may simply recall from the discussion in a previous section
that extraction region should be chosen beyond the range of the
interaction, where a free asymptote is reached. The very far
asymptote is neither suitable in numerical calculations, since the
CS wave function is very small and thus strongly affected by the
numerical inaccuracies.

On the other hand integral relations, like ones formulated in
eq.~(\ref{integral_2B}-\ref{integraal}), as well as CLD formalism
provide the ready recipes to extract  phaseshifts. Still accuracy
of the two methods is quite  different. In the
figure~\ref{fig_cld_vs_int} I present calculated S-wave
phaseshifts  for the potential consisting of two Yukawa functions
\begin{equation}
V(r)=678.097\frac{\exp(-2.54922r)}{r}-166.032\frac{\exp(-0.679864r)}{r},
\end{equation}
and aiming to describe n-$^3$H scattering, by setting a reduced
mass of the system to $\frac{\hbar^2}{2\mu}= 27.647$MeV.fm. In the
last expression the overall potential is expressed in units of
MeV, whereas the  distances are measured in fm units.

In these figures the absolute errors in the extracted phaseshifts
are presented as a function of Lagrange-Laguerre basis functions
used to realize the calculations. The scaling parameter for
Lagrange-Laguerre basis was optimized for CLD method and set to
$h=0.5$ fm (see section \ref{sec:ll_method}). One may see that
convergence of the phaseshifts obtained using CLD formalism are
somehow smoother. However results based on the integral relation
expression converge much faster and are systematically more
accurate than those based on CLD. Furthermore convergence of the
phaseshifts calculated by the CLD expression seem to saturate,
when accuracy of a few fractions in a degree is reached. I have
obtained very similar tendencies when employing other numerical
techniques or 2-body interaction models. This result is not
surprising however: the CLD formalism takes into consideration
only Hamiltonian eigenvalues, whereas integral expression involves
both eigenvalues and eigenvectors, thus absorbing richer
information on the original Hamiltonian.

Finally, the method based on the integral expressions provides
full scattering amplitude and thus S-matrix, not only phaseshifts.
Therefore at each calculation one may check how well unitarity of
the S-matrix is preserved. This verification provides also a good
indication of the accuracy of the calculated phaseshifts. In
particular, when keeping in mind that for the low energy
scattering problems it is more difficult to ensure unitarity of
the S-matrix than to obtain the accurate phaseshifts.

\bigskip

\subsubsection{CS transformation of the potential energy \label{sec:cs_pot_trnsf}}

Finally, one should discuss some technical aspects of the CS
method, which may hamper its successful implementation. As it has
been demonstrated in the previous sections, the  implementation of
the CS\ method is rather straightforward. This method may be
easily adapted to work in the majority of the existent bound state
codes. Still CS implies complex arithmetics and leads to
non-Hermitian matrices already for the problems involving only
binary scattering channels. Thus some of the linear-algebra
methods, which are limited to real Hermitian matrices, are
inappropriate. In particular, methods employed in bound state
calculations seeking for the extreme eigenvalues are not
applicable in CS problems. Indeed, the resonant states are
embedded between the continuum states and merely differentiate in
their real argument parts.

\begin{figure}[th]
\centering
\includegraphics[width=81mm] {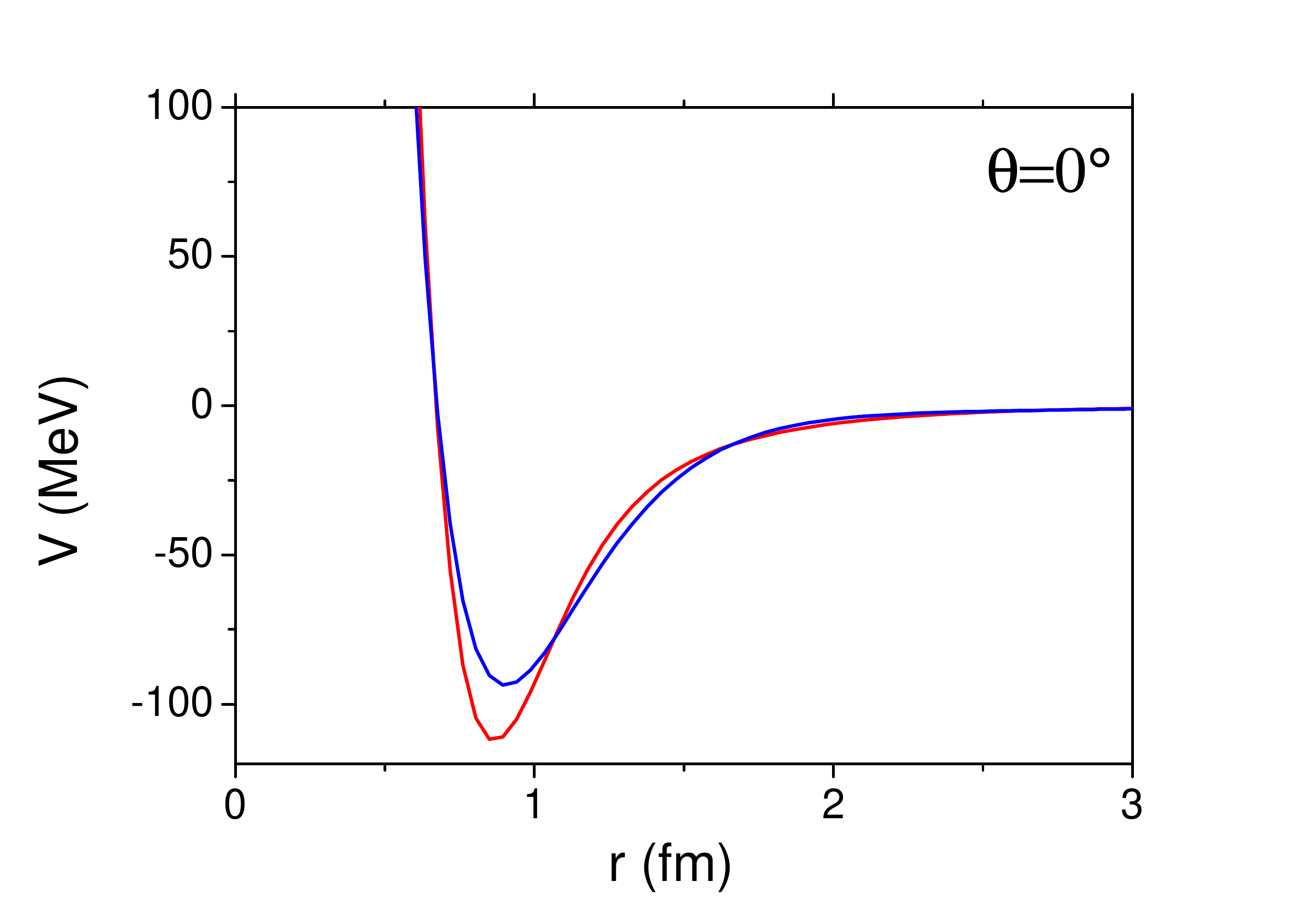} 
\includegraphics[width=81mm] {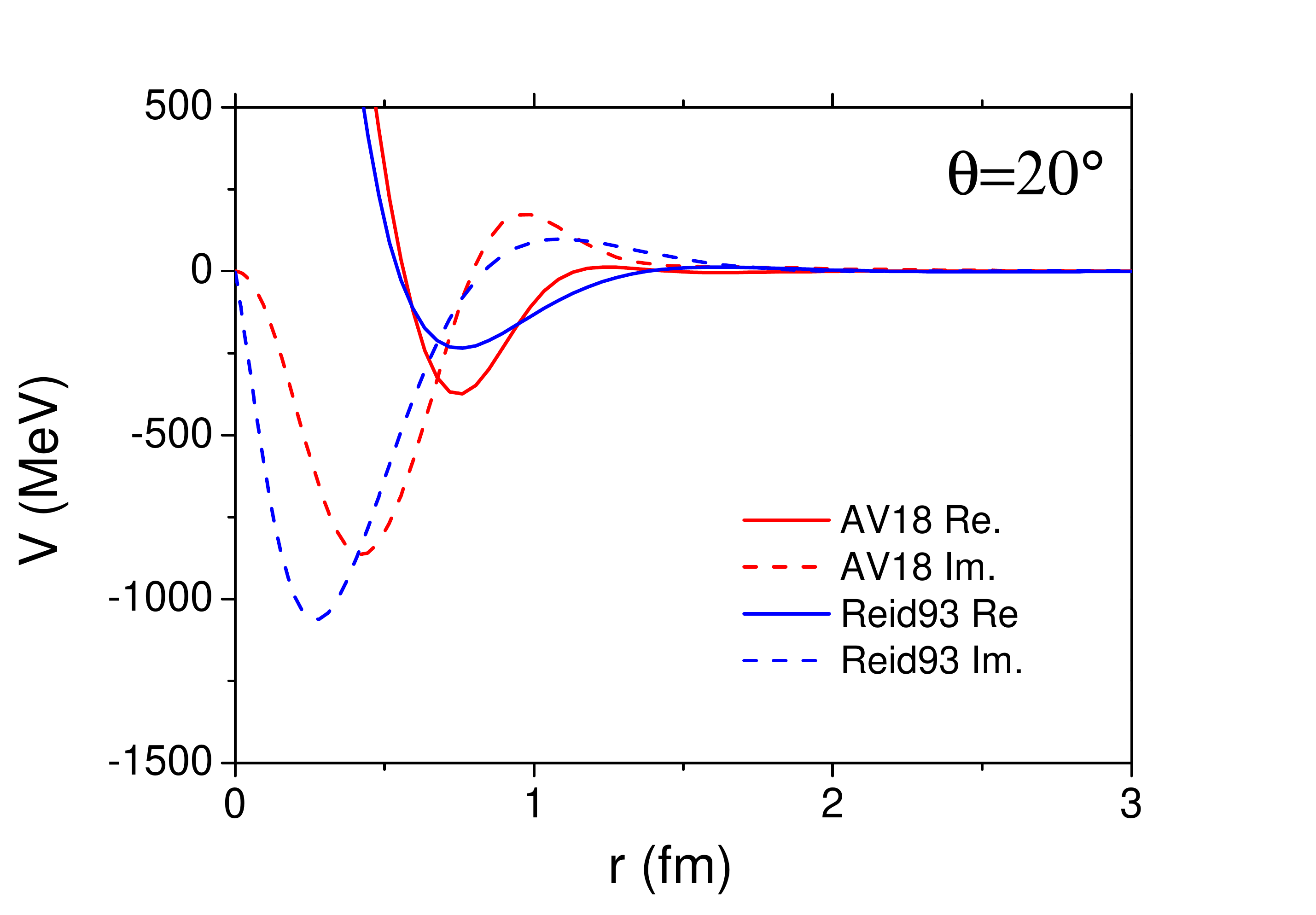} \includegraphics[width=81mm]
{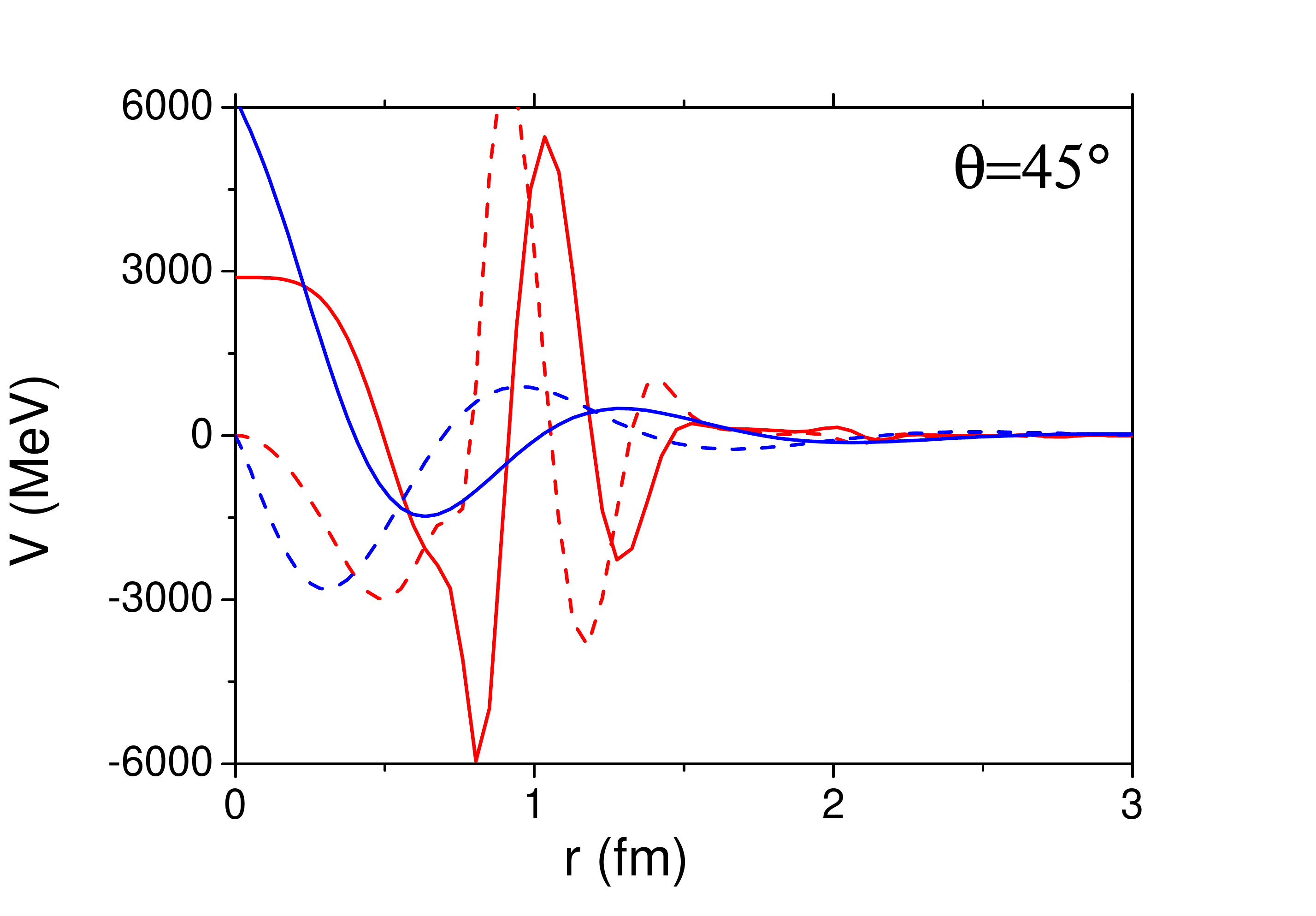}
\caption{Comparison of the CS potential energies for AV18 and Reid93 models,
when setting CS angle $\protect\theta$ to $0^\circ$ (non-scaled original
potentials), $20^\circ$ and $45^\circ$, as indicated in the insets of the
figures. Full lines represent real parts of the potentials, whereas dashed
lines imaginary ones. Av18 potential is plotted in red, whereas Reid93 in
blue.}
\label{fig_local_sc_pot}
\end{figure}

Other possible complication in implementing CS method are related
with the ability to calculate matrix elements of the potential
energy. In CS method one works with the analytical potentials
extended to the complex $r-$plane. However, as pointed out
in~\cite{REB81,Witala:3n,LC05_3n} not all the potentials comply
with the complex scaling. In particular, short-range potentials
may become strongly oscillatory and even start to diverge if large
value of the CS angle $\theta $ parameter is employed, see
figure~\ref{fig_local_sc_pot}. For example, if a potential
involves some exponential regulator in form of $f_{n}(r)=\exp
(-cr^{n})$, then after CS transformation this potential becomes
divergent for $\theta >\pi /2n$. Let us return to the
figure~\ref{fig_local_sc_pot}, both Reid93~\cite{nijm2} and
AV18~\cite{AV18} nucleon-nucleon interaction potentials have very
similar features. Nevertheless AV18 potential involves sharper
regulator with $n=2$ to rend potential repulsive close to the
origin. This results into very strong oscillations of the CS AV18
potential for $\theta =45^\circ$. Even more problematic turns to
be a case of chiral EFT nucleon-nucleon potentials, where one
employs high momenta regulators of the type $\exp (-cp^{n})$ with
$n=6-8$. Regardless the fact that these potentials are built in
momentum space, due to the equivalence of the CS transformation
reflecting $r\rightarrow re^{i\theta }$ in to $p\rightarrow
pe^{-i\theta },$ it turns to be that the same limitation applies
for the
momentum space regulators. As a consequence one has to limit $\theta <10%
{{}^\circ}%
$ when implementing CS method for the chiral EFT Hamiltonians.

To this respect it is of practical interest  to keep the angle
$\theta $ values small in numerical calculations, which would
guarantee smoothness of the potential after the
complex scaling and thus allows numerical treatability of the problem~\cite%
{Witala:3n,LC05_3n}. On the other hand the far asymptote of the
complex-scaled outgoing wave solution decays as $%
\exp(-k_{x}r_{x}\sin {\theta })$, where $k_{x}$ is a wave vector
corresponding to the last open-channel (channel with the lowest
free energy for the reaction products). To this aim,  large values
of the angle $\theta $  are preferred in order to damp efficiently
outgoing wave solution; and in particular if calculations are
performed close to the threshold (small $k_{x}$ value). The last
fact makes it
difficult to use CS at energies very close to the open thresholds. Condition~%
\label{cond_lim_nb} provides additional limit for angle $\theta $ to be used
when performing calculations at high energy. \bigskip

Regarding the practical calculation of the matrix elements, one
may remark that quite often the trial basis functions may have
better analytic properties than the potential energy. In this case
it is useful to employ the Cauchy theorem when estimating matrix
elements by deforming integration contour back to include real
$r-$axis. Due to the fact that the basis functions are square
integrable, integral over the radial contour at $\left\vert
r\right\vert =\infty $ vanishes, giving for the local potentials:
\begin{eqnarray}
V_{ij}^{\theta } &=&\int_{0}^{\infty }f_{i}(r)V(re^{i\theta })f_{j}(r)dr \\
&=&e^{-i\theta }\int_{0}^{\infty }f_{i}(re^{-i\theta
})V(r)f_{j}(re^{-i\theta })dr \label{eq:cauchy_vpot}
\end{eqnarray}
If the potential is non-local $V(r,r\prime )$:
\begin{eqnarray}
V_{ij}^{\theta } &=&\int_{0}^{\infty }e^{i\theta }f_{i}(r)V(re^{i\theta
},r^{\prime }e^{i\theta })f_{j}(r^{\prime })drdr^{\prime } \\
&=&e^{-i\theta }\int_{0}^{\infty }f_{i}(re^{-i\theta })V(r,r^{\prime
})f_{j}(r^{\prime }e^{-i\theta })drdr^{\prime }
\end{eqnarray}

\subsubsection{Complex scaling angle}

Complex scaling angle plays an important role for the successful
implementation of the CS technique. Naturally one would like to be
able to use the optimal values for this parameter, which would
allow to perform more accurate and faster converging calculations.
From one side large values of the CS angles allow to damp faster
the outgoing waves and thus should ensure convergence. However
quite often large CS angles involves much more complicated algebra
related with an emergence of the strongly oscillating wave
functions or potential energy terms.
 Moreover for certain problems there are some mathematical
limitations for the CS angle to be employed. Therefore in this
subsection I would like to overview this issue.

\vskip 0.5cm

First of all there is a natural limitation of the complex scaling
angle to be used, which is related with the ability to perform CS
transformation of the potential energy, discussed in  the previous
subsection. This issue is common to any problem treated by the CS
method -- bound, resonant states, or description of the diverse
reactions. Problem arises for the potentials, which involve an
exponential regulator in the form $f_{n}(r)=\exp (-cr^{n})$, such
potentials become divergent for CS transformations with
$\theta>\pi/2n$ and thus intractable numerically.

\vskip 0.5cm

According to the ABC theorem, see section \ref{sec:res_states}, to
determine position of a resonant state one should be able to
"expose" it by the cuts of the rotated continuum states. It means
that CS angle should satisfy, relation $\theta>-\frac{1}{2}Arg(E_{R})$, (see fig.~\ref%
{Fig_eigen_csm}), where energy of a resonant state $E_{R}$ is
calculated relative to the last open threshold.

\vskip 0.5cm

 There are two additional limitations for a CS angle
to be used arising when solving problems of particle collisions.
This feat is related to the fact that for this set of problems,
one should handle incoming particle(cluster) waves, whose wave
function becomes exponentially divergent after the CS transform.
To understand this issue let me briefly summarize the general
framework to treat collision of two multiparticle clusters. Lets
consider two clusters $a$ and $b$ formed by $N_{a}$ and $N_{b}$
particles (with $N_{a}+N_{b}=N$) whose binding energies are
$E_{a}$ and $E_{b}$ respectively. The relative kinetic energy of
the two clusters in the center of mass frame is
$E_{a,b}=E_{c.m.}-E_{a}-E_{b}=$ $\hbar ^{2}k_{a,b}^{2}/2\mu
_{a,b}$. Then the incoming wave takes the following form:
\begin{equation}
\Psi _{a,b}^{in}(\mathbf{k}_{a,b},\mathbf{r}_{i,a}\mathbf{,r}_{j,b}\mathbf{,r%
}_{a,b})=\psi _{a}(\mathbf{r}_{i,a})\psi _{b}(\mathbf{r}_{j,b})\exp (i%
\mathbf{k}_{a,b}\cdot \mathbf{r}_{a,b}),
\end{equation}
where $\psi _{a}(\mathbf{r}_{i,a})$ and $\psi
_{b}(\mathbf{r}_{j,b})$ represent
bound state wave functions of the clusters $a$ and $b$ respectively, with $\mathbf{%
r}_{i,a}(\mathbf{r}_{j,b})$ defining internal coordinates of the clusters, while $%
\mathbf{r}_{a,b}$ is a vector connecting the centers of mass of
the two clusters.

As previously, one is keen to write the Schr\"{o}dinger equation
in its inhomogeneous form and apply the complex scaling on all the
coordinates, getting:
\begin{equation}
\lbrack
E-e^{-i2\theta}\hat{H}_0-\sum\limits_{m<n}{V^\theta_{mn}}\left(
\mathbf{r}_{m}\mathbf{-r}_{\mathbf{n}}\right)]\overline{\Psi }%
_{a,b}^{sc}(\mathbf{r}_{i,a},\mathbf{r}_{j,b}\mathbf{%
,r}_{a,b})=\left[\sum\limits_{i\in a;j\in b}V^\theta_{ij}\left(
\mathbf{r}_{i}\mathbf{-r}_{\mathbf{j}}\right)\right]
(\Psi_{a,b}^{in})^\theta(\mathbf{r}_{i,a},\mathbf{r}_{j,b},\mathbf{%
r}_{a,b}). \label{Schr_dr_nbody}
\end{equation}
The term $\overline{\Psi}_{a,b}^{sc}(\mathbf{r}_{i,a},\mathbf{r}_{j,b},\mathbf{%
r}_{a,b})$ contains only complex-scaled outgoing waves in the
asymptote and thus is formally bound exponentially. Therefore, as
long as the right hand side of the last equation is bound, it
might be solved using a square  integrable basis set to express
the scattered part of the wave function
$\overline{\Psi}^{sc}_{a,b}(\mathbf{r}_{i,a}\mathbf{,r}_{j,b}\mathbf{,r}_{a,b})$.

\begin{figure}[tb]
\centering
\includegraphics[width=16 cm]{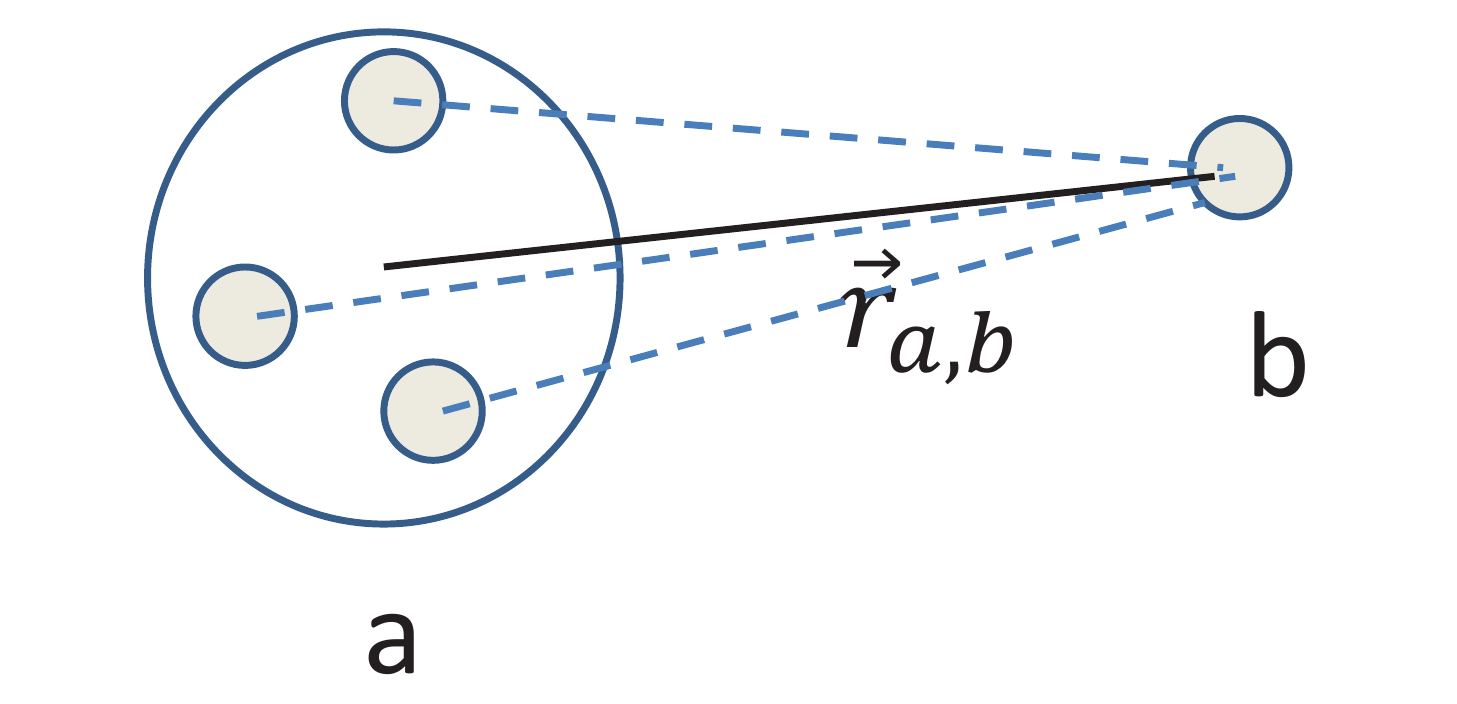}
\caption{Directions of the interaction terms between the two
multiparticle clusters (a and b) do not coincide exactly with the
wave vector $\vec{r}_{a,b}$ connecting their centers of
mass.\label{fig_cl_ab}}
\end{figure}

However the inhomogeneous term of eq.~(\ref{Schr_dr_nbody}) is not
necessarily exponentially bound even if all the interaction terms
are bound. The first issue is relevant for the problems governed
by the potentials containing exponentially decaying terms. Suppose
the slowest exponent for the potential describing interaction
between the particles $i \subset a$ and $j\subset b$ is:
\begin{equation}
V^\theta_{ij} (r\rightarrow \infty) \propto
\exp(-\nu^{min}_{ij}r),
\end{equation}
then inhomogeneous term will be convergent in the
$r_{a,b}\rightarrow \infty$  limit if:
\begin{equation}
\tan \theta <
\frac{\nu_{ab}\hbar}{\sqrt{E_{a,b}}}\sqrt{\frac{(M_a+M_b)}{2M_aM_b}}.
\end{equation}

For a two-body problem this condition translates into:
\begin{equation}
\tan \theta < \frac{\nu_{ab}}{k_{ab}};
\end{equation}
i.e. it may limit usage of the large CS angles for the
calculations involving large energies and slowly decaying
exponential potentials. It is not a very common issue in practice,
because in large $k_{ab}$ limit scattering observables are
dominated by the Born term, which may be estimated without use of
CS. Nevertheless for the scattering problems involving two heavy
clusters with a few light particles inside this condition is
strongly enhanced.

\vskip 0.5cm

Second limitation arises only in the scattering problems involving
more than two particles. In this case the vectors of the
interaction terms do not coincide with the wave vector connecting
the center of mass of the two clusters as shown in
Fig.~\ref{fig_cl_ab}. Still one may demonstrate that the
inhomogeneous term remains bound if an additional condition is
fulfilled~\cite{LC11}:
\begin{equation}
\label{cond_lim_nb}
\tan \theta <\min \left(\sqrt{\frac{B_{i\in a}}{E_{a,b}}\frac{m_{i}(M_{a}+M_{b})}{%
(M_{a}-m_{i})M_{b}}},\sqrt{\frac{B_{j\in b}}{E_{a,b}}\frac{m_{j}(M_{a}+M_{b})%
}{(M_{b}-m_{j})M_{a}}}\right),
\end{equation}
where $B_{i\in a}$ is the i-th particle removal energy from the
cluster $a$ and $M_{a}$ is a total mass of the cluster $a$. The
last condition implies additional limit on the complex scaling
angle $\theta $ to be used. For a system of equal mass particles
this limit does not have much effect and becomes important only
well above the break-up threshold $\left\vert E_{a,b}\right\vert
>>B_{i\in a}$ (or $\left\vert E_{a,b}\right\vert
>>B_{j\in b}$ respectively). Even at high energies
this limit is not so constraining,  since the exponent of the
scattered wave becomes proportional to $\sqrt{E_{a,b}}$ and
therefore one may achieve the same speed of convergence by
employing smaller complex scaling angle $\theta$ values. On the
other hand the condition in eq.~(\ref{cond_lim_nb})  may become
strongly restrictive for  the mass-imbalanced systems if one
considers light-heavy-heavy components.

\subsection{SRG transformation}

In the recent years the similarity renormalization group (SRG)
techniques became an indispensable part of the many-body structure
calculations. SRG is based on a smooth unitary transformations
that suppress off-diagonal matrix elements, gradually bringing the
Hamiltonian towards a band-diagonal form. SRG transformation are
used to soften Hamiltonians based on the interactions containing
repulsive cores (high-momenta components), they allow to greatly
improve convergence properties of the structure calculations, but
preserve the physical observables. It is of great interest if such
techniques might be beneficial when employed together with CS\
method. The answer is not obvious, since goals of the two
approaches are slightly different: SRG tries to soften
high-momenta components of the interaction, on the other hand
convergence of the CS method is related with the ability to
describe slowly dying asymptotes of the transformed wave
functions.

The SRG approach was developed independently by S.D.~Glazek and
G.~Wilson~\cite{PhysRevD.48.5863}  and by
F.~Wegner~\cite{Wegner_srg}. It resides on the similarity
transformation of the center-of-mass Hamiltonian $H=T+V$ by some
unitary operator $U(s)$:
\begin{equation}
H_{S}=U(s)HU^{\dagger }(s)=T+V_{s},
\end{equation}%
where $s$ is the flow parameter, whereas kinetic energy operator
is considered to be independent of $s$. Evolution of the
transformed Hamiltonian is determined by the flow equations:
\begin{equation}
\frac{dH_{s}}{ds}=\left[ \eta _{s},H_{s}\right] \label{eq:fl_eq1},
\end{equation}%
by selecting
\begin{equation}
\eta _{s}=\frac{dU(s)}{ds}U^{\dagger }(s)=-\eta _{s}^{\dagger }
\label{eq:fl_eq2},
\end{equation}%
which defines the SRG transformation.

Similar procedure can be applied to CS Hamiltonian. In this case
one has to choose between the two strategies: either first to
perform CS transformation and then SRG transformation of the CS
Hamiltonian, or first evolve initial potential with SRG and then
apply CS on the evolved Hamiltonian. The first procedure allows
one to use analytic properties of the initial potential when
performing CS transformation, thus leaving a choice how to
calculate the matrix elements of the potential energy via
eq.~(\ref{eq:me_cspot}) or via eq.~(\ref{eq:cauchy_vpot}). The
second strategy is applicable only if one performs SRG using a
basis of analytic functions, requiring to perform CS
transformation based on the relations obtained via Cauchy theorem.
This might be summarized as follows
\begin{eqnarray}
H_{S}^{\theta } &=&U^{\theta }(s)H^{\theta }U^{\theta ^{\dagger }}(s)=%
\widehat{S}U(s)\widehat{S}^{-1}\widehat{S}H\widehat{S}^{-1}\widehat{S}%
U^{\dagger }(s)\widehat{S}^{-1} \\
&=&\widehat{S}U(s)HU^{\dagger }(s)\widehat{S}^{-1}
\label{eq:sr_2_appr}
\end{eqnarray}%
here the first expression represents the first strategy, whereas
the last expression reflects the second strategy. If one is
capable to perform CS transformation
$\widehat{S}U(s)\widehat{S}^{-1}$ of the operator $U(s)$ the two
approaches become formally identical. However the full identity of
the two approaches is achieved only in the limit of the infinite
basis, employed in performing Hamiltonian transformations,  thus
numerical realization may highlight some differences.

Choosing $\eta_{s}$  specifies  the SRG transformation. Perhaps
the simplest and certainly the most popular choice
is~\cite{Wegner_srg}
\begin{equation}
\eta_{s}= \left[T,H_{s}\right].
\end{equation}
For this choice of transformation the flow parameter $s$ is
measured in units of $fm^4$ and thus is popularly quantified by a
parameter $\lambda\equiv s^{\frac{1}{4}}$ having dimensionality of
momenta.

Implementation of the SRG transformation requires solution of the
flow eq.~(\ref{eq:fl_eq1}-\ref{eq:fl_eq2}). Standard strategy to
solve these equations is based on discretizing Hamiltonian using
square integrable basis, leading to solve a set of first order
differential equations. In order to avoid numerical instabilities
differential equation solver of high quality is compulsory. The
fortran codes present in the publicly available
ODEPACK~\cite{hindmarsh1983odepack} library matches perfectly for
this task. In particular, Hermitian SRG flow equations (the second
approach)   might be comfortably solved employing DVODE code,
whereas for non-Hermitian flow equations (the first approach)
ZVODE code is appropriate.

The prove of principle for the second approach, the one
represented by eq.(\ref{eq:sr_2_appr}), has been already presented
a few years ago~\cite{PhysRevC.91.021001} employing realistic
nuclear Hamiltonians. I have also performed a few tests to
determine the relevance of the SRG transformation in the
calculations related with the complex scaling method.
\begin{figure}
\centering
\includegraphics[width=80mm] {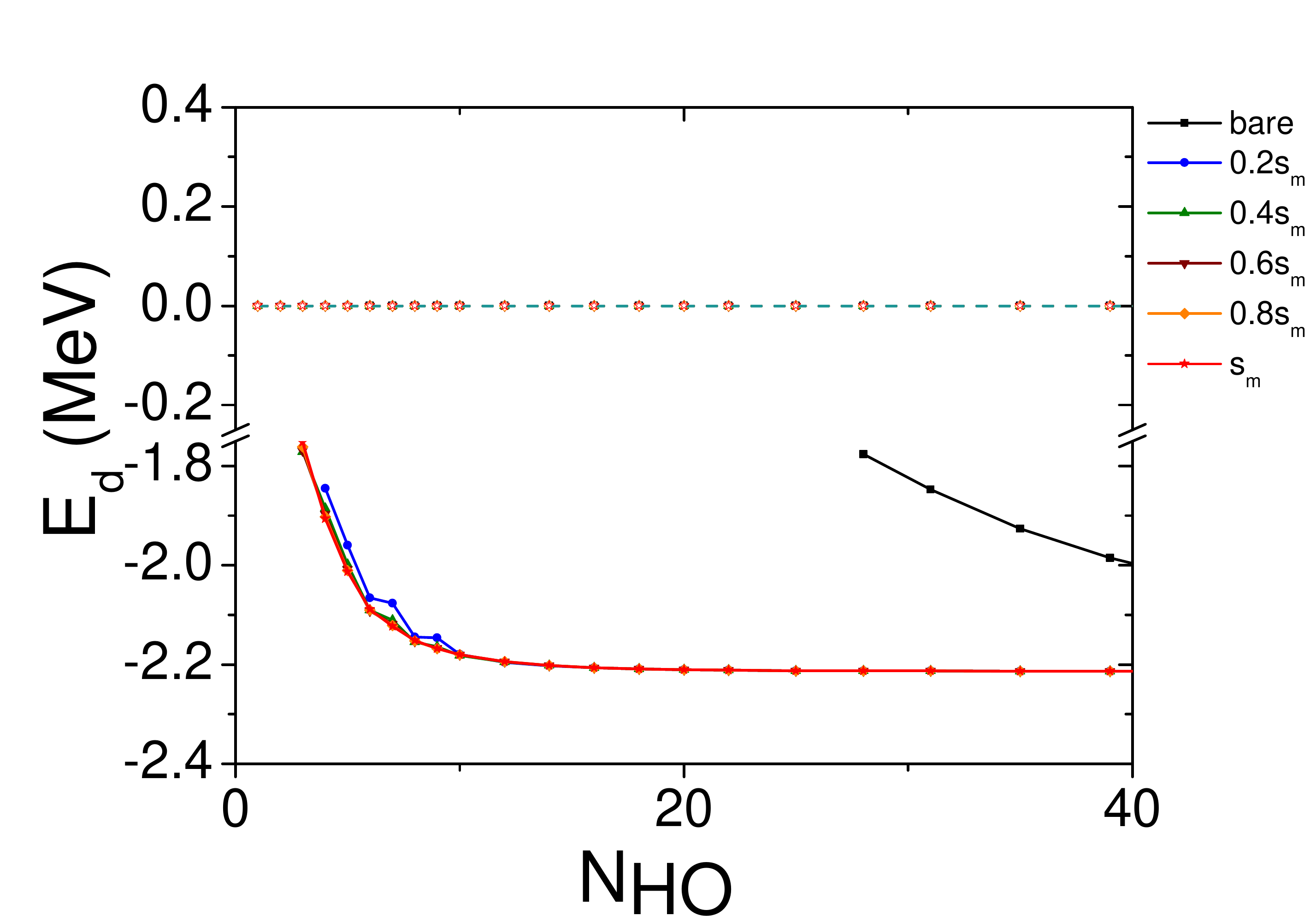} 
\includegraphics[width=80mm] {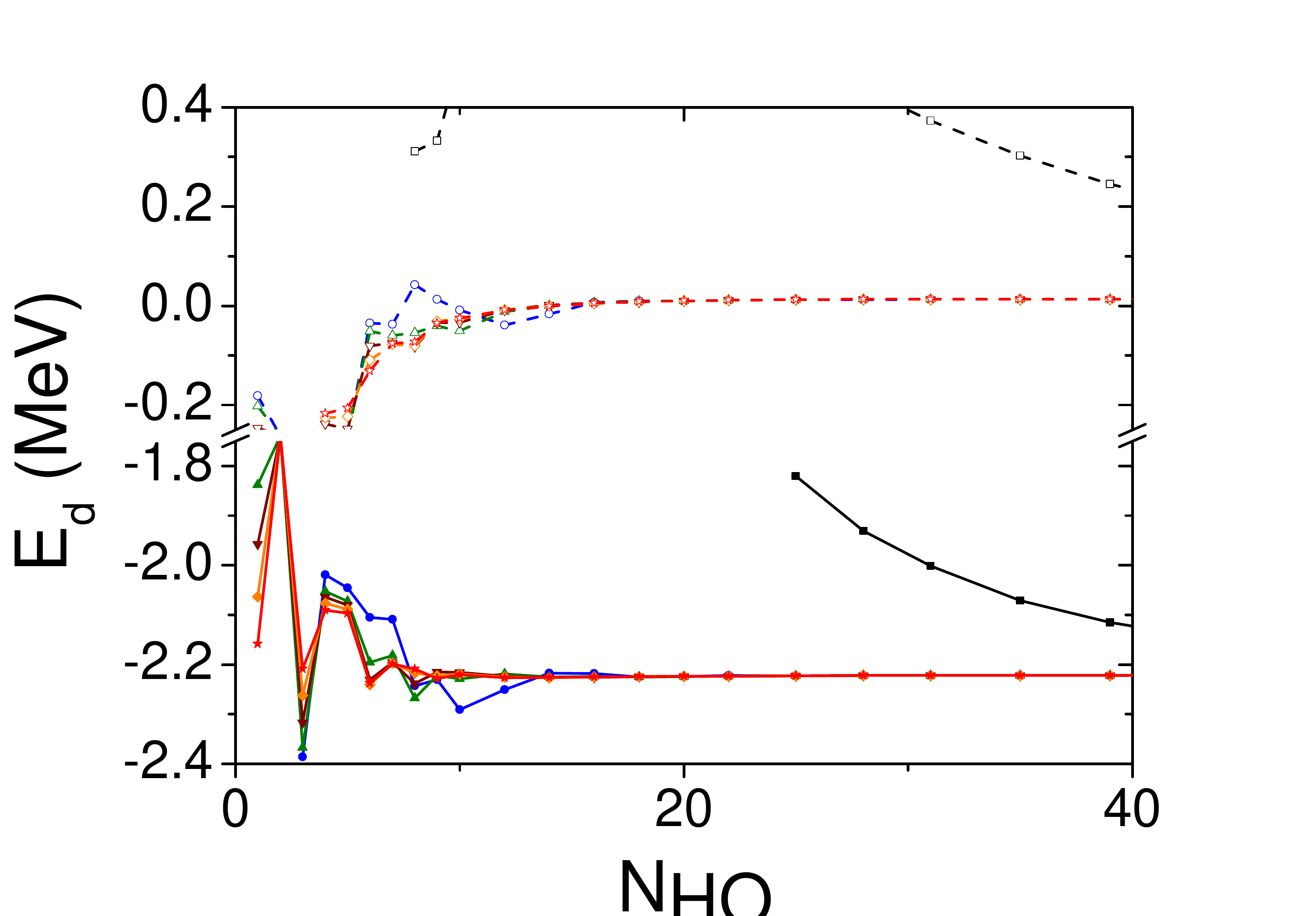}
\caption{ Convergence of the calculated deuterons binding energy
for MT I-III potential  as a function of the harmonic oscillator
(HO) basis functions used to discretize CS Hamiltonian. The
frequency of HO basis was $\hbar\omega=20$ MeV. The different
curves demonstrate convergence of the deuteron binding energy
(solid lines, bottom panel) and its spurious complex parts (dashed
lines, upper panel) with the flow parameter $s$. The parameter
$s_m$ corresponds to the transformation with a cutoff fixed at
$\lambda_m=2$ fm$^{-1}$. In the left panel results for non-rotated
Hamiltonian are presented, whereas in the right panel CS
Hamiltonian with $\theta=15^\circ$ is employed.}
\label{fig_srg_deut}
\end{figure}

Convergence of the calculated binding energies present very
similar features as ones realized by SRG evolution of the
Hermitian Hamiltonian. This feature is demonstrated in figure
\ref{fig_srg_deut}, where the deuteron binding energy convergence
is studied employing MT I-III potential both for the non-rotated
(left panel) and by $\theta=15^\circ$ rotated (right panel)
Hamiltonians. Convergence is sought by increasing size of harmonic
oscillator (HO) basis functions used to discretize CS Hamiltonian.
The frequency of HO basis was chosen to 20 MeV, whereas SRG
transformation has been realized within the basis of 150 HO
functions. The different curves demonstrate convergence of the
deuteron binding energy (solid lines, bottom panel) and its
spurious complex parts (dashed lines, upper panel) with the flow
parameter $s$. The parameter $s_m$ corresponds to the
transformation with a cutoff $\lambda_m=2$ fm$^{-1}$. CS
Hamiltonian results present less regular convergence pattern,
nevertheless the speed of convergence relative to the SRG
procedure is comparable. As expected SRG transformation softens
interaction by speeding-up convergence of the bound state energy
calculations within HO basis.

\bigskip
\begin{figure}[th]
\centering
\includegraphics[width=80mm] {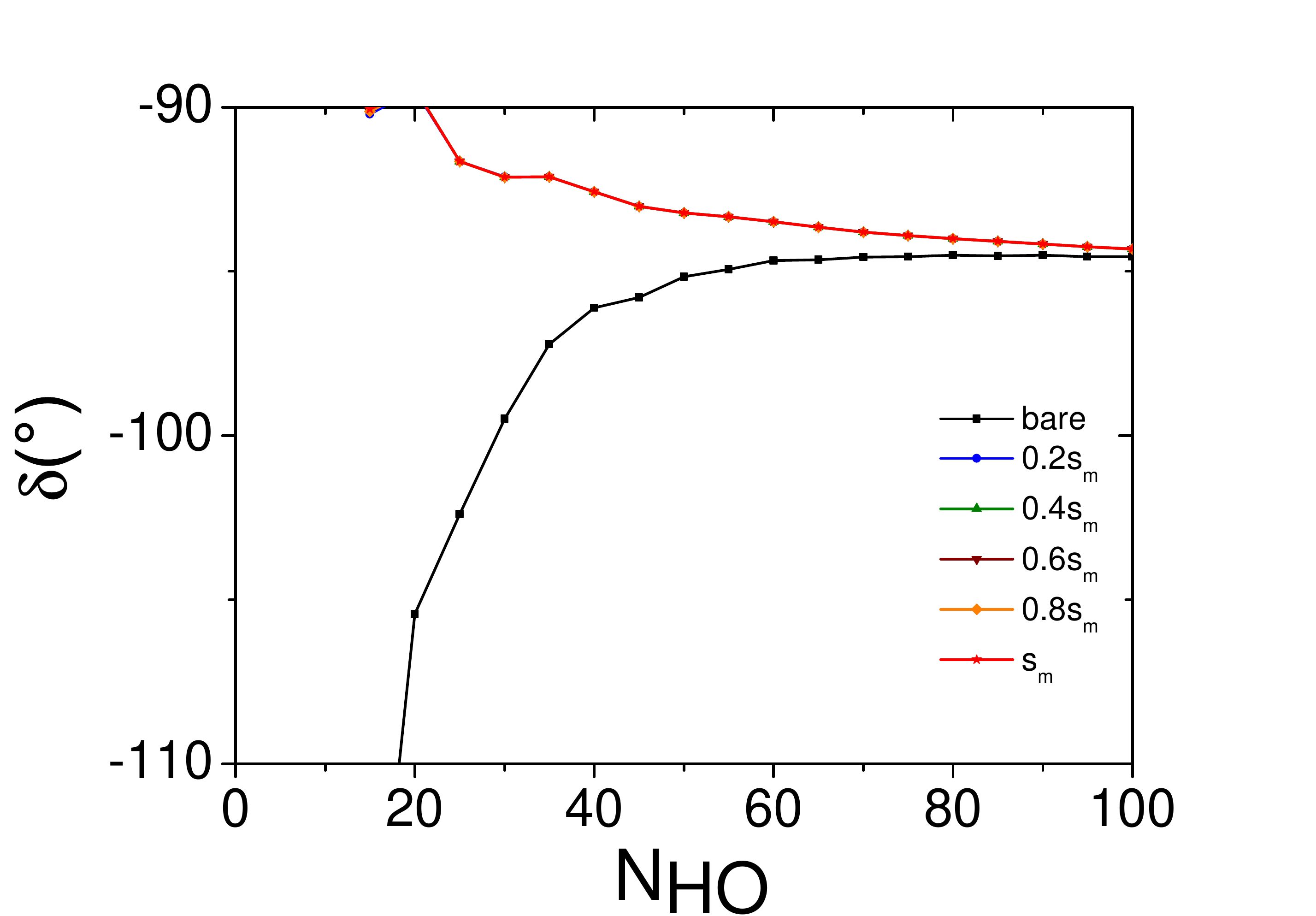} 
\includegraphics[width=80mm] {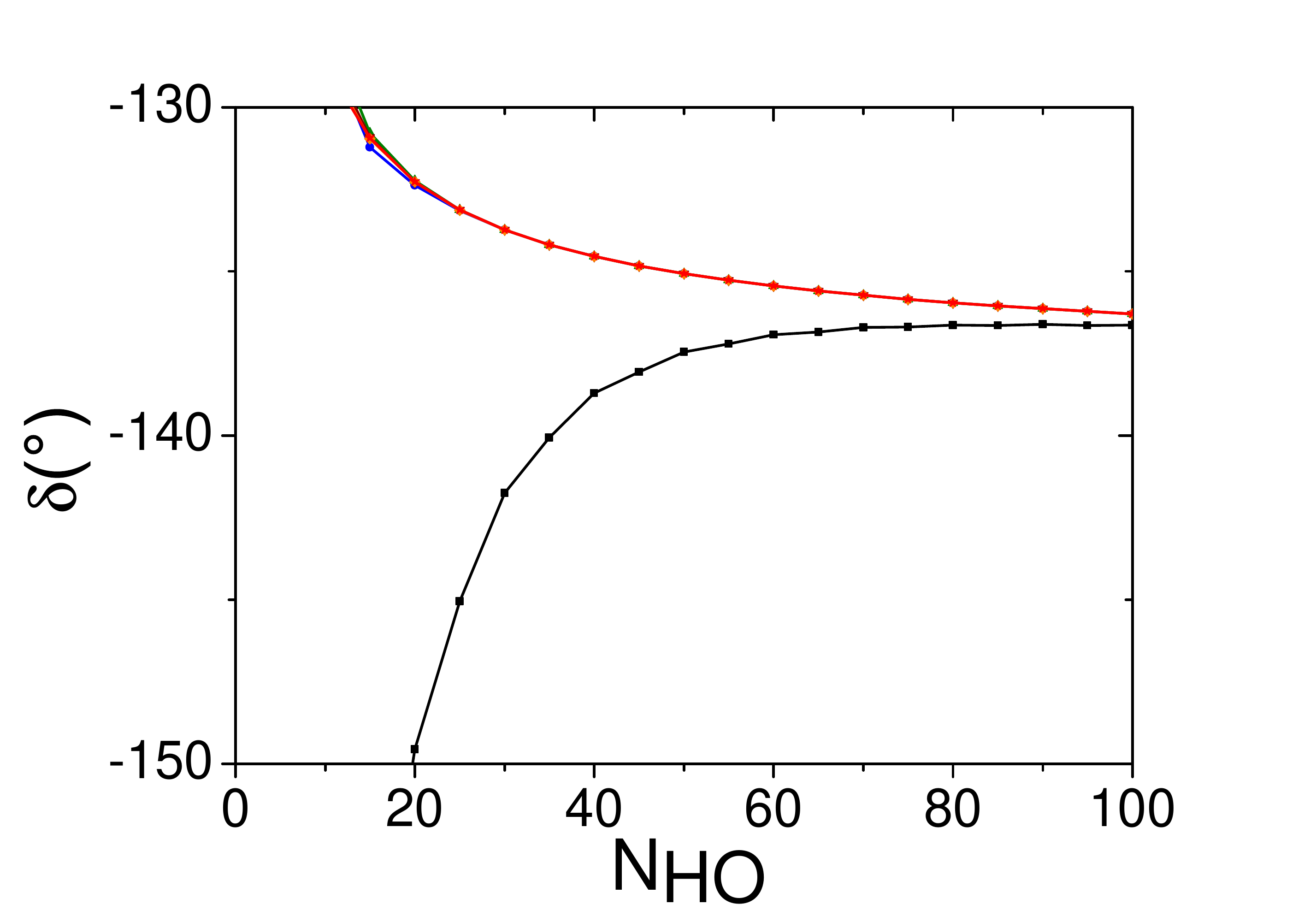}
\caption{Convergence of  the np doublet phaseshifts at E$_{cm}$=
10 MeV (left panel) and 50 MeV (right panel) realized SRG evoluted
CS Hamiltonians using CLD method. Calculations have been performed
for the same Hamiltonian as described in the inset of the figure
\ref{fig_srg_deut}.}
 \label{Fig_srg_phases}
\end{figure}

On contrary, the SRG effect on the convergence of the calculated
phaseshifts see figure~\ref{Fig_srg_phases} , extracted using CLD
technique, is strongly debatable. One can hardly see any evolution
apart from the first integration step from $s=0$ to $s=0.2s_m$ and
this is mostly due to the offset provided via
eq's.(\ref{eq:cld_b0}-\ref{eq:cld_b2}) by the inaccurately
reproduced deuterons binding energy, containing spurious imaginary
part and thus failing to shift a phaseshift by entire angle $\pi$
.

However one should not hastily discard the relevance of SRG in
relation with CS applications. This study has been done using CLD
to extract phaseshifts. However as demonstrated in the previous
section integral relations turns to be much more prominent
technique to extract the scattering observables. Unfortunately in
this pioneering study I have not managed to implement integral
relation method together with SRG for two-body problem. Actually I
have failed to evaluate numerically inhomogeneous term of
eq.(\ref{Schro_2B_cs}), which projected on HO basis becomes:
\begin{equation}
\left<\psi_{HO}(r) |V_S^{\theta }(\mathbf{r})\overline{\Psi }%
^{in,\theta }(\mathbf{r})\right>=\left<\psi_{HO}(r) |V_S^{\theta }(\mathbf{r})|\psi_{HO}(r)\right>\left<\psi_{HO}(r)|\overline{\Psi }%
^{in,\theta }(\mathbf{r})\right>.
\end{equation}
Via SRG procedure one readily disposes of matrix elements
$\left<\psi_{HO} |V_S^{\theta }(\mathbf{r})|\psi_{HO}(r)\right>$,
however one fails to get convergence then evaluating the full sum
-- due to diverging nature of the term $\overline{\Psi
}^{in,\theta }(\mathbf{r})$.

Effects in calculating scattering observables in $A>2$ systems
might be quite different. Estimation of the inhomogenious term
might not be so troublesome,  as discussed in the previous
section, one has to couple interaction terms with an incoming wave
whose space vectors does not coincide. Under certain (relatively
small CS angles employed) conditions the divergence of the CS
incoming wave is moderated by a faster converging bound state wave
functions describing compact colliding clusters. Secondly
successful calculation of the scattering observables in $A>2$
systems strongly resides on the accurate reproduction of the
thresholds, wave functions of the colliding clusters and
successful representation of the effective interaction between the
reaction products. SRG once properly implemented -- by including
induced many-body forces -- may clearly reduce the effort in
describing thresholds and cluster wave functions. Success in
description of the effective interaction is however less obvious
and requires more profound analysis, which is beyond my technical
baggage.

\section{Numerical methods}

In this study I will overview only two numerical techniques, which
were employed throughout this work: spline collocation method and
Lagrange-mesh method.

\subsection{Spline collocation\label{sec:spline_coll}}

For many applications, in particular ones related with description
of the complicated structural features, flexible bases are
required, which allow to highlight important space regions and if
necessary scan them with a denser distribution of points (basis
functions). Spline collocation method, widely employed in civil
engineering applications, is built for this purpose. In few-body
physics, it was introduced by G.L.~Payne~\cite{Payne_splines}.

The spline (or orthogonal collocation) method mathematical
foundations were laid by C.~de Boor and
B.~Swartz~\cite{Boor_Spl,Boor_book}. They showed that a basis of
piecewise polynomial functions of degree less than \textit{m+k}
with $m-1$ continuous derivatives can be used to approximate the
solution of \textit{m}-th order differential equation with an
error of $O(h^{m+k})$, where $h$ is the size of subintervals. One
should require that the differential equation is only exactly
satisfied at k Gauss quadrature points located in the
subintervals. The method consist of:

\begin{enumerate}
\item Subdividing the domain into a number of subintervals (a
grid) and associate to it a spline basis.

\item Expanding a wave function in the spline basis (in this work
piecewise Hermite polynomials are used).

\item Requiring the equation to be satisfied on a set of well-chosen points
(collocation points).
\end{enumerate}

This procedure leads to a finite-dimensional algebraic problem, which is
solved using linear algebra techniques.

Let us discuss the matter in more details. Suppose we want to solve
one-dimensional differential equation described by a linear operator $\hat{%
L}$, which is defined on a finite size domain $\Re \in \left[
r_{\min },r_{\max }\right] $:
\begin{equation}
\hat{L}\ast F(r)=0,  \label{Lin_op}
\end{equation}
with a solution $F$ satisfying some boundary conditions at $r=r_{\min }$ and
$r_{\max }$. To solve this system we divide $\Re $ in subintervals $%
r_{0}<r_{1}<r_{2}<...<r_{N}$ (for some finite grid $r_{0}=r_{\min },\
r_{N}=r_{\max }$). We search the solution $F$ in the form:
\begin{equation}
F(r)=\sum_{j=0}^{k(N+1)-1}C_{j}S_{j}(r),
\end{equation}
where $S_{j}$ are Hermite piecewise polynomials of
\textit{(2k-1)}-th order and where $C_{j}$ is a set of unknown
coefficients to be determined. Due to its linearity, operator
$\hat{L}$ of eq. (\ref{Lin_op}) acts only on known piecewise
functions $S_{j}$, and its action can be determined at any $r$
inside the domain $\Re $. In this way eq. (\ref{Lin_op}) becomes:
\begin{equation}
\sum_{j=0}^{k(N+1)-1}C_{j}\left[ \hat{L}\ast S_{j}(r)\right] =0.
\label{Lin_sys}
\end{equation}
We demand that this system of equations is satisfied on a number of
well-chosen points (collocation points, \textit{k} for each subinterval)%
\footnote{%
Knowing the properties of Gauss integral quadrature, it becomes rather
obvious \cite{Boor_book}, that if the exact solution can be extrapolated in
any subinterval by polynomials of order $\mathit{m=2k-1}$, then the
numerically obtained one would be exact if differential equations are
satisfied on only \textit{k} Gauss quadrature points of this subinterval.}.
Consequently we obtain $kN$ equations for $k(N+1)$ unknown coefficients $%
C_{j}$. We can as well implement $k$ different boundary conditions to have a
number of linear equations equal to the number of unknowns:
\begin{equation}
\sum_{j=0}^{k(N+1)-1}C_{j}\left[ \hat{L}\ast S_{j}(\tilde{r}_{i})\right]
=0\qquad i=1,2,\ ..\ ,k(N+1),  \label{Lin_sys_eq}
\end{equation}
where $\tilde{r}_{i}$ signifies \textit{i}-th collocation point.
\begin{figure}[th]
\centering
\includegraphics[width=80mm] {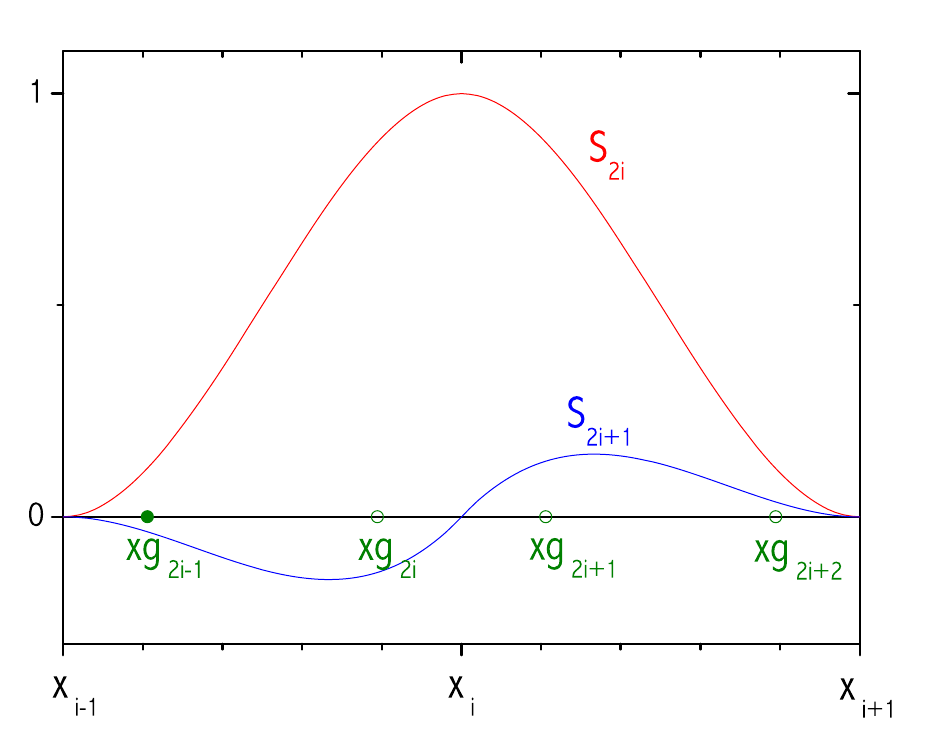} 
\includegraphics[width=80mm] {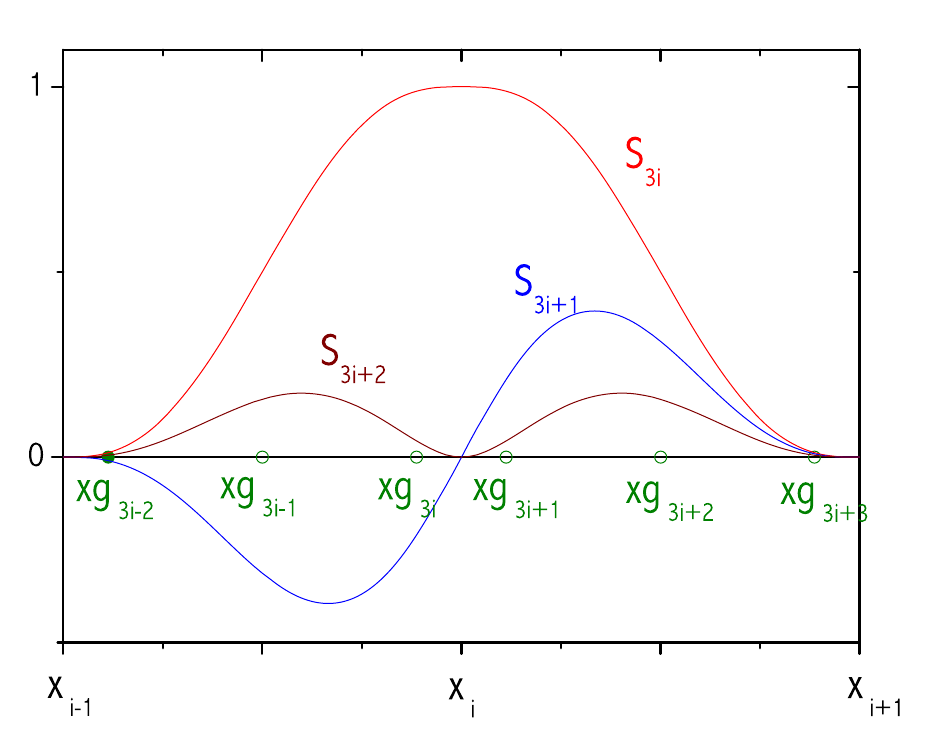}
\caption{The form of CHP (figure on the left) and QHP (figure on the right)
interpolants.}
\label{Fig_spline_func}
\end{figure}

In order for the $m$-th derivative to be continuous interpolant
polynomial functions should be of degree $m+1$ or higher. Since we
deal with a second-order differential equations, the spline
functions should have second order continuous derivatives. The
minimal order polynomials satisfying it are cubic ones. Therefore
we associate $k=2$ cubic Hermite polynomials (CHP) with each
breakpoint (see Fig. \ref{Fig_spline_func}), being defined as:
\begin{equation}
\begin{array}{cc}
\begin{array}{c}
\\
\text{for }X_{i-1}\leq x\leq X_{i} \\
\text{(with }r=\frac{x-X_{i-1}}{X_{i}-X_{i-1}}\text{)}%
\end{array}
& \left\{
\begin{array}{l}
S_{2i}(x)=r^{2}(3-2r) \\
S_{2i+1}(x)=-(X_{i}-X_{i-1})r^{2}(1-r)%
\end{array}
\right. \\
\begin{array}{c}
\\
\text{for }X_{i}\leq x\leq X_{i+1} \\
\text{(with }r=\frac{x-X_{i}}{X_{i+1}-X_{i}}\text{)}%
\end{array}
& \left\{
\begin{array}{l}
S_{2i}(x)=\left( 1-r\right) ^{2}(1+2r) \\
S_{2i+1}(x)=-(X_{i+1}-X_{i})r(1-r)^{2}%
\end{array}
\right.%
\end{array}%
\end{equation}

It turns to be an optimal choice~\cite{Boor_book}. However,
sometimes dealing with more acute wave functions or trying to
obtain better precision (especially, when expectation value of
kinetic energy is required), it is useful to use quintic Hermite
polynomials (QHP), having $k=3$ polynomials associated with each
breakpoint (see Fig. \ref{Fig_spline_func}):
\begin{equation}
\begin{array}{cc}
\begin{array}{c}
\\
\text{for }X_{i-1}\leq x\leq X_{i} \\
\text{(with }r=\frac{x-X_{i-1}}{X_{i}-X_{i-1}}\text{)}%
\end{array}
& \left\{
\begin{array}{l}
S_{3i}(x)=\left( 1-r^{3}\right) \left[ 1+3r\left( 1+2r\right) \right] \\
S_{3i+1}(x)=-(X_{i}-X_{i-1})r\left( 1-r^{3}\right) (1+3r) \\
S_{3i+2}(x)=\frac{1}{2}(X_{i}-X_{i-1})^{2}r^{2}(1-r)^{3}%
\end{array}
\right.%
\end{array}%
\end{equation}
\begin{equation}
\begin{array}{cc}
\begin{array}{c}
\\
\text{for }X_{i}\leq x\leq X_{i+1} \\
\text{(with }r=\frac{x-X_{i}}{X_{i+1}-X_{i}}\text{)}%
\end{array}
& \left\{
\begin{array}{l}
S_{3i}(x)=r^{3}\left[ 3(1-r)(3-2r)+1\right] \\
S_{3i+1}(x)=(X_{i+1}-X_{i})r^{3}(1-r)(4-3r) \\
S_{3i+2}(x)=\frac{1}{2}(X_{i}-X_{i-1})^{2}r^{3}(1-r)^{2}%
\end{array}
\right.%
\end{array}%
\end{equation}
Following the same procedure, one may easily construct polynomial
interpolants of even higher order (seventh, ninth,...), however
their relevance seems to be questionable. For the vast majority of
the applications related to the solution of second order
differential equations CHP interpolants are sufficient, also
proving to be the most efficient numerical procedure. Application
of the QHP interpolants is more expensive numerically, as it
results into denser matrices. In most cases QHP interpolants
provides only very moderate gain in accuracy compared to matrix
size equivalent CHP case.

Here I will figure out some useful properties of QHP and CHP interpolants.
First, one can notice that in each subinterval $i\equiv \left[ x_{i-1},x_{i}%
\right] $ there are only $2k$ non zero splines, therefore one
needs to sum at most $2k$ terms to reconstruct the functions value
at any given point:
\begin{equation}
f(x)=\sum\limits_{k\cdot (i-1)}^{k\cdot
(i+1)-1}C_{n}S_{n}(x)\qquad x\in \left[ x_{i-1},x_{i}\right].
\end{equation}
This feature turns to be very useful in numerical applications,
since it enables one to reduce number of arithmetic operations and
furthermore, when applied for solving systems of differential
equations, results in linear systems for sparse matrices. Sparse
matrices can be compactly stored, therefore considerably reducing
requirements of computer memory.

One can remark that it is easy to obtain the interpolated function
and its derivative values at the breakpoints, when QHP or CHP
interpolants are in use:
\begin{equation}
\begin{tabular}{cc}
$f(x_{i})=C_{k\cdot i}$ & $f^{\prime }(x_{i})=C_{k\cdot i+1}$ \\
$f^{\prime \prime }(x_{i})=C_{k\cdot i+2}$ & for QHP interpolants.%
\end{tabular}%
\end{equation}

These relations make implementation of the boundary conditions
rather straightforward. Furthermore, they can serve to interpolate
the functions, whose values and derivatives are known at the
selected breakpoints:
\begin{equation}
\begin{tabular}{cc}
$f(x)=\sum\limits_{i=0}^{N}\left[ f(x_{i})S_{2i}(x)+f^{\prime
}(x_{i})S_{2i+1}(x)\right] $ & for CHP. \\
$f(x)=\sum\limits_{i=0}^{N}\left[ f(x_{i})S_{3i}(x)+f^{\prime
}(x_{i})S_{3i+1}(x)+f^{\prime \prime }(x_{i})S_{3i+2}(x)\right] $ & for QHP.%
\end{tabular}%
\end{equation}

Beyond the flexibility to incorporate complicated boundary
conditions and manipulate the distribution of the breakpoints are
not the only assets of the spline collocation method. Spline
collocation method also offers possibility to factorize important
features of the described function. For instance, an unknown
function $F(r)$ might be approximated as:
\begin{equation}
F(r)=f(r)\sum_{j=0}^{k(N+1)-1}C_{j}S_{j}(r),
\end{equation}
where $f(r)$ is a chosen function intended to facilitate
interpolation of the function $F(r)$. In particular, when complex
scaling is in use the asymptote of the wave function is usually a
slowly decaying oscillating function, behaving as:
\begin{equation}
F(r\rightarrow\infty)\propto\exp(ikre^{i\theta})=\exp(ikr
cos\theta)\exp(-kr sin\theta).
\end{equation}
In this case it is very useful to factorize fast oscillating term, choosing
interpolation as:
\begin{equation}
F(r)=\exp(ikre^{i\theta})=\exp(ikr cos\theta)\sum_{j=0}^{k(N+1)-1}\tilde{C}%
_{j}S_{j}(r).
\end{equation}

\subsection{Lagrange mesh method \label{sec:ll_method}}

Gauss quadrature rules constitute one of the most popular and
efficient numerical technique to evaluate integrals. Gauss
quadrature is built for a specific interval (a,b) and for a
specific weighting function $w(x)$, by considering a family of the
orthogonal polynomials defined in this interval:
\begin{equation}
\int_{a}^{b}p_{k}(x)p_{n}(x)w(x)dx=\delta _{kn},
\end{equation}%
where $p_{k}(x)$ is an orthogonal polynomial of order $k$ with
respect to weighting function w(x) . A standard Gauss quadrature
constitutes of $N_{g}$ knots $x_{i}$ distributed within the
integration interval. These knots are the roots of the associated
orthogonal polynomial of order $N_{g}$; to each knot a weight
coefficient $w_{i}$ is associated. Such an quadrature is employed
to approximate the integral in the form:
\begin{equation}
\int_{a}^{b}h(x)w(x)dx\approx
\sum\limits_{i=1}^{N_{g}}w_{i}h(x_{i}). \label{eq:gauss_int}
\end{equation}%
It is easily demonstrated that if the function $h(x)$ is
polynomial of order $n\leq 2N_{g}-1,$ evaluation of the last
integral will be exact. Of the special importance are classical
Gauss-quadratures, summarized in table~\ref{tab:classGaussq}.
These quadratures are built for the so-called classical
polynomials, representing solutions of the self-adjoint
second-order differential equations. There are several assets to
employ classical quadratures; in particular that relates with the
simplicity to estimate positions of the quadrature knots and
associated weights. There also exist series of useful analytic
relations, which permits to calculate some important overlap
integrals. Nevertheless there exist vast potential to construct
non-classical quadratures, by selecting a smooth weighting
function $w(x)$ and an integration interval. A rich database of
such quadratures has been provided by W. Gautschi in his
repository at~\cite{PURR2399}.
\begin{table}[tbp]
\begin{center}
\caption{Definitions of the Gauss-quadratures based on the
classical polynomials.}
\label{tab:classGaussq}%
\begin{tabular}{|l|c|c|c|}
\hline
Type & w(x) & Interval & Limitations \\ \hline\hline
Gauss-Legendre & 1 & [-1,1] &  \\
Jacobi & $(1-x)^{\alpha }(1+x)^{\beta }$ & [-1,1] & $\alpha
>-1$; $\beta
>-1$ \\
\textrm{Generalized}\text{ }\textrm{{Laguerre}} & $x^{\alpha }\exp
(-x)$ &
[0,$\infty $) & $\alpha >-1$\\
\textrm{Generalized Hermite} & $x^{\alpha }\exp (-x^{2})$ &
($-\infty ,\infty$ ) & $\alpha >-1$ \\
\textrm{Exponential} & $x^{\alpha }$ & [-1,1] & $\alpha >-1$ \\
\textrm{Rational} & $x^{\alpha }(x+b)^{\beta }$ & [0,$\infty$ ) &
$\alpha
>-1$; $\beta +\alpha <-1$ \\
\textrm{Cosh} & $\frac{1}{\cosh (x)}$ & ($-\infty ,\infty$ ) &  \\
\hline
\end{tabular}%
\end{center}
\end{table}

Based on the ideas of Gauss quadrature and Lagrange interpolation
one can construct a very efficient numerical method to solve
integro-differential equations, popularly referred as Lagrange
mesh method \cite{Baye_bbl1,Baye_bible}. One may associate a
square-integrable basis with a Gauss quadrature defined in
eq.~(\ref{eq:gauss_int}), by:
\begin{equation}
f_{i}(x)=c_{i}\left( \frac{x}{x_{i}}\right) ^{n}\frac{L_{N_{g}}(x)}{(x-x_{i})%
}\sqrt{w(x)}, \label{eq:gauss_quad_lm}
\end{equation}%
with
\begin{equation}
L_{N_{g}}(x)=\prod\limits_{i=1}^{N_{g}}(x-x_{i}),
\end{equation}%
representing a characteristic polynomial of order $N_{g}$, built
for a weighting function $w(x);$ normalization coefficients
$c_{i}$ may be chosen to satisfy:
\begin{equation}
\int_{a}^{b}f_{i}(x)f_{i}(x)dx=1.
\end{equation}%
If required, basis functions might be regularized at the origin by
introducing a scaling factor $\left( \frac{x}{x_{i}}\right) ^{n}$.
One may
employ the same Gauss-quadrature\footnote{%
with the same number $N_{g}$ of knots and the same weighting function $w(x)$}%
$,$ to estimate a cross product of \ the basis functions :
\begin{equation}
\int_{a}^{b}f_{i}(x)f_{j}(x)dx\approx \sum\limits_{k=1}^{N_{g}}w_{k}\frac{%
f_{i}(x_{k})f_{j}(x_{k})}{w(x_{k})}=\delta _{i,j}w_{i}\left[ \frac{%
f_{i}(x_{i})}{\sqrt{w(x_{i})}}\right] ^{2}.
\end{equation}%
The last approximation becomes exact if $2N_{g}-1-2(N_{g}-1+n)\geq 0$; i.e. $%
n\leq 1/2$. For this case:
\begin{equation}
w_{i}=\left[ \frac{f_{i}(x_{i})}{\sqrt{w(x_{i})}}\right] ^{-2},
\end{equation}%
and the defined basis functions $f_{i}(x)$ are orthonormal in the
defined interval:
\begin{equation}
\int_{a}^{b}f_{i}(x)f_{i}(x)dx=\delta _{i,j}.
\end{equation}

\subsubsection{Evaluation of the matrix elements using the Langrange mesh
method}

In order to construct the matrix elements corresponding to some
local potential $V(x)$ one has to estimate:
\begin{equation}
O_{ij}=\left\langle f_{i}\left\vert \widehat{V}\right\vert
f_{j}\right\rangle =\int_{a}^{b}f_{i}(x)V(x)f_{j}(x)dx.
\end{equation}
This integral is conveniently realized if the same
Gauss-quadrature is employed to estimate the integral as one used
to construct basis functions. In this way:
\begin{eqnarray}
O_{ij} &=&\int_{a}^{b}f_{i}(x)V(x)f_{j}(x)dx \\
&\approx &\sum\limits_{k=1}^{N_{g}}w_{k}\frac{f_{i}(x_{k})\left[
V(x_{k})f_{j}(x_{k})\right] }{w(x_{k})}=V(x_{i})\delta
_{i,j}.\label{eq:V_matel_lm}
\end{eqnarray}

Projection of a given wave function $\phi (r)=F(r)/r$ on the
Lagrange-mesh basis is conveniently realized:
\begin{eqnarray}
F(r) &\approx &\sum\limits_{i=1}^{N_{g}}C_{i}f_{i}(r),  \\
C_{i} &=&\left\langle f_{i}\left\vert F\right. \right\rangle =\int_{a}^{b}%
\frac{F(r)}{r}\frac{f_{i}(r)}{r}r^{2}dr\approx \sum\limits_{k=1}^{N_{g}}w_{k}%
\frac{f_{i}(x_{k})F(x_{k})}{w(x_{k})}=w_{i}\frac{f_{i}(x_{i})F(x_{i})}{%
w(x_{i})}=\frac{F(x_{i})}{f_{i}(x_{i})}.\notag
\end{eqnarray}

 Expressions are slightly more complicated if the potential energy
operator is non-local. This situation arises when evaluating
matrix elements arising from non-local interactions. If
Gauss-quadrature rule is applied twice, one gets:
\begin{eqnarray}
V_{ij} &=&\int_{0}^{\infty }\left( r^{\prime }\right) ^{2}dr^{\prime
}\int_{0}^{\infty }\frac{f_{i}(r^{\prime })}{r^{\prime }}V(r^{\prime },r)%
\frac{f_{j}(r)}{r}r^{2}dr\approx \int_{0}^{\infty }\left( r^{\prime }\right)
^{2}dr^{\prime }\sum\limits_{k=1}^{N_{g}}\frac{w_{k}}{w(x_{k})}\frac{%
f_{i}(r^{\prime })}{r^{\prime }}V(r^{\prime },x_{k})f_{j}(x_{k})x_{k} \notag\\
&\approx &\sum\limits_{m=1}^{N_{g}}\frac{w_{m}}{w(x_{m})}f_{i}(x_{m})x_{m}%
\sum\limits_{k=1}^{N_{g}}\frac{w_{k}}{w(x_{k})}%
V(x_{m},x_{k})f_{j}(x_{k})x_{k}=\frac{\sqrt{w_{i}w_{j}}}{\sqrt{%
w(x_{i})w(x_{j})}}V(x_{i},x_{j})x_{i}x_{j}  \notag \\
&=&\frac{1}{f_{i}(x_{i})f_{j}(x_{j})}V(x_{i},x_{j})x_{i}x_{j}.
\end{eqnarray}

One of the greatest assets of Lagrange-mesh method is facility it
provides for estimating matrix elements. As follows from the
formulaes presented in this subsection matrix of the local
potentials are diagonal, while its estimation requires only values
of the potential energy at associated Gauss quadrature knots.
Potential matrix of non-local potentials is full, however
its estimation requires only knowledge of the potential energy values at
 $N_{g}\times N_{g}$ points of a double-quadrature mesh.

\pagebreak

\subsubsection{ Modifications of the Lagrange-mesh}

Lagrange-mesh method is very handy tool and rich variety of
different meshes might be constructed. All is needed is to
construct Gauss-quadrature for a chosen weighting function. The
rich database of Gauss-quadratures has been provided by W. Gautchi
in~\cite{PURR2399}. One may also construct his own quadrature
rule, to reflect better the problem at hand. One should mention
however that for a successful implementation of Lagrange-mesh
method Gausss quadrature knots and associated weights should be
determined very accurately, which turns to be far from trivial
numerical task. It requires very accurate estimation of the
polynomial integrals involving weighting function and may be
subject to possible numerical instabilities.

There exist other possibility to derive new-type of meshes by a
variable transformation. It is to construct basis $f(y)$, where y
is some variable obtained by smooth transformation of a variable
$x$ connected with one of the classical Lagrange-meshes.

\subsubsection{Solution of the Schr\"{o}dinger equation}

In order to solve complex-scaled radial Schr\"{o}dinger equation,
the radial functions $F^{(\theta )}(r)$, representing radial
dependence of the complex
scaled wave function are easily expanded using Lagrange-mesh basis functions~%
\cite{Baye_bible}:
\begin{equation}
F^{(\theta )}(r)=\sum_{i=1}^{N_{g}}C_{i}^{\theta }{f}_{i}\left(
r/h\right).
\end{equation}%
Since wave function $F^{(\theta )}(r)$ is complex, expansion
coefficients $C_{i}^{\theta }$ are complex numbers. To match
better the solution, radial scaling factor $h$ is introduced.

To solve radial Schr\"{o}dinger equation one needs to estimate
matrix elements of the kinetic energy $T_{ij}$, the potential
energy $V_{ij}$ as well as of the total energy $E_{ij}$. For this
problem it is practical to use Lagrange-meshes defined on the
infinite domain [0,$\infty$), like Lagrange-Laguerre one. For the
total energy, using Gauss-quadrature approximation with $N_{g}$
points, one gets:
\begin{equation}
E_{ij}=\int_{0}^{\infty }\frac{f_{i}(r/h)}{r}E\frac{f_{j}(r/h)}{r}%
r^{2}dr\approx h\sum\limits_{k=1}^{N_{g}}w_{k}\frac{f_{i}(x_{k})\left[
Ef_{j}(x_{k})\right] }{w(x_{k})}=h\delta _{i,j}E.
\end{equation}

The last relation is exact, as pointed out before, if
regularization factor n, defined in eq.~(\ref{eq:gauss_quad_lm}),
is chosen in the interval 0$\leq $n$\leq 1/2$. Otherwise norm
matrix is non-diagonal and if required might be estimated very
accurately using Gauss quadrature with a larger number of knots
$\widetilde{N}_{q}>N_{g}.$ For the kinetic energy operator exact
relation can be also derived based on Gauss-quadrature
approximation for many different meshes, if 0$\leq $n$\leq
3/2$~\cite{Baye_bible}. Otherwise one may also estimate it with an
ultimate accuracy using another Gauss quadrature constructed for a
much larger number of knots. For further convenience the basis
functions can be renormalized to get unit norm, by rescaling basis
coefficients $\widetilde{c}_{i}=c_{i}/\sqrt{h}$ from
eq.~(\ref{eq:gauss_quad_lm}).

\subsubsection{General remarks of practical interest}

\begin{figure}[h!]
\begin{center}
\includegraphics[width=14.cm]{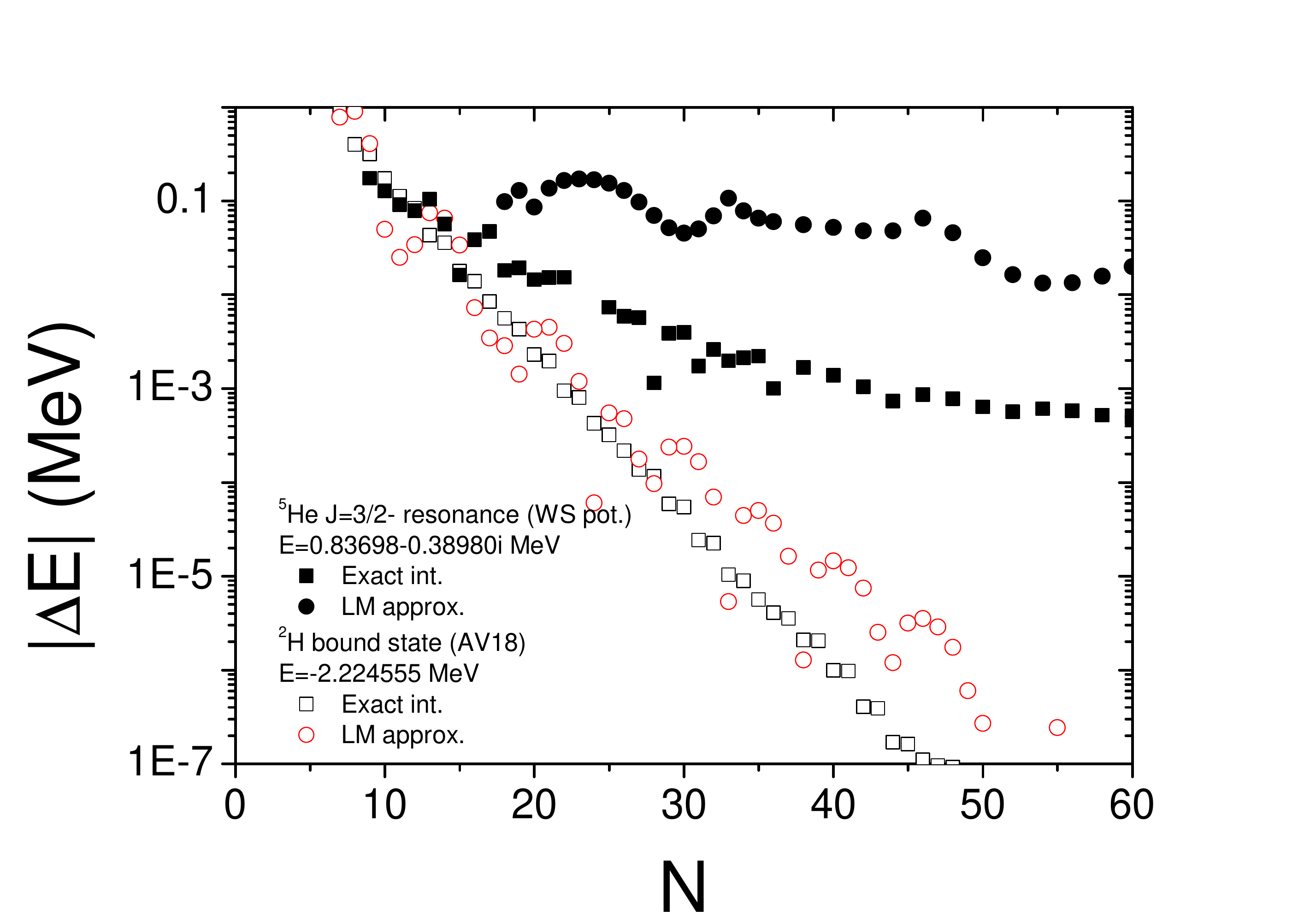}
\end{center}
\caption{Accuracy in calculated binding energy (resonance
positions) obtained using Lagrange-mesh functions. Two approaches
are compared:  when matrix elements of the potential matrix
calculated accurately (squares) and then estimated using
Lagrange-mesh approximation (circles). Two different physical
systems are also considered : binding energy of $^2$H nucleus
based on AV18 interaction, and  position of J$^\pi=3/2^-$
resonance in $^5$He based on Wood-Saxon potential from the
reference~\cite{bang1979}.} \label{Fig:2n_cs_accu}
\end{figure}

One of the assets of  Lagrange-mesh method is simple evaluation of
the matrix elements related with a potential energy. As could be
seen from eq.~(\ref{eq:V_matel_lm}), approximation based on
original Gauss quadrature (used to define Lagrange-mesh) gives
potential matrix in diagonal form. This turns to be very rough
approximation, however, as pointed out in~\cite{Baye_bible}, if
used in calculating binding energies delivers results of the
equivalent accuracy as a full variational method. This feature is
demonstrated in the figure~\ref{Fig:2n_cs_accu}, when comparing
two approaches to calculate binding energy of a deuteron  based on
AV18 nucleon-nucleon interaction~\cite{AV18}. Calculations using
diagonal potential matrix based on LM approximation (open circles)
are of equal accuracy to ones obtained using very accurate
estimation of the potential matrix (open squares), they also
provide very similar convergence pattern with respect to number of
Lagrange-Laguerre basis functions employed in the calculations.

Nevertheless this approximation starts faltering for the
applications related to the complex scaling method used for the
Hamiltonians based on short-range interactions. Indeed, as pointed
out in \ref{sec:cs_pot_trnsf} section, after the complex scaling
transformation short-range potentials start oscillating rapidly,
thus requiring stronger effort to evaluate their matrix elements.
This feat is demonstrated in the same figure~\ref{Fig:2n_cs_accu},
when comparing calculations of $^{5}$He J=3/2$^{-}$resonant state
position based on Wood-Saxon potential proposed by J. Bang and C.
Gignoux~\cite{bang1979}. Both methods converge when increasing
number of Lagrange-Laguerre basis functions employed in the
calculations. However convergence of the calculations based on
accurate estimate of the potential energy matrix (full squares) is
significantly faster and has much more regular pattern.

Based on the last observation for the applications related with
the complex scaling one should favor accurate estimation of the
matrix elements related to potential energy. Furthermore if the
employed Lagrange-mesh basis is analytic it is often beneficial to
use Cauchy theorem bringing the integral path along the contour
where argument of the potential energy is real (see
eq.~(\ref{eq:cauchy_vpot})).

Between different Lagrange-meshes, Lagrange-Laguerre quadrature
turns to be almost optimal choice for the complex scaling
applications. Generalized Lagrange-Laguerre basis is defined by
\begin{equation}
f_{i}(x)=c_{i}\left( \frac{x}{x_{i}}\right) ^{n}\frac{L_{N_{g}}^{\alpha }(x)%
}{(x-x_{i})}x^{\alpha /2}\exp (-x/2)
\end{equation}%
The power $\alpha $ should be chosen larger than $-1$. Exponential
factor provided by the weighting function of the Laguerre
polynomials is very well suited to describe asymptotic form of the
complex scaled wave functions. Indeed, one may demonstrate that
Lagrange-Laguerre mesh functions can be tuned to effectively
reproduce the shape of the complex-scaled outgoing free wave
$\exp(ikre^{i\theta })$ in its asymptote.

I have tested many different types of meshes in complex scaling
applications both classical as well as non-classical, but also
ones derived using variable-transformation. The bases which worked
the best were ones with the largest stretch of the knots
$s(N_{g})$, where stretch is for a given mesh is defined by the
ratio $s(N_{g})=x_{Ng}/x_{1}$. This is not surprising, for many
reasons and in particular related with difficulties in
transforming the potential energy -- the complex scaling angle
parameter usually should be kept small. This feat results the
complex scaled wave functions to be decaying very slowly $\psi
^{\theta }(r\rightarrow \infty )\propto \exp(-kr\sin \theta )$,
and thus requiring very extended meshes to encompass them. On the
other hand, if one works with the short-range potentials,
important density of the knots is required at the origin to follow
evolution of the potential energy. One may
easily see that for Lagrange-Laguerre mesh the stretch of the knots $%
s(N_{g})$ shrinks once increasing value of the power $\alpha$.
Therefore in complex-scaling applications it is beneficial to keep
$\alpha $ small, even negative, compensating regularization of the
systems wave function with the parameter $n$.

\bigskip

\chapter{Description of the resonant states}
\section{ Resonances in the e$^+$e$^-$p
system\label{sec_res_eep}\\ \footnotesize{\textit{(Results
presented in this section are based on the
study~\cite{Lazauskas_spec_iss})}}}

 There
is considerable speculation as to why the observable universe is
composed almost entirely of ordinary matter, as opposed to an
equal mixture of matter and antimatter. This asymmetry of matter
and antimatter in the visible universe is one of the great
unsolved problems in physics. There is therefore a natural
interest in producing and manipulating the simplest structures of
antimatter with an aim to compare their properties with an
ordinary matter.  In this line production of the
antihydrogen($\bar{\mathrm{H}}^+$) atoms presents a vital step.
Hot  antihydrogen has been produced and detected for the first
time in the 1990s. ATHENA collaboration produced cold antihydrogen
in 2002. For the first time it was  trapped by the Antihydrogen
Laser Physics Apparatus (ALPHA) team at CERN in 2010, allowing to
perform some measurements related to its structure and other
important properties. ALPHA, AEGIS, and GBAR plan to further cool
and study antihydrogen atoms.

Due to the extremely low yield of ($\bar{\mathrm{H}}$) atom
production, and high opportunity cost of using a particle
accelerator there is strong interest in optimizing experimental
conditions, in order to favor higher production yield. Good
knowledge of the reaction mechanism is essential. Production of
the antihydrogen is due to charge exchange three-body reaction
between antiprotons and positronium (Ps) atoms (hydrogen like atom
composed of electron and positron) :
\begin{equation}
\bar{p}+Ps^*\rightarrow e^- +\bar{H}^*, \label{eq:pps_eh}
\end{equation}
The positronium (and/or antihydrogen) might be produced both in
ground or one of the excited states, wherefore in the last
equation these atoms are denoted with asterisk.

By studying electron-Hydrogen scattering M. Gailitis and R.
Damburg pointed out existence of the oscillations in the
scattering cross section~\cite{Gailitis}, which are generated
close to each degenerate Hydrogen-atom threshold. These
oscillations are due to the rise of the long-range $1/R^2$
effective potential, which couples degenerate Hydrogen-atom
levels. Just below the degenerate threshold, oscillations are
caused by the presence of an infinite number of Feshbach
resonances, whose relative to threshold energies form a
logarithmic sequence. These resonances are common features for the
charged particle scattering on Hydrogen-atom like structures.
Resonances of this kind are also encountered in the
antiproton-positronium collisions, while their presence might turn
out to be important in boosting antihydrogen production cross
section, as originally pointed out in~\cite{Hu_PRL}.

Hamiltonian describing eq.~(\ref{eq:pps_eh}) is composed of the a
sum of the particle kinetic energies and the Coulomb potentials
\begin{equation}
H=H_{0}+\sum_{i<j}\frac{Z_{i}Z_{j}}{r_{ij}},
\end{equation}%
where $r_{ij}$ is the distance between the particles $i$ and $j$, while $%
Z_{i}$ indicates a charge of the particle $i$. Here, I use atomic
units setting $\hbar =e=m= 4 \pi \varepsilon_0 =1$. The perimetric
coordinates, introduced by James and Coolidge~\cite{Perim}, and
defined in eq.~(\ref{eq_perim_cor}) turn out to be a very
practical choice to express the system's wave function. By
limiting ourselves to the total angular momentum $L=0$ states
($S$-waves) the wave function of the system becomes independent of
the Euler's angles whereas the matrix elements of the kinetic
energy operator are expressed in eq.~(\ref{eq_h0_perim}).

In order to predict positions and widths of the resonant states we
employ the complex scaling method has been used. CS transformed
resonance wave functions $\widehat{S}\Psi(u,v,z)$ are
exponentially bound if the complex scaling parameter satisfies the relation $%
-\frac{1}{2} \mathrm{arg}(E_{res}-E_{th})<\theta<\pi/2$, where
$E_{th}$ denotes the closest threshold in the reaction
(\ref{eq:pps_eh}). The three-dimensional Schr\"{o}dinger equation
is solved using the Lagrange-mesh method~\cite{Baye_bible},
described in section \ref{sec:ll_method}. The necessary integrals,
involved in estimating matrix elements of the potential energy,
were estimated by using Gauss approximation associated with a
chosen mesh. The three-dimensional wave function function is
discretized as:
\begin{equation}
\Psi(u,v,z)=\sum_{i=1}^{N_i}\sum_{j=1}^{N_j}\sum_{k=1}^{N_k}
C_{ijk} f_i(u/h_u)f_j(v/h_v)f_k(z/h_z),
\end{equation}
where $C_{ijk}$ represent the expansion coefficients, $h_u,h_v$
and $h_z$ are scaling parameters. The basis functions are defined
on a grid based on Lagrange-Laguerre quadrature
\begin{equation}
f_i(x)=(-1)^i c_i (x_i)^{1/2}\frac{L_N(x)}{x-x_i} e^{-x/2}\;,
\end{equation}
where $L_N(x)$ is a $N$ degree Laguerre polynomial, whereas $x_i$,
as usual, denotes its roots.

Eigenvalues, representing $S$-wave resonant states, of the
$\mathrm{e}^+\mathrm{e}^-\mathrm{\overline{p}}$ system are
summarized in Table~\ref{tab:resen}. These values were
 calculated
using the complex scaling method. I compare obtained results with
the most accurate values found in the literature. As
aforementioned, the Feshbach resonances in this system can be
grouped into families, each of them being associated to each
degenerate atom-charged particle threshold. Furthermore,
approximate discrete symmetry indicates that the positions and
widths of the resonances in each family should approximately
satisfy the discrete scaling invariance:
\begin{equation}
\frac{\mathrm{Re}(E_f^i)-E_f^{thr}}{\mathrm{Re}(E_f^{i+1})-E_f^{thr}}\approx\frac{\mathrm{Im}(E_f^i)}{%
\mathrm{Im}(E_f^{i+1})}\approx d_f,
\end{equation}
where $E_f^{thr}$ is the position of the threshold $f$ whereas
$E_f^i$ is an eigenvalue of the $i^{th}$ resonance belonging to
the family-$f$. Discrete scaling coefficient $d_f$ is related to
the dipole-coupling strength between degenerate channels and can
be determined analytically. The value of this coefficient is
usually much larger than 1. Therefore, numerically, one is able to
identify only a few resonances in each sequence. Other resonances
have very extended wave-functions and are situated too close to
the threshold to be determined numerically. On the other hand
resonances situated very close to the threshold should disappear
once relativistic corrections are taken into account and the
degenerate thresholds become separated.

\begin{figure}
\begin{center}
\includegraphics[scale=0.44]{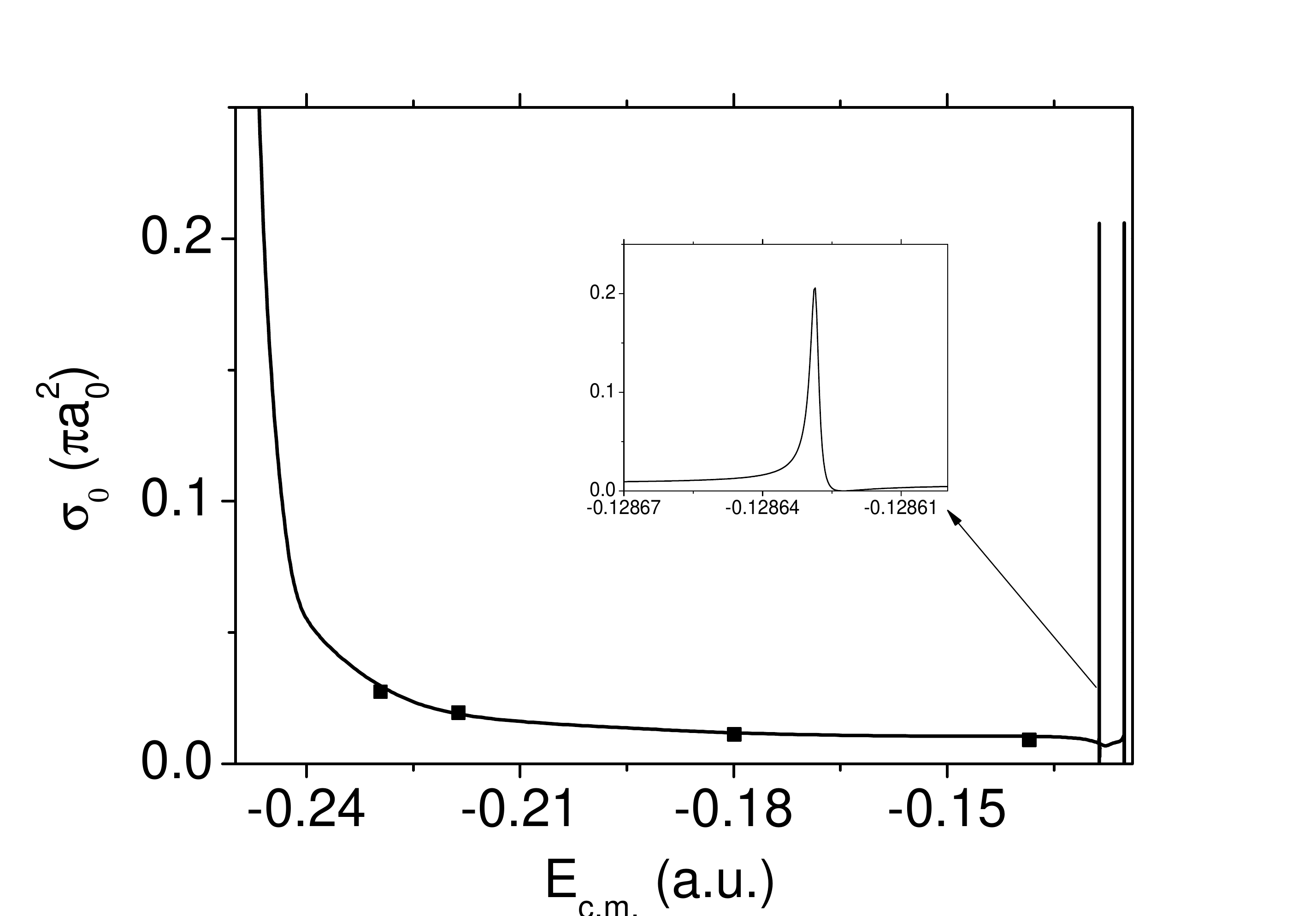}
\normalsize \caption{\label{fig:h_prod_cs} $S$-wave antihydrogen
formation cross section for antiproton-positronium collisions in
the Ore gap. The present results are compared to the ones obtained
by Hu~\cite{Hu_frs} that are depicted by the full squares.}
\end{center}
\end{figure}

\begin{table}[tbp]
\caption{$L=0$ resonances of
$\mathrm{e}^+\mathrm{e}^-\mathrm{\overline{p}}$ system and their
respective thresholds. The notation $a[b]$ means $a\times10^b$.}
\label{tab:resen}%
\vspace{0.25cm}
\begin{center}
\begin{tabular}{|l|ll|lll|}
\hline
& \multicolumn{2}{|c|}{This work} & \multicolumn{3}{c|}{Literature} \\
Threshold & -$\mathrm{Re}(E_{res})$ & $\Gamma$/2 &
-$\mathrm{Re}(E_{res})$ & $\Gamma$/2 & Ref.
\\ \hline\hline
$\mathrm{\overline{H}}(n=2)$ & 0.128622631 & 3.3283[-5] & 0.128623 & 3.33[-5] & \cite{Umair} \\
0.124932 & 0.1251318 & 1.82[-6] & 0.125132 & 2.50[-6] & \cite{Umair} \\
\hline $\mathrm{Ps}(n=2)$ & 0.07513977 & 1.67290[-4] & 0.075140 &
1.67[-4] & \cite{Umair}
\\
{0.0625} & 0.0658293 & 8.127[-5] & 0.065830 & 8.06[-5] & \cite{Umair} \\
& 0.0633866 & 2.494[-5] & 0.063387 & 2.48[-5] & \cite{Umair} \\
& 0.06274 & 6.9[-6] & 0.0627218 & 6.89[-6] & \cite{Hu_jmp} \\
\hline
$\mathrm{\overline{H}}(n=3)$ & 0.05802577 & 3.1057[-4] & 0.058059 & 2.86[-4] & \cite{Umair} \\
{0.055525} & 0.0560311 & 6.399[-5] & 0.056034 & 6.40[-5] & \cite{Umair} \\
& 0.05564 & 8.77[-5] & 0.055571 & 9.45[-5] & \cite{Umair} \\
\hline
$\mathrm{\overline{H}}(n=4)$ & 0.03853098 & 2.3837[-5] & 0.038536 & 2.50[-5] & \cite{Umair} \\
0.031233 & 0.03393264 & 2.3938[-5] & 0.033942 & 2.8[-5] & \cite{Umair} \\
& 0.032244 & 8.08[-6] & 0.032294 & 1.29[-5] & \cite{Umair} \\
& 0.03184 & 2.45[-5] & 0.031843 & 2.58[-5] & \cite{Umair} \\
& 0.031649 & 1.6[-6] & 0.031617 & 2.32[-5] & \cite{Umair} \\
\hline
\end{tabular}%
\end{center}
\end{table}
It is natural to ask the question if these resonances may have
non-negligible impact on the antiproton capture cross section. In
~\cite{Hu_PRL,Hu_jmp} it has been demonstrated that the Gailitis
oscillations lead to a rapid rise of the cross section just above
the $\mathrm{Ps}(n=2)$ threshold, whereas the same authors
concluded that the resonances situated below $\mathrm{Ps}(n=2)$
threshold had small effect on the total antiproton capture cross
section. In figure \ref{fig:h_prod_cs} the $S$-wave antihydrogen
production cross section is depicted in the Ore gap\footnote{The
energy interval between the positronium formation threshold and
the first excitation threshold of the target atom, is referred to
as the Ore gap.} region, calculated as described in reference
\cite{Lazauskas_spec_iss}. One may clearly identify two narrow
resonances situated just below the $\overline{\mathrm{H}}(n=2)$
threshold, whose position and width coincide well with the values
provided in the Table~\ref{tab:resen}. The $S$-wave antihydrogen
production cross section is enhanced by a factor 20 at the
resonance, reaching a value of $\sim0.2\pi a^2_0$. One may argue
however that this effect is largely due to the smallness of the
$S$-wave cross section. Indeed, antihydrogen production cross
section in the Ore gap region is relatively large $\sim4\pi a^2_0$
and is dominated by the contribution from the higher partial
waves~\cite{Hu_frs}. Therefore, the $S$-wave resonances have a
limited practical impact. Nevertheless, a very similar behavior is
expected for the resonances in higher partial waves and thus, one
may expect a very sizeable impact of the latter on the cross
section.

\section{Three-neutron resonant states \\ \footnotesize{\textit{(Results presented in this section are based on
the study~\cite{LC05_3n})}}}

 Possible existence of the pure
neutron nuclei is a long standing ambiguity in nuclear physics.
Neutron-neutron (\emph{nn}) scattering length is negative and
rather large $a_{nn}=-18.59\pm0.40$ fm~\cite{nn_scl4}, indicating
that this system is almost bound in $^{1}\emph{S}_{0}$ state. It
still contains a signature of a bound state - a virtual one -- just $\approx$%
100 keV above the threshold. Then it is expected that adding a few
additional neutrons one can finish by binding multineutron, as it
happens in other pure-fermionic system, namely clusters of He
atoms~\cite{Guardiola_00}. This is a reason for from time to time
rising turmoils in the community of nuclear
physics~\cite{Oglobin,Marques,KisamoriPH,Kisamori_PRL}.
Nevertheless, weakness of nuclear interaction in higher partial
waves (namely \emph{P} and \emph{D}), in comparison with
centrifugal energy terms they bring with, excludes the theoretical
mechanism of binding `virtual' dineutrons
together~\cite{These_Rimas_03}. Non-existence of small bound
multineutron clusters seems to be settled out
theoretically~\cite{These_Rimas_03,Pieper,Bertulani,Natasha}.
Still the existence of resonant states in such nuclei, which can
have observable effects, can not be straightforwardly eliminated
and continue to provoke some controversial debates
\cite{LC05_3n,PhysRevC.72.034003,Shirokov,Plosczaj,Arnas_3n}.

In spite of the numerous experimental and theoretical studies that
exploit different reactions and methods, the situation concerning
few-neutron resonances is not firmly established. One does not
have clear ideas even for the simplest case: three-neutron
compound. A nice summary on the three-neutron system status up to
1987 can be found in~\cite{3N_analysis}. A few more recent
experimental studies have not provided any conclusive results
either. In~\cite{He_3n_97} analyzing the process
$^{3}He(\pi^{-},\pi^{+})3n$ no evidence of a three-neutron
resonant state has been found. The claims~\cite{3n_res_exp_70} to
explain differential cross sections of double charge exchange
process in $^{3}$He by the existence of a broad $E=(2-6i)$ MeV
three-neutron resonance were recently criticized by a more
thorough experimental study~\cite{3n_res_exp_86}. Nevertheless
this study further
suggested existence of a wide resonance at even larger energies with $%
E_{r}\approx(20-20i)$ MeV.

There were several theoretical efforts to find $^{3}$n and $^{4}$n
resonances. A variational study based on complex-scaling and
simplified nucleon-nucleon (\emph{NN}) interaction was carried
through in~\cite{Csoto} with the prediction of $^{3}$n resonance
at $E=(14-13i)$ MeV for a $J^{\pi }=3/2^{+}$ state. On the other
hand no real $^{3}$n, and even $^{4}$n, resonances were found by
Sofianos et al.~\cite{Sofianos} using MT I-III potential model;
only existence of some broad subthreshold resonances was pointed
out. Realistic interaction models however can provide different
conclusions. These models contain interaction in \emph{P}- and
higher partial waves; due to the necessity of antisymmetric wave
functions -- a crucial ingredient in binding pure fermionic
systems. The only study performed in part using realistic
potentials was carried by Gl\"{o}ckle and Wita\l
a~\cite{Witala:3n}. These authors were not able to find any real
three-neutron resonances. However due to some numerical
instabilities full treatment of $^{3}$n system has not been
accomplished and conclusions have
been drawn basing only on phenomenological Gogny interaction model~\cite%
{Gogny}. The reference~\cite{Glockle_Hem} is probably the most
complete study of three neutron system. In this work full
trajectories for $^{3}$n states with $|J| \leqslant $3/2, obtained
by artificially enhancing nn interaction to bind three-neutron,
have been traced. Though once again simplified to finite rank NN
interaction model have been used. In this section I explore all
$^{3}$n quantum states upto $|J| $=5/2 and this time fully relying
on realistic \emph{NN} interactions.


Before analyzing three neutron system it is useful to discuss the
basic properties of dineutron and \emph{nn} interaction in
general. As mentioned above, dineutron is almost bound in
$^{1}\emph{S}_{0}$ state, one should enhance nuclear potential
only by the factor $\gamma\sim1.08$ to make it bound, see
Table~\ref{gamma_2N}. However spherical symmetry of this state
determines that when reducing $\gamma$ to 1 (i.e. to real value of
the potential) the bound state pole moves further down, staying on
the imaginary k axis, and thus becomes a virtual state and not a
resonance. The approximate position of this virtual state can be
already evaluated from the \emph{nn} scattering lengths by using
relation $E_{virt}\approx\frac{\hbar^2}{ma^2}$: these approximate
and exactly calculated virtual state energies are summarized in
Table~\ref{long_de_diff}. One has very good agreement for the
enhancement factors $\gamma$, as predicted by different local
\emph{NN}-interaction models. Only AV14 result slightly deviates
from the other model predictions, which is caused by charge
invariance assumption in this model. This potential being adjusted
to
reproduce neutron-proton (\emph{np}) scattering data, ignores the fact that experimental $^{1}\emph{%
S}_{0}$ \emph{nn} scattering length is smaller in magnitude than
\emph{np} one~\cite{nn_scl4}.

In fact, multineutron physics, being in low energy regime, is
dominated by large \emph{nn} scattering length ($a_{nn}$). Systems
wave function has only small part in the interaction region
($r_0<<a_{nn}$) and therefore marginally depends on a particular
form of \emph{nn} potential in $^{1}\emph{S}_{0}$ waves can take,
provided $r_0$ and $a_{nn}$ are fixed~\cite{These_Rimas_03}. On
the other hand $r_0$ is controlled by the theory (the pion-range),
whereas $a_{nn}$ is constrained by experiment. These effective
range theory arguments~\cite{Platter} shows that one should not
count on the modifications of $^{1}\emph{S}_{0}$ waves in order to
favor existence of bound or resonant multineutron  states.
\begin{figure}
\begin{center}
\includegraphics[width=14. cm]{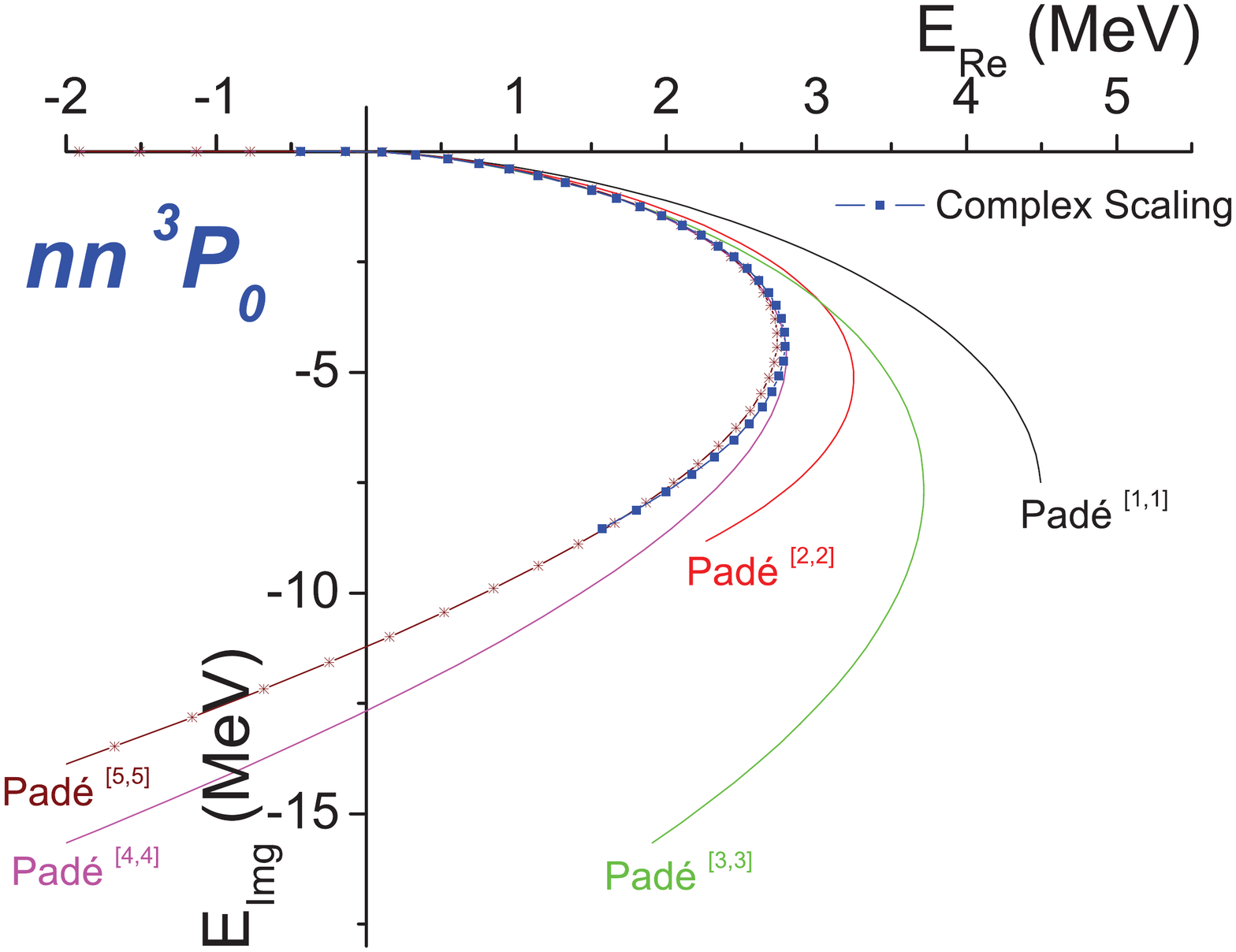}
 \caption{ \label{Fig:2n_ACCC_CS}Comparison of ACCC and CS method results for
$^{3}\emph{P}_{0}$ \emph{nn} resonance trajectories. ACCC results
with various order Pad\'e extrapolants extending from $\gamma$=6.1
to 1.0 are presented by solid lines. CS values are presented by
large distinct points, obtained by reducing  enhancement factors
$\gamma$ from 6.1 to 2.7 in step of 0.1. Small snowflake-like
points correspond to [5,5] Pad\'e extrapolation used in ACCC for
$\gamma$ ranging from 6.1 in step of 0.1. These points are already
very close to CS ones, whereas adding few additional terms in
extrapolation perfect agreement between ACCC and CS results can be
obtained (see next figure).}
\end{center}
\end{figure}

\begin{figure}[h!]
\begin{center}
\includegraphics[width=14.cm]{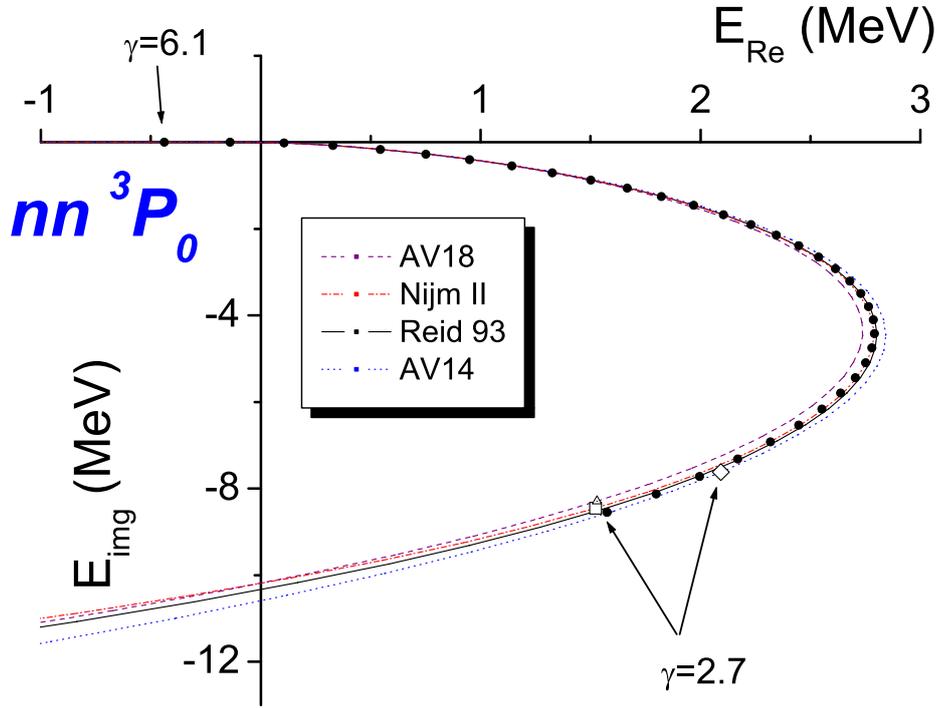} \vspace{-1.2 cm}
\end{center}
\caption{Dineutron $^{3}\emph{P}_{0}$ resonant state trajectories
in complex energy plane for  AV14, NijmII, Reid 93 and AV18
\emph{nn} interactions. Different points in
Reid 93 potential curve correspond to different values of enhancement factors $%
\protect\gamma$ changing from 6.1 to 2.7 in step of 0.1 obtained
with CS method. Continuous lines represent ACCC results. ACCC and
CS results superimpose up to $\gamma$=2.7 point, limit of CS
methods applicability.} \label{Fig:2n_res_3P0}
\end{figure}

\begin{figure}[h!]
\begin{center}
\includegraphics[width=14.cm]{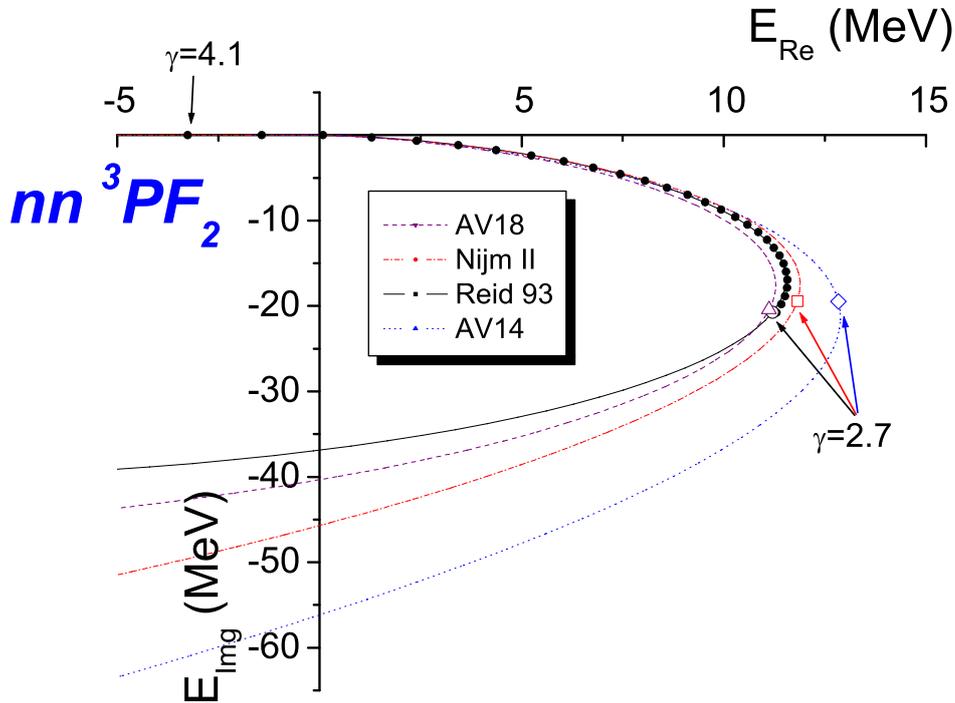} \vspace{-1.2 cm}
\end{center}
\caption{Dineutron $^{3}\emph{P}_{2}-^{3}\emph{F}_{2}$ resonant
state trajectories in complex energy plane for AV14, AV18 and Reid
93 \emph{nn} interactions. Different points in Reid 93 potential
curve correspond different values of enhancement factors
$\protect\gamma$ changing from 4.1 to 2.7 in step of 0.05.
Resonance trajectories beyond $\gamma$=2.7 point are presented
using ACCC method results, while for $\gamma\geqslant$2.7 CS and
ACCC calculated values are in full agreement.}
\label{Fig:2n_res_3PF2}
\end{figure}

\begin{table}[tbp]
\begin{center}
\caption{Critical enhancement factors $\protect\gamma$ required to
bind dineutron in various states and for different \emph{NN}
realistic interaction models in use.}
\label{gamma_2N}%
\vspace{0.5 cm}
\begin{tabular}{|l|llll|}
\hline & Nijm II & Reid 93 & AV14 & AV18 \\ \hline\hline
$^{2}n(^{1}S_{0})$ & 1.088 & 1.087 & 1.063 & 1.080 \\
$^{2}n(^{3}P_{0})$ & 5.95 & 5.95 & 5.46 & 6.10 \\
$^{2}n(^{3}PF_{2})$ & 3.89 & 4.00 & 4.30 & 4.39 \\
$^{2}n(^{1}D_{2})$ & 9.28 & 9.22 & 9.54 & 10.20 \\ \hline
\end{tabular}%
\end{center}
\end{table}

\begin{table}[tbp]
\caption{Nuclear model predictions for \emph{nn} scattering length
(in fm) as well as corresponding virtual state (in MeV), evaluated
from scattering length and calculated exactly.}
\label{long_de_diff}%
\vspace{0.5 cm}
\begin{center}
\begin{tabular}{|l|llll|}
\hline & Nijm II & Reid 93 & AV14 & AV18 \\ \hline\hline
$a_{nn}(^{1}S_{0})$ & -17.57 & -17.55 & -24.02 & -18.50 \\
${\hbar^2}/(ma^{2}_{nn})$ & 0.134 & 0.135 & 0.072 & 0.121 \\
$E_{virt}(^{1}S_{0})$ & 0.1162 & 0.1165 & 0.0647 & 0.1055 \\
\hline
\end{tabular}%
\end{center}
\end{table}

\begin{table}[h]
\caption{Enhancement factors $\gamma'$ at which dineutron
resonances  become subthreshold ones. Values in MeV of imaginary
energy for which such transition is effected E$_{img}(\gamma')$
and subthreshold resonance position E$_{res}$ for real \emph{nn}
interaction (i.e. at $\gamma=1.0$). These results are obtained
using ACCC method.}
\label{gamma_res_2N}%
\begin{center}
\footnotesize{
\begin{tabular}{|r|cccc|cccc|}
\hline

    & \multicolumn{4}{c|}{$^{3}P_{0}$} & \multicolumn{4}{c|}{$^{3}PF_{2}$} \\ 

                   & Nijm II & Reid 93 & AV14 & AV18  & Nijm II & Reid 93 & AV14    & AV18 \\ \hline\hline
$\gamma'$          & 2.27    & 2.26   & 2.08 & 2.24  & 1.64    &  1.71   &  1.46   & 1.73  \\
E$_{img}(\gamma')$ & -10.2   & -10.3  & -10.6& -10.2 & -45.6    & -36.9   & -56.2   & -40.3   \\
E$_{res}(1.0)$&-14.1-17.2\textit{i}&-14.2-18.5\textit{i}&-10.3-18.1\textit{i}&-12.1-18.0\textit{i}
           & -20.5-64.8\textit{i}& -15.9-39.9\textit{i}& -17.9-80.1\textit{i}&-34.1-45.4\textit{i}\\
\hline
\end{tabular}}
\end{center}
\end{table}

\emph{P}-waves of \emph{nn} interaction are extremely weak, this
turns to be a major reason why multineutrons are not
bound~\cite{These_Rimas_03}. Neutron-neutron interaction in
$^{3}\emph{P}_{1}$ channel is even repulsive, whereas potentials
in $^{3}\emph{P}_{2}-^{3}\emph{F}_{2}$ and $^{3}\emph{P}_{0}$
channels should be multiplied by considerable factors
$\gamma=[3.9-4.4] $ and $[5.5-6.1]$ respectively (see
Table~\ref{gamma_2N}), to force dineutron's binding. Presence of
centrifugal terms in these channels results that these
artificially bound states turn into resonances when factor
$\gamma$
 is slightly reduced from the critical values presented above.

Calculations employing the CS method for the realistic NN
potentials may be successful only for relatively narrow
resonances. As explained in section~\ref{sec:cs_pot_trnsf},
presence of short-range regulators in these potentials make these
potential divergent once large CS angles are employed. This was
also the reason to concentrate on Reid93 interaction in this
study, which turns to be the most compliant to CS transformation.
In order to explore the broader structures Analytic Continuation
in the Coupling Constant (ACCC) method is
used~\cite{KK77,Kukulin_PL78,Kukulin_book}, which allows via
Pad\'e extrapolation to extend trajectories of S-matrix poles
emerging from the bound state region. For the details about
implementation of ACCC method one may refer to the more complete
description of this study~\cite{LC05_3n}.

Resonance (S-matrix pole) trajectories for the dineutrons,
obtained when combining CS and ACCC methods, are traced in
figures~\ref{Fig:2n_res_3P0} and~\ref{Fig:2n_res_3PF2}  for the
Nijm II, Reid 93, AV14 and AV18 models. In fact, using high order
Pad\'e extrapolants and accurate input of ${^2n}$ binding energies
for ACCC method we obtain perfect agreement between two different
techniques. CS method, due to requirement to scale with an
increasing angle $\theta$, was applied only up to enhancement
factor $\gamma=2.7$ values. This value corresponding resonance
positions are marked in figures. One should note that resonance
trajectories have very similar shapes. First, when reducing
$\gamma $ (being close to one binding dineutron) imaginary energy
part of the resonance speeds-up and then continues to fall
linearly with the enhancement $\gamma$. On the other hand real
energy part of the resonance first grows linearly with enhancement
factor being reduced from its critical value (the one binding
three-neutron). Afterwards it temporary saturates reaching its
maxima. Further reducing enhancement factor real energy part of
the resonance quickly vanishes and becomes negative. ACCC method
provided resonance trajectories were extended up to $\gamma=1.0$
points; for these values dineutron resonances are already deep
subthreshold ones, whereas transition to third energy quadrant
happens well before the enhancement $\gamma$ turns to 1.  Some
resonance trajectory properties obtained using ACCC method are
summarized in Table~\ref{gamma_res_2N}.

Therefore existence of observable $\emph{P}$-wave dineutron
resonances should be excluded: only subthreshold ones with large
widths persist, making such structures physically of little
interest. Still we would like to remark that some few-nucleon
scattering calculations indicate that for a good description of
$3N$ and $4N$
scattering observables stronger \emph{NN} $\emph{P}$-waves are required~\cite%
{P_waves_Gloe,P_waves_Pisa,These_Rimas_03}. However these
discrepancies can be removed by modifying \emph{NN} \emph{P}-waves
by less than 10\%; nevertheless within such enhancements dineutron
resonances always remain in the subthreshold region.

One should quote the astonishing similarity for the \emph{P}-wave
dineutron resonance trajectories, when different realistic
\emph{NN}-interaction models with quite different shapes (see
Fig.~\ref{fig_local_sc_pot}) are  in use. $^{3}\emph{P}_{0}$
resonance curves for all three interaction models superimpose,
whereas in  $^{3}\emph{P}_{2}-^{3}\emph{F}_{2}$ case they separate
only when very large resonance energies are reached. Enhancement
factors employed in tracing these curves are unphysically large
and produce very
broad resonances: ($^{3}\emph{P}_{0}$ resonance slips into adjacent energy quadrant at $%
E_{img}\sim 10$ $MeV$, while in
$^{3}\emph{P}_{2}-^{3}\emph{F}_{2}$ case this value explodes
beyond 30 $MeV$).

Two neutron system, when having orbital angular momentum $\ell$=2,
can be realized only in singlet state ($^{\emph{1}}$\emph{%
D}$_{\emph{2}}$). This state is dominated by large centrifugal
terms; enhancement factors $\gamma$ for this wave should be
considerably large in order to overcome these terms and bind
dineutron, see the last line of Table~\ref{gamma_2N}. Effective
potentials, containing centrifugal energy,  in this and higher
angular
momentum \emph{nn} partial waves:%
\begin{equation*}
V_{eff}(r)=V_{nn}(r)+\frac{\hbar^{2}}{m_{n}}\frac{\ell(\ell+1)}{r^{2}}
\end{equation*}
are smoothly decreasing functions, without any dips. This is a
crucial fact, why dineutron can not be resonant in
$\ell\geqslant$2 states.


\bigskip

The spline collocation method employed here to solve Faddeev
equations describing 3n systems leads to solution of a large scale
linear algebra problem, well beyond the outreach of direct linear
algebra methods. We were unable to invert directly 3n matrices in
order to obtain all the eigenvalue spectra. Only a few specific
eigenvalues of the discretized 3n Hamiltonian could be extracted
when applying iterative linear algebra methods. These techniques
do not allow to separate a-priori  eigenvalues related to the
resonances from the spurious ones related to the rotated continuum
in CS method. In order to force numerical process converge to the
resonance position one should provide for it a rather accurate
guess value. This feat obliged me to follow the procedure employed
in~\cite{Witala:3n}: first three-neutron is bound artificially by
making \emph{nn} interaction stronger, and then gradually removing
additional interaction follow the trajectory of this state. Note,
that in bound state calculations one can use linear algebra
methods determining extreme eigenvalues of the spectra (as Lanczos
or Power-method), whereas resonance eigenvalue is not anymore an
extreme one in CS matrices.

 By enhancing \emph{nn}-potential in $^{\emph{1}}$\emph{S}$%
_{\emph{0}}$ channel one is not able to bind three neutrons
without first binding dineutron. On the other hand, as quoted
before, this wave is controlled together by theory and experiment,
whereas modification of its form can not affect multineutron
physics. Three-neutron can neither be bound if we keep
$^{\emph{1}}$\emph{S}$_{\emph{0}}$ interaction unchanged, whereas
multiply all \emph{nn} \emph{P}-waves with the same enhancement
factor. In this case dineutron is first bound in
$^{3}\emph{P}_{2}\emph{-}^{3}\emph{F}_{2}$ channel. Then we tried
to enhance only one of $\emph{P}$ channels, whereas keeping the
natural strengths for the other ones. The $^{3}\emph{P}_{1}$
channel is purely repulsive and the enhancement of this wave can
not give any positive effect. The enhancement of
$^{3}\emph{P}_{0}$ wave gives null result as well: dineutron is
always bound before any of $^{3}n$ states is formed. By enhancing
$^{3}\emph{P}_{2}\emph{-}^{3}\emph{F}_{2}$ channel we managed to
bind $^{3}n$ only in $\frac{3}{2}^{-}$ state, without first
binding dineutron. These tendencies have been found to be general
for four realistic interactions (AV14, Reid 93, Nijm II and AV18)
we have used. Very similar observations have been made also for
non-local interaction models in a very recent
study~\cite{Arnas_3n}.

Critical enhancement factors required to bind ${^3n}$ are
summarized in Table~\ref{gamma_3n}, they are so large that
dineutron is already resonant in
$^{3}\emph{P}_{2}\emph{-}^{3}\emph{F}_{2}$ state; the
critical factors corresponding dineutron resonance positions are
 summarized in the bottom line of Table~\ref%
{gamma_3n}. Once again one should remark rather good agreement
between the different model predictions. The latter fact as well
as similarity of dineutron predictions suggest that different
realistic local-interaction models have qualitative agreement in
multineutron physics as well. Therefore in further analysis of
three-neutron resonances I decided to rely on single interaction
model. In this scope Reid 93 model is the most suited, since it
possess the best analytical properties and consequently provides
the most stable numerical results for CS method.

\begin{table}[tbp]
\begin{center}
\caption{Critical enhancement factors $\protect\gamma$ required for $^{3}%
\emph{P}_{2}\emph{-}^{3}\emph{F}_{2}$ \emph{nn} channel to bind
$J^{\protect\pi}=\frac{3}{2}^{-}$ three-neutron and these factors
corresponding $J^{\protect\pi}=2^{-}$ dineutron resonances in
MeV.}
\label{gamma_3n}%
\vspace{0.5cm}
\begin{tabular}{|l|llll|}\hline
 & Nijm II & Reid 93 & AV14 & AV18 \\ \hline\hline
$\gamma({^{3}n})$ & 3.61 & 3.74 & 3.86 & 3.98 \\
E($^{2}n$) MeV & 5.31-2.41$i$ & 5.41-2.52$i$ & 5.20-2.49$i$ &
4.83-2.31$i$ \\ \hline
\end{tabular}%
\end{center}
\end{table}

\begin{figure}[h!]
\begin{center}
\includegraphics[width=14.cm]{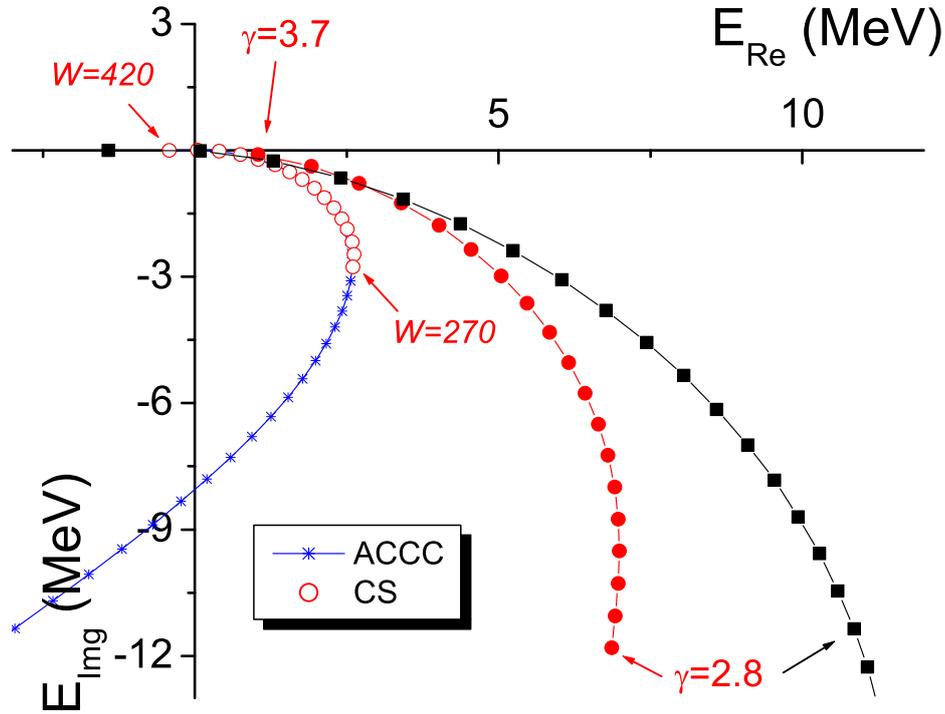} \vspace{-1.2 cm}
\end{center}
\caption{$J^\protect\pi=3/2^-$ three-neutron state resonance
trajectory obtained when reducing the strength $W$ of
phenomenological Yukawa-type force (open circles for CS and solid
line+snowflake points for ACCC methods). Trajectory
depicted by full circles represents one obtained using CS,
when reducing enhancement factor $\gamma$ for $%
^{3}\emph{P}_{2}-^{3}\emph{F}_{2}$ \emph{nn} interaction.
Trajectory
depicted by full squares is dineutron resonance path in $^{3}\emph{P}%
_{2}-^{3}\emph{F}_{2}$ channel, obtained by enhancing
\emph{nn}-interaction in these waves. Presented results are based
on Reid 93 model.} \label{Fig:3_2_nnn_traj}
\end{figure}

As mentioned above only $\frac{3}{2}^{-}$ three-neutron state can
be bound by enhancing single \emph{NN} interaction channel,
without first binding dineutron. In Fig.~\ref{Fig:3_2_nnn_traj}
with full circles the $^3n$ resonance trajectory is traced for
this state when reducing enhancement factor in
$^{3}\emph{P}_{2}\emph{-}^{3}\emph{F}_{2}$ channel from 3.7 to 2.8
with step of 0.05 obtained by CS calculations. Extension of CS
calculations to smaller $\gamma$ values was causing numerical
instabilities, which set for broad resonances due to necessary
scaling of Faddeev equations with ever increasing $\theta$ value.
However it can be seen that this trajectory bends faster than
analogous one for the dineutron in $^{3}\emph{PF}_{2}$ state,
therefore indicating that it will finish in third energy quadrant
with Re(E)$<$0.

Still three-neutron can be bound in states $\frac{3}{2}^{+}$ and $\frac{1}{2}%
^{-}$ by combining together enhancement factors for $^{3}\emph{P}_{2}\emph{-}%
^{3}\emph{F}_{2}$ and $^{3}\emph{P}_{1}$ waves, however such
binding is result of strongly resonant dineutrons in both
mentioned waves. These resonances are very sensitive to the
reduction of enhancement factor and thus quickly vanish leaving
only dineutron ones.

In order to explore all the three-neutron states for a presence of
resonance systematically it has been decided to keep the
\emph{NN}-interaction unchanged, whereas force three-neutron
binding by means of the phenomenological attractive three-body
force, expressed by means of Yukawa function:
\begin{equation}
V_{3n}=-W\frac{e^{-\rho/\rho_{0}}}{\rho},\text{ with }\rho=\sqrt{%
x_{ij}^{2}+y_{ij}^{2}}   \label{Yukawa_3NF}
\end{equation}
and fixing $\rho_{0}=2$ fm. In this way we hold dineutron physics
not affected.

\begin{table}[tbp]
\begin{center}
\caption{Critical strengths $W_0$ in MeV*fm of the
phenomenological Yukawa-type force of eq. (\ref{Yukawa_3NF})
required to bind three-neutron in various states. Parameter
$\protect\rho_{0}$ of this force was fixed to 2 fm. $W' $  are the
values at which three-neutron resonances become subthreshold ones,
whereas $B_{trit}$ are such 3NF corresponding triton binding
energies in MeV.} \label{Tab:Crit_val_w0} \vspace{0.5cm}
\begin{tabular}{|l|llllll|}
\hline
$J^{\pi}$ & $\frac{1}{2}^{+}$ & $\frac{3}{2}^{+}$ & $\frac{5}{2}^{+}$ & $%
\frac{1}{2}^{-}$ & $\frac{3}{2}^{-}$ & $\frac{5}{2}^{-}$ \\
\hline\hline
$W_0 $      & 307   & 1062 & 809 & 515 & 413 & 629 \\
$W' $       & 152   &    - & 329 & 118 & 146 & 277 \\
$B_{trit} $ & 21.35 &    - &  44.55 & 17.72 & 20.69& 37.05
\\\hline
\end{tabular}%
\end{center}
\end{table}

\begin{figure}[h!]
\begin{center}
\includegraphics[width=14.cm]{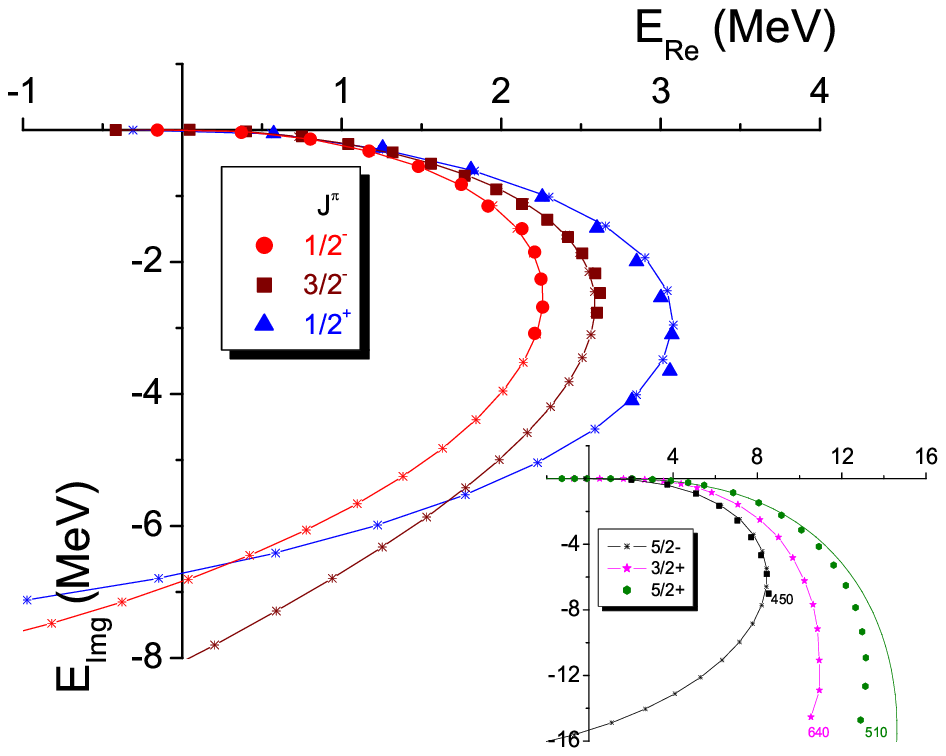} \vspace{-1.2 cm}
\end{center}
\caption{Three-neutron resonance trajectories obtained when
varying the strength $W$ of the phenomenological Yukawa-type 3NF.
Results obtained using CS method are presented by distinct solid
points. For $J^\protect\pi=1/2^-$ state the value of $W$ was
reduced from 520 MeV*fm
to 300 MeV*fm in step of 20 MeV*fm, for the  $J^\protect\pi%
=3/2^-$ state it was $[420,270,10]$ and for the
$J^\protect\pi=1/2^+$ state $[310,210,10]$. Other three-neutron
states require considerably stronger 3NF to be bound and thus
results large values for the resonance energies, their
trajectories are depicted in a smaller figure. For
$J^\protect\pi=5/2^-$ state $W$ was changed $[610,450,10]$; for
$J^\protect\pi=5/2^+$ state $W$ was first reduced with
$[810,750,10]$ and then $[750,510,20]$, whereas
$J^\protect\pi=3/2^+$ trajectory is plotted using $[1060,800,20]$
and then $[800,640,40]$ phenomenological 3NF strengths. ACCC
method results are presented by solid lines, supported by
snowflake-like points.} \label{Fig:3n_res_traj}
\end{figure}

In table~\ref{Tab:Crit_val_w0}  the critical values $W_{0}$ of the
parameter $W$ are summarized for which three-neutron is bound in
different states. Corresponding resonance trajectories obtained by
gradually reducing parameter $W$ are traced in
Fig.~\ref{Fig:3n_res_traj}. As previously CS results are presented
by separate solid points, whereas ACCC ones using continuous line
and snowflake-like points. One has very nice agreement between two
methods except for the $J^{\pi}=\frac{5}{2}^{+}$ three-neutron,
where discrepancy between two methods sets in for large energy
resonances. This is probably an artifact of very strong 3NF used.
Such 3NF confines three-neutron inside $\approx$1.4 fm box (well
inside the range of its action) and starts to compete against
repulsive part of the \emph{nn} interaction, making ACCC method
badly convergent for broad resonances. For
$J^{\pi}=\frac{3}{2}^{+}$ state due to requirement to perform
calculations  with even more brutal $W$ values ACCC method have
not been used.

One can perceive that resonance trajectories have similar shapes
for all three-neutron states, while the resonance poles tend to
slip into adjacent quadrant with Re(E)$<$0 well before $W$ turns
to 0 (i.e. when additional 3NF is removed and only
\emph{NN}-interaction remains). In Table~\ref{Tab:Crit_val_w0}
estimated $W'$ values are presented, obtained using ACCC method,
at which resonance trajectory cross imaginary energy axis and thus
${^3n}$ resonances become subthreshold ones. These values are
still rather large, strongly exceeding ones that could be expected
for a realistic 3NF. To  demonstrate how strongly such 3NF
violates nuclear properties -- we present triton binding energies,
which are obtained supposing that the same 3NF with $W'$ acts in
the \emph{nnp} compound. These energies are expected to be even
larger for more realistic 3NF models, since in our model to favor
extended three-neutron structures we have permitted for this
interaction to have rather long range.

Presented results demonstrate that realistic \emph{NN}-interaction
models exclude the
existence of observable three-neutron resonances. In~\cite{Csoto} ${^3n}$ resonance in $%
\frac{3}{2}^{+}$ state was claimed at $E=(14-13i)$ MeV for
non-realistic Minnesota potential. Our results using realistic
\emph{nn} interaction however contradict existence of such
resonance. Very strong additional interaction is required to bind
three-neutron in $J^\pi=\frac{3}{2}^{+}$ state, whereas removing
this interaction imaginary part of the resonance grows very
rapidly. On the other hand the real energy part of the resonance
saturates
rather early -- it reaches its maximal value when $W$ is reduced from $%
\approx$1060 MeV*fm to $\approx$ 720 MeV*fm. Then, once the
maximal value for its real part is reached, resonance trajectory
have to move rapidly into 3-rd quadrant.

In figure \ref{Fig:3n_res_traj} $^3n$ resonance trajectories are
presented only partially without following them to their final
positions, when additional interaction is completely removed. The
reason is that these positions are very far from bound region,
requiring many terms in Pad\'e expansion to attain accurate ACCC
predictions. Then one can imagine a hypothetical scenario that
these trajectories turn around and return to positive real parts;
although I have never encountered such trajectories in practical
calculations it is ignored if such trajectories can be in
principal excluded by rigorous mathematical arguments.
Nevertheless I would like to stress that such development is very
unlikely, in particular due to the fact that one manipulates with
purely attractive external force. Furthermore in order to get back
to fourth-energy quadrant resonance trajectory should exhibit very
sharp behavior after leaving it -- from
Table~\ref{Tab:Crit_val_w0} one can see that  larger part of
trajectory is already depicted in 4-th quadrant -- in contrary
these trajectories continue smoothly gaining in energy and do not
show any signs of turning around after passing to third quadrant.

\begin{figure}[h!]
\begin{center}
\includegraphics[width=14.cm]{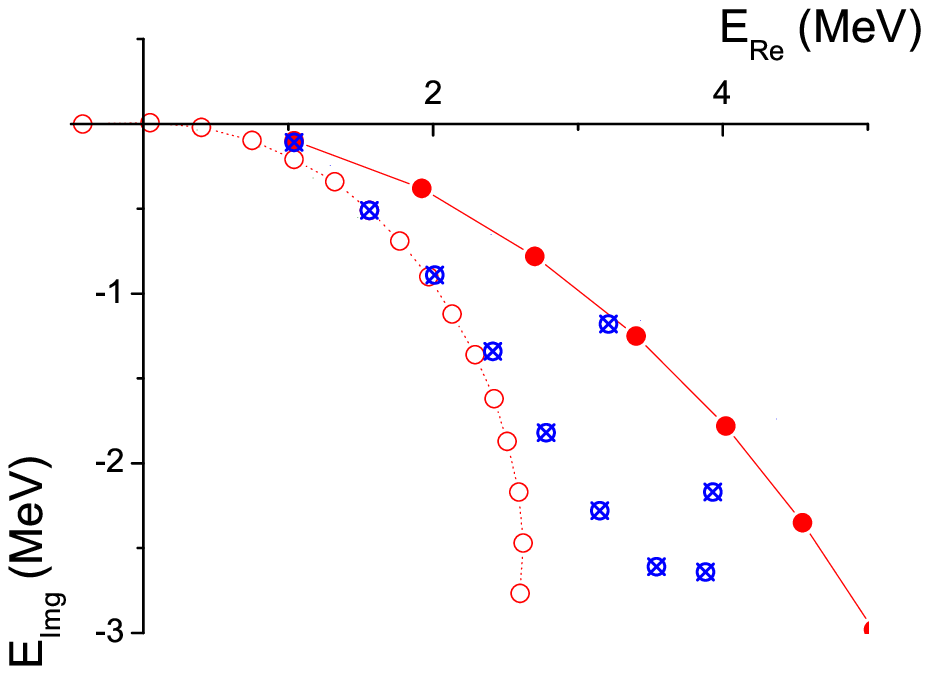} \vspace{-1.2 cm}
\end{center}
\caption{$J^\protect\pi=3/2^-$ three-neutron resonance
trajectories obtained when reducing the strength $W$ of the
phenomenological
Yukawa-type 3NF (open circles) and enhancement factor $\protect\gamma$ for $%
^{3}\emph{P}_{2}-^{3}\emph{F}_{2}$ \emph{nn} interaction (full
circles). Crossed circles indicate resonance path, which is
obtained, when at point $W_{0}=360$ MeV*fm phenomenological 3NF is
gradually removed, however at the same time increasing enhancement
of $^{3}\emph{P}_{2}\emph{-}^{3}\emph{F}_{2}$ \emph{nn} channel
$\protect\gamma$ from 1 to 3.7. } \label{Fig:3_2_nnn_identity}
\end{figure}

Finally, one can expect that enhanced (artificial) bound state -
resonance pole relation is not unique. I.e. some resonance can
exist due to continuation of a bound state of the other symmetry,
which is for some reason is less affected by the modifications of
the interaction in the former
calculations. To investigate such a possibility I have chosen a resonance in $%
J^{\pi}=\frac{3}{2}^{-}$ state, obtained using help of
phenomenological 3NF force eq.(\ref{Yukawa_3NF}) having $W=360$
MeV*fm. Then we gradually reduce $W$ to zero, whereas at the same
time at each step increasing the enhancement factor for the
$^{3}\emph{P}_{2}\emph{-}^{3}\emph{F}_{2}$ channel from 1 to 3.7.
Obtained trajectory of the resonance is traced in
Fig.~\ref{Fig:3_2_nnn_identity} (circles with the crosses)
together with the resonance curves obtained with additional 3NF
(open circles) and when enhancing \emph{nn} interaction in
$^{3}\emph{P}_{2}\emph{-}^{3}\emph{F}_{2}$ channel (full circles).
Once 3NF was completely removed the resonance pole
rejoined the curve obtained by enhancing $^{3}\emph{P}_{2}\emph{-}^{3}\emph{F%
}_{2}$ channel. Note, that structure of bound state obtained with
3N force and enhancement of $\emph{P}$-waves are quite different.
3NF requires very dense and spherical symmetric neutron wave
functions, this is the reason why the $\frac{1}{2}^{+}$ state is
more favorable than $\frac{3}{2}^{-}$ (see
Table~\ref{Tab:Crit_val_w0}) to bind three-neutron with such
additional force.

In this section the results obtained more than ten years ago in
\cite{LC05_3n} have been summarized. Recent study by
Deltuva~\cite{Arnas_3n}, using very different technique based on
solution of the AGS equations in conjunction with complex energy
method and Padé extrapolation technique fully confirmed the
presented results. Presence of independent resonant structures,
which do not evolve from a bound state, have not been  observed in
the work of Deltuva~\cite{Arnas_3n} nor in our more recent study
where full diagonalisation of CS 3n  Hamiltonian have been
achieved~\cite{Arnas_3n}.

\section{Four-neutron resonant states \\
\footnotesize{\textit{(Results presented in this section are based
on the study \cite{PhysRevC.93.044004})}}}

A recent experiment on the $^4{\rm He}(^8{\rm He},\mbox{$^8{\rm
Be}$})4n$ reaction generated an excess of $4n$ events with low
energy in the final state. This observation has been associated
with a possible existence of4$n$ resonance with an estimated
energy $E_R=0.83\pm0.65\pm1.25$ MeV above the $4n$ breakup
threshold and an upper limit of width $\Gamma=2.6$
MeV~\cite{KisamoriPH,Kisamori_PRL}. Low statistics, however, have
not allowed one to extract the spin or parity of the corresponding
state. It is worth noting that a further analysis of the
experimental results of  Ref.~\cite{Marques} concluded that the
observed (very few) events  were also compatible with a $E_R=0-2$
MeV tetraneutron resonance~\cite{Marques:2005vz}.

Ten years ago in collaboration with Jaume Carbonell I have
demonstrated that existence of the observable four-neutron
resonances are incompatible with the present understanding of the
nucleon-nucleon interaction~\cite{PhysRevC.72.034003}. This work
has been mostly realized using ACCC technique. Two important
questions have not been explored in that paper: possible existence
of four-neutron resonances, which does not evolve form the bound
state and the possible impact of the three-neutron force on such a
resonance.

 In view of the obvious tension  between the theoretical
predictions and the last experimental results, we believed that it
would be of  some interest  to reconsider this problem by putting
emphasis on the two aforementioned aspects. This time we have
decided to employ the complex scaling method, in conjunction with
our preferred method based on solution of Faddeev-Yakubovsky
equations but also to cross-check obtained results by employing
Gaussian-expansion
method~\cite{Kami88,Kame89,Hiya03,Hiya12FEW,Hiya12PTEP,Hiya12COLD},
developed by E.~Hiyama and M.~Kamimura, who have joined our team.

As explained in a previous section two-neutron system is resonant
in the $^1S_0$ partial wave. From the perspective of the S-wave
interaction $0+$ tetraneutron state is an ideal system to comply
with the effective field theory predictions in the unitary limit.
Indeed, neutron-neutron interaction length turns to be much larger
than the interaction range and thus for such an extended system
short-range details of the short range interaction does not
matter. Indeed, EFT in the unitary limit predict strong repulsion
between the loosely bound difermion pairs. Thus any arbitrary
enhancement of the $^1S_0$ cannot benefit $4n$ system, due to the
Pauli principle the effective interaction between dineutrons in a
relative S-wave remains mostly repulsive
 and thus  $^1S_0$ partial wave does not contribute much
 in building attraction between the dineutron pairs.

In contrast,   the Pauli principle  does not prevent contributions
from $P$- and higher partial waves to increase the attraction
between a dineutron and another neutron. As aforementioned
$P$-waves are subject of a long standing controversy in nuclear
physics~\cite{EpelKam,Wood,Dolesch},  and some few-nucleon
scattering observables (as analyzing powers) would favor stronger
$P$-waves. Nevertheless the discrepancies with scattering data
might be accounted for   a small  variation of the $nn$ $P$-waves,
of the order of 10\%. In fact, some previous
studies~\cite{These_Rimas_03} showed that, in order to bind the
tetraneutron, the  attractive $nn$ $P$-waves should be multiplied
by a factor $\eta\sim4$\footnote{As it has been demonstrated the
numerical value of this factor depends on the interaction model,
however qualitatively all the models present the same physical
features relative to modification of nn P-waves.}, rending the
dineutron strongly resonant in these $P$-waves. In order to create
a narrow 4$n$ resonance, a slightly weaker enhancement is
required, but still this enhancement factor remains considerable,
 $\eta\gtrsim3$. Therefore such a modification strongly
contradicts the nature of the nuclear interaction, which respects
rather well the isospin conservation.

Finally, as  noticed in Ref.~\cite{pieper:01a}, a three-neutron
force might make a key contribution in building the additional
attraction required to generate resonant multineutron clusters.
The presence of an attractive $T=3/2$  component in the 3$N$ force
is clearly suggested in the studies based on the best $NN$ and
$T=1/2$ $3N$ potentials,  which often underestimate the binding
energies of the neutron-rich systems. Furthermore the contribution
of such a force should rise quickly with the  number of neutrons
in the system, and we will indeed demonstrate this feature when
comparing 3$n$ and 4$n$ systems.


 In our  previous
studies~\cite{These_Rimas_03,LC05_3n} we have employed different
realistic $NN$ interaction models (Reid93, AV18, AV8$'$, INOY) in
analyzing multineutron systems  and found that they provide
qualitatively the same results. For all these reasons  led us to
focus on the modification of the 3$N$ force in the total isospin
$T=3/2$ channel. To this aim we have fixed the $NN$ force with a
realistic interaction  and introduce a simple isospin-dependent
$3N$ force acting in both isospin channels. Its $T=1/2$ part was
adjusted to describe some $A=3$ and $A=4$ nuclear states, while
the $T=3/2$ one was tuned until a $^4n$ resonance had manifested.
The exploratory character of this study, as well as the final
conclusions, justify the simplicity of the phenomenological force
adopted here.

\subsection*{Hamiltonian}\label{H}

We started with a general nonrelativistic  nuclear Hamiltonian
\begin{equation}
H=T +\sum_{i<j} V_{ij}^{NN} + \sum_{i<j<k}V_{ijk}^{3N},
\end{equation}
where $T$ is a four-particle kinetic-energy operator,
$V_{ij}^{NN}$ and $V_{ijk}^{3N}$ are respectively two- and
three-nucleon potentials. In this study the AV8$'$
version~\cite{GFMC_97} of the $NN$ potentials has been used,
derived by the Argonne group. This model describes well the main
properties of the $NN$ system and it is well suited to be handled
by the Gaussian expansion method. The main properties of this
interaction are outlined in the benchmark calculation of the
$^4$He ground state~\cite{Kama01}.

As  most of $NN$ forces,  AV8$'$ fails to reproduce binding
energies of the lightest nuclei, in particular ones of $^3$H,
$^3$He and $^4$He. A $3N$ interaction is required and we have
therefore supplemented AV8$'$ with a purely phenomenological $3N$
force which is assumed to be isospin-dependent and  given by a sum
of two Gaussian terms:
\begin{equation}\label{V3NT}
V_{ijk}^{3N}=\!\!\sum_{T=1/2}^{3/2}\: \sum_{n=1}^2
W_n(T)e^{-(r_{ij}^2+r_{jk}^2+r_{ki}^2)/b_n^2} \, {\cal
P}_{ijk}({T})\; .
\end{equation}
where ${\cal P}_{ijk}({T})$ is a projection operator on the total
three-nucleon isospin $T$ state. The parameters of this force --
its strength $W_n$ and range $b_n$ -- were adjusted to reproduce
the phenomenology.

In the case of $T=1/2$  they were fixed in
Ref.~\cite{Hiya04SECOND} when studying $J^\pi=0^+$ states of
$^4$He nucleus. They are:
\begin{equation}\label{Para_T1/2}
\begin{array}{rclcl}
W_1(T=1/2)&=& -2.04 \;{\rm MeV},    &&  b_1=4.0\,\; \; {\rm fm},
\cr W_2(T=1/2)&=& +35.0 \;{\rm MeV},   &&  b_2=0.75\, {\rm fm}.
\end{array}
\end{equation}
Using this parameter set, in addition to the AV8$'$ and Coulomb
interaction, one obtains  the following binding energies:
$^3$H=8.41 (8.48) MeV, $^3$He=7.74 (7.72) MeV, $^4$He ($0^+_1$)=
28.44 (28.30) MeV and the excitation energy of
$^4$He$(0^+_2)$=20.25 (20.21) MeV~\cite{Hiya04SECOND}, where the
experimental values are shown in parentheses. Furthermore, this
parameterization allows to reproduce the observed transition form
factor  $^4{\rm He}(e,e')^4{\rm He}(0^+_2)$ (cf. Fig.~3 of
Ref.~\cite{Hiya04SECOND})\footnote{Although $^3$H and $^3$He
nuclei contain in their wave functions small admixture of isospin
$T=3/2$ configurations, the last calculations have been performed
by neglecting it as it is a case in most of the few-nucleon
calculations.}.

\begin{figure}[h!]
\begin{center}
\includegraphics[width=12.cm]{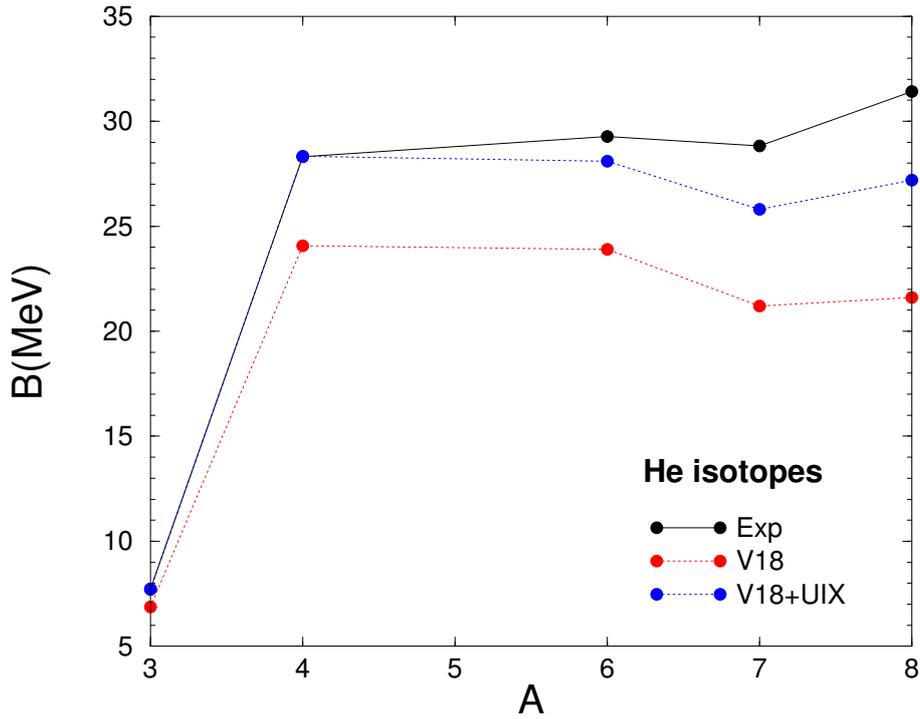}
\end{center}
\caption{(color online) Experimental binding energies of He
isotopes  compared with the predictions of  AV18 $NN$ potential
and a purely $T=1/2$  $3N$ force (UIX), taken
 from Table II of Ref.~\cite{pieper:01a}.}
\label{He_Isotopes}
\end{figure}

$4n$ is only sensitive to  $T=3/2$ component of the $3N$
interaction. This component has almost no effect in proton-neutron
balanced nuclei but it manifests clearly  in the series of He
isotopes, where the purely $T=1/2$ $3N$ force, adjusted to
reproduce well the $^4$He, fails to describe the increasingly
neutron-rich He isotopes. This can be illustrated with the results
of the GFMC calculations, Table II of Ref.~\cite{pieper:01a},
which are displayed in Fig.~\ref{He_Isotopes}.

This situation was dramatically improved in
Ref.~\cite{pieper:01a}, where several $3\leq A \leq 8$ nuclei were
used to fix the parameters of a new series of spin-isospin
dependent Illinois $3N$ forces (IL1$-$IL5) which reproduces well
the experimental data in Fig.~\ref{He_Isotopes}. It is worth
noting however that, from the results~in~Fig.~\ref{He_Isotopes},
the effect of the $T=3/2$ component of the $3N$ force remains
inferior to the $T=1/2$ one.

All along the present section, the attractive strength parameter
of the $T=3/2$ component, $W_1(T=3/2)$, will be considered as a
free  parameter and varied in order to  analyze the existence of a
possible tetraneutron resonance. The other parameters retain the
same value of the $T=1/2$ case, that is we use:
\begin{equation}\label{Para_T3/2}
\begin{array}{rclcl}
W_1(T=3/2)&=& \;\; {\rm free}, \;      &&  b_1=4.0\,\;  {\rm fm},
\cr W_2(T=3/2)&=& +35.0 \;{\rm MeV},   &&  b_2=0.75\, {\rm fm} .
\end{array}
\end{equation}
We will explore in parallel  an effect of such a force on the
$A=4$ nuclei that could be sensitive to the $T=3/2$ component,
that is: $^4$H, $^4$He and $^4$Li, in  states with total isospin
$T=1$ and angular momentum $J^{\pi}=1^-$ and $2^-$.

\begin{figure}[h!]
\centering
\includegraphics[width=81mm]{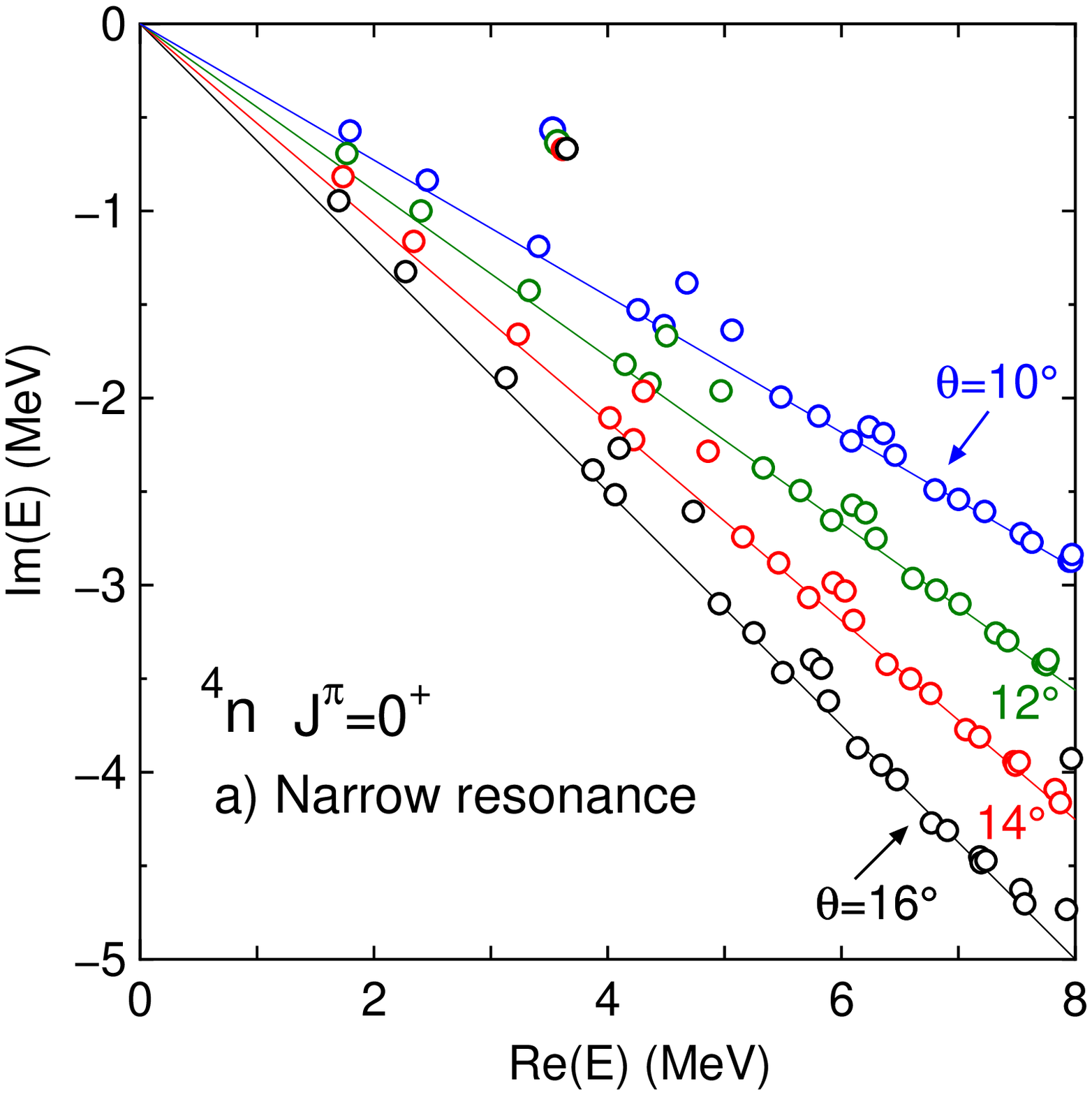}
\includegraphics[width=81mm]{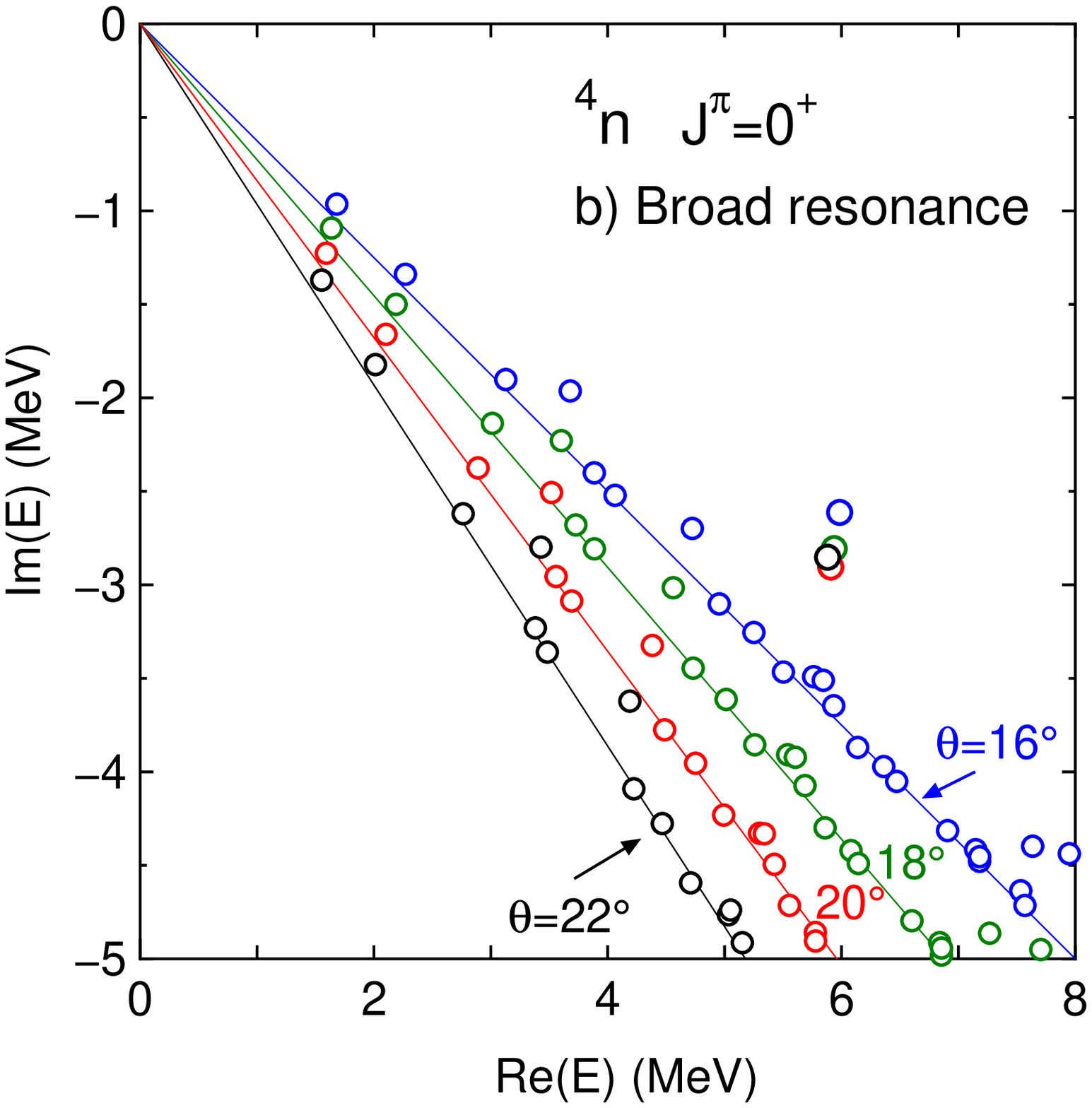}
\caption{Dependence of eigenenergy distribution on the complex
scaling angle $\theta$ for  $^4n$ system with $J^\pi=0^+$. Two
different cases are considered a) presence of a narrow resonance
at $E_{\rm res}=3.65-0.66i$ MeV for $W_1(T=3/2)=-28$ MeV and b)
presence of a broad resonance at $E_{\rm res}=5.88-2.85i$ MeV for
$W_1(T=3/2)=-21$ MeV. } \label{fig:pole-example}
\end{figure}

\subsection*{Results and Discussion}\label{Results}

The Gaussian expansion method allows to achieve numerical
convergence by solving considerably smaller linear algebra
problems than ones required to achieve comparable accuracy by FY
equations method. Furthermore it turns to be possible to perform a
full diagonalization of the CS Hamiltonian matrix for a $4n$
system built by the Gaussian expansion method, thus obtaining full
spectra. Such a spectra is demonstrated in
Fig.~\ref{fig:pole-example} for the $J^\pi=0^+$ state of the
tetraneutron. We were able to check that unless strong attractive
three-neutron force is employed no narrow resonances are observed,
thus denying the hypothesis about the possible presence of narrow
tetraneutron resonant states, which does not evolve into the
lowest bound state once some strong auxiliary interaction is
added.

\subsection*{4n bound state}\label{4n_bs}

Our primary goal was to determine the most favorable tetraneutron
configurations to support narrow resonances.
 For this purpose, we calculate a critical strength of the
attractive 3$N$ force $W_1(T=3/2)$, defined by Eq.~(\ref{V3NT}),
to make different $4n$ states bound at $E=-1.07$ MeV. This energy
corresponds to the lowest limit value compatible with the RIKEN
data~\cite{Kisamori_PRL}. The calculated results, denoted as
$W_1^{(0)}(T\!=\!3/2)$,  are given in
Table~\ref{table:critical-strength}.

\begin{table}[tbp]
\begin{center}
\caption{Critical strength   $W_1^{(0)}(T=3/2)$ (MeV)  of the
phenomenological $T=3/2$ $3N$ force required to bind the $4n$
system at $E=-1.07$ MeV, the
 lower bound of the experimental value~\cite{Kisamori_PRL}, for different states as well
  as the probability (\%) of their four-body  partial waves.}
\label{table:critical-strength} \vspace{0.5cm}
\begin{tabular}{|c|cccccc|} \hline
 $J^{\pi}$  & $0^+$  &$1^+$  &$2^+$  &$0^-$  &$1^-$  &$2^-$ \\ \hline
 $W_1^{(0)}(T\!=\!\frac{3}{2})$ & $-36.14$ & $-45.33$ & $-38.05$ &
  $-64.37$ & $-61.74$ & $-58.37$  \\ \hline\hline
   $S$-wave & $93.8$ &
$0.42$  & $0.04$ &
           $0.07$ & $0.08$ & $0.08$  \\ $P$-wave & $5.84$ & $98.4$  & $17.7$ &
          $99.6$ & $97.8$ & $89.9$  \\
$D$-wave & $ 0.30$ & $1.08$  & $82.1$ &
          $0.33$ & $2.07$ & $9.23$  \\
$F$-wave & $ 0.0$ & $0.05$  & $0.07$ &
           $0.0$ & $0.10$ & $0.74$  \\ \hline
\end{tabular}
\end{center}
\end{table}

As one can see from this table, the smallest critical strength is
$W_1^{(0)}(T\!=\!3/2)=-36.14$ MeV and corresponds to the $J=0^+$
state.  It is consistent with a result reported in
Ref.~\cite{PhysRevC.72.034003}, where tetraneutrons binding was
forced using artificial four-body force in conjunction with Reid93
$nn$ potential. Next most favorable configuration is established
to be $2^+$ state, which is bound by $1.07$ MeV for 3NF strength
of $W_1^{(0)}(T\!=\!3/2)$. The calculated level ordering is
$J^{\pi}=0^+, 2^+, 1^+,  2^-, 1^-, 0^-$. The level ordering
calculated in Ref.~\cite{PhysRevC.72.034003} is $J^{\pi} =0^+,
1^+, 1^-, 2^-, 0^-, 2^+$. These differences are related to the
different binding mechanism of four-nucleon force used in
Ref.~\cite{PhysRevC.72.034003}.

It should be noted that,  in comparison with $W_1(T\!=\!1/2)=-2.04
$ MeV established for the $T=1/2$ $3N$ force, we need extremely
strong $T=3/2$ attractive term to make the 4$n$ system weakly
bound; when the $J=0^+$ state is at $E=-1.07$ MeV with
$W_1(T\!=\!3/2)=-36.14$ MeV, the expectation values of the kinetic
energy, $NN$  and $3N$ forces are  $+67.0, -38.6$ and $-29.5$ MeV,
respectively. We see that the expectation value of $3N$ is almost
as large as one of $NN$ potential. The validity of this strongly
attractive $T=3/2$ $3N$ force will be discussed after presenting
results of $4n$ resonant states.


\subsection*{4n resonances}\label{4n_res}

 After determining critical strength of $W_1(T=3/2)$
required to bind tetraneutron we gradually released this parameter
letting  4$n$ system to move into  continuum. In this way we
follow complex-energy trajectory of the $^4n$ resonances for
$J=0^+, 2^+$ and $2^-$ states. We remind that these trajectories
are controlled by a single parameter $W_1(T=3/2)$, whereas other
parameters remain
 fixed to the values given in Eq.(\ref{Para_T1/2}) and Eq.(\ref{Para_T3/2}).

\begin{figure}[h!]
\centering
\includegraphics[width=81mm]{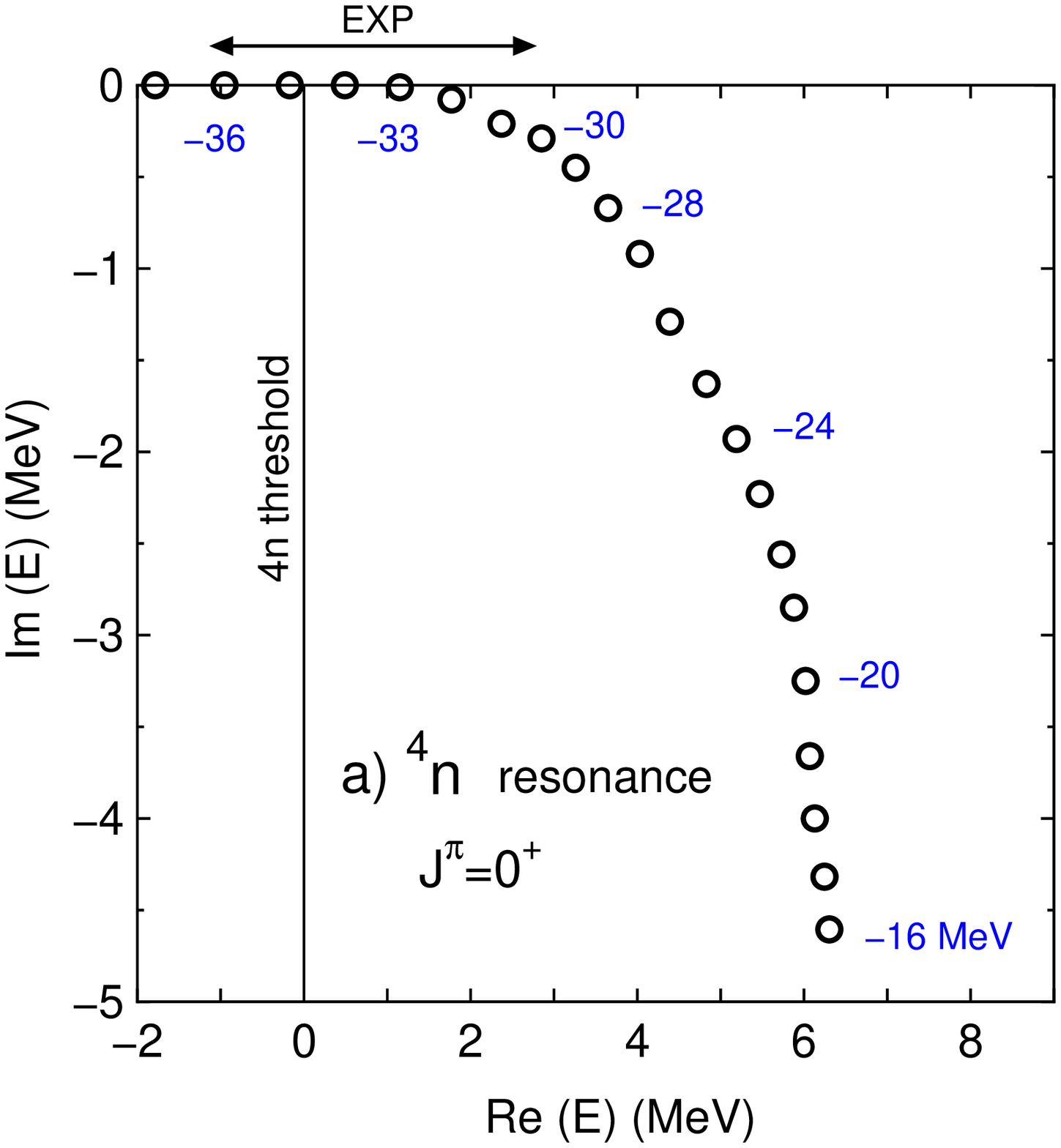}
\includegraphics[width=81mm]{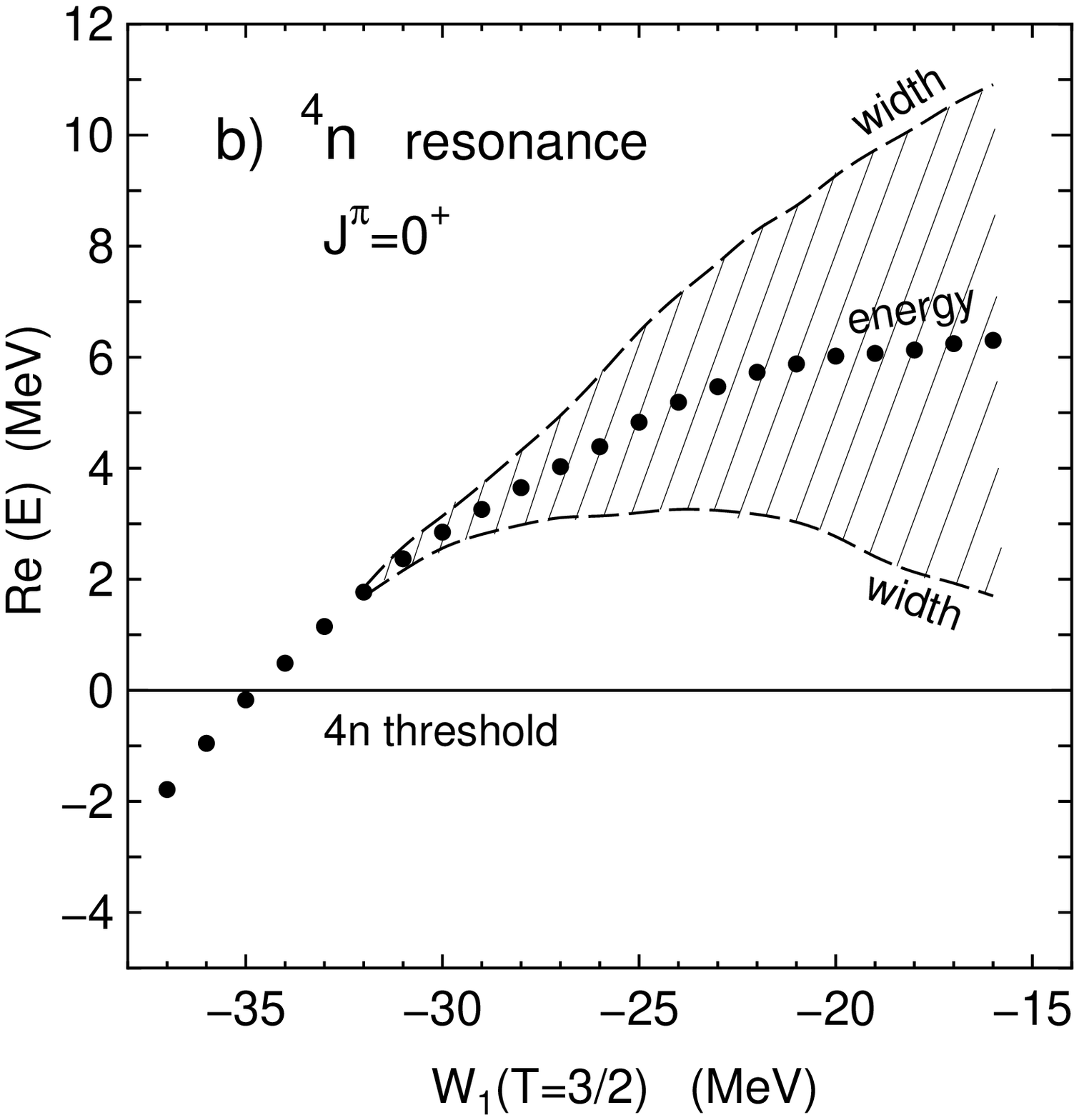}
\caption{a) Tetraneutron resonance trajectory for $J^\pi=0^+$. The
circles correspond to resonance positions for $W_1(T=3/2)$ values
from $-37$ to $-16$ MeV in step of 1 MeV. The observed resonance
energy Re($E_{\rm res})$ including the error is indicated by the
arrow at the top, and upper limit of the observed width
$\Gamma~(=-2\,{\rm Im}(E_{\rm res}))$ is 2.6
MeV~\cite{Kisamori_PRL}. b) The same resonance energy (closed
circles) and width (shadowed area) as those in the upper panel but
explicitly shown with respect to $W_1(T=3/2)$. }
\label{fig:nnnn-trajectry}
\end{figure}

In Fig.~\ref{fig:nnnn-trajectry}a, we display the $^4n$ S-matrix
pole (resonance) trajectory for $J=0^+$ state by reducing the
strength parameter
 from $W_1(T\!=\!3/2)=-35$ to $-16$ MeV in step of $1$ MeV.
We were unable to continue the resonance trajectory beyond $W_1
(T=3/2) =-16$ MeV value with CSM, resonance becoming too broad to
be separated from the non-resonant continuum.
To guide the eye, at the top of the same figure, we presented an
arrow to indicate the $^4n$ real energy range suggested by the
recent measurement~\cite{Kisamori_PRL}. In that range the maximum
value of calculated decay width $\Gamma$ is 0.6 MeV, which is to
be compared with the observed upper limit width $\Gamma=2.6$ MeV.
 In Fig.~\ref{fig:nnnn-trajectry}b the
contents of Fig.~\ref{fig:nnnn-trajectry}a are illustrated in a
different manner to display explicitly the resonance energy and
width versus $W_1(T\!=\!3/2)$. The real energy of the resonances
reaches its maximal value of ${\rm Re}(E_{\rm res})\sim\! 6$ MeV.
Once its real energy maxima is reached the width starts quickly
increasing as the strength $W_1(T\!=\!3/2)$ is further reduced.

\begin{figure}[h!]
\vskip 0.5cm
\begin{center}
\includegraphics[width=12.cm]{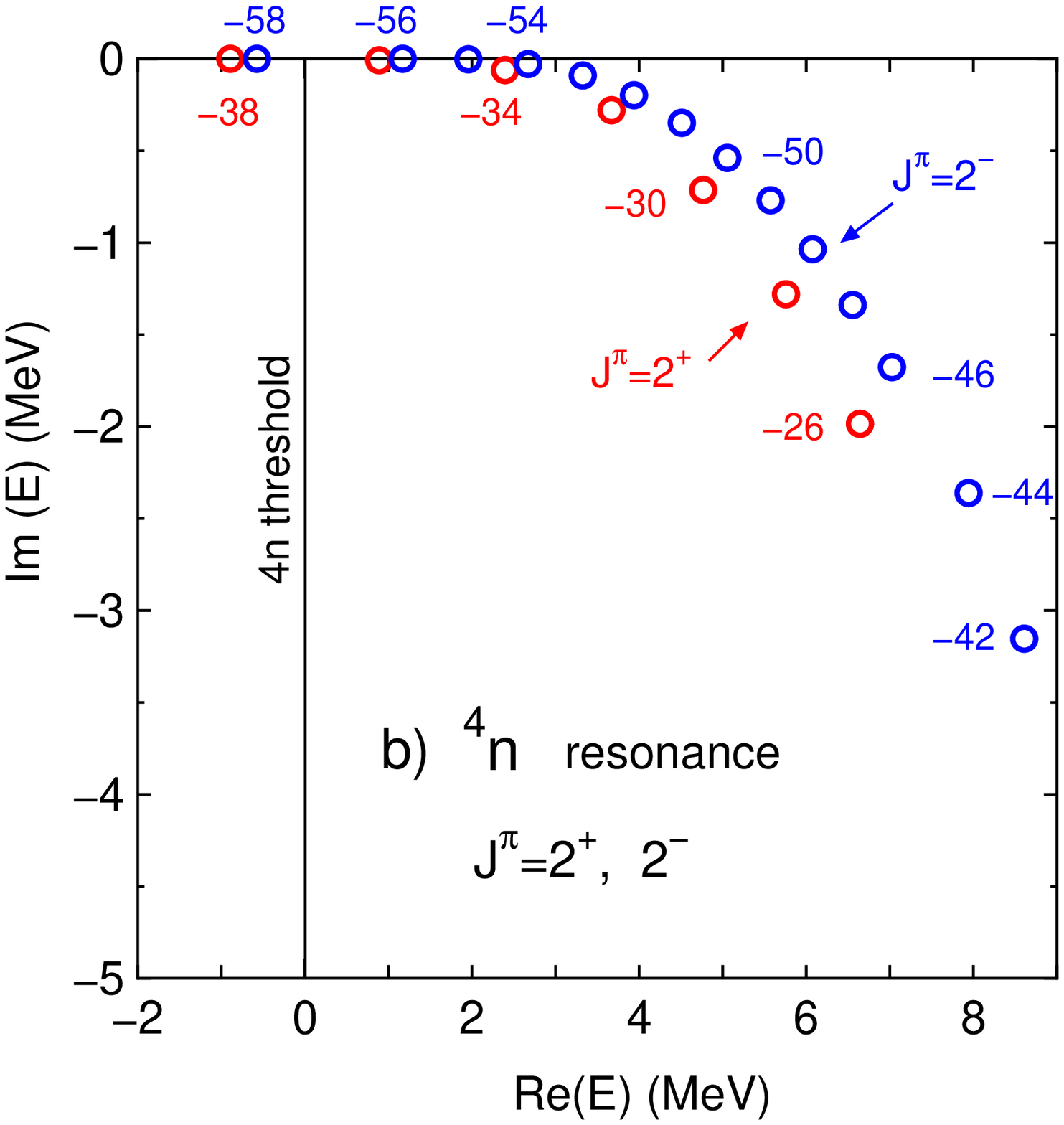}
\end{center}
\caption{Tetraneutron  resonance trajectories for $J^\pi=2^+$ and
$2^-$ states for $W_1(T=3/2)$ values from $-38$ to $-26$ MeV and
from $-58$ to $-42$ MeV, respectively. }
\label{fig:nnnn-width-avoid}
\end{figure}

In Fig.~\ref{fig:nnnn-width-avoid}, we present calculated $^4n$
resonance trajectories for $2^+$ and $2^-$ states. The $J=2^+$ is
the next most favorable configuration to accommodate a bound
tetraneutron, whereas $J=2^-$ is the most favorable negative
parity state, see Table~\ref{table:critical-strength}. The
trajectory of $2^+$ state is very similar to that of the $0^+$
state.
 On the other hand in order to bind  or even to hold a resonant $J=2^-$  state, in the
region relevant for a physical observation, attractive
three-nucleon force term $W_1(T\!=\!3/2)$ should be almost twice
as large as one for $J=0^+$  state. The strength of
$W_1(T\!=\!3/2)$ required to produce resonant $4n$ system in any
configuration, which could produce pronounced experimental signal,
is much larger than $W_1(T\!=\!1/2) (-2.04$ MeV) required to
settle the binding energies of $^3$H, $^3$He and $^4$He nuclei.

As was expected, based on our experience from previous studies on
multineutron systems~\cite{These_Rimas_03,LC05_3n}, tetraneutron
trajectory turns to be very rigid with respect to the employed NN
interaction model, provided this model is capable to reproduce NN
scattering data. To demonstrate this feature the $^4n$ resonance
trajectory for $J=0^+$state based on INOY04(is-m) NN model has
been calculated~\cite{Dolesch}. This semi-realistic interaction
model strongly differs from the other ones in that it contains
fully phenomenological and strongly non-local short range part in
addition to the typical local long range part based on one
pion-exchange. Furthermore this model reproduce triton and
alpha-particle binding energies without contribution from a 3NF
force in $T=1/2$ channel. Finally, P-waves of this interaction are
slightly modified in order to match better low energy scattering
observables in 3N system. Regardless mentioned qualitative
differences for INOY04(is-m) interaction with respect to AV8' one
the obtained results for $^4n$ resonance trajectory are
qualitatively the same and demonstrate only minor quantitative
differences see fig.~\ref{fig:nnnn-trajectry}a.

In order to approve(disprove) possible existence of the observable
tetraneutron resonances, we should consider validity of the
strongly attractive  3$N$ force in isospin $T=3/2$ channel. One
should mention that parametrization of the phenomenological $3NF$
adapted in this study is very favorable for dilute states, as
expected for the tetraneutron resonances. Attractive $3NF$ term
has larger range than one obtained from the pion-exchanges.
Furthermore tetraneutron states, unlike compound $^4He$ or $^3H$
ground states, do not feel repulsive core contribution.

As pointed out already, the GFMC calculation for $3\leq A \leq 8$
suggested existence of a weaker $T=3/2$ $3NF$ component than the
$T=1/2$ one~\cite{pieper:01a,Pieper2002}. From the same study it
follows that the binding energies of neutron-rich nuclei are
described without notable contribution of $T=3/2$ $3NF$. A similar
conclusion was reached in neutron matter calculations, where the
expectation values of the $T=3/2$ $3NF$ are always inferior to
$T=1/2$ one~\cite{nucl_matt}.

 These features are in full agreement with the
considerations of EFT, which  asserts $T\!=\!3/2$ $3N$ force to be
of the subleading order compared to $T\!=\!1/2$ one \cite{VNNN}.
In this way, we find  no physical reason for the fact that the
$T\!=\!3/2$ term to be order of magnitude more attractive than the
$T=1/2$ one, which turned to be necessary to form observable
tetraneutron states as one suggested by a recent interpretation of
the experimental results of the $^4$He($^8$He,$^8$Be)4n
reaction~\cite{Kisamori_PRL}.

\subsection*{T=1 states in $^4$H, $^4$He and $^4$Li}\label{A=4}

\begin{table} [tbp]
\begin{center}
\caption{Observed energies $E_R$ and widths $\Gamma$ (in MeV) of
the $J^\pi=2_1^-$ and $1_1^-$ states in $^4$H,  $^4$He $(T=1)$ and
$^4$Li, \mbox{$Et_R$ being  measured} from the $^3$H$+n$,
$^3$H$+p$ and $^3$He$+p$ thresholds,
respectively~\cite{Tilley1992}. } \label{table:Exp-A=4}
\vspace{0.5cm}
\begin{tabular}{|c|cccccc|}\hline
  & & \multicolumn{1}{c}{$^4$H} & & \multicolumn{1}{c}{$^4$He $(T=1)$}   & & \multicolumn{1}{c|}{$^4$Li} \\\cline{2-7}

 $J^{\pi}$  & & $E_R \, (\Gamma)$  & & $E_R \, (\Gamma)$
& & $E_R \, (\Gamma)$  \\\hline \hline
$2_1^-\;$ & $\;$ & 3.19 (5.42)  & $\;$ & 3.52 (5.01) &$\;$  & 4.07 (6.03) \\
$1_1^-\;$ & & 3.50 (6.73)  & & 3.83 (6.20) & & 4.39 (7.35)\\\hline
\end{tabular}
\end{center}
\end{table}

\begin{figure}[h!]
\centering
\includegraphics[width=81mm]{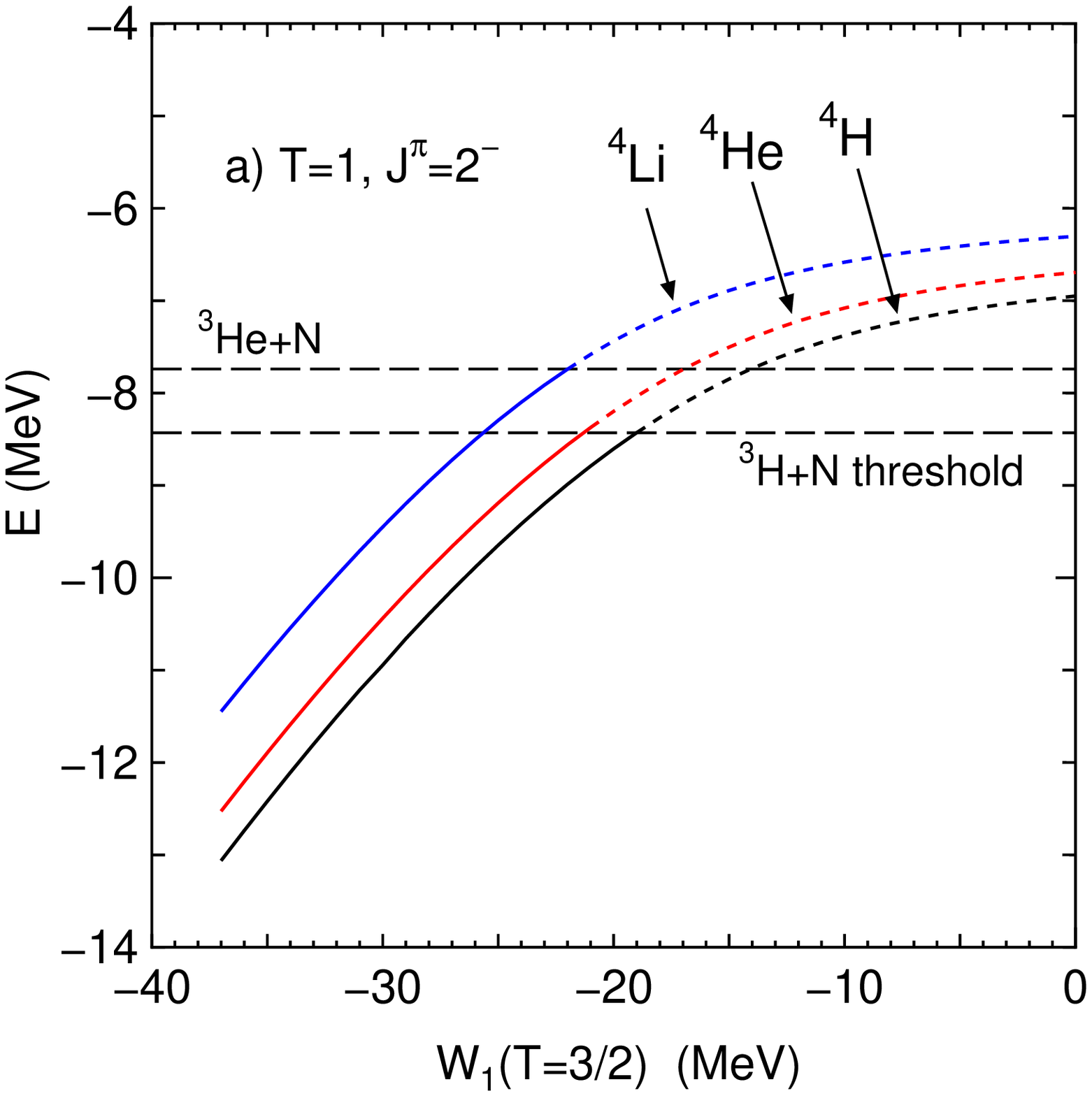}
\includegraphics[width=81mm]{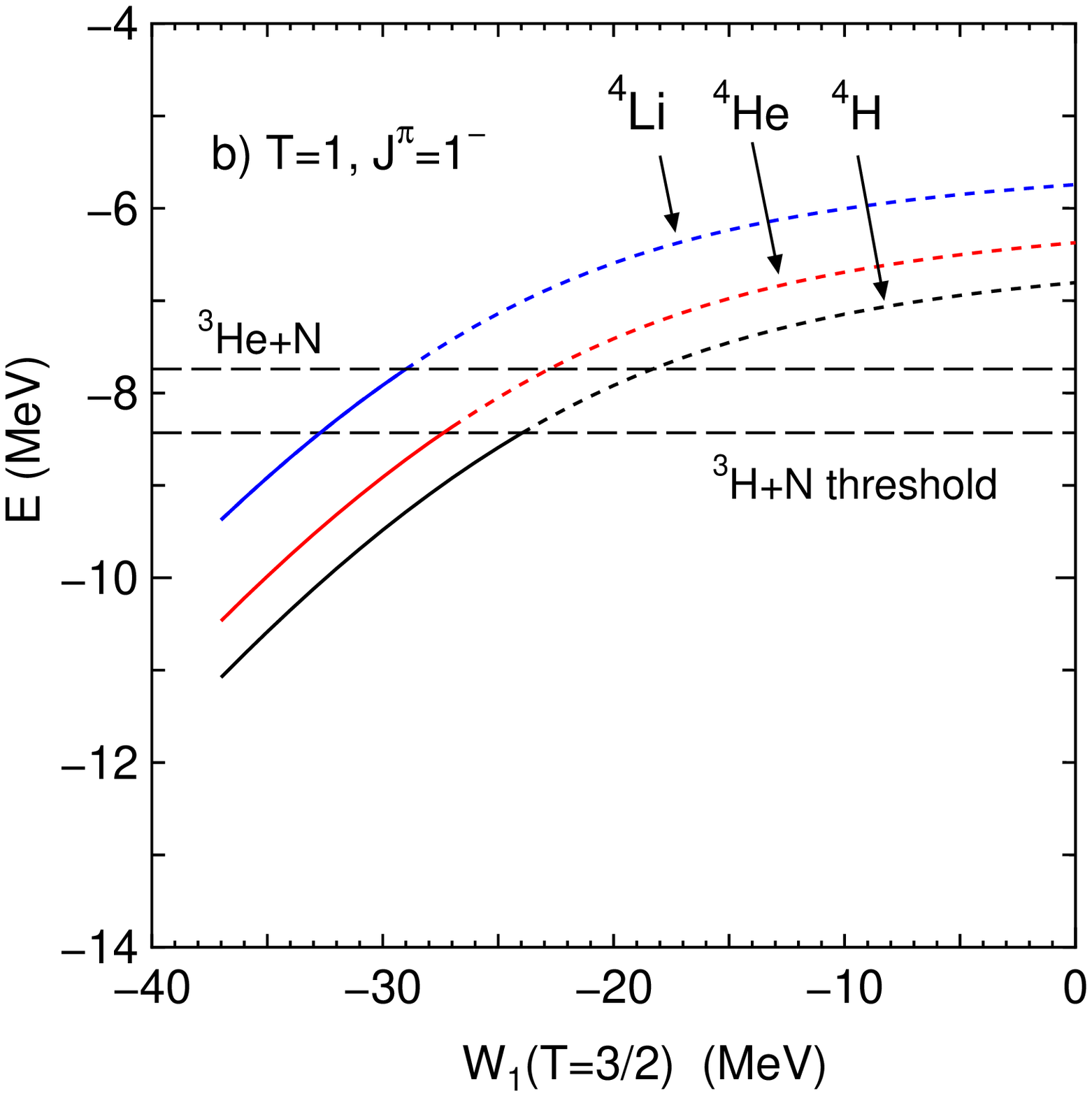}
\caption{ a) Calculated energies of the lowest $T=1, J^\pi=2^-$
states in $^4$H, $^4$He and $^4$Li with respect to the strength of
$T=3/2$  $3N$ force, $W_1(T=3/2)$. \mbox{b) The same} but for
$T=1, J^\pi=1^-$ states. The horizontal dashed lines show the
$^3{\rm He}+N$ and $^3{\rm H}+N$ thresholds. The solid curve below
the corresponding threshold indicates evolution of a bound state,
while the dotted curve above the threshold stands approximately
for the evolution of a resonance obtained by the diagonalization
of $H(\theta=0)$ with the $L^2$ basis functions. }
\label{fig:W1-dependence}
\end{figure}

In the following we would like to investigate the consequences of
a strongly attractive 3NF component in the isospin $T\!=\!3/2)$
channel. It is clear that such a force will have the most dramatic
effect on nuclei with a large isospin number, i.e. neutron (or
proton) rich ones as well as on infinite neutron matter.
Nevertheless this includes mostly nuclei with $A>4$, not within
our current scope.
Still we were able to investigate effect on other well known
states of $A=4$ nuclei, namely negative parity, isospin ($T=1$)
states of $^4$H, $^4$He  and $^4$Li. These structures represent
broad resonances~\cite{Tilley1992} (see Table~\ref{table:Exp-A=4})
established in nuclear collision experiments. Calculated energies
of those states are shown in Fig.~\ref{fig:W1-dependence} with
respect to increasing $W_1(T=3/2)$ from $-37$ to \mbox{0 MeV}. The
solid curve below the corresponding threshold indicates evolution
of a bound state, whereas the dotted curve above the threshold
stands approximately for the trajectory of a resonant state
obtained within a bound state approximation, that is, by
diagonalizing $H(\theta=0)$ using the $L^2$ basis functions of the
Gaussian-expansion method.

As demonstrated in Fig.~\ref{fig:W1-dependence}, values of an
attractive 3NF term in the range of $W_1(T\!=\!3/2) \simeq -36$ to
$-30$ MeV, which is compatible with a reported $^4n$ resonance
region in Ref.~\cite{Kisamori_PRL}, gives rise to the appearance
of bound $J=2^-$ and $J=1^-$ states in $^4$H, $^4$He($T=1$) and
$^4$Li nuclei. Unlike observed in the collision experiments, these
states become stable with respect to the $^3$H $(^3$He$)+N$ decay
channels. This means that such a strong 3NF force in $(T=3/2)$ has
already dramatic consequences for the lightest nuclei, like $^4$H,
$^4$He ($T=1$) and $^4$Li and is expected to have even more
catastrophic consequences on heavier neutron (or proton)
unbalanced nuclei.

In contrast, it is interesting to see the energy of 4n system when
we have just unbound states for  $^4$H, $^4$He ($T=1$) and $^4$Li
in Fig.~\ref{fig:W1-dependence}a. Use of $W_1(T=3/2)=-19$ MeV
gives rise to an unbound state of $J=2^-$ in $^4$H with respect to
the disintegration into  $^3{\rm H}+N$. However, using this
strength of $W_1(T=3/2)$, we have already a very broad $^4n$
resonant state at Re($E_{\rm res})=6$ MeV with $\Gamma=7.5$ MeV,
see Fig.~\ref{fig:nnnn-trajectry}a, which is inconsistent with the
recent experimental claim~\cite{Kisamori_PRL} of resonant $^4n$.
Moreover the value of $W_1(T=3/2)$ that reproduces the observed
broad resonance data of $2^-$ in $^4$H should be much less
attractive than $-19$ MeV.

\begin{figure}[h!]
\begin{center}
\includegraphics[width=12.cm]{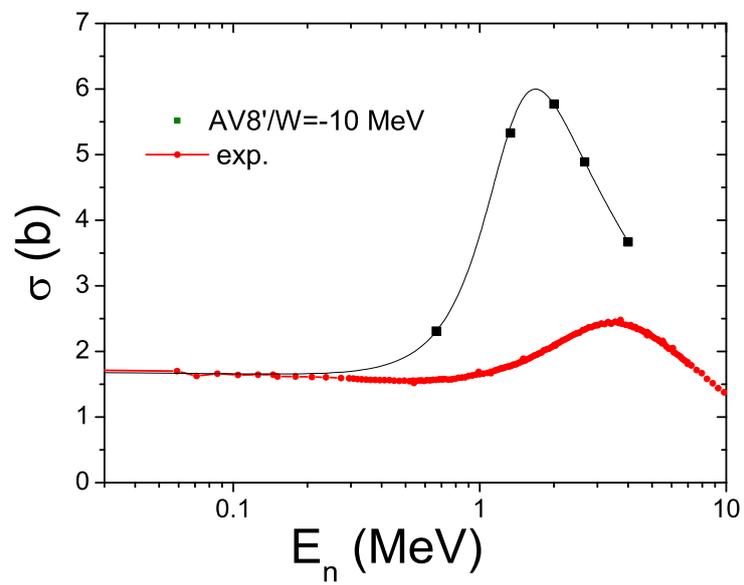}
\caption{The calculated total cross section of $^3$H$+n$ in black
solid line using $W_1(T$=3/2)=$-10$ MeV. The experimental
data~\cite{phillips:80} is illustrated in red solid line.}
\label{fig:cross-section}
\end{center}
\end{figure}

Results presented in Fig.~\ref{fig:W1-dependence}a, however, give
little insight to the properties of $^4$H, once it becomes a
resonant state for $W_1(T=3/2)>-19$ MeV. Moreover it is well
known, as explicitly written in~\cite{Tilley1992}, that for broad
resonances the structure given by the S-matrix poles may be
different from that provided by an R-matrix analysis. Therefore,
it makes much more sense to compare directly the calculations with
the measurable $^3$H$ + n$ data, namely scattering cross sections.
In Fig.~\ref{fig:cross-section} we present $^3$H$ + n$ total cross
section calculated for a value of $W_1(T=3/2)=-10$ MeV. The total
cross section is clearly dominated by a pronounced negative-parity
resonances in $^4$H system. These resonances contribute too much
in the total cross section, resulting in the appearance of a
narrow peak shifted significantly to the lower-energy side.
Furthermore, in order to reproduce the shape of the experimental
$^3$H$ + n$ cross section, a very weak $3NF$ is required in the
isospin $T\!=\!3/2$ channel. From this fact, we conclude that even
$W_1(T\!=\!3/2)=-10$ MeV value renders 3NF to be excessively
attractive.

In conclusion, as far as we keep the consistency with the observed
low-lying energy properties of the  $^4$H, $^4$He ($T=1$) and
$^4$Li nuclei, it is difficult to produce an observable $^4n$
resonant state.

\chapter{Reactions induced by the perturbations}

\section{Tetraneutron response functions\\ \footnotesize{\textit{(Results presented in this section are based on
the study \cite{doi:10.1093})}}}

As mentioned in the previous section, a recent experiment at RIKEN
\cite{KisamoriPH,Kisamori_PRL} observed the sharp structure in
$^4$He($^8$He,$^8$Be)4n reaction cross section near the $4n$
threshold,  suggesting the existence of a narrow resonant state of
tetraneutron ($4n$).

The dineutron-dineutron correlation has sometimes been invoked as
a possible enhancement mechanism, due to the large value of the
scattering length~\cite{LF_PAN_2008,Bertulani_Nature}. However
previous
calculations~\cite{These_Rimas_03,PhysRevC.72.034003,Bertulani}
indicated that the
 interaction between two (artificially bound) di-neutron
was repulsive and so the probability to find four neutrons at the
same point of the phase space is very weak.  A similar conclusion
was reached   in the framework of the Effective Field Theories
(EFT) for a more general case of fermionic systems close to the
unitarity limit~\cite{dd_EFT_Petrov_2004,dd_EFT_PLB_2017}. Their
conclusions are model independent and rely only on the fact that
the fermion-fermion scattering length is much larger than the
interaction range, which is the case of the neutron-neutron
system. In view of these results, and contrary to some theoretical
claims, it seems very unlikely that the tetraneutron system could
manifest a nearthreshold resonant state.

Finally theoretical results presented in the previous section,
demonstrate the difficulty to accommodate such a near-threshold
resonance of the $4n$ system without dramatically disturbing the
well established neighboring nuclear chart. However as pointed out
in a previous section some reaction mechanism,  being able to
produce an enhancement of the cross section at small energy,
should be investigated. It is indeed well known that without a
presence of S-matrix poles there exist other possibilities to
generate sharp structures in a reaction cross
section~\cite{Calucci} and even in simple fully repulsive
systems~\cite{flugge}.

It is of great interest to study a possibility to observe the
sharp $4n$ response functions, without a presence of associated
S-matrix poles. Unfortunately reaction $^4$He($^8$He,$^8$Be)4n ,
studied experimentally, turns to be too complicated to be
addressed with an accurate theoretical model. Nevertheless one may
try to construct a simplistic approach, which may mimic the gross
features of the aforementioned reaction.

The experiment held by Kisamori et al. used 186 MeV/u $^8$He beam
to bombard  $^4$He. The reaction $^4{\rm He}(^8{\rm He},^8{\rm
Be})^4n$ has been studied in a very particular
 kinematical conditions, where most of the kinetic energy of the  projectile has been transferred to $^8$Be nucleus.
 The decay products of $^8$Be, namely the two alpha particles, were detected in order to reconstruct the kinematics.
The particular
  kinematics employed in this
 experiment suggests to use approximate methods in order to estimate the possible  response.

 The principal reaction mechanism is  a double charge exchange with little energy transfer to the target  nucleus $^4{\rm He}$ ,
 which transforms it into a tetraneutron.
 The transition amplitude for such a process might be split  in two pieces
 \begin{equation}
A\approx\langle ^4n |\hat{O}_1|^4{\rm He}\rangle \langle ^8{\rm
Be} |\hat{O}_2|^8{\rm He}\rangle  \; \label{eq:amp},
\end{equation}
where $\hat{O}_i$ are some transition operators. These two factors
correspond respectively to the ''fast''  process   $\langle ^8{\rm
Be}|\hat{O}_2|^8{\rm He}\rangle$ carrying  the bulk of the 186
MeV/u kinetic energy of the projectile and a ''slow'' one
$\langle^4n |\hat{O}_1|^4{\rm He}\rangle$  constituent of the
charge exchange reactions and which remains practically static.

Total reaction cross section takes then the form:
\begin{equation}
\sigma_{tot}(E)\propto|\langle ^4n |\hat{O}_1|^4{\rm He}\rangle
\langle ^8{\rm Be}
 |\hat{O}_2|^8{\rm He}\rangle |^2 \delta(E_i-E_f),
 \label{eq:cs}
\end{equation}
We are interested in the first term $\langle ^4n |\hat{O}_1|^4{\rm
He}\rangle$ of the last expression, since this term should bring
into  evidence any resonant features of the tetraneutron or any
alternative mechanism for enhancing the cross section (if at all).
The other term, related with a rapid process and involving large
momenta, may affect the overall size of the total cross section,
but should not have significant influence on the low-energy
distribution of $^4n$ system.

On the other hand, the features of $\langle ^4n |\hat{O}_1|^4{\rm
He}\rangle$ matrix element will critically depend on the
particular transition operator $\hat{O}_1$, which is unknown. In
this work  the most probable operator form is assumed. Since
$^4{\rm He}$ and $^4n$ wave functions are coupled with little
momenta transfer, the corresponding transition operator should
contain only low order momenta terms and thus its space-spin
structure should have quite a simple form. Furthermore, it is
assumed that both $^4$He and $^4n$ wave functions are $J=0^+$
states since, as pointed out in our previous studies
\cite{PhysRevC.72.034003,PhysRevC.93.044004}, this state is the
most favorable tetraneutron configuration  revealing resonant
features. The transition operator $\hat{O}_1$ should be therefore
a scalar.

One possibility could be  $E_0$ or $\sigma_i.\sigma_j$ operators.
However the effect of these operators would be strongly suppressed
by the spatial orthogonality between the $^4{\rm He}$ and $^4n$
wave functions. This follows from the shell model representation
of $^4{\rm He}$ and $^4n$ wave functions with   s-wave protons
replaced by p-wave neutrons. The second operator
{$\sigma_i.\sigma_j$ term}  implies correlated double-charge
exchange, but since exchange of the nucleons takes very short time
uncorrelated process is expected to dominate. The simplest
operator allowing such a transition might be represented as a
double spin-dipole term:
\begin{equation}
\hat{O}_1=(\sigma_i.r_i)(\sigma_j.r_j)\tau_i^-\tau_j^-
 \label{eq:op},
\end{equation}
In the last expression $\tau_i^-$ isospin reduction operators are
added which enable charge exchange, i.e. replace a proton by
neutron.

\bigskip
Once fixed the transition operator we are interested in evaluating
the response (or strength) function, given  by
\begin{equation}\label{S_E}
S(E)=\sum_{\nu }\left\vert \left\langle \Psi _{\nu }\left\vert
\widehat{O}_1\right\vert \Psi _{0}\right\rangle \right\vert
^{2}\delta (E-E_{\nu }),
\end{equation}
 where $\Psi _{0}$ represents the ground state wave
function of the $^4$He nucleus, with ground-state energy $E_{0}$,
and  $\Psi_{\nu }$ represents the wave function of the $^4n$
system in the continuum with an energy $E_{\nu }$. Both wave
functions are solutions of the four-nucleon Hamiltonian $H$. The
energy is measured from some standard value, e.g. a particle-decay
threshold energy.

The Strength function (\ref{S_E}) may be calculated within the
formalism explained in the section \label{sec:react_exprb}, using
complex scaling method applied to four-body Faddeev-Yakubovsky
equations as explained in section \ref{sec:FY_eq}.

The nuclear Hamiltonian considered in this study coincides with
one of the previous section and
reference~\cite{PhysRevC.93.044004}, consisting of the Argonne
AV8' two-neutron interaction~\cite{GFMC_97}  plus three-nucleon
forces in both T=1/2 and T=3/2 total  isospin channels, as
explained in the previous section and the
works~\cite{Hiya04SECOND}.

In that concerns the numerical calculations, for FY equations
partial-wave basis has been limited to angular momenta
max$(l,L,\lambda)\leq7$, providing total of 1541 partial
amplitudes. Furthermore $25^3$ Lagrange-mesh points were used to
describe radial dependence of Faddeev-Yakubovsky components,
resulting into linear-algebra problem of $2.4\times10^7$
equations. Such a large basis size ensured accurate results, which
can be traced by comparing FY calculation with Gaussian expansion
method in Table~\ref{Table_1}. Even for a very shallow
tetraneutron state of $\sim$1 MeV difference in calculated binding
energy was less than 20 keV, whereas expectation values differed
by less than 1\%.

\begin{figure}[h!]
\begin{center}
\includegraphics[width=12.cm]{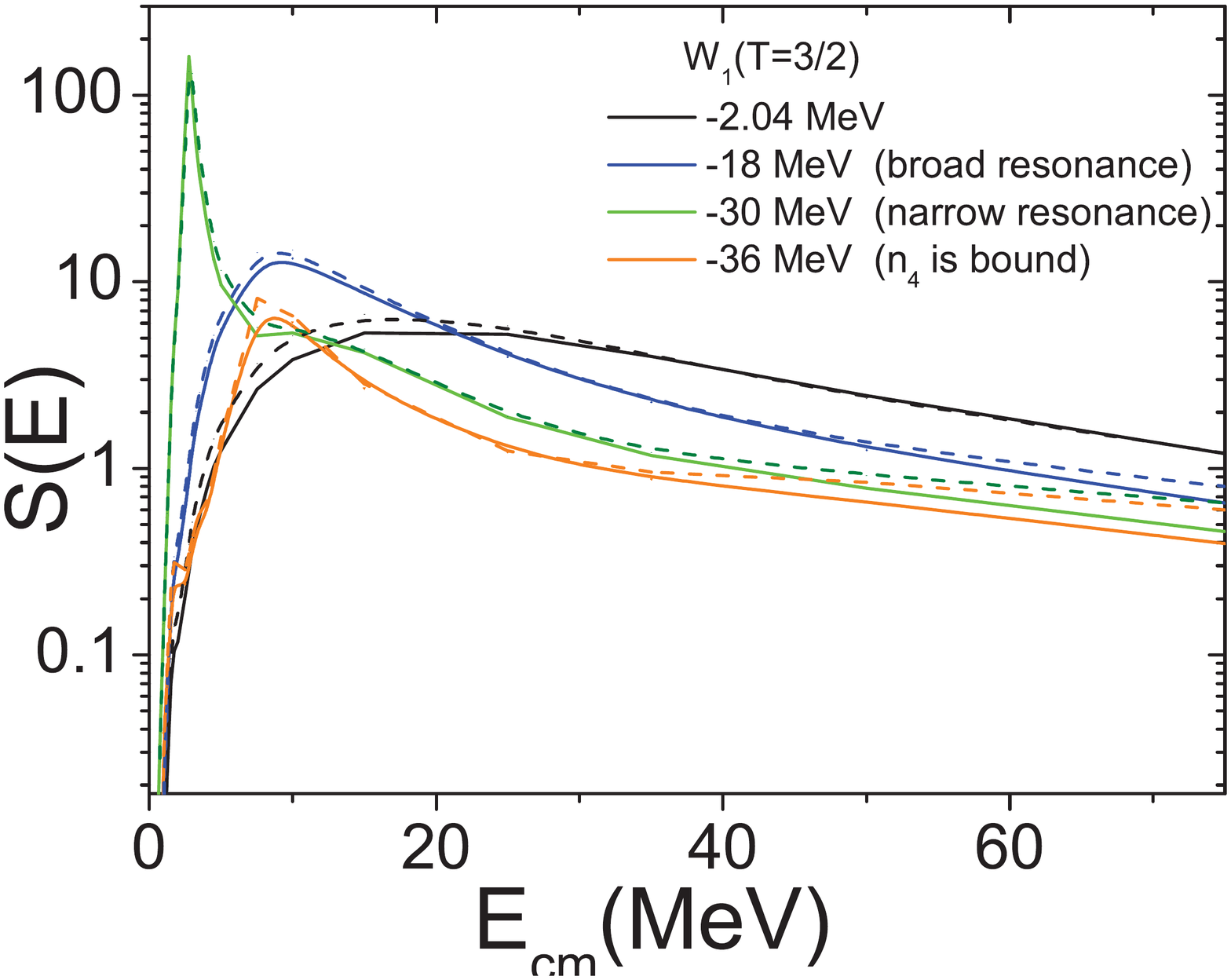}
\end{center} 
\caption{ \label{fig:ptx}  Response function for tetraneutron
production from $\alpha$  particle due to double-dipole charge
exchange operator.}
\end{figure}

Obtained results are concluded in the figure~\ref{fig:ptx}. The
black curve corresponds to the nuclear Hamiltonian, based on
isospin independent three nucleon force. In this case, the
response function is flat without any near-threshold sharp
structure. By increasing the attractive part of the T=3/2
contribution, a resonant peak appears. For $W_1(T=3/2) =-18$ MeV
(blue curve), still far from the values compatible with the RIKEN
result,  the underlying structure is already visible, although
quite broad. It becomes  sharper and sharper by further increasing
the attraction and by moving the resonant pole close to the
threshold.

For  $W_1(T=3/2)=-30$ MeV (green curve),  the tetraneutron
resonance parameters are provided in the inset figure
\ref{fig:nnnn-trajectry} are $E_R=2.8$ MeV  and $\Gamma=0.7$ MeV.
In the vicinity of this value the corresponding response function
takes the usual Breit-Wigner form.

When further increasing the attraction the resonance becomes a
bound state (orange curve, corresponding to $W_1(T=3/2)=-36$ MeV
). The response function, which  has a pole at negative energy,
displays  also some pronounced structure at positive energies
although with reduced strength.

\begin{table} [tbp]
\begin{center}
\caption{Two- and three-body contribution to the potential energy
of the $4n$  system in a $J^{\pi}=0^+$ state as a function  of
$W_1(T=3/2)$ (all units are in MeV). Results denoted by $^4$n'
correspond to the bound state approximation and $^4$n to the
continuum resonant states. The results  are compared with the
$^4$He ground and first excited state with the physical strength
$W_1(T=1/2)=-2.04$. The T=3/2 contribution in 4n required to
accommodate a resonant state is more than one order of magnitude
larger than the T=1/2 (see the rightmost  column).}
\label{Table_1} \vspace{0.5cm}
\begin{tabular}{|l|cccccr|} \hline

                    &   $ W_1(T=3/2) $   &     E             &    $ \langle T \rangle $       &    $\langle V_{2N} \rangle $    &      $\langle V_{3N}\rangle$      &     $\langle V_{3N} \rangle\over\langle V_{2N} \rangle $ \\   \hline \hline
$^4$n'    &     -36                 &   -1.00          &      67.02      &     -38.58         &         -29.52         &          76.5 \%  \\
                    &     -33                 &   +1.18        &     46.67       &     -28.13         &       -17.35           &          61.7  \%  \\
                    &     -30                 &   +2.70        &      29.11       &    -18.36         &          -8.05           &   43.8    \%  \\
                    &     -27                 &   +4.70        &     25.20       &  -15.03            &    -5.48                 &     36.5  \%  \\
                    &    -24                  &   +5.18        &    19.83        &  -11.98            &   -2.66                  &   22.2  \%  \\  \hline
                                       & -36                &     -0.98          &  66.79          &           -38.47 &
                    -29.31          &  76.2\% \\
$^4$n    &     -30                  & +2.84-0.33i  &     -               &   -26.7+6.5i    &  -10.1+4.4i          &   40.1 \%  \\
                   &    -24                   & +5.21-1.88i  &     -              &    -19.3+8.8i    &  -2.3+5.4i           &  27.7\%  \\\hline
$^4$He   &    -2.04               &     -28.44     &     106.12 &
-131.17         &         -3.50          & 2.59 \%  \\
\hline $^4$He*   &                           &     - 8.13 & 49.36
&     - 56.71         &          -0.78         & 1.38 \%      \\
\hline
\end{tabular}
\end{center}
\end{table}

It worths to emphasize  that the presented results are essentially
independent of the nuclear Hamiltonian and the mechanism
considered to artificially produce the $4n$ bound or resonant
state. Several two- and three- and even four-nucleon interactions
have been indeed examined in  previous calculations
\cite{These_Rimas_03, PhysRevC.72.034003,PhysRevC.93.044004} and
led to very similar results. The underlying reason is  that, when
any ad-hoc mechanism is considered to enhance the 4n attraction in
order to accommodate a resonant state, this state is in fact,
essentially supported by an artificial  binding mechanism adjusted
to this aim: the details of the remaining nucleon-nucleon
interaction are residual.

This fact is  illustrated in Table \ref{Table_1} where we have
compared the contributions of the two- and three-nucleon force
(averaged values of the corresponding potential energies) both for
the $^4$He and the 4n system, for several values of the strength
parameter $W_1(T)$. As one can see from the results of this table
the $V_{2n}$ and $V_{3n}$, the contributions to the $^4$n state in
the resonance region  are of the same order and its ratio (the
rightmost column) remains in any case more than one order of
magnitude larger than for the T=1/2 case in $^4$He, the contrary
of one could expect from physical arguments.

\chapter{Description of a few particle collisions}

\section{Nucleon scattering on deuteron\\ \footnotesize{\textit{(Results presented in this section are based on
the study~\cite{LC11})}}}

As a first non-trivial application of the complex scaling method
in describing particle collisions we have considered the
nucleon-deuteron (N-d) $L=0$ scattering in spin-doublet ($S=1/2$)
and spin-quartet ($S=3/2$) states. For this pioneering study, the
interaction between the nucleons have been described by a
phenomenological MT I-III potential~\label{MT13}, defined in
section \ref{sec:2b_sc_splines}. This work has been realized
employing spline collocation method. Calculations have been
performed both below and above the three-particle breakup
threshold. Below the breakup threshold, the results are stable and
independent of the scaling angle,  in a similar  way as for  the
two-body case. Phaseshifts might be accurately extracted using
either differential or integral expressions.

The application of the differential relations for extracting
scattering phaseshifts and inelasticities above the breakup
threshold   does not lead to a very convincing results. It is
always a difficult task to find the stability domain. We have
therefore  employed integral expressions, obtained using the
Greens theorem, which once again proved their worth. We have
summarized some obtained results in Tables~\ref{tab_3Bnd_brca} and
~\ref{tab_3Bpd_brca}, respectively for n-d and p-d scattering
above the breakup threshold. Very accurate results are obtained
for both the phaseshifts and the inelasticity parameters, once the
complex scaling angle is chosen in the interval
$[4^\circ,12.5^\circ]$ for incident neutron with energy
E$_{lab}$=14.1 MeV and in the range $[3^\circ,7.5^\circ]$ at
E$_{lab}$=42 MeV. A stability of the final result within at least
three significant digits is assured, providing an excellent
agreement with the benchmark calculations
of~\cite{Friar_nd_bench,Arnas_bench}. The calculated integral
gradually ceases to converge on the finite domain for the
calculations when larger complex scaling angles are chosen. This
is related to the failure to damp inhomogeneous term involving
diverging CS incoming wave, finally leading to the CS angle
limiting conditions discussed in section \ref{SA_GF}.

We have displayed   in table~\ref{tab_3B_brca},  the $^3S_1$ n-d
breakup amplitude as a function of the breakup angle $\vartheta$,
which defines the pair and spectator wave numbers via k=K
cos($\vartheta$) and q= 2K sin($\vartheta$)/$\sqrt{3}$
respectively. A nice agreement is obtained with the benchmark
calculation of~\cite{Friar_nd_bench}. Some small discrepancy
appears only for the $\vartheta$ values close to 90$^\circ$, which
corresponds to a geometric  configuration where, after the
breakup, one pair of particles  remains at rest. This is due to
the slow convergence of the integral relation for the breakup
amplitude for $\vartheta\rightarrow90^\circ$ in y-direction. A
special procedure must be undertaken in this particular case to
evaluate the contribution of the slowly convergent integral
outside the border of resolution domain limited by y$_{max}$.

\begin{table}[tbh]
\caption{Neutron-deuteron scattering phaseshift and inelasticity
parameter as a function of the complex rotation angle
$\protect\theta $ compared with benchmark results
of~\cite{Friar_nd_bench,Arnas_bench}. Our calculations has been
performed by setting y$_{max}$=100 fm.} \label{tab_3Bnd_brca}
\vspace{0.5cm}
\begin{center}
\begin{tabular}{|ccccccccc|}\hline
& 3$^\circ$ & 4$^\circ$ & 5$^\circ$ & 6$^\circ$ & 7.5$^\circ$ &
10$^\circ$ & 12.5$^\circ$ & Ref.~\cite{Friar_nd_bench,Arnas_bench}
\\\hline\hline
 \multicolumn{9}{c}{nd doublet at E$_{lab}$=14.1 MeV } \\\hline
Re($\delta )$ & 105.00 & 105.43 & 105.50 & 105.50 & 105.50 &
105.49 & 105.48 & 105.49 \\
 $\eta $ & 0.4559 & 0.4638
& 0.4653 & 0.4654 & 0.4653 & 0.4650 & 0.4649 & 0.4649 \\\hline
 \multicolumn{9}{c}{nd doublet at E$_{lab}$=42 MeV } \\\hline
Re($\delta )$ & 41.71 & 41.63 & 41.55 & 41.51 & 41.45 & 41.04 &  & 41.35 \\
$\eta $ & 0.5017 & 0.5015 & 0.5014 & 0.5014 & 0.5015 & 0.5048 &  &
0.5022 \\\hline
 \multicolumn{9}{c}{nd quartet at E$_{lab}$=14.1 MeV } \\\hline
Re($\delta )$ & 68.47 & 68.90 & 68.97 & 68.97 & 68.97 & 68.97 &
68.97 & 68.95
\\
$\eta $ & 0.9661 & 0.9762 & 0.9782 & 0.9784 & 0.9783 & 0.9782 &
0.9780 & 0.9782 \\\hline
 \multicolumn{9}{c}{nd quartet at E$_{lab}$=42 MeV } \\\hline
Re($\delta )$ & 37.83 & 37.80 & 37.77 & 37.77 & 37.74 & 38.06 &  & 37.71 \\
$\eta $ & 0.9038 & 0.9034 & 0.9032 & 0.9030 & 0.9029 & 0.8980 & &
0.9033
\\\hline
\end{tabular}
\end{center}
\end{table}

\begin{table}[tbh]
\caption{Proton-deuteron scattering phaseshifts and inelasticity
parameters as a function of the complex rotation angle
$\protect\theta $ compared with benchmark values
of~\cite{Arnas_bench}. Our calculations has been performed by
setting y$_{max}$=150 fm. }
\label{tab_3Bpd_brca}%
\begin{center}
\vspace{0.5cm}
\begin{tabular}{|ccccccccc|}\hline
& 3$^\circ$ & 4$^\circ$ & 5$^\circ$ & 6$^\circ$ & 7.5$^\circ$ &
10$^\circ$ & 12.5$^\circ$ & Ref.~\cite{Arnas_bench} \\\hline\hline
 \multicolumn{9}{c}{pd doublet at E$_{lab}$=14.1 MeV } \\\hline
Re($\delta )$ & 108.46 & 108.43 & 108.43 & 108.43 & 108.43 &
108.43 & 108.42
& 108.41[3] \\
$\eta $ & 0.5003 & 0.4993 & 0.4990 & 0.4988 & 0.4986 &  0.4984 &
0.4981  & 0.4983[1] \\\hline
 \multicolumn{9}{c}{pd doublet at E$_{lab}$=42 MeV } \\\hline
Re($\delta )$ & 43.98 & 43.92 & 43.87 & 43.82 & 43.78 & 44.83 & -
& 43.68[2]
\\
$\eta $ & 0.5066 & 0.5060 & 0.5056 & 0.5054 & 0.5052 & 0.5488 & -
& 0.5056\\\hline
 \multicolumn{9}{c}{pd quartet at E$_{lab}$=14.1 MeV } \\\hline
Re($\delta )$ & 72.70 & 72.65 & 72.65 & 72.64 & 72.64 & 72.63 &
72.62 & 72.60
\\
$\eta $ & 0.9842 & 0.9827 & 0.9826 & 0.9826 & 0.9826 & 0.9828 &
0.9829 & 0.9795[1] \\\hline
 \multicolumn{9}{c}{pd quartet at E$_{lab}$=42 MeV } \\\hline
Re($\delta )$ & 40.13 & 40.11 & 40.08 & 40.07 & 40.05 & 40.35 & -
& 39.96[1]
\\
$\eta $ & 0.9052 & 0.9044 & 0.9039 & 0.9036 & 0.9034 & 0.9026 & -
& 0.9046
\\\hline

\end{tabular}
\end{center}
\end{table}
\bigskip
\begin{table}[tbp]
\caption{Neutron-deuteron $^3S_1$ breakup amplitude calculated at
$E_{lab}$=42 MeV as a function of the breakup angle $\vartheta$.}
\label{tab_3B_brca}%
\begin{center}
\vspace{0.5cm} \footnotesize{
\begin{tabular}{|c|cccccccccc|}\hline
& 0$^\circ$  & 10$^\circ$  & 20$^\circ$  & 30$^\circ$   &
40$^\circ$   & 50$^\circ$  & 60$^\circ$  & 70$^\circ$  &
80$^\circ$  & 90$^\circ$    \\\hline\hline This work
Re($^{3}$S$_{1}$) & 1.49[-2] & 8.84[-4] & -3.40[-2] & 3.33[-2] &
7.70[-2] &
2.52[-1] & 4.47[-1] & 6.47[-1] & 6.30[-1] & -1.62[-1]   \\
This work Im($^{3}$S$_{1}$) & 1.69[0] & 1.74[0] & 1.87[0] &
1.92[0] & 1.80[0] & 1.68[0] & 1.70[0] & 1.96[0] & 2.23[0] &
3.17[0]\\\hline Ref.~\cite{Friar_nd_bench} Re($^{3}$S$_{1}$) &
1.48[-2] & 9.22[-4] & -3.21[-2] & 3.09[-2] & 7.70[-2] &
2.52[-1] & 4.51[-1] & 6.53[-1] & 6.93[-1] & -1.05[-1]   \\
Ref.~\cite{Friar_nd_bench} Im($^{3}$S$_{1}$) & 1.69[0] & 1.74[0] &
1.87[0] & 1.92[0] & 1.80[0] & 1.67[0] & 1.70[0] & 1.95[0] &
2.52[0] & 3.06[0] \\\hline
\end{tabular}}
\end{center}
\end{table}

\section{Three-body scattering including optical potentials\\ \footnotesize{\textit{(Results presented in this section are based on
the study~\cite{DFL12})}}}

The three-nucleon system is the only nuclear three-particle system
that may be considered as a realistic in the sense that the
interactions are given by high precision potentials valid over a
broad energy range. Nevertheless, in the same way one considers a
nucleon as a single particle by neglecting its inner quark
structure, in a further approximation one can consider a cluster
of nucleons (composite nucleus) to be a single particle that
interacts with other nucleons or nuclei via effective potentials
whose parameters are determined from the two-body data.
 A classical example is the $\alpha$ particle, a tightly
bound four-nucleon cluster. As demonstrated in
Ref.~\cite{deltuva:06b}, the description of the
 $(\alpha,p,n)$ three-particle system with real potentials
is quite successful at low energies but becomes less reliable with
increasing energy where the inner structure of the $\alpha$
particle cannot be neglected anymore. At higher energies the
nucleon-nucleus or nucleus-nucleus interactions are modelled by
optical potentials (OP) that provide quite an accurate description
of the considered two-body system in a given narrow energy range;
these potentials are complex to account for the inelastic
excitations not explicitly included in the model space. The
complex scaling method built on Faddeev-Merkuriev equations can be
applied also in this case, however, the potentials within the
pairs that are bound in the initial or final channel must remain
real.

In the past, the description of three-body-like nuclear reactions
involved a number of approximate methods that have been developed.
Well-known examples are the distorted-wave Born approximation
(DWBA), various adiabatic approaches \cite{johnson:70a}, and
continuum-discretized coupled-channels (CDCC)
method~\cite{austern:87}. The first fully rigorous solution of
this problem has been realized in in Ref.~\cite{deltuva:06b} by
solving Alt, Grasseberger and Sandhas equations
(AGS)~\cite{alt:jinr} formulated in momentum space. These
equations are formally equivalent to the 3-body Faddeev equations.
The comparison of the two methods based:  solution of the AGS and
complex scaled Faddeev-Merkuriev equations will be performed in
the next section~\ref{sec:compare} for a chosen 3-body problem
involving OP.

Compared to DWBA or CDCC, the present methods based on exact
Faddeev or AGS equations, being  more technically and involved,
have some disadvantages. Namely, their application in the present
technical realization is so far limited to a system made of two
nucleons and one heavier cluster. The reason is that the
interaction between two  heavy clusters involves very many angular
momentum states and the partial-wave convergence becomes very
slow. The comparison between traditional nuclear reaction
approaches and momentum-space Faddeev/AGS methods for various
neutron + proton + nucleus systems has been realized by A. Deltuva
\textit{et al.}
in~\cite{PhysRevC.76.064602,PhysRevC.79.064610,PhysRevC.83.054613,PhysRevC.85.054621}.

On the other hand, the  Faddeev and AGS  methods may be more
flexible with respect to dynamic input and thereby allows to test
novel aspects of the nuclear interaction not accessible with the
traditional approaches.

\subsection*{Numerical comparison of AGS and FM methods} \label{sec:compare}

As a test case, the $n+p+^{12}C$ system is considered. For the
$n$-$p$ interaction we use a realistic AV18 model~\cite{AV18} that
accurately reproduces the available two-nucleon scattering data
and deuteron binding energy. To study not only the $d+{}^{12}$C
but also $p+{}^{13}$C scattering and transfer reactions we use a
$n$-$^{12}$C potential that is real in the $^2P_\frac{1}{2}$
partial wave and supports the ground state of $^{13}C$ with 4.946
MeV binding energy; the parameters are taken from
Ref.~\cite{nunes:11b}. In all other partial waves we use the
$n$-$^{12}$C optical  potential from Ref.~\cite{CH89} taken at
half the deuteron energy in the $d+{}^{12}$C channel. The
$p$-$^{12}$C optical potential is also taken from
Ref.~\cite{CH89}, however, at the proton energy in the
$p+{}^{13}$C channel. We admit that, depending on the reaction of
interest, other choices of energies for OP may be more
appropriate, however, the aim of the present study is comparison
of the methods and not the description  of the experimental data
although the latter are also included in the plots.

\begin{figure}[h!]
\includegraphics[width=14.cm]{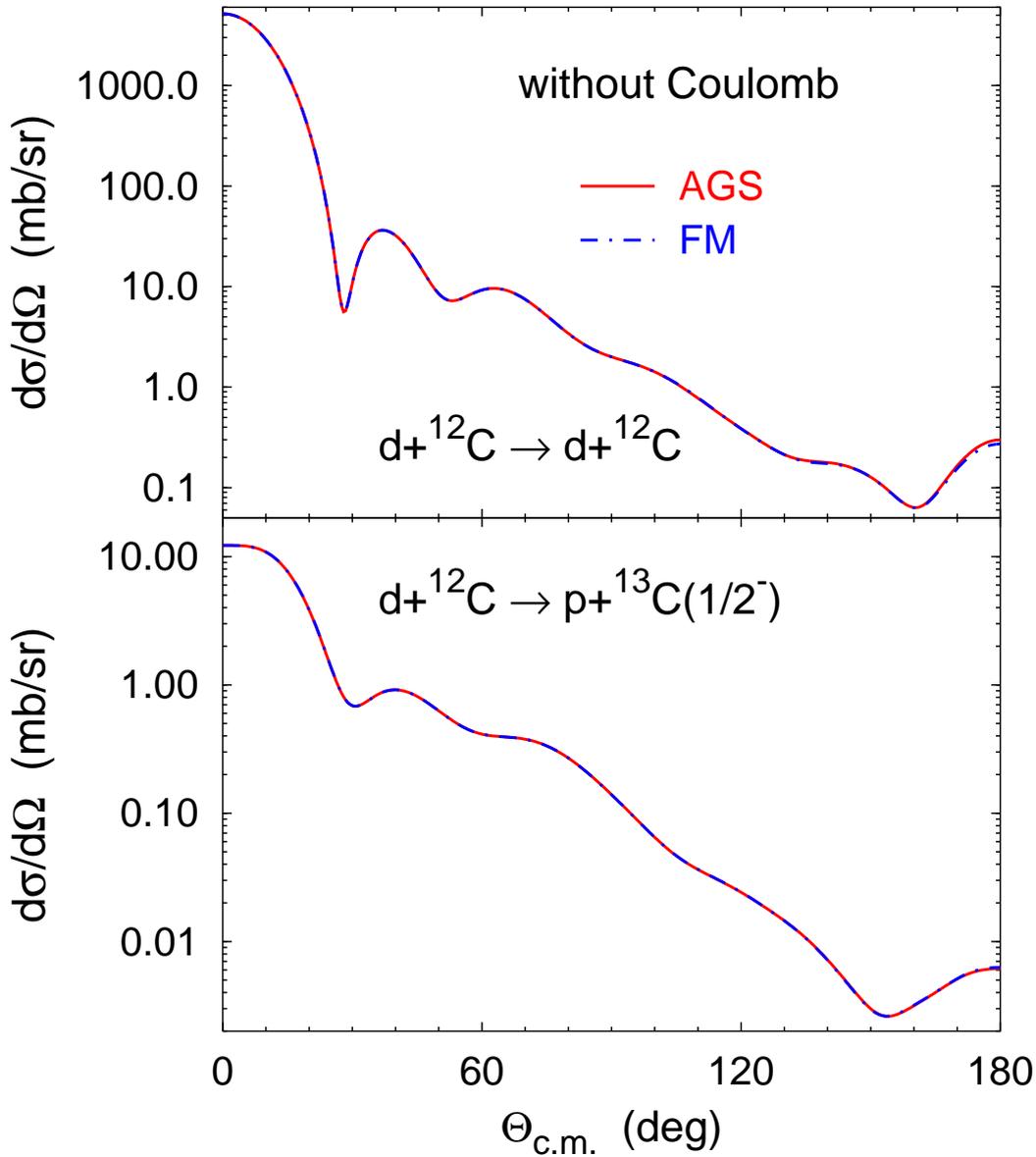}
\caption{ Comparison of momentum- (solid curves) and
configuration-space (dashed-dotted curves) results for the
deuteron-${}^{12}$C scattering at 30 MeV deuteron lab energy.
Differential cross sections for elastic scattering and stripping
are shown neglecting the Coulomb interaction.} \label{fig:dC-noC}
\end{figure}

We consider $d+{}^{12}$C scattering at 30 MeV deuteron lab energy
and  $p+{}^{13}$C scattering at 30.6 MeV proton lab energy; they
correspond to the same energy in c.m. system. First we perform
calculations by neglecting the $p$-$^{12}$C Coulomb repulsion. One
observes a perfect agreement between the AGS and FM methods.
 Indeed, the calculated S-matrix elements in each three-particle
channel considered (calculations have been performed for total
three-particle angular momentum states up to $J=13$) agree  within
 three digits. Scattering observables  converge quite slowly with $J$
as different angular momentum state contributions cancel each
other at large angles. Nevertheless, the results of the two
methods are practically indistinguishable as demonstrated in
Fig.~\ref{fig:dC-noC} for $d+{}^{12}$C elastic scattering and
transfer to $p+{}^{13}$C.

\begin{figure}[h!]
\includegraphics[width=14.cm]{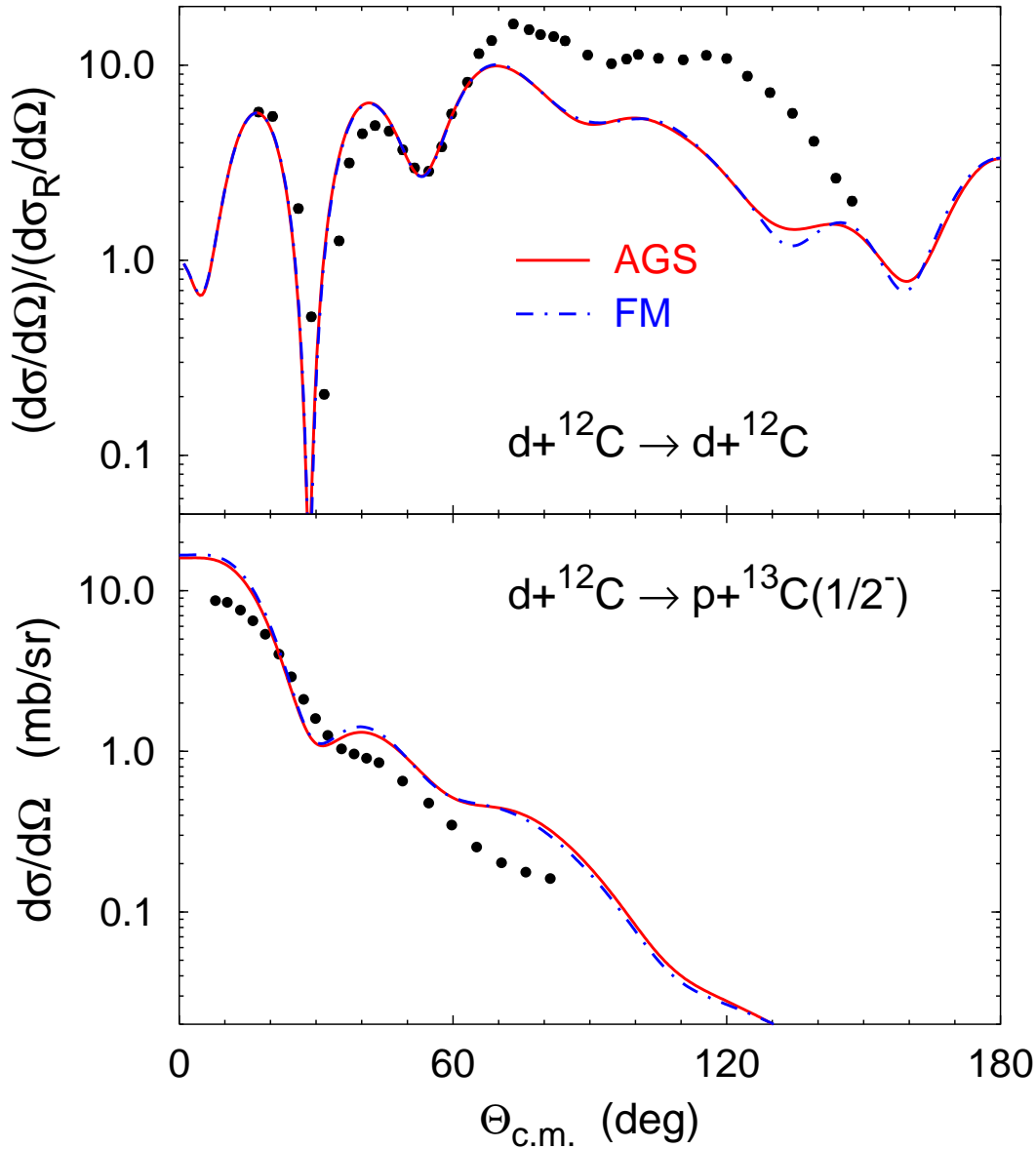}
\caption{ Comparison of momentum- (solid curves) and
configuration-space (dashed-dotted curves) results for the
deuteron-${}^{12}$C scattering at 30 MeV deuteron lab energy.
Differential cross sections for elastic scattering and stripping
are shown, the former in ratio to the Rutherford cross section
$d\sigma_R/d\Omega$. The experimental data are from
Refs.~\cite{perrin:77,dC30p}.} \label{fig:dC}
\end{figure}

\begin{figure}[h!]
\includegraphics[width=14.cm]{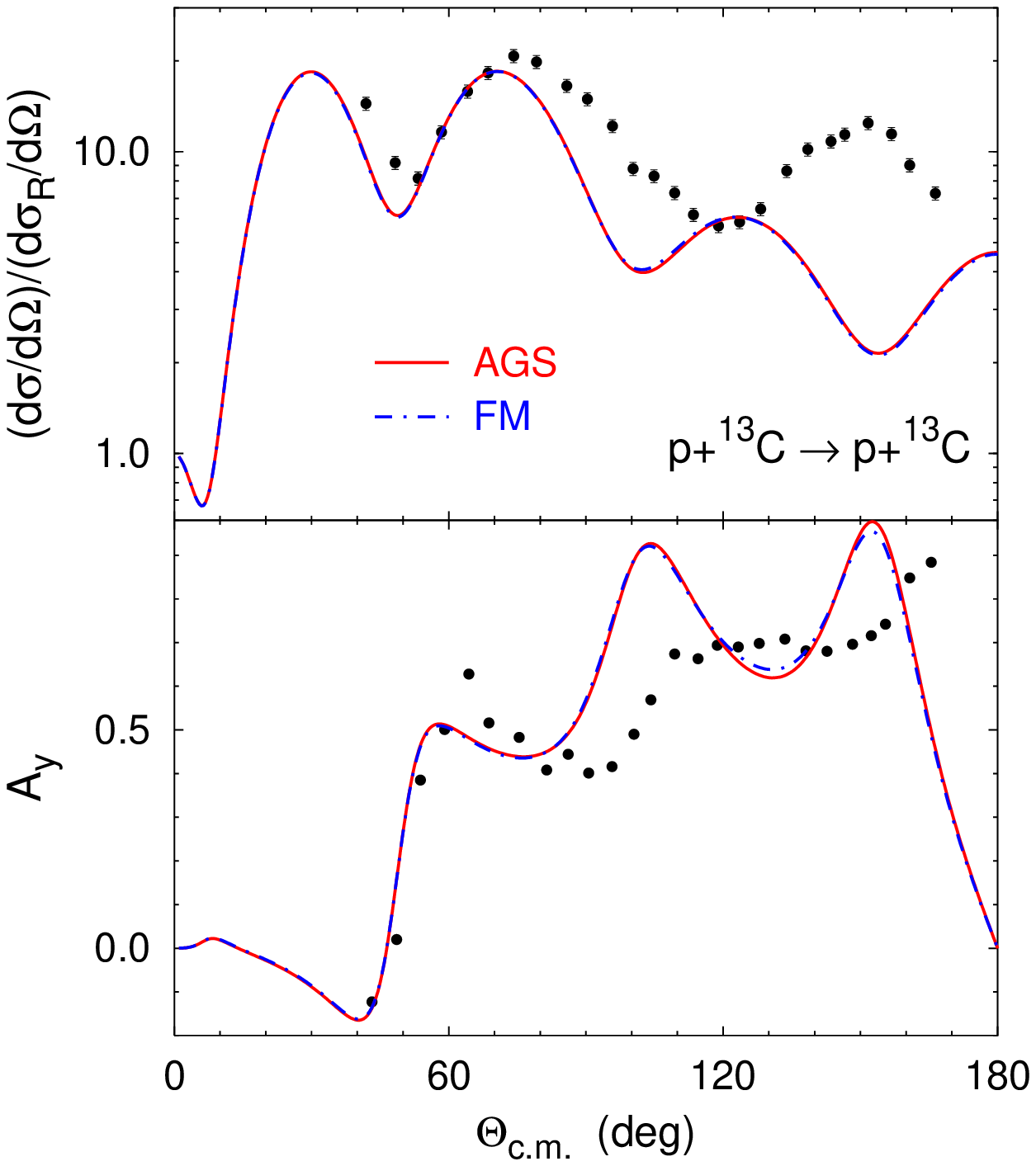}
\caption{ Comparison of momentum- (solid curves) and
configuration-space (dashed-dotted curves) results for the
proton-${}^{13}$C elastic scattering at 30.6 MeV proton lab
energy. Differential cross section divided by the Rutherford cross
section and proton analyzing power are shown. The experimental
data are from Ref.~\cite{pC30}.} \label{fig:pC}
\end{figure}

Next we perform the full calculation including the $p$-$^{12}$C
Coulomb repulsion;  we note that inside the nucleus the Coulomb
potential is taken as the one of a uniformly charged
sphere~\cite{deltuva:06b}. Once again we obtain good agreement
between the AGS and FM methods. However, this time small
variations up to the order of 1\% are observed when analyzing
separate $S$-matrix elements, mostly in high angular momentum
states.
 This leads to small differences in some scattering observables, e.g.,
differential cross sections for $d+{}^{12}$C elastic scattering
(at large angles where the differential cross section is very
small) and for the deuteron stripping reaction $d+{}^{12}$C $ \to
p+{}^{13}$C shown in Fig.~\ref{fig:dC}. The  $p+{}^{13}$C  elastic
scattering observables presented in Fig.~\ref{fig:pC} converge
faster with $J$. As a consequence, the results of the two
calculations are indistinguishable
 for the $p+{}^{13}$C elastic cross section
and only tiny differences  can be seen for the proton analyzing
power at large angles. In any case, the agreement between the AGS
and FM methods exceeds both the accuracy of the data and the
existing discrepancies between theoretical predictions and
experimental data.

\section{Four-nucleon scattering using phenomenological interactions\\ \footnotesize{\textit{(Results presented in this section are based on
the study~\cite{La12})}}}

As discussed above, the MT I-III potential turns to be very well
adapted to perform various tests in studying few-nucleon systems.
Therefore this model has been chosen for the first implementation
of the CS method in describing four-nucleon reactions. Within this
model, the nuclear interaction turns out to be isospin independent
and thus nucleonic systems conserve the total isospin ($T$). In
addition, due to the S-wave limitation of the  MT I-III potential,
nucleonic systems separately conserve the total spin and the
orbital angular momentum. This potential is fitted to reproduce
the correct binding energies of the deuteron ($^2H$) and the
triton ($^3H$), at -2.230 \textrm{MeV} and -8.535 \textrm{MeV}
respectively. However, the absence of the Coulomb interaction
makes $^3He$ ground state to be located at the same energy as the
$^3H$ ground state. Two-cluster collisions are available in $T=1$
and $T=0$ channels, which will be discussed further on.

The calculations were performed by employing spline collocation
method, described in the section \ref{sec:spline_coll}, and based
on numerical techniques developed
in~\cite{These_Rimas_03,Lazauskas_4B,LC11}.
 50 discretization points in each direction (x,y,z) have been used.
The complex scaling angle was fixed at $\theta=$9$^\circ$.
Vanishing boundary conditions for FY partial amplitudes were
imposed at the borders of the discretized grid, which was varied
from 35 to 50 fm. The results have been tested to be stable when
modifying the scaling angle and the grid parameters. Basically,
the extracted amplitudes turn out to be accurate to 3-digits,
which guarantees the 3-digit accuracy for the extracted
phaseshifts. Nevertheless, this method is slightly less accurate
for the inelasticity parameter, especially once its value is very
close to 1. Due to the S-wave limitation of the interaction model,
partial amplitudes with $l_x\neq0$ do not contribute in solving FY
eq.~(\ref{FY_drive}), however, one must include these amplitudes
in evaluating the integrals of eq.~({\ref{integral_2B}). The
expansion into tripolar harmonics was limited by the
$max(l_x,l_y,l_z)\leq3$ condition. The results are converged to
four significative digits with  respect to the partial angular
momentum basis.

First of all, I present the results for the $T=1$ case, which well
reflects the reality of the $n-^3H$ collisions. The values of the
calculated phase shifts and the inelasticity parameters are
summarized in the table~\ref{tab_4nt_ps}. The phase-shifts are
obtained with very high accuracy, a variation is observed only in
the third digit. The variation of the inelasticity parameter is of
the order 0.005, which looks as a rather accurate result.
Nevertheless, since the values of the inelasticity parameter  are
very close to unity such accuracy might be critical in determining
the small value of the total break-up cross-section.

In the table~\ref{tab_4nt_cs}, the calculated total elastic
cross-sections are compared with the experimental values. One may
notice a rather good agreement. These calculations have been
performed for total orbital momentum states $L\leq3$ and seem to
be converged to this respect.
 In figure~\ref{fig:r_dep} we present the comparison of the differential elastic cross-sections,
calculated for the incident neutron at lab. energy 14.4 MeV (left
pane) and 22.1 MeV (right pane), with the experimental values. One
may notice that a rather good agreement is also obtained  in this
case. Only at the minimum region, for the 14.4 MeV neutrons, the
theoretical results underestimate the experimental values.
Nevertheless, the overall agreement remains very good and is far
beyond expectations for such a simplistic interaction model as MT
I-III. It proves that the $n-^3H$ cross-sections at higher energy,
beyond the resonance region, are rather insensitive to the details
of the nucleon-nucleon interaction. As has been shown
recently~\cite{Arnas_cem:2012} the realistic interaction models
further improve description of $n-^3H$ elastic cross-sections,
providing almost perfect agreement with the data also in the
minimum region.

Next we consider the total isospin $T=0$ case. This isospin
channel is a very rich one, combining the $d+d$, $n-^3H$ and
$p-^3He$ binary scattering modes in addition to 3- and 4-particle
break-up ones. Due to the absence of the Coulomb interaction,  the
$n-^3H$ and $p-^3He$ thresholds coincide in our calculations. The
soundness of these calculations is further shrouded by the fact
that we neglect the Coulomb interaction in the asymptotes of the
open channels. Therefore, there is not much sense in comparing the
obtained results compiled in the table~\ref{tab_20_5} with the
experiment. One may notice, see table~\ref{tab_3B_uzu}, that our
obtained values are also rather different from the ones calculated
for $J^\pi=0^+$ case by Uzu et al.~\cite{Uzu_prc68:2003}, who have
used the same assumptions as in the present paper but employed a
separable Yamaguchi interaction.  The last fact indicates the
strong sensitivity of the $T=0$ channel to the details of the
nucleon-nucleon interaction. However, this sensitivity is not
surprising, as the $T=0$ channel is strongly attractive and
contains the series of  resonances also above the four-particle
break-up threshold. It is also confirmed by rather large inelastic
cross-sections (inelasticity parameters).

\begin{figure}[h!]
\centering
\includegraphics[width=81mm]{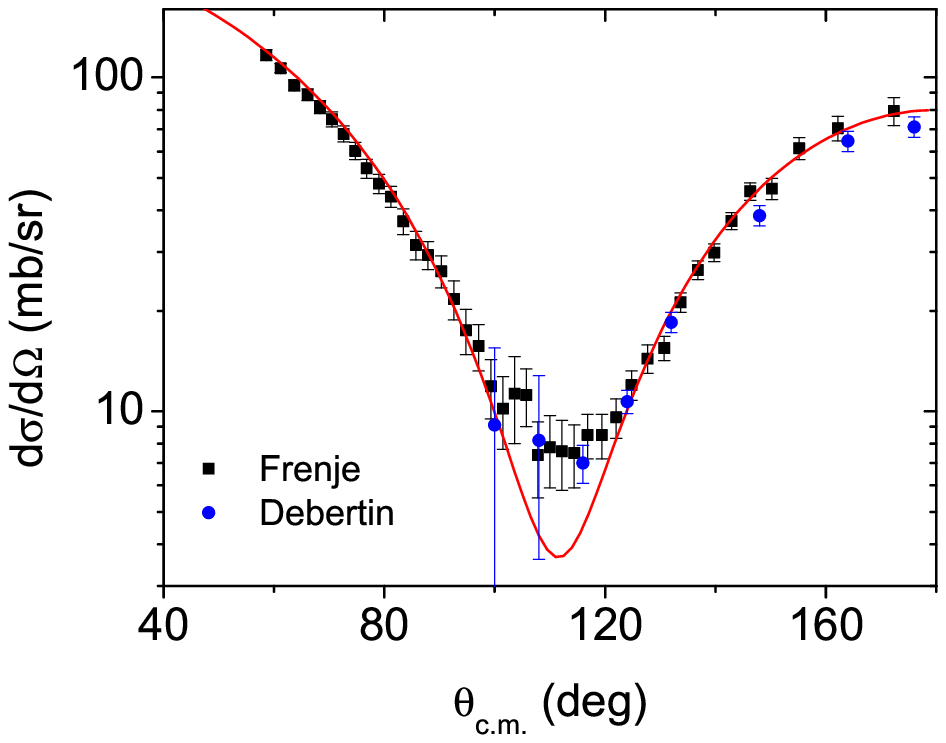}
\includegraphics[width=81mm]{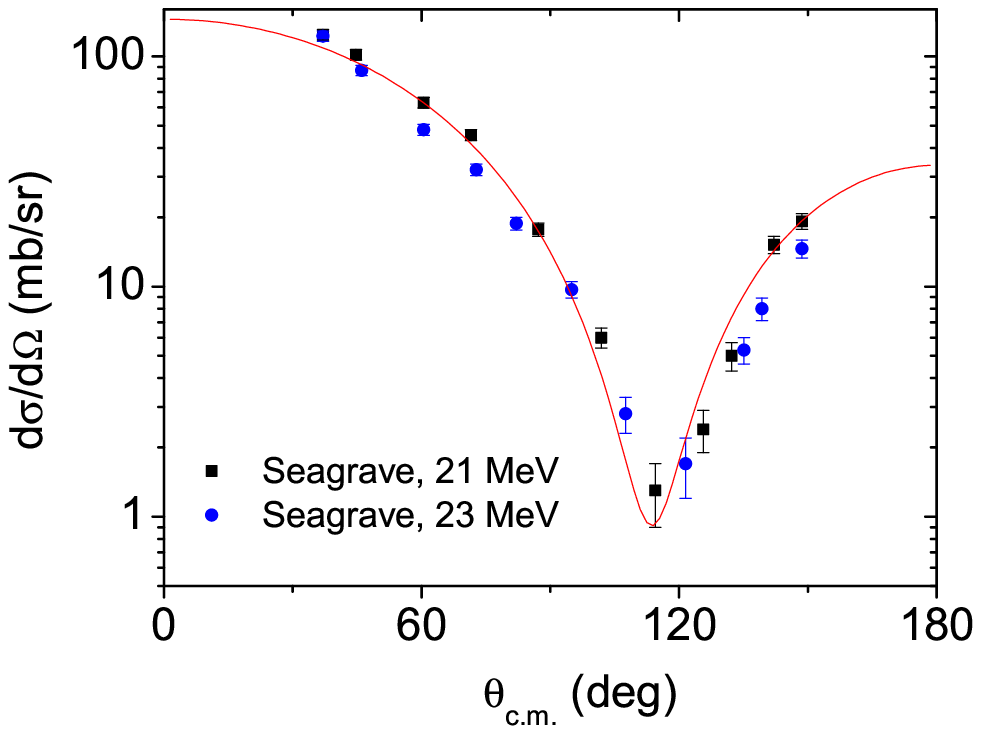}

\caption{Calculated $n-^3H$ elastic differential cross-sections
for neutrons of lab. energy 14.4 \textrm{MeV} (left panel) and
22.1 \textrm{MeV} (right panel) compared with the experimental
results of Frenje et al.~\cite{Frenje_nt}, Debretin et
al.~\cite{Debertin} and Seagrave et al.~\cite{Seagrave}.}
\label{fig:r_dep}
\end{figure}

\begin{table}[tbp]
\caption{Neutron-triton scattering phaseshifts (in degrees) and
inelasticity parameters. Accuracy for calculated phaseshifts is
about 0.1 \textrm{deg}, whereas inelasticity parameter has
accuracy around 0.005. }
\label{tab_4nt_ps}%
\vspace{0.5cm}
\begin{center}
\begin{tabular}{|c|cc|cc|cc|}\hline

 $E_{\textrm{lab.}}$              & \multicolumn{2}{c|}{$L=0$} &   \multicolumn{2}{c|}{$L=1$} &   \multicolumn{2}{c|}{$L=2$} \\
 (MeV) & $S=0$ & $S=1$ & $S=0$   & $S=1$ & $S=0$ & $S=1$ \\ \hline\hline
14.4   & 72.7    & 81.2  & 40.0  & 57.4  & -3.92 & -2.45  \\
       & 0.993   & 0.988 & 0.988 & 1.00  & 0.999 & 0.988 \\\hline
18.0   & 65.5    & 74.4  & 38.8  & 55.4  & -3.24 &  -1.98\\
       & 0.990   & 0.984 & 0.968 & 0.983 & 0.995 & 0.973\\\hline
22.1   & 58.4    & 67.4  & 37.1  & 53.0  & -2.40 & -1.21\\
       & 0.988   & 0.983 & 0.944 & 0.952 & 0.988 & 0.955\\\hline
\end{tabular}
\end{center}
\end{table}

\begin{table}[tbp]
\caption{Neutron-triton elastic ($\sigma_e$), inelastic
($\sigma_b$) and total ($\sigma_t$) scattering cross-sections (in
units of \textrm{mb}) for the selected neutron \textrm{lab.}
energies (in units of \textrm{MeV}) compared with the experimental
data. Calculations has been limited to the maximal total orbital
angular momentum states $L\leq3$. }
\label{tab_4nt_cs}%
\vspace{0.5cm}
\begin{center}
\begin{tabular}{|c|ccc|cc|}\hline
 $E_{\textrm{lab.}}$ &   \multicolumn{3}{c|}{MT I-III} &   \multicolumn{2}{c|}{ Exp. } \\
  (MeV)  & $\sigma_e$ & $\sigma_b$ & $\sigma_t$ & $\sigma_t$ & [Ref.]  \\
  \hline\hline
14.4           & 922    & 11  & 933 &978$\pm$70 &\cite{Battat}  \\
18.0           & 690    & 25  & 715 &750$\pm$40 &\cite{Battat}  \\
22.1           & 512    & 38  & 550 &620$\pm$24
&\cite{phillips:80}\\\hline
\end{tabular}

\end{center}

\end{table}

\begin{table}[tbp]
\caption{Nucleon-trinucleon scattering phaseshifts (in degrees)
and inelasticity parameters calculated for the center of mass
energy of 20.5 MeV and 30 MeV, nucleon lab. energies of 27.3 MeV
and 40 MeV respectively.}
\label{tab_20_5}%
\vspace{0.5cm}
\begin{center}
\begin{tabular}{|cc|cc|cc|}\hline

 & &  \multicolumn{2}{c|}{$E_{\textrm{c.m.}}=20.5$  MeV} & \multicolumn{2}{c|}{$E_{\textrm{c.m.}}=30$  MeV} \\
 &  & $\delta$ (deg) & $\eta$ & $\delta$ (deg) & $ \eta$ \\ \hline\hline
 L=0 & S=0 & -56.6 & 0.650 & -81.0 & 0.618 \\
     & S=1 & 68.8 & 0.947 & 56.9 & 0.882 \\\hline
  L=1 & S=0 & -85.3 & 0.945 & 78.9 & 0.918 \\
     & S=1 & 64.9 & 0.886 & 52.8 & 0.843 \\ \hline
   L=2 & S=0 & 47.1 & 0.678 & 44.7 & 0.720 \\
     & S=1 & 1.09 & 0.896 & 4.49 & 0.851 \\\hline
\end{tabular}
\end{center}
\end{table}

\begin{table}[tbp]
\caption{Nucleon-trinucleon scattering phaseshifts (in degrees)
and inelasticity parameters for $J^\pi=0^+$ case and at the chosen
center of mass projectile energies (in units of MeV). The results
of this manuscript using MT I-III interaction are compared with
the ones of ref.~\cite{Uzu_prc68:2003}, who employed the Yamaguchi
potential.}
\label{tab_3B_uzu}%
\vspace{0.5cm}
\begin{center}
\begin{tabular}{|c|cc|cc|}\hline
 $E_{\textrm{c.m.}}$& \multicolumn{2}{c|}{MT I-III (this work)} & \multicolumn{2}{c|}{Yamaguchi (ref.~\cite{Uzu_prc68:2003})}
 \\
         & $\delta$ (\textrm{deg}) & $\eta$ & $\delta$ (\textrm{deg}) & $ \eta$ \\ \hline\hline
 7.3  & -4.46 & 0.988 & -5.51 & 0.899 \\
 20.5  & -56.6 & 0.650 & -61.7 & 0.746 \\\hline
\end{tabular}
\end{center}
\end{table}

\section{Four-nucleon scattering using realistic interactions\\ \footnotesize{\textit{(Results presented in this section are based on
the study~\cite{PhysRevC.91.041001})}}}

In the previous section CS method has been already applied to
study n-$^3$H scattering above the breakup threshold~\cite{La12}.
In that work, realized in collaboration with J.~Carbonell, due to
large numerical costs we were obliged to use the simplistic
S-waves nucleon-nucleon interaction model. More recently spline
collocation method, employed previously to discretize radial
dependence of FY amplitudes, has been replaced by the
Lagrange-mesh technique. This modification allowed to improve
significantly numerical accuracy and challenge realistic
description of the four-nucleon reactions above the three and
four-fragment breakup thresholds. As the first step of the longer
program intended to cover fully four-nucleon continuum  I have
realized calculations of neutron scattering on $^3$H nucleus.
Calculations presented in this section have been performed using
three formally and structurally different realistic nuclear
Hamiltonians: INOY04~\cite{Dolesch}, $\chi$N3LO~\cite{xN3LO} and
AV18~\cite{AV18}.

Some years ago pioneering realistic calculation on n-$^3$H system
above the breakup threshold has been  undertaken  by A.~Deltuva et
al.~\cite{Arnas_cem:2012}. In his works A.~Deltuva employs
momentum space formulation of the complex-energy
method~\cite{MDN69,PPNP}. In Table~\ref{tab_nt_phs} the
phaseshifts and the inelasticity parameters obtained in this study
are compared with the ones published by A.~Deltuva for INOY04
model. An excellent agreement is obtained between the two
calculations reaching three-digit accuracy. The largest
discrepancy of 0.5\% is observed for the inelasticity parameter in
$^1S_0$ channel, which is due to the fact that this parameter is
very close to unity.

\begin{table}[tbp]
\caption{ Some calculated phaseshifts
 $\delta $ and inelasticity
parameters $\eta $ for 22.1 MeV neutron scattering on triton using
INOY04 potential. This work results are compared with the ones
from ref.~\cite{Arnas_cem:2012}.} \label{tab_nt_phs}
\vspace{0.5cm}
\begin{center}
\begin{tabular}{|l|ll|ll|}\hline
&\multicolumn{2}{c|}{ $\eta$ }& \multicolumn{2}{c|}{ $\delta$ (deg.) }\\
PW& This work & Ref.~\cite{Arnas_cem:2012}& This work & Ref.~\cite{Arnas_cem:2012}
\\\hline\hline
$^1S_0$ &0.985 & 0.990 & 62.74 & 62.63\\
$^3P_0$ & 0.959&  0.959& 43.07 &43.03\\
$^3P_2$ &0.949& 0.950 &65.25& 65.27 \\\hline

\end{tabular}
\end{center}
\end{table}

\begin{table}[tbp]
\begin{center}
\caption{ Integrated elastic ($\sigma_{el}$), breakup
($\sigma_{b}$) and total ($\sigma_{t}$) cross sections for neutron
scattering on $^3$H. Calculations have been performed using INOY04
NN potential model. This work results are compared with the ones
from ref.~\cite{Arnas_cem:2012} and experimental values
from~\cite{Battat,phillips:80}.} \label{tab_nt_cs} \vspace{0.5cm}
\vspace{0.5cm}
\begin{tabular}{|l|lll|lll|l|}\hline

& \multicolumn{3}{c|}{ This work } & \multicolumn{3}{c|}{Ref.~\cite{Arnas_cem:2012} } & Exp. \\

$E_n$ (MeV) & $\sigma_{el}$ (mb) & $\sigma_{b}$ (mb)&
$\sigma_{tot}$ (mb)& $\sigma_{el}$ (mb) & $\sigma_{b}$ (mb)&
$\sigma_{tot}$ (mb)& $\sigma_{tot}$ (mb) \\\hline\hline

14.1 & 927 & 19 & 947 & 928 & 19 & 947 & 978$\pm$70 \\
18.0 & 697 & 42 & 739 & 697 & 41 & 738 & 750$\pm$40 \\
22.1 & 535 & 61 & 596 & 536 & 61 & 597 & 620$\pm$24 \\\hline

\end{tabular}
\end{center}
\end{table}

Excellent agreement between the two calculations is also obtained
for the integrated cross sections, see Table~\ref{tab_nt_cs}.
These calculations includes all the scattering states with total
angular momentum J$\leq$5. Including more partial waves yields no
change for the elastic cross section and only entirely
insignificant changes for the breakup one. The total cross
sections are also in good agreement with the experimental data
from M.~E. Battat~\cite{Battat} and T.~W.
Phillips~\cite{phillips:80} -- they fall within experimental
error-bars but favors slightly lower values than the experimental
centroid.

 In figure~\ref{fig:nt_dcs} the elastic
differential cross section  as well as the neutron analyzing power
$A_y$ are presented
 for 22.1 MeV neutron scattering on triton. In this figure results
 obtained using three different realistic nuclear Hamiltonians,
 namely
 INOY04~\cite{Dolesch}, $\chi$N3LO~\cite{xN3LO} and
AV18~\cite{AV18}, are presented. Before discussing agreement with
the experimental data, one should notice  that
 not all of the employed Hamiltonians are equally
 successful in describing bound state properties of the $^3$H
 (i.e. target nucleus). It is commonly accepted that most
 of the nuclear interaction
models require three-nucleon force to provide extra binding for
the trinucleon. The INOY04, $\chi$N3LO and AV18 models produce the
tritons with binding energy of 8.48, 7.85 and 7.62 MeV
respectively, and thus with exception of the INOY04 model, they
underbind triton (experimental binding energy of the triton is
8.482 MeV ). However correct positioning of the thresholds are
crucial in describing low-energy scattering cross sections. In
vicinity of a threshold, due to the kinematical form factor, the
breakup cross section increases with the available kinetic energy.
This feature is clearly demonstrated in the figure~\ref{4b_summ},
where total cross sections  provided by four different realistic
nucleon-nucleon interaction models are plotted against the binding
energy of $^3$H~\footnote{CD-Bonn model result is taken from the
ref.~\cite{Arnas_cem:2012}.}. On the other hand the total elastic
cross section has opposite behavior -- it increases with the
binding energy of the triton compensating effect from the breakup
cross section. One may observe linear correlation pattern for both
cross sections. Existence of such a correlation indicates that at
these energies the neutron cross sections are not very sensitive
to the off-shell structure of a nuclear Hamiltonian, being
determined by the on-shell properties of the 2-nucleon system and
the binding energy of the triton. It is expected that once
three-nucleon force is introduced to correct the binding energy of
the trinucleons, different realistic nuclear Hamiltonian
predictions should align with a result of INOY04 model. While
extensive model dependence of the n-$^3$H cross sections has been
performed only for 22.1 MeV neutrons, our other calculations
suggest that this tendency should remain valid for the broader
energy range above the three- and four-nucleon breakup thresholds.
On the other hand this tendency is clearly broken below the
three-nucleon breakup threshold, where four pronounced neutron
resonances are present~\cite{Lazauskas_4B,viviani:11a}.


The same correlation pattern is also observed for the differential
elastic cross section, see Fig.~\ref{fig:nt_dcs}. Elastic cross
section increases with the trinucleon binding energy, which is the
most pronounced at the cross-sections minima. Cross sections
provided by the INOY04 model, which must stand as a reference for
any realistic Hamiltonian calculation with correct trinucleon
threshold, provides the worst agreement with the experimental data
of~\cite{Seagrave} at the cross sections minima. On the other
hand, as demonstrated in~\cite{Arnas_cem:2012}, the calculated
cross sections at E$_n$=18 MeV lie in the middle between data sets
of~\cite{Seagrave} and~\cite{Debertin}. Thus one might expect a
lack of reliability for the data from ref.~\cite{Seagrave}. As
disagreement is due to  the cross sections minima underestimation
of the experimental error-bars might be the reason of this
discrepancy. New precise measurements are required to resolve this
discrepancy.

Agreement between the theoretical and the experimental neutron
analyzing powers is not perfect, however is much improved compared
to one obtained for slower neutrons. In particular it contrast
with the existence of the well known Ay-puzzle for p-$^3$He
scattering below p+p+d breakup threshold~\cite{Viviani_PRL:2013}.
\begin{figure}
\begin{center}
\hspace{-0.5cm}\includegraphics[width=85mm]{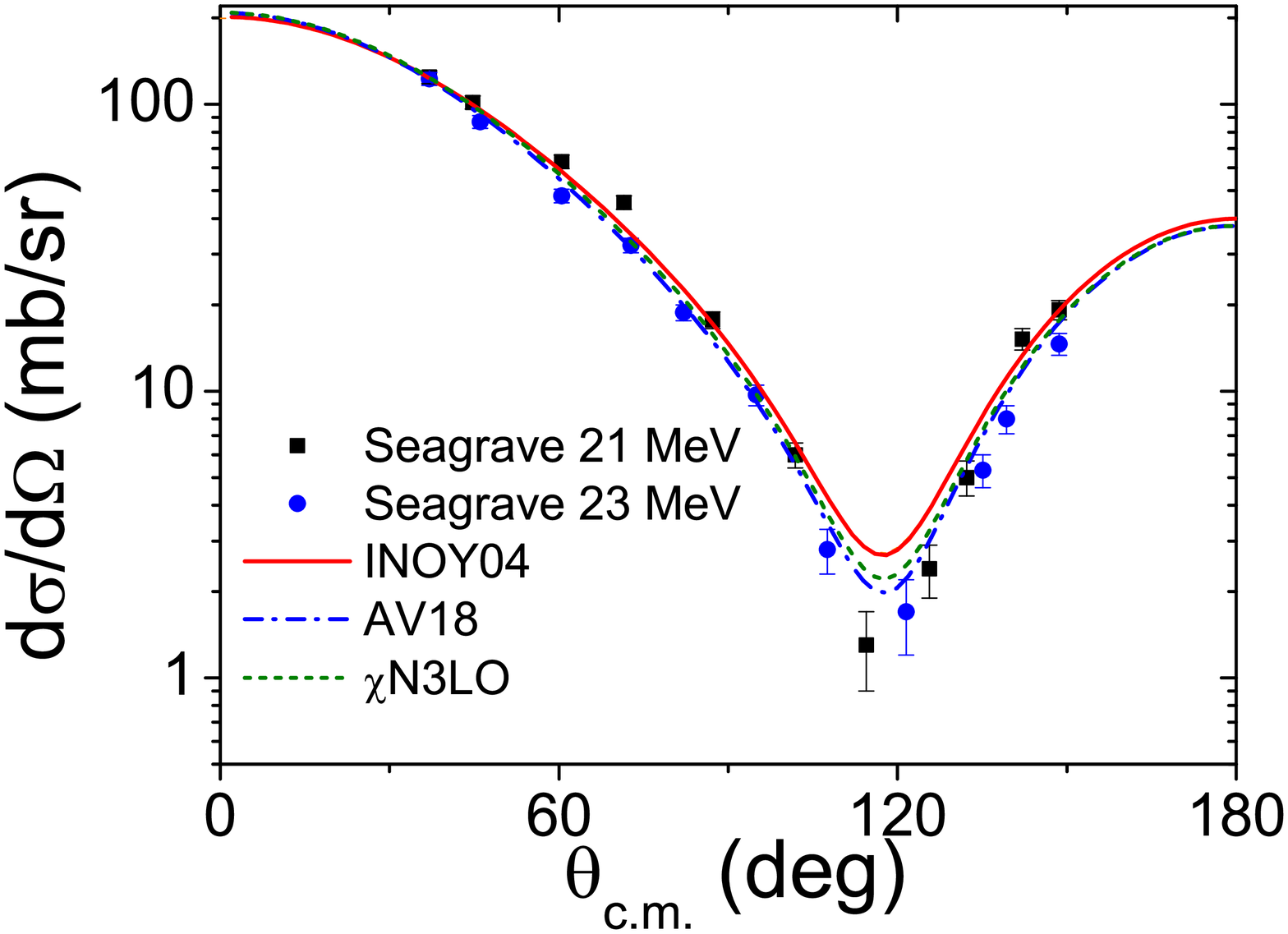}\hspace{-0.5cm}
\includegraphics[width=85mm]{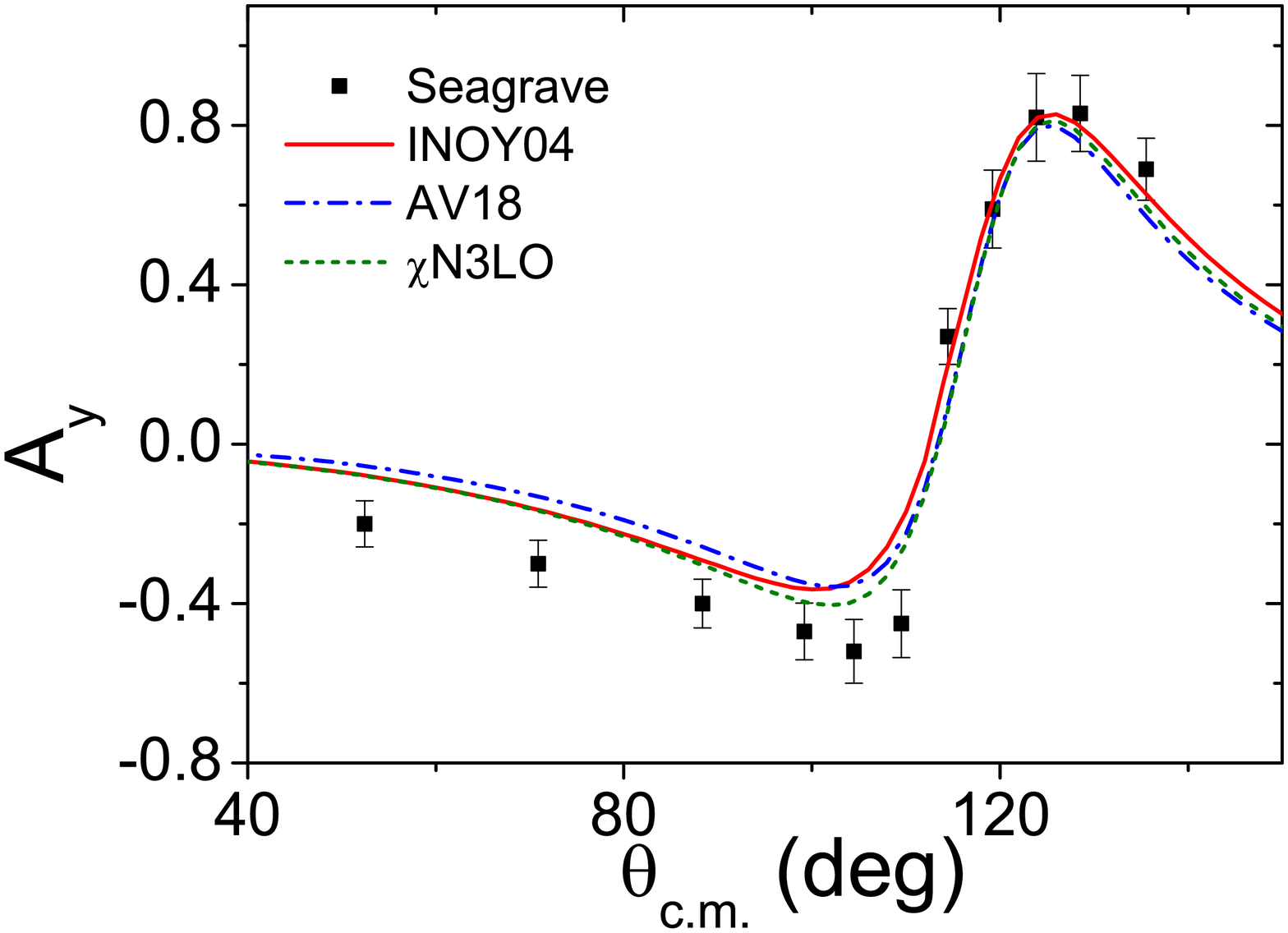}
\caption{\label{fig:nt_dcs} Calculated n-$^3$H elastic
differential cross sections (left panel) and neutron analyzing
power $A_y$ (right panel) for incident neutrons at laboratory
energy 22.1 MeV. Calculated values are compared with the
experimental results of J.~Seagrave et al.~\cite{Seagrave}. }
\end{center}
\end{figure}
\begin{figure}[h!]
\begin{center}
\includegraphics[width=14cm]{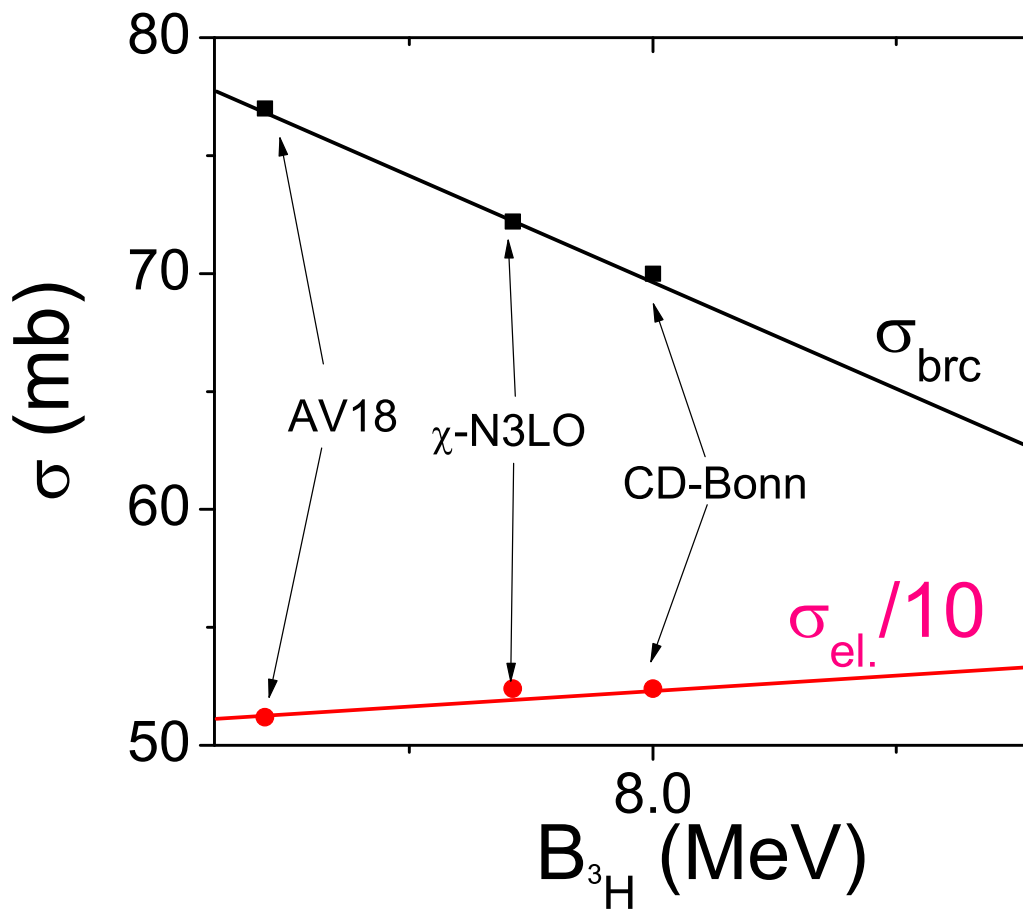}
\end{center}
\caption{Dependence of the calculated  n-$^3$H total elastic and
inelastic (breakup) cross sections on the triton binding energy.
Calculations have been performed for  neutrons with laboratory
energy of 22.1 MeV.} \label{4b_summ}
\end{figure}

\section{Three-body Coulomb scattering\\ \footnotesize{\textit{(Results presented in this section are based on
the study~\cite{Laz_csm_atomic})}}}

 The unique asset of the Complex scaling
method is that it presents an unified formalism enabling to treat
bound, resonant and scattering states. Nevertheless, an originally
formulated~\cite{NC69} smooth complex scaling method is not
directly applicable in solving scattering problems with the
long-range interactions. For this purpose exterior complex scaling
method has been proposed~\cite{ecsm_SIMON1979211}. This method has
been successfully implemented in describing scattering of
electrons on the Hydrogen
atoms~\cite{Rescigno2474,0953-4075-37-17-R01,Bartlett_ecsm} and
recently for describing fully elastic scattering in the systems of
three different charged particles~\cite{Volkov_ecsm}. Nevertheless
exterior complex scaling method contains several drawbacks from
the formal as well as practical point of view, which stalls its
further developments. Formally, exterior complex method:
\begin{itemize}
    \item is limited to a case of central and local interaction
    \item is difficult to use together with the partial-wave expansion
    \item is difficult to generalize for N$\geq$3 particle system
\end{itemize}
Alternatively, the smooth complex scaling method is not affected
by the aforementioned complications. In this section is
demonstrated that the smooth complex scaling method can be
successfully employed in describing Coulombic three-body
collisions, thus overcoming its original limitation to the
scattering dominated by the short-range interactions.

Technically the most advanced methods of the atomic collisions
evolved from the close-coupling (CC) expansion introduced by
Massey and Mohr~\cite{Massey289}, which is based on the expansion
of the system's wave function in terms of the eigenstates of the
target atom. The success of these techniques relies on the
simplicity of the Hydrogenic wave functions and the ability to
find the  analytical expression for the matrix elements involved
in the numerical solution. Techniques based on the close-coupling
(CC) expansion have proved to be also successful in solving e-H
scattering problem in a wide range of energies, which also allows
to evaluate ionization cross
sections~\cite{PhysRevA.53.1525,PhysRevLett.69.53}. For this aim
the positive-energy pseudostates of H atom should be included in
the wave functions expansion~\cite{PhysRevLett.69.53}. Given a
sufficiently large basis and successful parameterizations of the
pseudostates, these methods become very efficient and provide very
accurate solutions.  Clear advantage of the pseudostate methods is
due to the fact that they allow to perform calculations by keeping
the analytical part of the problem almost unchanged.

Presence of three different charged particles reveal more severe
formal difficulties for the conventional CC methods. For this aim
two-center CC expansions has been introduced~\cite{Mitroy}. The
drawback of the last method is that two-center CC basis becomes
overcomplete and results into mathematically ill-conditioned
problem~\cite{PhysRevA.93.012709}. Nevertheless by well mastering
parameter space this method may turn into the very efficient tool.

The complex scaling technique by itself is just a tool, which
allows one to avoid complications related with the complex wave
function behavior in the far asymptotes, and might be used
successfully in conjunction with CC expansion. In this study I
have however decided to apply the complex scaling method in
conjunction with the Faddeev-Merkuriev (FM) equations, described
in section~\ref{sec_FM_equations}. The FM equations present a
mathematically rigorous formulation of the three-body Coulombic
problem. When exploring potential of the complex scaling technique
this presents one clear advantage, since mathematically
well-conditioned formulation of the problem should guarantee
convergence of the basis expansion, regardless to the fact that a
chosen basis is not optimized. The price to pay for using FM
equations, compared to the CC approaches, is in appearance of some
complicated integrals, which are not possible to perform
analytically and require numerical approximations to be used.

\subsection{Bound state input}

The first step in performing any many-body scattering calculations
is to determine projectile (target) bound state wave functions
from which the free-wave solutions are constructed.  Clearly
accuracy of any scattering calculation critically depends on this
input. Lagrange-Laguerre quadrature, being based on Laguerre
polynomial basis, is naturally well fitted to describe Hydrogenic
wave functions. Numerous calculations exist proving accuracy of
this method in solving Coulombic bound state
problems~\cite{ISI_BAYE}. Nevertheless it is not obvious how this
basis complies with the complex scaling transformation.

 In Table \ref{tab:ps-bs}  accuracy  in determining
 Positronium binding energies are presented.
 The parameters of the Lagrange-Laguerre quadrature and complex scaling angles
 are chosen to comply with the parameters of the  three-body scattering calculations
 (presented in the following subsections). These parameters
 were not optimized to reproduce excited states of the
 Positronium. Due to the complex scaling operation binding energies
 are obtained as the complex numbers, contaminated by a small imaginary part -- reflecting
 numerical artifacts of the CS transformation.
 As one can see ground state binding energy of the
 Positronium is quasi-exact already when using a modest quadrature
 of 15 points. The inaccuracy of the calculated ground state energy
 is due to the dominance of the machine round off error, rather than numerical method.
 Accuracy of the excited states is also quite satisfactory and
 is improving  systematically with a number of the quadrature points
 (basis size). Naturally more accurate values are obtained when
 small complex scaling angles are employed, resulting
 in a weaker overall effect of the CS transformation on Positronium's wave function.

\begin{table}
\begin{center}
\caption{Relative error $(E_{exact}-E_{calc})/E_{exact}$ of the
calculated binding energies ($E_{calc}$) of the S-wave n=1 and n=5
Positronium
  states. Values are tabulated as a function of number ($N_x$) of  Lagrange-Laguerre quadrature points used in calculation. Two sets of
  calculations respectively
  for the complex scaling angle of $\theta=6^\circ$ and of  $\theta=8^\circ$ are
  compared. The
notation $x [y]$ means $x10^y$.
  }\label{tab:ps-bs}
  \vspace{0.5cm}
\begin{tabular}{|c|c|c|c|c|}\hline
 &\multicolumn{2}{|c|}{ $\theta=6^\circ$ }&\multicolumn{2}{c|}{ $\theta=8^\circ$ }\\
 \cline{2-5}
  $N_x$ & n=1 &  n=5 &n=1 &n=5\\
  \hline\hline
  15 & -9.6[-17]$-$ 4.0[-15]i& 1.0[-2]$-$ 1.4[-2]i& 1.7[-16]$+$ 8.0[-16]i& 5.7[-3]$-$ 1.8[-2]i \\
  20 & 1.6[-16]$+$ 4.0[-15]i & -1.4[-5]$-$ 7.5[-4]i & 7.8[-16]$+$ 4.0[-15]i & -1.3[-3]$-$ 3.2[-3]i \\
  30 & -1.0[-15]$-$ 2.0[-13]i & -3.5[-6]$-$ 9.6[-7]i & -3.4[-16]$-$ 1.9[-13]i & -6.6[-6]$+$ 4.8[-6]i \\
  40 & -1.9[-15]$-$ 1.5[-13]i & -3.3[-10]$+$ 7.1[-10]i& -2.7[-15]$-$ 1.5[-13]i & -1.1[-9]$+$ 8.1[-10]i \\
  \hline

\end{tabular}
\end{center}
\end{table}

 There is no point in repeating the same analysis for the Hydrogen
 atom, since its bound state wave functions coincide with the Positronium ones
 after a trivial coordinate scaling.

 Positions of three-body bound and resonant states
 influence strongly the scattering observables. Ability of the CS method to reproduce
 resonant states has been already demonstrated in section~\ref{sec_res_eep}. In order to demonstrate the
 level of accuracy
 of the numerical technique used in this work in Table~\ref{tab:ps-bs}
 convergence of the ground state of the Positronium ion
 ($e^+e^-e^-$) is
 presented. Positronium's ion is relatively weakly bound
 structure and therefore is suitable as a testground for the three-body calculations.
 Calculations presented in Table~\ref{tab:ps-bs} were performed using the same
 configuration as in the scattering calculations of the next
 subsection. The partial wave expansion has been limited to
 max$(l_x,l_y)\leq9$, whereas convergence has been studied as a function
 of Lagrange-Laguerre quadrature size ($N=N_x=N_y$) employed in
 expanding radial parts of the FM components.
 One may see that already a moderate basis of 20$\times$20 points (functions)
 provides accuracy of six significant digits, further improvement
 of the calculation is stalled and would require enlargement of the
 partial-wave basis\footnote{It is well known that convergence of the
 partial-wave series is slow for the Coulombic problems due to
 the awkward
 "cusp" behavior at the two-particle collision points.}. Presence
 of the complex scaling transformation has only the limited impact
 on the calculated binding energies.
 The smallness of the spurious imaginary part of the binding energy as
 well as
 weak deterioration of the calculated values when increasing
 complex scaling parameter $\theta$ proves this point.

\begin{table}[tbp]
\begin{center}
\caption{ Calculated Positronium ion ($e^+e^-e^-$) binding
energies as a function of the number ($N=N_x=N_y$) of
Lagrange-Laguerre quadrature points used in the calculation. Two
sets of
  calculations, respectively
  for the complex scaling angle set to $\theta=6^\circ$ and to $\theta=8^\circ$ are compared.
 In the last line reference value of one of the most accurate variational calculations
 is provided.  }\label{tab:ps-bs}
 \vspace{0.5cm}
\begin{tabular}{|c|c|c|}\hline
 $N$ & $\theta=6^\circ$ &$\theta=8^\circ$ \\
 \hline\hline
  15 & 0.26200102+2.64$\times10^{-4}i$ & 0.26200001+2.63$\times10^{-4}i$  \\
  20 & 0.26200597+4.02$\times10^{-5}i$ & 0.26200597+3.73$\times10^{-5}i$\\
  30 & 0.26200533-6.73$\times10^{-6}i$ & 0.26200533-4.71$\times10^{-6}i$ \\
  40 & 0.26200543-2.12$\times10^{-7}i$ & 0.26200543-2.56$\times10^{-7}i$\\
  \hline
  Ref.~\cite{PhysRevA.57.2436} &\multicolumn{2}{|c|} {0.2620050702329757} \\
  \hline
\end{tabular}
\end{center}
\end{table}
\subsection{e+Ps(n=1) scattering}

Electron scattering on positronium constitutes probably the
simplest realistic Coulombic three-body system. This system has
been well explored at low energies, below the first positronium
excitation
threshold~\cite{Ward_JPB,PhysRevA.50.1924,PhysRevA.61.032710,PhysRevA.72.062507,PhysRevA.92.032713,Gilmore2004124}.
Above the positronium excitation threshold only the calculations
based on close-coupling method are
available~\cite{Gilmore2004124}, which if properly parameterized
may provide very accurate results but in general are not
constrained to provide an unique physical solution.
\begin{figure}
\centering
\hspace{-0.5cm}\includegraphics[width=85mm]{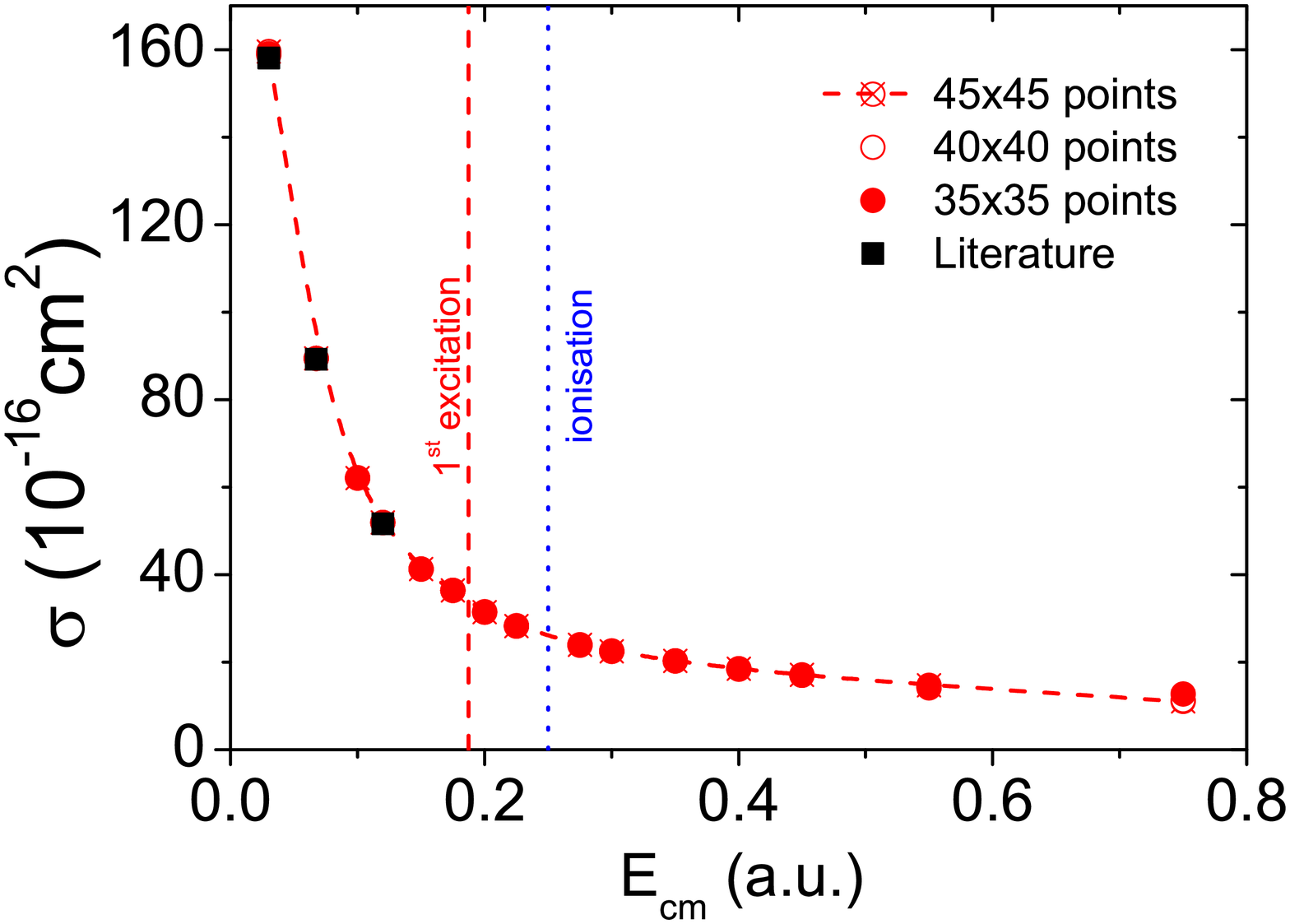}
\hspace{-0.5cm}\includegraphics[width=85mm]{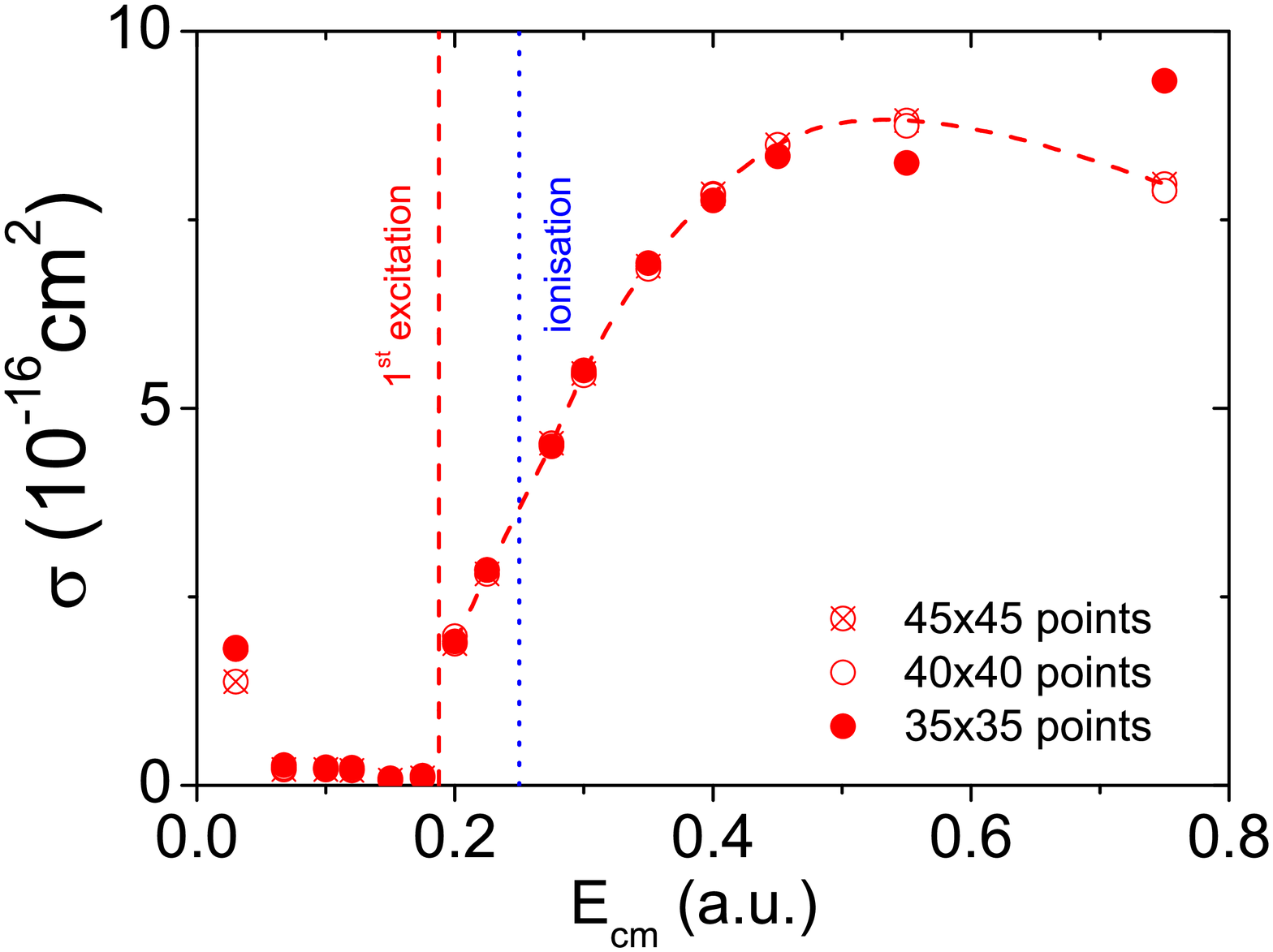}
\caption{\label{fig:eps_cs} Calculated  e+Ps(n=1) total (left
panel) and inelastic (right panel) cross sections in a wide energy
range. The dashed lines connecting calculated points are just
drawn to guide an eye.
 Current calculations are well converged, which becomes clear
by increasing the size of Lagrange-mesh basis. Below the Ps(n=2)
excitation threshold the calculated total cross sections are
compared with the ones compiled from the
literature~\cite{Ward_JPB,PhysRevA.61.032710} and represented by
full black squares. }
\end{figure}

In figure~\ref{fig:eps_cs}  calculated cross sections of electron
scattering on the ground state of positronium (Ps(n=1)) are
presented. These calculations cover a broad energy region,
starting with a purely elastic case and spreading well above the
positronium ionization threshold. Below the positronium excitation
threshold results of this work are compared with the most accurate
values from literature, summarized in Table.I of
ref.~\cite{PhysRevA.61.032710}.

Present calculations have been performed by considering free
e+Ps(n=1) waves to represent the incoming wave function in
eq.~(\ref{eq_inwave_sep}-\ref{Sc_eq_aux_pot}). The inhomogeneous
term arising from the incoming wave has been screened in
eq.~(\ref{eq_FM_inh}) for e+Ps(n=1) separations exceeding
$y_{eps}=35$ a.u. Partial wave expansion has been limited to
max$(l_x,l_y)\leq9$ and proved to be sufficient to get well
converged results. Calculations were also limited to total angular
momentum states $L\leq$5.

 As can be seen in figure~\ref{fig:eps_cs}, a basis of 35x35 Lagrange-Laguerre
 mesh functions is sufficient to describe radial dependence of the FM amplitudes
 and  to get converged results in a broad energy region.
 Only well beyond the positronium ionization threshold a basis of 35x35 functions
 turns to be insufficient in describing inelastic cross section, nevertheless
 convergence is reached by increasing basis to 40x40 functions.

 As discussed in section \ref{SA_GF}, large complex scaling angles are not suited
 to perform scattering calculations in $A>2$ particle systems.
 This work reconfirmed this feat. In this work complex scaling parameter
 has been limited to $\theta<10^\circ$, with $\theta=7-8^\circ$
 representing an optimal choice. Regardless simplicity of the
 employed approach calculations turn to be very accurate and are in
 line with the most accurate published values. The phaseshifts calculated below
 the Ps(n=2) threshold differ from ones reported in~\cite{Ward_JPB,PhysRevA.61.032710} by less than
 0.5$\%$. This proves that the elastic differential cross sections,
 which are usually determined from the calculated phaseshifts, are well
 reproduced as well.

 As it is well known, complex scaling operation breaks Hermiticity
 of the Hamiltonian. Consequently the unitarity of the S-matrix is not
 provided by the symmetry properties of the CS equations.
 This is the reason why using complex scaling it is more difficult
 to attain  the unitarity of  S-matrix  than to
 get highly accurate phaseshifts.  Regardless this fact the
 unitarity of  S-matrix in presented e+Ps(n=1) calculations is
 assured with a three-digit accuracy once electron impact energy exceeds
 0.03 a.u. This is clearly demonstrated by
 analyzing inelastic e+Ps(n=1) cross section, extracted
 relying on the unitarity property of the S-matrix. In particular,
 inelastic cross sections are consistent with a zero value
 in the purely elastic region, below the Ps(n=2) threshold.
 Accurate description of the nearthreshold collisions is naturally the most
 problematic case for the complex-scaling method. After the complex scaling
 operation outgoing waves converge with an exponential factor -$k r sin\theta$,
 where $k$ is
 a relative momenta of the  scattered clusters and $r$ is a target-projectile separation distance. This exponent
 vanishes at low impact energies and therefore  approximation of the outgoing waves by
 using the square-integrable basis functions becomes inefficient.

\subsection{$e^-$+H scattering}
\begin{figure}[h!]
\centering \hspace{-0.5cm}\includegraphics[width=85mm]{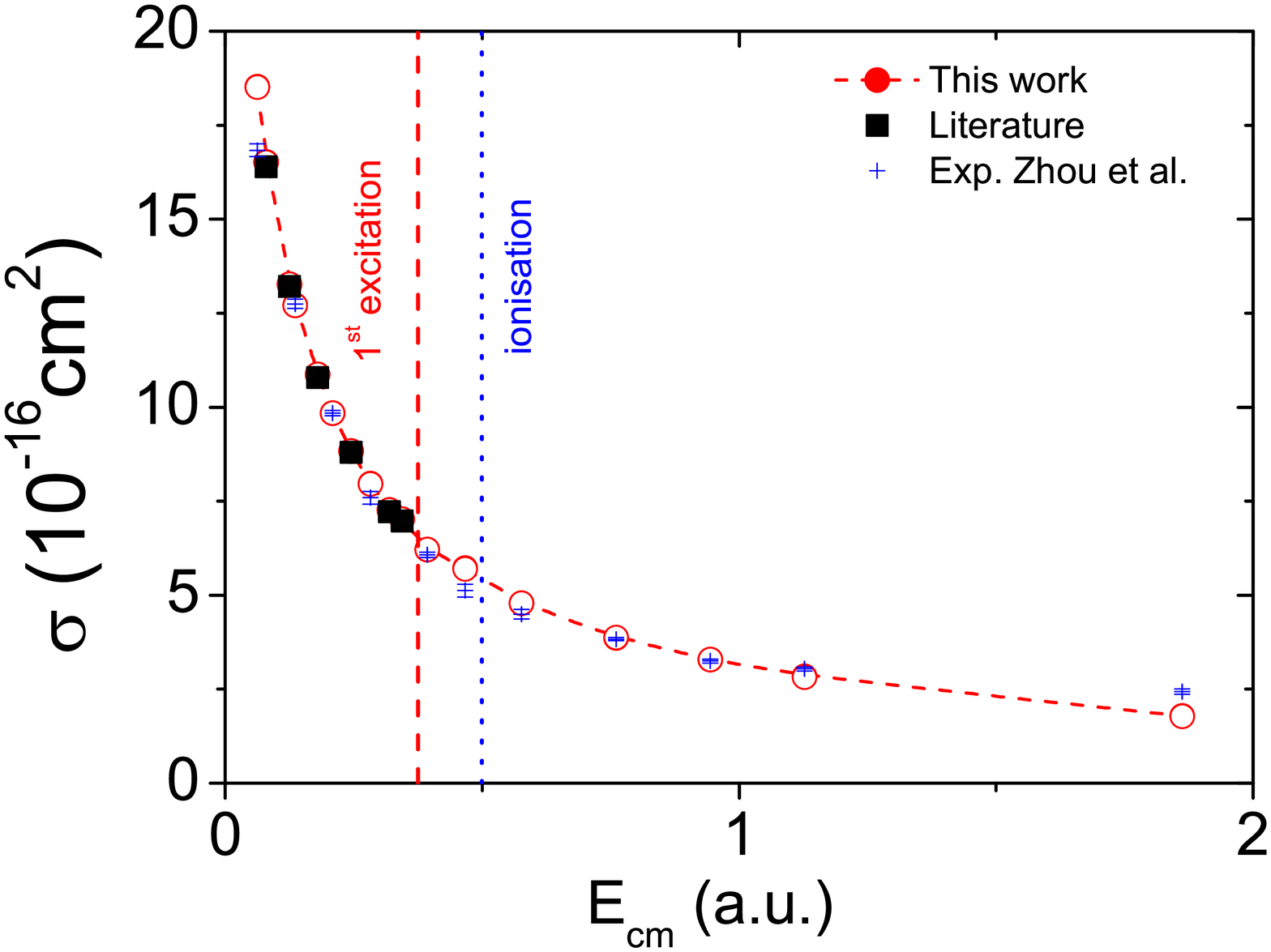}
\hspace{-0.5cm}\includegraphics[width=85mm]{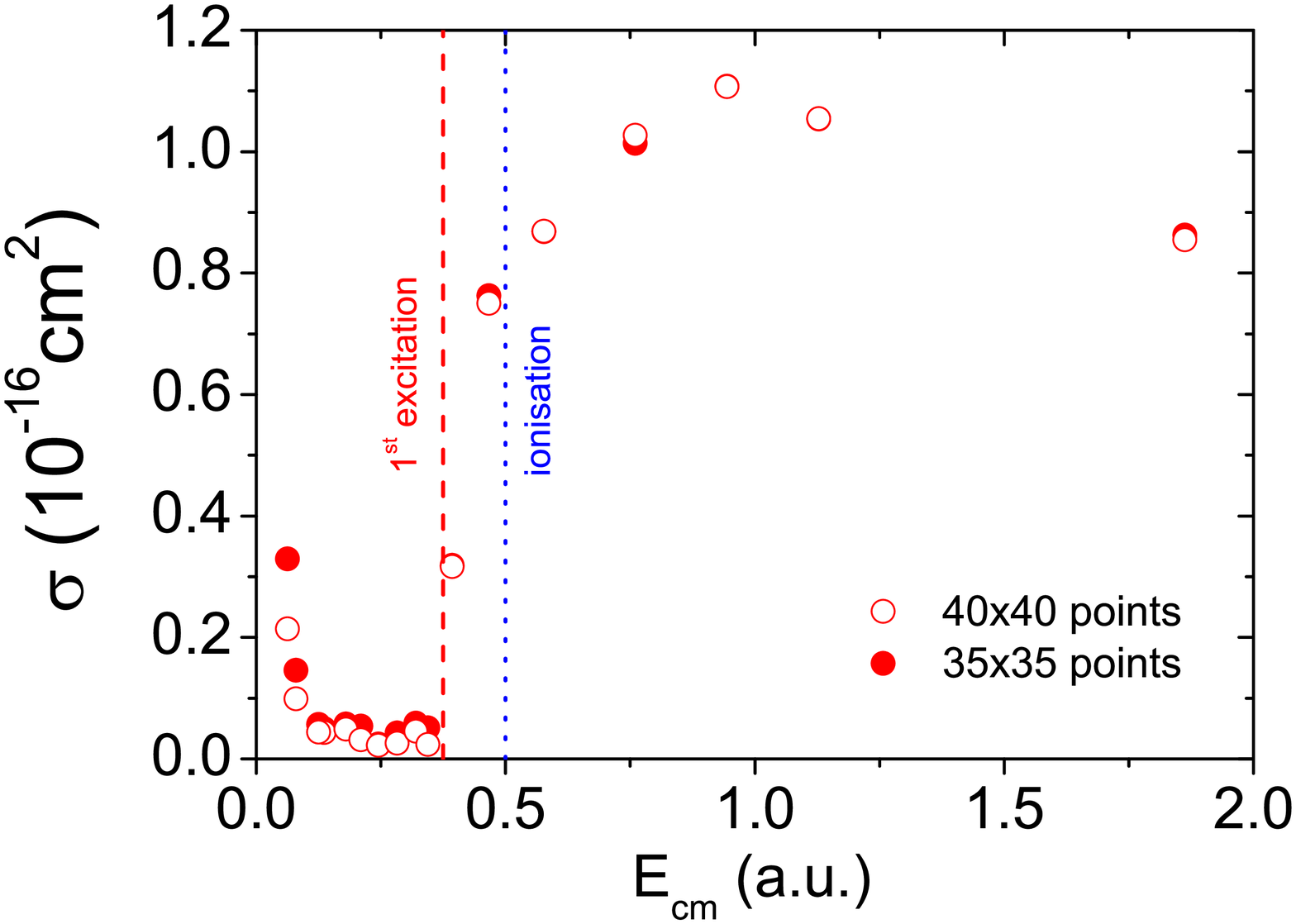}
\caption{\label{fig:eH_cs} The same as in Fig.~\ref{fig:eps_cs}
but for  $e^{-}$+H(n=1) scattering. Calculated values for the
total cross section are compared with the experimental data of
Zhou et al.~\cite{PhysRevA.55.361}. Below $H(n=2)$ excitation
threshold the calculated total cross sections are compared with
ones compiled from the literature~\cite{Gien_JPB} and represented
by full black squares. }
\end{figure}

Electron collisions with the atomic hydrogen is well studied
problem, presenting probably the most popular benchmark for a
three-body Coulombic scattering problem. This system has been
considered by several different techniques, finally giving rise to
public access databases~\cite{Bartlett_ecsm,Bray1995209}, as well
as public access codes~\cite{Benda20142903}.

\begin{figure}[h!]
\centering
\includegraphics[width=14cm]{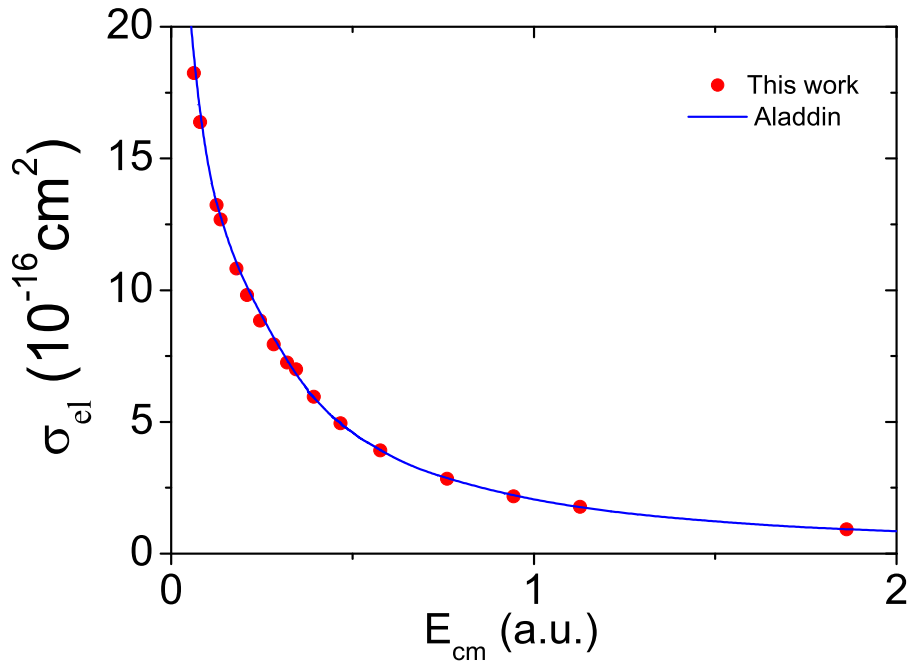}
 \caption{Calculation of the total  elastic cross section of
 electron scattering on atomic Hydrogen. Results of this work are compared
 with the values of the Aladdin database, based on the computation by Bray and
  Stelbovics~\cite{Bray1995209} using converged close coupling method.}
 \label{Fig_H_totel}
\end{figure}
In the figure~\ref{fig:eH_cs} calculations of the electron
scattering on the ground sate of Hydrogen atom are presented.
Present calculations have been performed using the same setup as
for the e+Ps(n=1) case, described in the last subsection. A free
e+H(n=1) wave  is considered when separating inhomogeneous term in
eq.~(\ref{eq_inwave_sep}-\ref{Sc_eq_aux_pot}).  35-40
Lagrange-mesh functions were employed for discretizing radial
dependence of the FM amplitudes in $x$ and $y$ directions and
proved to be enough to get the converged results. The calculated
values agree perfectly with the ones found in
literature~\cite{Gien_JPB,PhysRevA.66.052714} as well as with the
experimental data of Zhou et al.~\cite{PhysRevA.55.361}. Only the
last point (at 1.87 a.u.) seems to underestimate the experimental
total cross section. The total cross section in this energy
region, well above the Hydrogen atom ionization threshold, has
non-negligible contribution of high angular momentum states
(beyond $L$=5)\cite{Benda20142903}, which have not been included
in a present calculation.

 The phaseshifts calculated below
 the $H(n=2)$ threshold agree perfectly well with the most accurate calculations
 found in the literature.
 All the phaseshifts fall within the limits defined by the values compiled in the
 references~\cite{Gien_JPB,PhysRevA.66.052714}.
 Calculated total elastic cross sections, see Fig.~\ref{Fig_H_totel}, both below as well as above ionization threshold
 perfectly agrees with the
 published values in the Aladdin database,  based on the computation by Bray and Stelbovics~\cite{Bray1995209} using
 converged close coupling method.

As pointed out in the previous section, presenting the e+Ps(n=1)
scattering, complex scaling technique turns to be the most
difficult to apply at very low energies, close to the threshold.
By reducing energy it turns increasingly difficult to preserve the
unitarity of the calculated S-matrix. This feat is best
demonstrated by the deviation from the zero-value of the inelastic
e$^{-}$+H(n=1) cross section close to H(n=1) threshold (see two
lowest energy points, situated at $E_{cm}=0.0624$ and $0.08$ a.u.
respectively). Naturally the unitarity of the calculated S-matrix
improves once number of basis functions is increased, nevertheless
at very low energies this convergence turns to be rather slow.

\subsection{$e^+$-H(n=1)$\leftrightarrows$p+Ps(n=1) scattering}

\begin{figure}[h!]
\centering
\hspace{-0.5cm}\includegraphics[width=85mm]{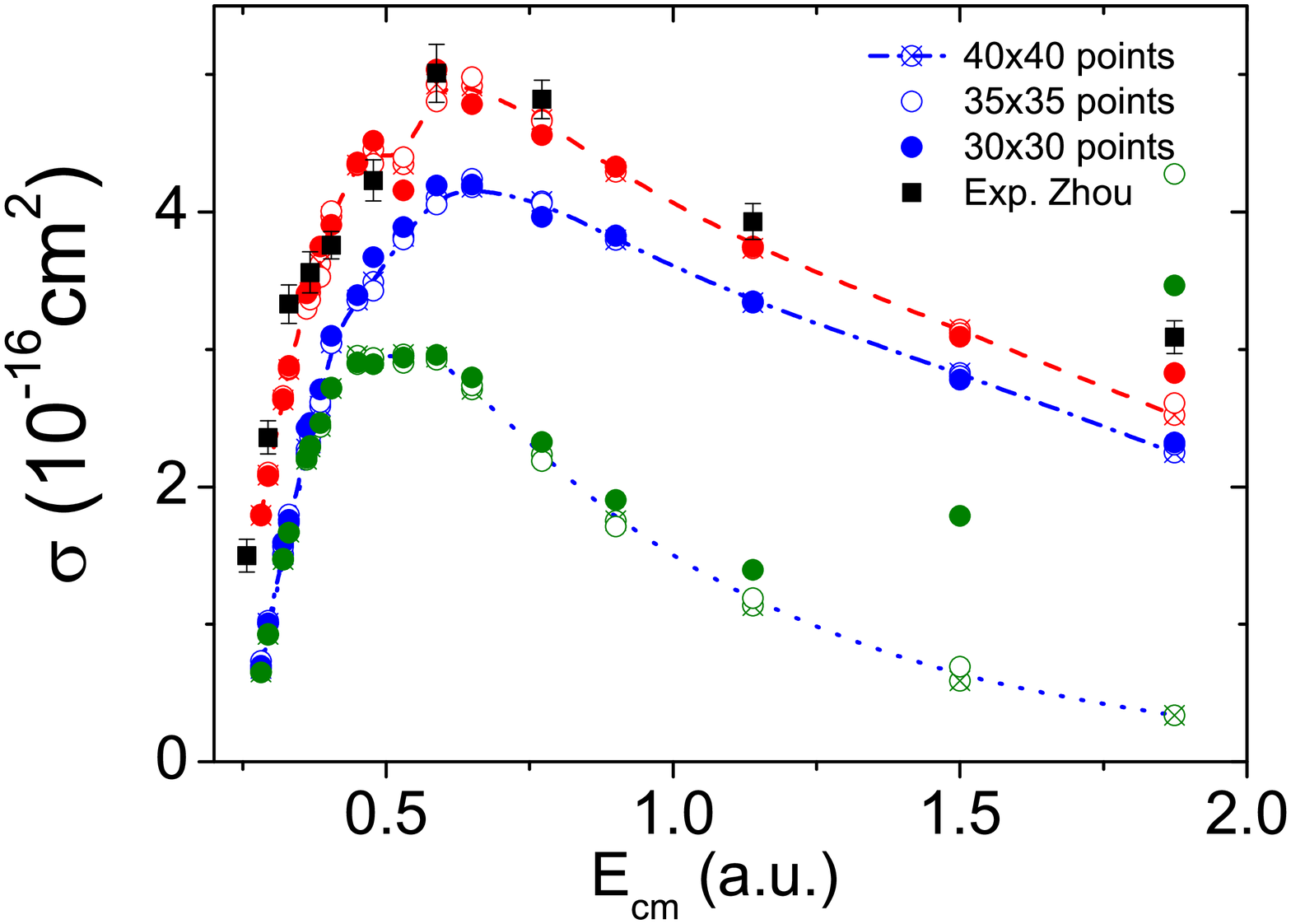}
\hspace{-0.5cm}\includegraphics[width=85mm]{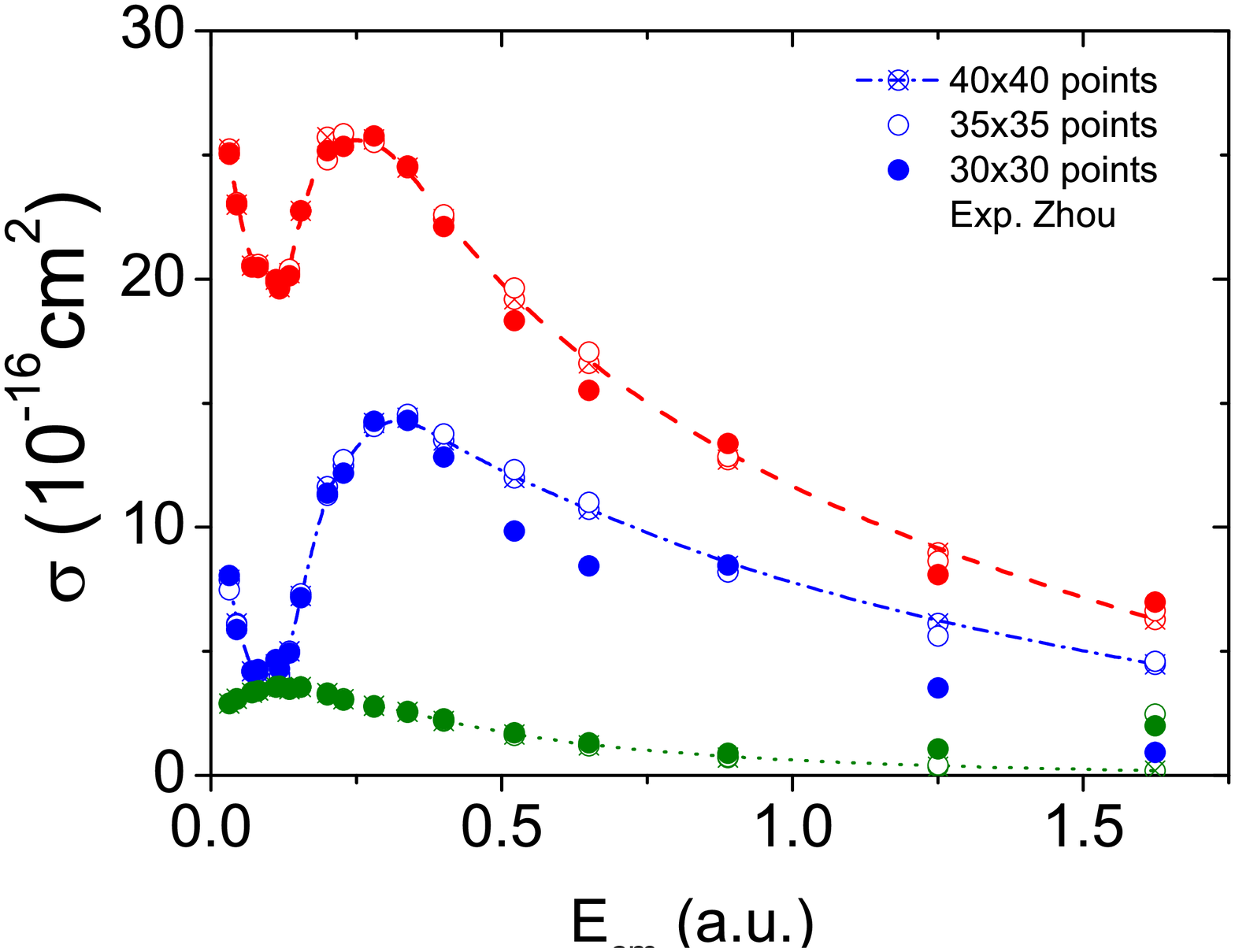}
\caption{\label{fig:pPS_cs} Study of e$^+$+H(n=1) (left panel) and
p+Ps(n=1) (right panel) collisions. Energy evolution of the total
(red), the total inelastic (blue) and the
$e^+$+H(n=1)$\leftrightarrows$ p+Ps(n=1) (olive) cross sections
presented. Three sets of the calculations performed using
different radial basis sizes. The calculated total cross section
for a e$^+$+H(n=1) process is compared with the experimental data
of Zhou et al.~\cite{PhysRevA.55.361}. }
\end{figure}

There is an increased interest in studying
(anti)proton-positronium collisions in view of the possible
production of antihydrogen atoms. This system was mostly explored
using different variations of the close coupling
method~\cite{PhysRevA.93.012709,Mitroy}; There also exist
calculations based on
Hyperspherical-Harmonics~\cite{PhysRevA.50.232},variational
method~\cite{PhysRevA.9.219,0953-4075-30-10-020} as well as
Faddeev-Merkuriev
equations~\cite{Hu_PRL,PhysRevA.59.4813,0953-4075-28-2-015,PhysRevA.63.062721}
 but these limited to the energy region of a few lowest energy excitations
of either the Hydrogen or the Positronium atom.

Elastic e$^+$+H(n=1) collisions below the positronium excitation
threshold does not present any new features compared to the
e$^-$+H(n=1) or e+Ps(n=1) elastic scattering, discussed in two
previous subsections. Therefore I will concentrate on the energy
region above the p+Ps(n=1) production threshold. In
figure~\ref{fig:pPS_cs}  the calculations performed by considering
only a free  e$^+$+H(n=1) (left panel) or p+Ps(n=1) (right panel)
waves to separate inhomogeneous term in
eq.~(\ref{eq_inwave_sep}-\ref{Sc_eq_aux_pot}).

Calculations considering the $e^+$+H(n=1) entrance channel are
well converged for a moderate basis of 30x30 Lagrange-mesh
functions and does not depend on the variation of the CS parameter
in the range $\theta=5-10^\circ$. Results of the present work
agree perfectly with other theoretical calculations as well as
with the experimental data of Zhou et al.~\cite{PhysRevA.55.361}.
 The experimental total cross section is only underestimated for
the highest energy point, which has still to non-negligible
contribution from large total angular momentum states not included
in a present calculation. For this point contribution of the $L$=7
state, the largest total angular momentum state considered in this
calculation, still accounts for $\approx10\%$ of the total cross
section, whereas this state has negligible contribution at lower
energies. The unitarity of the S-matrix is well preserved, which
is demonstrated by the feat that below the H(n=2) excitation
threshold the inelastic cross section agrees with the Ps(n=1)
production one (at these energies Ps(n=1) production represents
the only inelastic channel).
\begin{figure}[h!]
\centering
\includegraphics[width=14. cm]{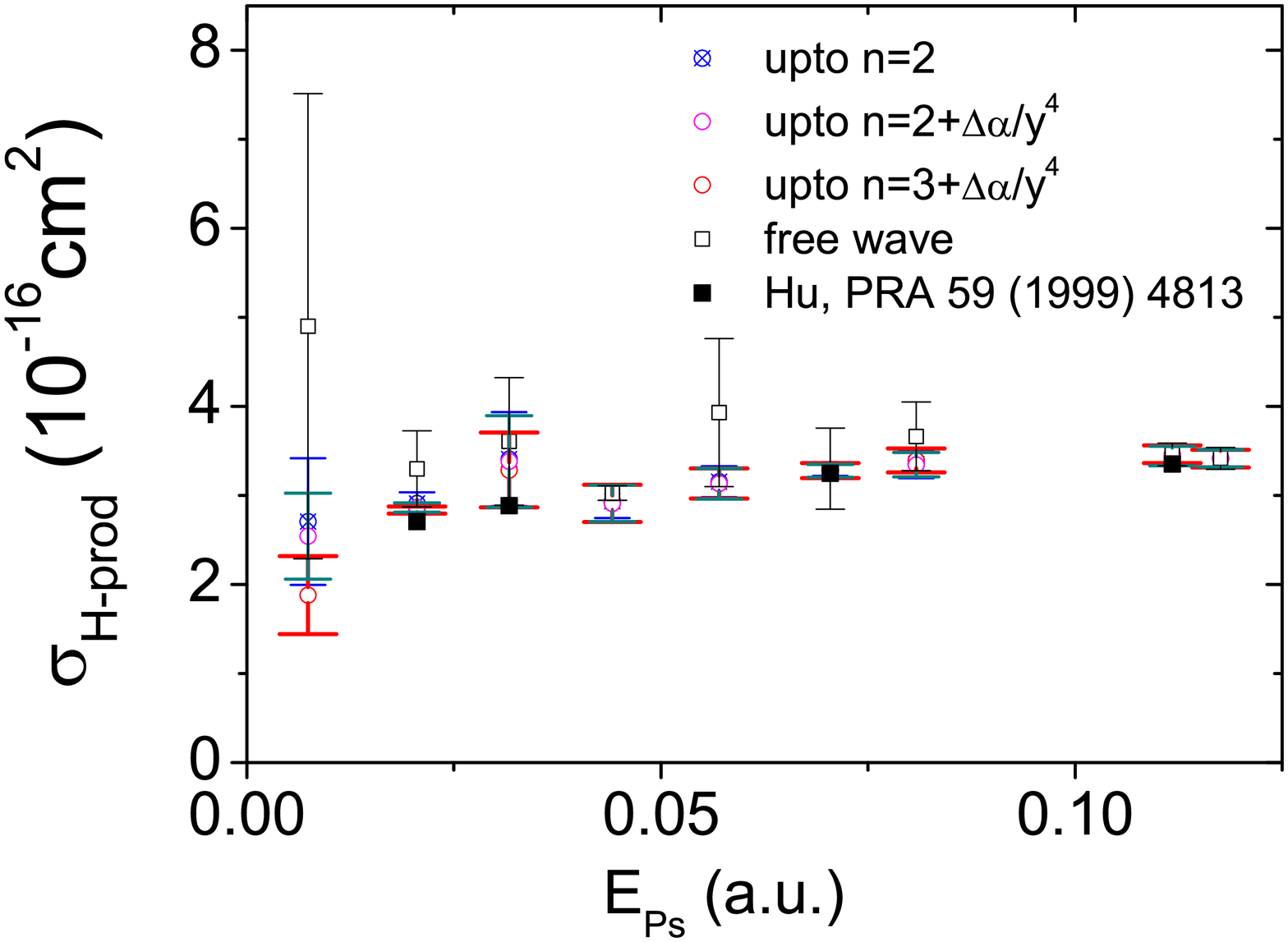}
 \caption{Calculation of the Hydrogen production cross section for the p+Ps(n=1) collisions in the Ore gap region.
 Calculations using different assumptions for the inhomogeneous term, based on distorted incoming wave of  p+Ps(n=1), were performed.
 These results are compared with the calculations of Hu~\cite{PhysRevA.59.4813}, using conventional boundary condition method.}
 \label{Fig_H_prod_err}
\end{figure}

The calculations considering p+Ps(n=1) entrance channel  turns to
be less accurate. In particular, problematic are the calculations
performed in the Ore gap region\footnote{Ore gap is the energy
region between the positronium formation threshold and the first
excitation of the target atom (in our case Hydrogen).} and
dominated by the relatively low proton(positronium) impact
energies. In this region the inelastic p+Ps(n=1) cross section,
extracted using the unitarity property of S-matrix, is visibly not
converged and improves only moderately when increasing the size og
Lagrange-mesh basis. On the other hand,  the Hydrogen production
cross sections calculated from the non-diagonal S-matrix element
coupling $e^+$+H(n=1) and p+Ps(n=1) channels turns to be accurate
and well converged even at very low energies.

Even though the low energy region is not the most relevant region
to use the complex scaling method -- it is still worthy  to pay
more attention to the Ore gap region, where p+Ps(n=1) cross
sections converge slowly. In order to improve the convergence I
have constructed the inhomogeneous term in
eq.~(\ref{eq_inwave_sep},\ref{Sc_eq_aux_pot}) based on the
distorted waves instead of the simple free waves used before.
Effect of the choice of the inhomogeneous term is studied in
figure~\ref{Fig_H_prod_err} by comparing the inelastic p+Ps(n=1)
cross sections in the problematic  Ore gap region. These
calculations were performed using a basis of
 30x30 Lagrange-mesh functions, with the CS parameter set to $\theta=5^\circ$ and total angular momentum
 expansion limited to $L$=3\footnote{This limitation have been used in order to compare the results with ones from the ref.~\cite{PhysRevA.59.4813}}.
 Four types of the distorted waves, based on the choice of long-range potential in eq.~(\ref{eq_aux_pot}), have been used:
 \begin{itemize}
    \item distorted wave by considering long-range dipole coupling of Ps(n$\leq$2) states,
    with  $\lambda _{ab}(y_{\alpha })=-C_{\alpha }\langle\varphi _{b,l_{x}^{(b)}}(\vec{x}_{\alpha })|\vec{x}_{\alpha }|\varphi _{a,l_{x}^{(a)}}(\vec{x}_{\alpha })\rangle/\tilde{y_{ab}}_{\alpha }^2$\footnote{Expression $\tilde{y_{ab}}_{\alpha }=y_{\alpha }+a_0*n_a^3*n_b^3/y^2_{\alpha }$ has been used to regularize former potential at the origin.
Coefficient $C_{\alpha }$ is a result of the presence of mass
scaling factors, present in a definition of Jacobi coordinates
$x_{\alpha },y_{\alpha }$.}
    \item considering long-range dipole coupling of the Ps(n$\leq$2) states together with a residual p+Ps(n=1) polarization potential
    \item dipole coupling of the Ps(n$\leq$3) states together with a residual p+Ps(n=1) polarization potential
    \item inhomogeneous term based on a free wave
 \end{itemize}
 In the figure~\ref{Fig_H_prod_err}  the calculated p+Ps(n=1)$\leftrightarrows e^+$+H(n=1) reaction cross section is presented as a range,
  obtained by comparing three different values: cross sections calculated from
 the non-diagonal S-matrix elements  ($S_{e^++H(n=1),p+Ps(n=1)}$ and $S_{p-Ps(n=1),e^++H(n=1)}$) as well as cross section
 extracted from the diagonal S-matrix element $S_{p-Ps(n=1),p-Ps(n=1)}$ via
 unitarity condition\footnote{In the Ore gap region relation $|S_{p+Ps(n=1),e^++H(n=1)}|^2=1-|S_{p+Ps(n=1),p+Ps(n=1)}|^2$ should hold}.
 It is clear that the distorted waves improve considerably accuracy of the
 calculated cross sections even at very low
 energies.
 Inclusion of the dipole coupling of the Ps(n$\leq$2) states is already enough to get rather accurate results,
 in agreement with the ones from ref.~\cite{PhysRevA.59.4813},
 obtained employing the conventional boundary condition approach.
 By considering the more complete residual p+Ps(n=1) interaction to determine the distorted incoming wave allows to
  improve the accuracy of the results even further.

\bigskip
\bibliographystyle{plain} 
\bibliography{Bib_Rimas}

\end{document}